%% file: main.tex
\documentclass{article}

\usepackage[T1]{fontenc}
\usepackage{geometry}
\usepackage[utf8]{inputenc}
\usepackage{sansmath}

\usepackage{authblk}
\usepackage{graphicx}
\usepackage[resetlabels]{multibib}
\newcites{si}{Supplementary References} 
\usepackage{booktabs}
\usepackage{amssymb}
\usepackage{threeparttable}
\usepackage[section]{placeins}
\usepackage{pdflscape}

\usepackage{stackengine}
\usepackage[version=4]{mhchem}
\usepackage{siunitx}
\usepackage{textgreek}

\DeclareSIUnit\angstrom{\text{Å}}
\DeclareSIUnit\bar{\text{bar}}
\usepackage{changepage}
\let\oldfigure\figure
\let\endoldfigure\endfigure
\let\oldtable\table
\let\endoldtable\endtable
\expandafter\let\expandafter\oldfigurestar\csname figure*\endcsname
\expandafter\let\expandafter\endoldfigurestar\csname endfigure*\endcsname
\renewenvironment{figure*}{%
   \oldfigurestar
   \begin{adjustwidth}{-1cm}{-1cm}%
}{%
   \end{adjustwidth}%
   \endoldfigurestar
}
\renewenvironment{figure}[1][htbp]{%
   \oldfigure[#1]
   \begin{adjustwidth}{-1cm}{-1cm}%
}{%
   \end{adjustwidth}%
   \endoldfigure
}
\renewenvironment{table}[1][htbp]{%
   \oldtable[#1]
   \begin{adjustwidth}{-1cm}{-1cm}%
}{%
   \end{adjustwidth}%
   \endoldtable
}

\usepackage{caption}
\usepackage{url}
\usepackage{hyperref}

\DeclareUnicodeCharacter{0226}{\.{A}}
\DeclareUnicodeCharacter{03BC}{\textmu}
\DeclareUnicodeCharacter{03B2}{\textbeta}
\DeclareUnicodeCharacter{03B3}{\textgamma}

\usepackage{xltabular}
\usepackage{tabularx}
\usepackage[table]{xcolor}

\newcolumntype{R}{>{\raggedleft\arraybackslash}X}
\newcolumntype{L}{>{\raggedright\arraybackslash}X}

\newcommand{\mytitle}{Enhanced sampling and cryo-EM data resolve magnesium binding to RNA}
\title{\mytitle}

\newcommand{\motifX}{$\text{G}^\text{257}_\text{OP1}$,\allowbreak$\text{U}^\text{258}_\text{OP2}$,\allowbreak$\text{C}^\text{260}_\text{O2}$,\allowbreak$\text{U}^\text{305}_\text{OP2}$-\underline{Cis}-3$\text{O}_\text{ph}$$\text{O}_\text{b}$}
\newcommand{\motifXVII}{$\text{G}^\text{257}_\text{OP1}$,\allowbreak$\text{U}^\text{258}_\text{OP2}$,\allowbreak$\text{U}^\text{305}_\text{OP2}$-\textit{mer}-3$\text{O}_\text{ph}$}
\newcommand{\motifV}{$\text{A}^\text{184}_\text{OP1}$,\allowbreak$\text{A}^\text{186}_\text{OP1}$,\allowbreak$\text{A}^\text{187}_\text{OP2}$-\textit{mer}-3$\text{O}_\text{ph}$}
\newcommand{\motifXXXVII}{$\text{A}^\text{184}_\text{OP1}$,\allowbreak$\text{A}^\text{186}_\text{OP1}$,\allowbreak$\text{A}^\text{187}_\text{OP2}$,\allowbreak$\text{G}^\text{188}_\text{OP2}$-\underline{Trans}-4$\text{O}_\text{ph}$}

\begin{document}

\author[a,1]{Olivier Languin-Cattoën}
\author[a]{Elisa Posani}
\author[a,1]{Giovanni Bussi}

\affil[a]{Scuola Internazionale Superiore di Studi Avanzati (SISSA), Molecular and Statistical Biophysics Group, Trieste 34136, Italy}

\date{}
\maketitle
\begin{abstract}
  Magnesium ions are essential for RNA structure but difficult to model due to slow binding kinetics and experimental limitations. We present an enhanced-sampling strategy that accelerates \ce{Mg^{2+}} inner-shell binding by orders of magnitude, enabling quantitative exploration of ion-binding motifs in a large ribozyme. The method combines a barrier-flattening bias with Hamiltonian replica exchange to efficiently sample multiple equivalent binding sites, and builds on an approach that achieved top performance in the CASP16 blind assessment of RNA solvation structure.
  Using cryo-electron microscopy maps for validation, we introduce a local analysis framework that infers the population of individual binding motifs from their agreement with experimental density, enabling site-by-site validation. We find that insufficient sampling of inner-shell binding leads to significantly poorer agreement with experiment, whereas force fields predicting different inner/outer binding equilibria remain largely indistinguishable at the current experimental resolution.
  These results highlight the dominant role of sampling in modelling divalent ion binding and provide a general strategy for integrating simulations with experimental data in complex biomolecular systems.
\end{abstract}

Magnesium(II) (\ce{Mg^{2+}}) is among the most abundant divalent atomic cations in living cells and is essential to a wide range of biological processes. \ce{Mg^{2+}} is specially relevant to the biochemistry of nucleic acids, where it plays a crucial role due to its high affinity for the poly-anionic phosphate backbone of both RNA and DNA molecules \cite{bowman_cations_2012}. Due to its plasticity, RNA is particularly affected by ionic conditions, that have been shown to modulate folding stability and conformational ensembles \cite{yamagami_functional_2021}.
Conceptually, magnesium interacts with RNA in two ways. First, it participates with other electrolytes in a diffuse cloud stabilizing the negatively charged backbone, also termed ion atmosphere. Second, it attaches to RNA at specific binding sites, through direct (inner-sphere) coordination or solvent-mediated (outer-sphere) interaction \cite{draper_guide_2004, hud_complexes_2008, martinez-monge_measurement_2022}.

Experimental study of RNA--\ce{Mg^{2+}} interactions is challenging. Ion counting experiments allow direct measurement of the excess or deficit of ion species recruited around an RNA molecule, with respect to bulk concentration \cite{jacobson_counting_2016, gebala_quantitative_2019}. While very useful to investigate various phenomena such as competition between cationic species---and how they relate to RNA dynamics, these techniques offer little insight on individual ion localization and binding modes. On the other hand, spatially resolved experiments have their own shortcomings. X-ray crystallography is notoriously troublesome for RNA molecules, suffers from crystallization artifacts and is often performed using artificial buffers that preclude studying biological ionic conditions \cite{jackson_general_2023}. Another promising approach is single-molecule cryo-electron microscopy (cryo-EM), whose tremendous progress in the past years has allowed to reach atomistic resolution \cite{chari_prospects_2023}. However, ion assignment has remained a challenging task due to the insufficiency of distinctive features in the density map, and lack of automatized validation tools \cite{auffinger_deflating_2021, naleem_cat_wiz_2026}.
In addition, spatial averaging of RNA conformational heterogeneity often results in poorly resolved regions, for which single-structure refinement procedures fail to appropriately model the observed density. In such cases, one might require integrative modeling techniques to infer molecular ensembles. This applies even more strongly to ion binding and solvent shell organization, which are inherently dynamic processes. As a result, ion identification in RNA structures deposited on the Protein Data Bank (PDB) has been inconsistent and error-prone \cite{leonarski_principles_2025}.

Molecular dynamics (MD) simulations stand out as an appealing approach to complement experimental data for its ability to capture dynamics at an atomistic level of detail. Several studies have successfully shown the importance of modeling cations to understand nucleic acid plasticity \cite{languin-cattoen_rna_2026}. However, it has been acknowledged that \ce{Mg^{2+}} poses a singular difficulty because of its slow binding kinetics, that defy typical all-atom MD simulation time spans. Water-water exchange dynamics in the inner shell of \ce{[Mg(H2O)6]^{2+}} has been measured in the microsecond range \cite{bleuzen_water_1997}, and inner-sphere binding to RNA moieties is expected to reach equilibrium on even slower time scales. Hence, plain MD simulations often reveal inadequate if one desires to sample the ion distribution at the surface of nucleic acid molecules, and more advanced techniques are called for.

Recently, the ability of the modeling community to describe the Ångström-scale solvent and ion structure around an RNA molecule was tested for the first time in one target (R1260) of the CASP16 (Critical Assessment of Structure Prediction) blind experiment \cite{kretsch_blind_2026}. The participants were given the task to submit a structural ensemble for the already-known \textit{Tetrahymena thermophila} group I intron ribozyme fold \cite{su_cryo-em_2021}, including its solvent layer, and the results were assessed using then-unpublished, high resolution cryo-EM density maps \cite{kretsch_complex_2025}. Our group submitted two MD-based predictions, one using conventional MD with strong restraints on the RNA native fold (\textit{bussilab\_plainmd}) and another employing weaker restraints, but an enhanced-sampling technique designed to allow equilibration of inner-bound \ce{Mg^{2+}} (\textit{bussilab\_replex}). Both ranked first in at least one of the various evaluation metrics, and in the top-7 at worst. Despite these successes, the top scores remained well below the theoretical limit imposed by experimental uncertainty. Aside from the complexity of the assessment process itself, that involves local alignment and back-mapping of the density from an ensemble of atomic coordinates, the assessors emphasized the large number of variables worthy of investigation such as RNA restraints and sampling strategy. For example, in absence of a more systematic analysis, it remains unclear to what extent enhanced sampling of \ce{Mg^{2+}}, prior knowledge on binding sites, or the use of restraints on the RNA fold affected the final results.

In this work we present in details a refined version of our enhanced-sampling MD strategy to accelerate magnesium inner-shell binding to its most common RNA binding sites, backbone phosphates. The improved method allows forecasting complex binding motifs involving multiple phosphate groups chelating a central cation, a recurrent pattern found in RNA structures. We characterize with good convergence the ion distribution around the \textit{Tethrahymena} ribozyme, as predicted by state-of-the-art force fields for RNA and ions, and validate the results against cryo-EM data. We show that accelerated methods sample a much wider range of significant binding motifs and are required for de novo prediction of already-known sites. Direct comparison to experimental maps allows to discriminate between setups and highlight the dominant role of restraints on the RNA dynamics, presence of magnesium ions and prior knowledge of keystone motifs.

\section{Results}

\begin{figure*}
  \begin{minipage}{0.67\linewidth}
    \stackinset{l}{1em}{t}{0.5em}{\textsf{\textbf{A}}}{
      \includegraphics[width=\linewidth]{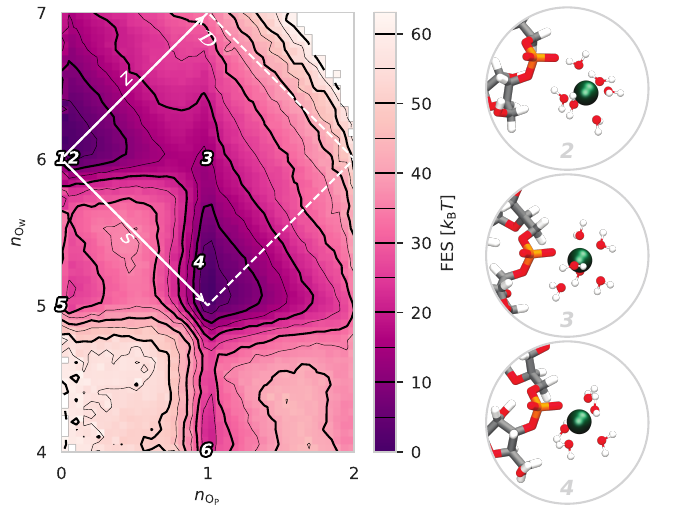}
    }
  \end{minipage}
  \hfill
  \begin{minipage}{0.31\linewidth}
    \stackinset{l}{1em}{t}{0.5em}{\textsf{\textbf{B}}}{
      \includegraphics[width=\linewidth]{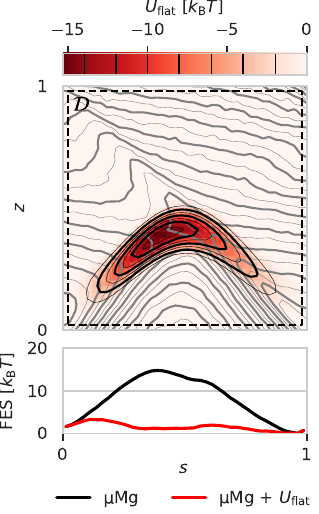}
    }
  \end{minipage}
  \caption{Designing a barrier-flattening bias. \textit{(A)} The free energy landscape of \ce{Mg^{2+}} binding to diuridine phosphate oxygen atoms is obtained performing well-tempered metadynamics on the two-dimensional space formed by the coordination number of \ce{Mg^{2+}} with water oxygen atoms ($n_{\text{O}_\text{W}}$) and phosphate OP1/OP2 ($n_{\text{O}_\text{P}}$). Numbers indicate states shown on the right and in Fig.~\ref{sfig:uu_snapshots}. \textit{(B)} A barrier-flattening bias $U_\text{flat}$ is defined on a two-dimensional domain $\mathcal{D}$ (top) and fitted so as to reduce the effective free-energy barrier projected along the binding coordinate $s$ (bottom).}
  \label{fig:bias}
\end{figure*}

\subsection{Mechanism for \ce{Mg^{2+}}--phosphate binding in molecular dynamics simulations}

Binding and unbinding \ce{Mg^{2+}} and RNA phosphate oxygen require a concerted rearrangement of the ion coordination sphere \cite{neumann_artificial_2022}. In MD simulations, the exact mechanism and corresponding free-energy barrier depend on the force-field parametrization \cite{falkner_kinetic_2021}.
We base our approach on the microMg (μMg) magnesium model, whose Lennard-Jones parameters use specific combination rules with RNA atom types in order to match geometric, kinetic and thermodynamic observables such as bond length, water exchange rate and solvation free energy \cite{grotz_optimized_2021}. μMg shares with most Mg models slow binding kinetics that prevent satisfactory sampling of the RNA's inner-shell ion distribution using conventional MD simulations. Its authors also derived a similar model, nanoMg (nMg). Despite sharing similar equilibrium properties, nMg features exchange dynamics that are two orders of magnitude faster, serving as a reference for an alternative sampling strategy.

To accelerate μMg inner-sphere binding dynamics in general RNA systems, we first use enhanced sampling to characterize the free-energy landscape of the binding process to non-bridging phosphate oxygen atoms (OP1, OP2) in a model system, namely
diuridine (UpU) with one \ce{Mg^{2+}} ion in a water box. Taking inspiration from previous work \cite{cunha_unraveling_2017}, we used well-tempered metadynamics (WT-MetaD) \cite{barducci_well-tempered_2008} on the two-dimensional space formed by the metal's coordination number to water oxygen ($n_{\text{O}_\text{W}}$) and phosphate oxygen atoms ($n_{\text{O}_\text{P}}$).
We note that previous works used as a reaction coordinate the distance between the divalent ion and
either the phosphorus \cite{koca_findik_binding_2024}
or the phosphate oxygen atoms \cite{cunha_unraveling_2017, grotz_optimized_2021}.
We employ a coordination number instead, to facilitate the extension of the method
to binding sites of higher denticity, as discussed below.

The resulting landscape (Fig.~\ref{fig:bias}a and \ref{sfig:uu_snapshots}) includes two main minima: one for the hexahydrated, unbound/outer-bound state (states \textbf{\textit{1}}/\textbf{\textit{2}}) and one for the pentahydrated, monodentate inner-bound state (\textbf{\textit{4}}). These two states are separated by a free-energy barrier with a saddle point located near a heptacoordinated transient state (\textbf{\textit{3}}), indicating an interchange associative ($\mathrm{I_a}$) mechanism, as previously observed for the μMg model \cite{falkner_kinetic_2021}. Note that this is at odds with the interchange dissociative ($\mathrm{I_d}$) or dissociative (D) mechanism suggested by the positive activation volume measured for first-shell water exchange \cite{bleuzen_water_1997} and observed in ab initio calculations \cite{ferretti_accurate_2025}. WT-MetaD also transiently explores low-population metastable states with apparently lower total coordination (Fig.~\ref{fig:bias}a, \textbf{\textit{5}} and \textbf{\textit{6}}) which correspond in fact to hexacoordinated complexes with alternative RNA moieties such as base or ribose oxygen atoms (see Fig.~\ref{sfig:uu_snapshots}).

Bidentate coordination to both OP1 and OP2 of the same phosphate is unfavorable, as shown by the absence of a local minimum at $n_{\text{O}_\text{P}} \approx 2$. By contrast, WT-MetaD simulation on a triuridine (UpUpU) system (Fig.~\ref{sfig:uuu_snapshots}) demonstrates the stability of adjacent-phosphate clamps, or “10-membered rings”,  which constitutes a recurrent motif in RNA structures \cite{petrov_bidentate_2011,zheng_magnesium-binding_2015, cowan_understanding_2024}. In fact, using as collective variables two coordination numbers reveals the similarity of the free-energy barriers in going from unbound to monodentate and from
monodentate to bidentate complexes, which allows us in principle to accelerate any number of exchanges of water ligands with phosphate oxyanions moieties, as explained hereafter.

\subsection{A flattening potential enables sub-nanosecond time scale for phosphate binding}

We next use the UpU--\ce{Mg^{2+}} free-energy landscape to design a potential energy bias with the goal of lowering the barrier's height, as shown in Fig.~\ref{fig:bias}b. Importantly, we enforce the following requirements: \textit{(i)}~the bias cancels out in the unbound state and the bound state, acting only near the barrier saddle point, \textit{(ii)}~the barrier height is lowered as much as feasible and \textit{(iii)}~the resulting ensemble is minimally perturbed. We do so by hand-crafting a crescent-shaped functional form, whose three parameters are fitted numerically to minimize barrier height (see Materials and Methods).

\begin{figure*}
  \centering
  \begin{minipage}{0.49\linewidth}
    \stackinset{l}{1em}{t}{0.5em}{\textsf{\textbf{A}}}{
      \includegraphics[width=\linewidth]{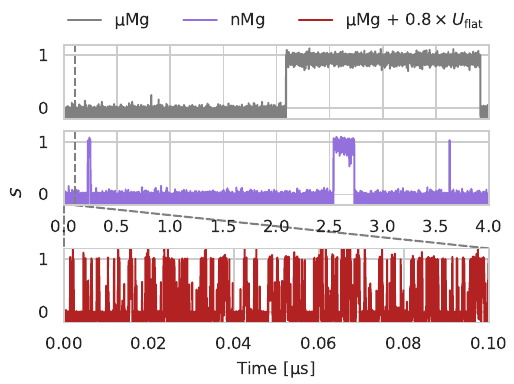}
    }
  \end{minipage}
  \begin{minipage}{0.49\linewidth}
    \stackinset{l}{1em}{t}{0.5em}{\textsf{\textbf{B}}}{
      \includegraphics[width=\linewidth]{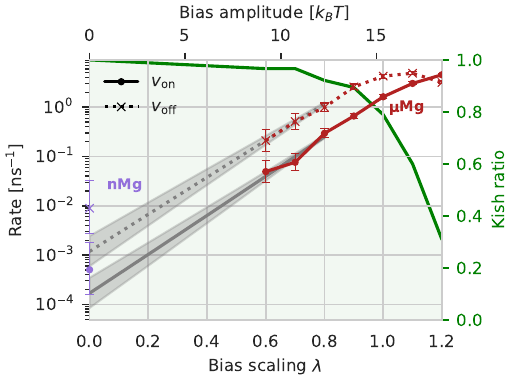}
    }
  \end{minipage}
  \caption{Testing the efficacy of the flattening bias on UpU. \textit{(A)} Timeseries of the binding coordinate $s$ for diuridine--\ce{Mg^{2+}} simulations using μMg (top, gray), nMg (middle, violet) and biased μMg (bottom, red) simulations. Without bias, a single binding event is observed within \qty{4}{\micro\second} of simulation and nMg results in a few exchanges. The flattening bias, here scaled with a prefactor 0.8, allows many exchanges between the unbound and bound states on the hundreds-of-nanosecond time scale. \textit{(B)} Estimate of on- and off-rates as a function of $U_\text{flat}$ for the UpU-\ce{Mg^{2+}} simulation box with biased μMg (red) or nMg (violet). Markers and 95~\% confidence intervals indicate $\lambda$-wise estimates from independent simulations. Gray curves and shaded areas show the best fit and block-bootstrap 95~\% confidence interval for the extrapolated μMg rates, based on the low-scaling/high-barrier regime ($\lambda \in \{0.6, 0.7, 0.8\}$).
  Weighting efficiency---or Kish ratio---is also reported (green curve).}
  \label{fig:bias_kinetics}
\end{figure*}

Simulating the UpU--\ce{Mg^{2+}} system with increasing bias strength demonstrates its ability to accelerate binding and unbinding events (Fig.~\ref{fig:bias_kinetics}a). On- and off-rates show an exponential dependence to the bias strength $\lambda$ at lower values, then saturate and decline at higher amplitudes (Fig.~\ref{fig:bias_kinetics}b) as over-stabilized intermediate states negatively impact the kinetics (Fig.~\ref{sfig:bias_timeseries}). The extrapolated on- and off-rates at zero bias, $v^0_\text{on} = \qty{0.19}{\per\micro\second}$ (95\% CI \numrange{0.10}{0.35}) and $v^0_\text{off} = \qty{1.3}{\per\micro\second}$ (\numrange{0.7}{2.4}) are consistent with a unique binding event observed in a 4-\unit{\micro\second} unbiased trajectory. By contrast, full bias ($\lambda = 1$) leads to sub-nanosecond time scales, $v_\text{on} = \qty{1.6}{\per\nano\second}$ (\numrange{1.4}{1.9}) and $v_\text{off} = \qty{6.2}{\per\nano\second}$ (\numrange{5.4}{7.2}). This also compares favorably to the nMg model, which completes only three binding events in a similar 4-\unit{\micro\second} simulation, $v_\text{on} = \qty{0.5}{\per\micro\second}$ (\numrange{0.2}{1.8}) and $v_\text{off} = \qty{8.9}{\per\micro\second}$ (\numrange{2.8}{32.2}).

Because the system spends most of the time in the unbound or bound states and only resides transiently on the barrier, the bias value remains close to zero for most of the simulation frames (Fig.~\ref{sfig:bias_timeseries}). This allows straightforward Boltzmann re-weighting of the trajectory to the unbiased conformational statistics with a good effective sample size (ESS)---that is, the number of unweighted samples that would provide the same level of statistical precision as the re-weighted ones. This can be measured by the Kish ratio, or weighting efficiency \cite{kish_survey_1995}
\begin{equation}
  \text{Kish ratio} = \frac{\text{ESS}}{n} = \frac{\left(\sum_i^n w_i\right)^2}{n\sum_i^n w_i^2},
\end{equation}
\noindent with $w_i$ the unnormalized weights of the $n$ sample frames, which exceeds \qty{80}{\percent} for scalings up to $\lambda = 1$ (Fig.~\ref{fig:bias_kinetics}b).

Conveniently, the barrier-flattening bias for monodentate OP1/OP2 binding generalizes to higher denticities without reparameterization, and is shown to adequately flatten the mono-to-bidentate transition observed in the UpUpU system (Fig.~\ref{sfig:uuu_snapshots} and Fig.~\ref{sfig:uuu_bias_timeseries}). We make use of this property to sample complex binding motifs in a structured RNA system.

\subsection{Replica-exchange sampling of the \textit{T. thermophila} ribozyme ion distribution}

The \textit{Tetrahymena thermophila} group I self-splicing intron (Apo L-21 ScaI) has 386 phosphate groups and requires a box with hundreds of \ce{Mg^{2+}} ions to be simulated in the proper conditions. Despite the locality of the previous barrier-flattening bias, a direct re-weighting approach similar to the one tested on the UpU-Mg system does not scale to such a large system, because the ESS is expected to shrink exponentially with the number of occupied binding sites. Indeed, assuming $N$ independent subsystems akin to the UpU--Mg setup, the Kish ratio for the whole system would factorize as $K_N = K_1^N$, where $K_1$ is the Kish ratio for reweighting a single subsystem. Given the Kish ratios reported in Figure \ref{fig:bias_kinetics}, this back-of-the-envelope calculation for a simulation biased with a scaling factor 0.8 yields a vanishing weighting efficiency $< \qty{1}{\percent}$ for $N > 43$ sites. Inspired by previous works \cite{curuksu_enhanced_2009, gil-ley_enhanced_2015}, we therefore introduce a replica-exchange procedure where the flattening potential is gradually activated across the ladder (see Materials and Methods), and the first replica samples the unbiased Boltzmann distribution.

The initial structure (PDB: \textsc{7ez0} \cite{su_cryo-em_2021}) contained 27 magnesium ions, from which we could identify 14 plausibly inner-bound to RNA ($r < 2.8$~Å), see Table~\ref{stable:motif_table}. Nearly all were bound to non-bridging phosphate oxygen atoms (OP1, OP2), including 8 monodentate, 3 bidentate and 2 tridentate sites. A single one (507MG in the original structure) was bound to a base carbonyl (O6). Multiple ions laid in the exclusion zone (\qtyrange{2.2}{3.8}{\angstrom}) suggested by curated databases and ab initio simulations \cite{leonarski_mg2_2017, kolev_sodium_2022}, indicating either a misassignment (e.g. instead of \ce{Na+}) or smearing artifacts due to local dynamics. It is worth mentioning that minimization and short NPT equilibration of the solvated native structure including its magnesium ions leads to differences in the binding motifs despite using positional restraints on the ions and heavy atoms of the RNA. Most notably, several ions previously in the exclusion zone or outer sphere end up inner-bound (Table~\ref{stable:motif_table}). This is in part due to the standard solvation protocol, that fails adding water molecules in between the outer-bound ions and RNA atoms, resulting in a cavity that is only resolved through direct binding. Since our scope is precisely to overcome slow binding kinetics through enhanced sampling, we did not further investigate initial placement of inner-bound ions. Extra attention should be paid in absence of an adequate sampling strategy.

In order to assess our ability to predict binding sites \textit{de novo}, we carried out two sets of simulations, either keeping or ignoring the PDB ion positions. Extra \ce{Na+}, \ce{Cl-} and \ce{Mg^{2+}} ions were added so as to approximate the experimental buffer ionic conditions of the validation cryo-EM data used hereafter (10~mM \ce{MgCl2}, 50~mM Na--HEPES modelled here with 50~mM NaCl, see Methods and Figure~\ref{sfig:bulk_concentrations}). Importantly, these ions were randomly distributed in the bulk using different seeds to generate the initial configuration of each replica. Relatively strong conformational restraints were used to keep the RNA near its input conformation. Each replica-exchange simulation used 8 replicas with increasing bias strength from $\lambda = 0$ to $\lambda = 0.87$ and was performed in two independent replicates. As a basis for comparison, we ran traditional MD simulations of the system under three conditions: μMg, nMg, and another nMg using softer conformational restraints on the RNA, with 8 replicates each.

\begin{figure*}
  \centering
  \begin{minipage}{0.49\linewidth}
    \stackinset{l}{1em}{t}{0.5em}{\textsf{\textbf{A}}}{
      \includegraphics[width=\linewidth]{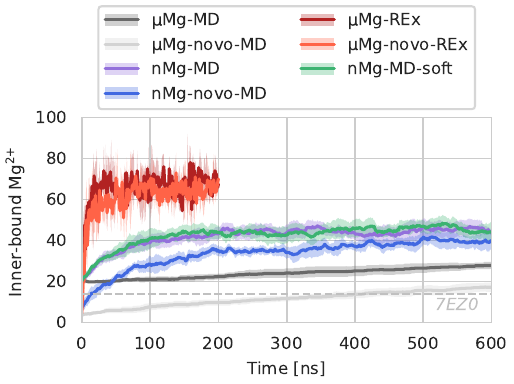}
    }
  \end{minipage}
  \begin{minipage}{0.49\linewidth}
    \stackinset{l}{1em}{t}{0.5em}{\textsf{\textbf{B}}}{
      \includegraphics[width=\linewidth]{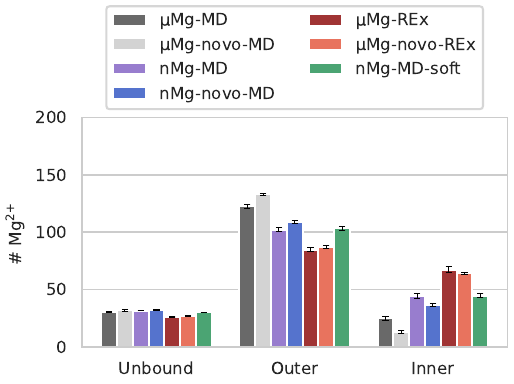}
    }
  \end{minipage}
  \caption{Magnesium binding to Apo L-21 ScaI. \textit{(A)} Time evolution of the number of inner-bound ($d \le \qty{2.8}{\angstrom}$) \ce{Mg^{2+}} for the two models microMg (μMg) and nanoMg (nMg) using conventional Molecular Dynamics (MD) or Replica Exchange with accelerating bias (REx). Curves are averages over 8 (MD) or 2 (REx) replicates, shaded area show point-wise 95\% CI. The dashed line indicates the number of inner-bound ions in the input PDB: \textsc{7ez0}. \textit{(B)} Number of \ce{Mg^{2+}} ions in the inner-sphere, outer-sphere ($\qty{2.8}{\angstrom} < d \le \qty{5}{\angstrom}$) and unbound ($d > \qty{5}{\angstrom}$), discarding the first \qty{25}{\%} of each trajectory. Error bars indicate \qty{95}{\%} CI estimated with Bayesian bootstrap on blocks.}
  \label{fig:bound_mg}
\end{figure*}

In the unbiased replica of replica-exchange simulations, the number of inner-bound \ce{Mg^{2+}} (using a \qty{2.8}{\angstrom} cutoff, see Fig.~\ref{sfig:rdfs}) quickly increases within the first 50~ns before reaching a plateau around 70 (Fig.~\ref{fig:bound_mg}a).
By contrast, conventional MD simulations using μMg fails to reach a plateau and shows a slow and steady increase, with only a few binding events within 600~ns. This demonstrates the comparative advantage of our enhanced sampling method.
On the other hand, the faster nMg model allows reaching a plateau after 200~ns of conventional MD, although it converges to a smaller value of about 45 inner-bound \ce{Mg^{2+}}. This discrepancy corresponding to a difference of \qty{-0.4}{k_BT} inner-sphere binding free energy in favor of μMg is compatible with previously reported values (μMg: $\Delta G^0 = \qty{-0.6 +- 0.6}{k_BT}$; nMg: $\Delta G^0 = \qty{-0.3 +- 0.1}{k_BT}$) \cite{grotz_optimized_2021}. This is partly compensated by a gap in the number of observed outer-bound \ce{Mg^{2+}} (Fig.~\ref{fig:bound_mg}b, see Fig.~\ref{sfig:bound_mg} for timeseries). Importantly, such differences are undetectable in the overall number of excess cations and default anions around the RNA, as one would measure in ion counting experiments (see preferential interaction coefficients, Fig.~\ref{sfig:bulk_concentrations}). Note also that \ce{Na+} distribution converges very quickly and is hardly affected by \ce{Mg^{2+}} inner-binding dynamics (Fig.~\ref{sfig:bound_na}), supporting the idea that ion competition is insensitive to binding mode.

\subsection{Global Cryo-EM validation}\label{sec:cryo}

We note that a validation of predicted ensembles against deposited PDB entries can be affected by modeling inaccuracies and compromises related to single-structure refinement protocols.
Therefore, we assess the accuracy of the simulation approaches presented so far by directly comparing
them with the cryo-EM density maps.
To do so, we back-calculate cryo-EM density maps from our trajectories using a forward model based on scattering factors \cite{peng_robust_1996, hoff_accurate_2024} and compare them to experimental densities. Following the protocols discussed in the CASP16 experiment \cite{kretsch_blind_2026}, we compare simulations based on the intron structure derived from the first experiment (PDB: \textsc{7ez0}, EMD: 31385) to the structure and higher-resolution map obtained from the most recent experiment (PDB: \textsc{9cbu}, EMD: 42499). The trajectories are aligned either globally, through root-mean-square-deviation (RMSD) minimization to the reference, or with residue-wise alignment of local environments (see Methods). Global alignment is akin to the process of density map reconstruction from a set of 2D projections from individual particles, possibly from slightly different conformational states. On the other hand, local alignment discards part of the RNA flexibility that may be incorrectly modeled by the chosen force field, and enforces closer adherence to the reference native structure. This approach was preferred in the CASP16 assessment to emphasize differences in modeling the solvent shell in lieu of the RNA dynamics.

\begin{figure*}
  \centering
  \begin{minipage}[t]{\linewidth}
    \stackinset{l}{1em}{t}{0.5em}{\textsf{\textbf{A}}}{
      \includegraphics[width=\linewidth]{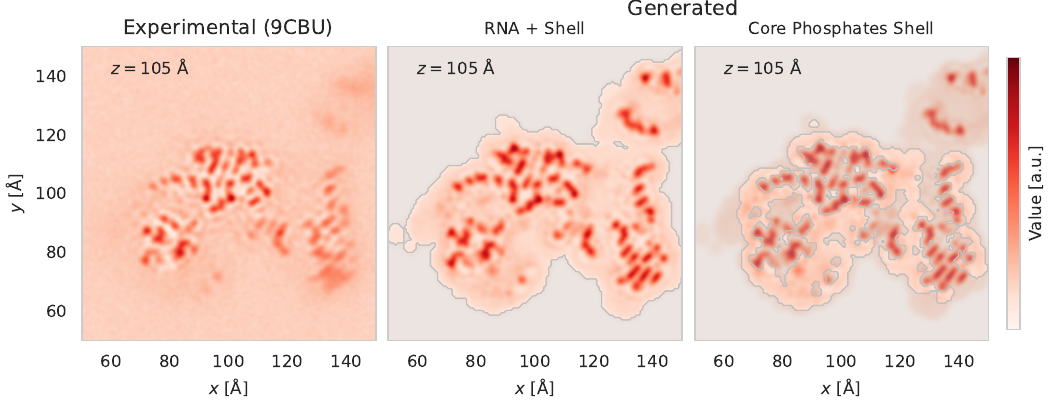}
    }
  \end{minipage}

  \begin{minipage}[t]{0.44\linewidth}
    \raisebox{-\height}{%
      \stackinset{l}{1em}{t}{0.5em}{\textsf{\textbf{B}}}{
        \includegraphics[width=\linewidth]{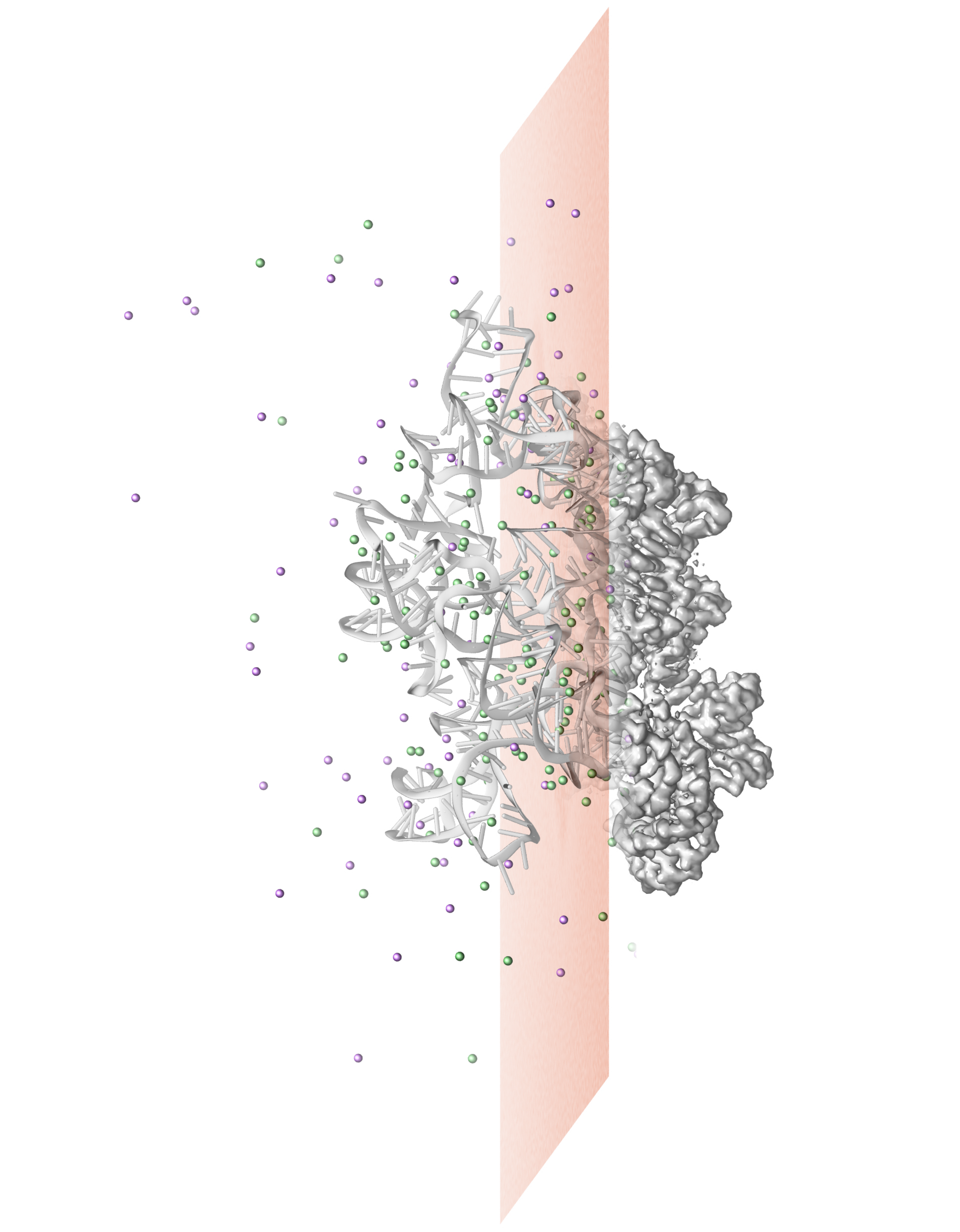}
      }
    }%
  \end{minipage}%
  \hfill%
  \begin{minipage}[t]{0.54\linewidth}
    \raisebox{-\height}{%
      \stackinset{l}{1em}{t}{0.5em}{\textsf{\textbf{C}}}{
        \includegraphics[width=\linewidth]{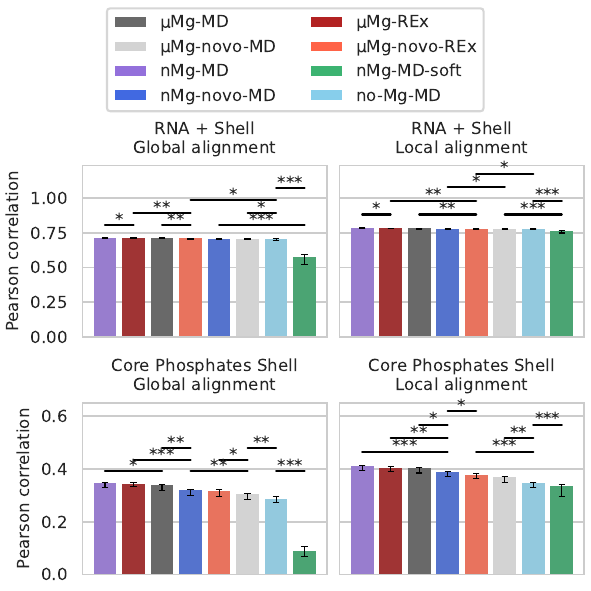}
      }
    }%
  \end{minipage}
  \caption{Validation of simulation setups against experimental cryo-EM maps. \textit{(A)} Slices of the reference map (PDB: \textsc{9cbu}, EMD-42499) and globally aligned generated map (here μMg-REx). Two voxel masks are shown (shaded area). RNA + Shell: whole RNA including its solvent shell ($d_\text{P} < \qty{6}{\angstrom}$). Core Phosphates Shell: solvent shell around phosphates from the well-resolved core of the RNA scaffold ($d_\text{RNA} < \qty{6}{\angstrom}$ and $d_\text{RNA} > \qty{1.8}{\angstrom}$). \textit{(B)} Molecular rendering illustrating: on the left, a snapshot of the intron with surrounding \ce{Mg^{2+}} (green) and \ce{Na+} (purple) cations; on the right, iso-surface of the back-calculated density from the trajectory; in-between, the $z = \qty{105}{\angstrom}$ slice displayed in panel A. \textit{(C)} Cross correlation computed for the RNA + Shell and Core Phosphates Shell voxel selections, with either global or local alignment. Bars indicate 95\% confidence interval computed from Bayesian bootstrap on blocks. Stars indicate the significance level for the first significantly lower-performing setup. $\ast$:~$p < 0.05$. ${\ast}{\ast}$:~$p < 0.01$. ${\ast}{\ast}{\ast}$:~$p < 0.001$.}
  \label{fig:correlation}
\end{figure*}

Maps are generated by averaging the contribution of all heavy atoms for subsampled trajectory frames, on a grid identical to that of the reference map. To account for finite-sample artifacts, generated maps are smoothed with B-factor values maximizing agreement with experiment (Fig.~\ref{sfig:tempfactor_scan}). We compare the Pearson correlation coefficients, or cross-correlation (CC), between predicted and observed voxel values for voxel selections based on the reference PDB coordinates (\textsc{9cbu}) (Fig.~\ref{fig:correlation}). Note that we use the raw, unsharpened half maps as reference instead of the sharpened, main deposited map. The sharpened reference systematically leads to lower CC, and sometimes negative CC values, without improving discriminatory power between setups (Fig.~\ref{sfig:tempfactor_scan_sharpened}).

When including the whole molecule and its solvent shell ($d < 6$~Å) with global alignment, all setups reach comparable CC values around 0.7, except for the soft-restrained one, which performs significantly lower ($p < 10^{-3}$) at 0.57. Among the tightly restrained simulations, small but statistically significant differences can be observed. The best performing setups are the ones informed with initial PDB ions, nMg-MD, μMg-REx and μMg-MD, with a low-significance advantage ($p < 0.05$) for nMg-MD. This is expected, and indirectly confirms the consistency between the two experimental datasets, as the initial ions are taking from the older experiment and the validation is done against the newer experiment.
These predictions are underperformed with moderate significance ($p < 0.01)$ by those from setups without initial ions, nMg-novo-MD, μMg-novo-REx, μMg-novo-MD and no-Mg-MD. Somewhat surprisingly, at this scale, there is no significant difference between enhanced sampling setups (μMg-REx and μMg-novo-REx) and their kinetically trapped equivalent (respectively μMg-MD and μMg-novo-MD) despite considerable differences in the number of inner-bound ions, as previously shown.

To increase the sensitivity of the comparison with respect to phosphate binding mode, we restrict the voxel selection to the solvent shell around phosphate atoms ($d < \qty{6}{\angstrom}$) of the core fold, excluding poorly defined peripheral helices (excluding residues 63--88, 229--245, 284--295, 332--335 and 364--402 as in \cite{kretsch_blind_2026}), and excluding RNA itself ($\qty{1.8}{\angstrom} < d$). A slightly different picture emerges, with now μMg-novo-MD ranking significantly below μMg-novo-REx and nMg-novo-MD, although the ranking between μMg-REx and μMg-MD remains below significance.
Together, these results are indicative of the ability of the employed analysis protocol to discriminate features of the ion-solvent shell induced by different sampling strategies, although with moderate effect size. The principal impacting factors on the ranking are, by order of importance: \textit{(i)} the strength of conformational restraints on the RNA, \textit{(ii)} the use of adequate \ce{[Mg^{2+}]} \textit{(iii)} the inclusion of prior ions, \textit{(iv)} the sampling strategy together with the Mg model and \textit{(v)} the kind of trajectory alignment.
Importantly, the ranking of nMg-MD and μMg-REx is not robust to a change in alignment and voxel selection.
Using softer restraints on the RNA leads to a decrease in CC, both for the shell-only and shell-RNA selections. This is indicative that RNA flexibility is a strong component of the agreement between the back-calculated density and the experimental maps, although the impact is alleviated with local alignment, as expected. Overall, it is likely that the observed variations in density are dominated by how cations organize RNA local dynamics rather than direct features of the solvation shell, with the exception of a few highly structured keystone motifs. For that reason, it is also possible that the relatively strong conformational restraints we employed by default partially obscure otherwise detectable differences between force fields and sampling strategies.

\subsection{Analysis of predicted binding motifs}

In all setups, nearly all inner-bound \ce{Mg^{2+}} interacts with non-bridging phosphate oxygen, with a slight (60:40) preference for OP2 versus OP1 (Fig.~\ref{sfig:site_types}),
in qualitative agreement with previous analysis of structural databases \cite{zheng_magnesium-binding_2015} and computational study \cite{cunha_unraveling_2017}. This contrasts with \ce{Na+}, which exhibits a wider range of partners including base nitrogen (N1/N3/N7) and base oxygen (O2/O4/O6). Binding is unevenly distributed along the sequence (Fig.~\ref{sfig:residue_sites}).

\begin{figure*}
  \centering
  \includegraphics[width=\linewidth]{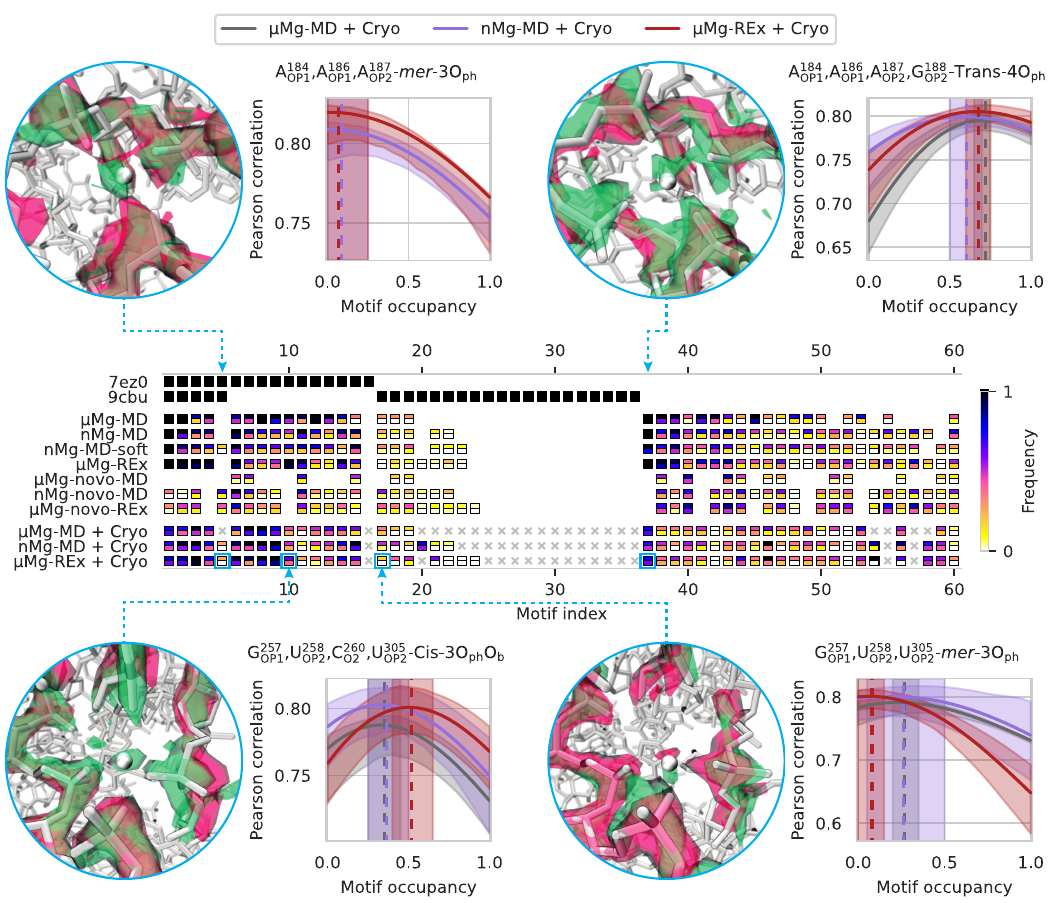}
  \caption{Prediction of \ce{Mg^{2+}} binding motif frequency in simulations and after integrating cryo-EM data. Black squares indicate either the presence in reference PDB structures \textsc{7ez0} and \textsc{9cbu}. Bi-colored squares (lower half and upper half show the 2.5th and 97.5th percentiles) indicate the empirical frequency of significant motifs (frequency $\ge 0.05$ observed in at least one setup) or occupancy that maximizes correlation with experimental cryo-EM map (+~Cryo). Absence of square indicate no detection in that setup. For cryo-EM-informed prediction, crosses indicate no prediction could be made--rather than prediction of absence. Only 60 of the most relevant motifs are shown, ranked by presence in PDB and average frequency in simulations. The complete list can be found in Table~\ref{stable:motif_table}. Four motifs are highlighted, showing a molecular rendering of the motif together with isosurfaces for the μMg-REx local density map computed with the motif present (green) and the absent (red). The local cross-correlation to experimental map is plotted as a function of the proportion of present-motif map in the generated map mixture, and it maximum yields our best estimate for occupancy (vertical dashed lines). Shaded areas correspond to bootstrap \qty{95}{\%} CI.}
  \label{fig:motifs}
\end{figure*}

Due to its strict hexacoordination, \ce{Mg^{2+}} forms highly structured motifs with RNA moieties, many of which have been identified in deposited structures and can be characterized by the set of coordinated atoms and a given stereochemistry \cite{zheng_magnesium-binding_2015, cowan_understanding_2024}. The simplest and most frequent kind of motif involves one cation coordinating a single RNA site and 5 water molecules. More complex motifs with higher denticities are not uncommon and play a key role in RNA folding. Most notable ones include Magnesium clamps and 10-member rings, that respectively bridge consecutive or distant phosphates from the backbone \cite{petrov_bidentate_2011}.

We analyse the frequency of such motifs in our simulation setups. As expected, monodentate interactions dominate, with several bidentate and only a few tri- and tetradentate sites counted at equilibrium (Fig.~\ref{sfig:denticities}). The total number of significant motifs (frequency $> \qty{5}{\%}$) explored in each setup demonstrates a comparative advantage of REx, which exceeds 295 in all setups tested, while nMg with traditional MD tops at 230 at best (Table~\ref{stable:motif_count}), for similar computational cost (Table~\ref{stable:simulations}). Note that those numbers exceed by up to 5 times the equilibrium counts (Fig.~\ref{fig:bound_mg}), due both to the aggregation of multiple trajectories/replicas and to the transient nature of many binding sites. As expected, μMg-MD and μMg-novo-MD shows limited sampling and only reach about 80 unique motifs.

Figure~\ref{fig:motifs} show the empirical frequency of significant motifs across setups. When comparing them to motifs identified in the PDB, 14 out of 16 motifs present in the initial structure (\textsc{7ez0}) are retained in simulations including them (μMg, nMg). More strikingly, fast-binding simulations without prior \ce{Mg^{2+}}, μMg-novo-REx and nMg-novo-MD, predict 12 (resp. 10) motifs out of 16. Consistently with the low binding rate, most of μMg-MD motifs are retained from post-equilibrated configurations and only a few (4) are found in μMg-novo-MD.

An important number (12) of sites only identified in the recent \textsc{9cbu} are not found in simulations. However, most of these involve non-phosphate moieties (base oxygen, base nitrogen or ribose oxygen), for which (a) enhanced sampling was not applied and (b) the \ce{Mg^{2+}} force field was not explicitly parameterized.

Higher denticity motifs are particularly challenging. We note two tetradentate motifs. \motifX (motif \#10) is suggested in \textsc{7ez0} but not \textsc{9cbu}, with an ambiguous involvement of the base O2 at a distance \qty{2.67}{\angstrom} and non-octahedral geometry in the former, and a clear outer-bound configuration in the latter (\qty{3.86}{\angstrom} with a mediating structural water molecule). In simulations, equilibration leads to a well defined tetradentate geometry, but $\text{C}^{260}_\text{O2}$ is partly lost in production in favor of its tridentate counterparts, \motifXVII and -\textit{fac}-3$\text{O}_\text{ph}$ (motifs \#17 and \#84), which could indicate it was purely an artifact from the solvation protocol. Another example is the tridentate \motifV (\#5), clearly identified in both \textsc{7ez0} and \textsc{9cbu}, that becomes tetradentate \motifXXXVII (\#37) during system preparation and is retained in production. None of the \textit{de novo} setups is able to reach 4-ligand configurations, probably because of insufficient simulation time.

\subsection{Integration of Cryo-EM data and validation of motif populations}

Correlation coefficients such as those discussed in Section \ref{sec:cryo} only provide a global assessment but do not validate individual motifs.
We now show cryo-EM data can be effectively integrated \cite{bernetti_integrating_2023} with simulated ensembles to improve their accuracy and to alleviate issues related to limited sampling.

In standard reweighting protocols, the weight of each analyzed snapshot would depend on the degree of agreement of that specific snapshot with experiment. In our case, the agreement might result from a mixture of quasi-independent binding sites, only a fraction of which would be in agreement with experiment. To overcome the difficulties of such a global reweighting procedure,
we adopt a local strategy in which the contribution of each binding motif is assessed independently to infer its effective population under experimental conditions.

For each of the identified motifs, we separate simulation frames where the motif is present from those where it is absent, and compute density maps of the motif neighborhood in those two extreme cases after local alignment to the reference structure. Note that such an analysis is only feasible when both the "present" and "absent" states were sampled. Therefore, data from both PDB-informed and \textit{novo} systems are combined to maximize the number of motif that get representative frames from both states. This results in three conformational ensembles, for the μMg-MD, nMg-MD and μMg-REx setups. We then compute the neighborhood's cross-correlation with the reference map for varying mixtures of the present/absent maps, and record the occupancy that maximize agreement as our best estimate for how populated is the motif in experimental conditions. Neighborhoods with low maximum cross-correlation ($\text{CC}_\text{max} < 0.7$) predominantly come from poorly resolved regions outside the intron core (Fig.~\ref{sfig:ccmax_pcryo}), including flexible peripheral helices such as P6b or P9.2, and are discarded from this analysis since they tend to give unreliable results. Visual representation of the motif occupancy determined from cryo-EM-informed reweighting is shown in Figure~\ref{fig:motifs} (+~Cryo) and detailed in Table~\ref{stable:motif_table}.

As a general trend, and despite marked differences in empirical frequencies, reweighted occupancies agree across conformational ensembles, showing that the approach is robust to force-field and sampling discrepancies.

Our results validate the motifs from \textsc{7ez0}, attributing them high occupancies, with a few interesting exceptions. One case is that of the tridentate \motifV (motif \#5), present in both \textsc{7ez0} and \textsc{9cbu}. System preparation leads the neighboring $\text{G}^{188}_\text{OP2}$ atom to form a tetradentate \underline{Trans}-4$\text{O}_\text{ph}$ (motif \#37) site shown to be very stable in both μMg and nMg simulations. The tridentate motif is also spontaneously sampled in \textit{novo} simulations, but not the tetradentate one, leaving a doubt on the latter being purely artefactual. However, the motif occupancy that maximizes agreement with the experimental map clearly favors the tetradentate state ($p_\text{cryo} =$ \numrange{0.6}{0.8}) with respect to the tridentate one (\numrange{0.0}{0.2}), suggesting a suboptimal refinement procedure in both PDB structures, which failed to report it.

Similarly, the ambiguous participation of the distal C260 base oxygen in \motifX\slash \motifXVII (motifs \#10\slash \#17, present in \textsc{7ez0} and \textsc{9cbu}, respectively) is resolved by acknowledging that agreement is maximized for intermediate occupancy ($p_\text{cryo} \approx 0.5$) of the tetradentate state and non-zero occupancy of the tridentate one. This example highlights how two experiments with different map resolutions can lead to different compromises in single-structure coordinate assignment, that propagate downstream to MD. It is worth mentioning that neither μMg nor nMg were optimized to faithfully capture binding to nucleobase sites.

\section{Discussion}

In this work, we (a) developed an enhanced sampling protocol to accelerate Mg--RNA binding and (b) we used it to compare different force fields and simulation settings, by assessing them in their capability to reproduce cryo-EM maps. The calculations are designed similarly to those
reported in a recent CASP experiment \cite{kretsch_blind_2026}, where our group presented highly ranked predictions using a preliminary version of the method used in the current paper.

The enhanced sampling approach presented here combines the specificity of CV-based methods, that focus on selected kinetic bottlenecks, and the power of replica exchange to accelerate concurrently a large number of degrees-of-freedom \cite{henin_enhanced_2022}. As such it takes inspiration from Replica Exchange with Collective-Variable Tempering \cite{gil-ley_enhanced_2015}, with several key differences. First, similarly to \cite{curuksu_enhanced_2009}, it implements a fixed bias instead of an adaptive one, which allows to fine-tune an efficient and unobtrusive barrier-flattening potential. Second, the same bias is applied to a multitude of equivalent CV spaces. This is possible because all \ce{Mg^{2+}} ions are equivalent, and because the shape of the free-energy barrier for binding $\text{O}_\text{ph}$ is only marginally affected by the local environment. For the same reason, the barrier is similar regardless of the motif's current denticity, which allows us to reuse the bias functional form to tackle polydentate motifs. Another recently proposed sampling strategy, termed fast-ion replica exchange (FIREX) \cite{kantin_firex_2026}, also employs Hamiltonian replica exchange to accelerate \ce{Mg^{2+}} inner-binding dynamics, although it rescales the ion charge and Lennard-Jones parameters rather than biasing a selected set of CVs. This choice avoids the overhead associated with computing coordination number-based CVs and can be readily used with standard molecular dynamics topologies. On the other hand, the limited flexibility of the perturbation used in the FIREX method may restrict the achievable acceleration, and its non-local nature is likely to require a larger number of replicas when scaling to large systems. In its current implementation, FIREX has been shown to be compatible with solute tempering (REST2) \cite{wang_replica_2011}, enabling simultaneous enhancement of RNA conformational sampling. Likewise, similar hybrid strategies where two tempering factors are simultaneously used to accelerate two types of processes can be used with PLUMED-based biases such as the one introduced in this work, as enabled by the generalized replica-exchange framework \cite{bussi_hamiltonian_2014} and exemplified in previous works \cite{appadurai_high_2021, bellucci_fibrillation-prone_2017}. Such a combination is particularly attractive for systems in which REST2 can be applied to a relatively small region.

We believe the presented methodology could be adapted to other magnesium models if needed. Possible improvements involve the inclusion of minor binding sites ($\text{O}_\text{r}$, $\text{O}_\text{b}$, $\text{N}_\text{b}$) and automatized, on-the-fly fitting of the accelerating bias for diverse atom types.
More generally, this strategy should be transferable to systems in which slow dynamics arise primarily from repeated local activated events that can be described by compact collective variables. In RNA and DNA, for instance, nucleobase pairing and stacking transitions may not represent the only relevant barriers, but they often constitute a dominant kinetic bottleneck and can be described by relatively local descriptors \cite{bottaro_role_2014}. Other examples include proton-transfer events in reactive molecular dynamics models, where local bonding or proton-sharing coordinates define the elementary event; proline cis–trans isomerization in proteins containing multiple prolines, where the peptide-bond dihedral provides a natural local variable; lipid flip-flop in membranes, where headgroup position, hydration, and membrane insertion describe rare interleaflet transfer; and ion transport or ligand-exchange processes in energy materials, including solid-state ion conductors, battery electrolytes, ionic liquids, and deep eutectic solvents, where local coordination numbers capture hopping or solvation-shell exchange. These examples share the key features exploited here: a high local kinetic barrier, a relatively simple transition coordinate, and many equivalent or quasi-equivalent copies of the same microscopic process within a larger system.

A similar principle underlies the local reweighting strategy used here, where the contribution of individual binding sites is assessed independently to circumvent the intractability of global reweighting. This decomposition into approximately independent local observables is conceptually related to earlier approaches that exploit locality to address complex biomolecular systems, such as piecewise or iterative force-field refinement schemes based on experimental data \cite{li_iterative_2011}, or reweighted hierarchical chain growth \cite{stelzl_global_2022}. While the objectives and implementations differ, these approaches share the idea that large systems can be effectively analyzed through a set of weakly coupled local contributions.

The generated ensembles are then validated against cryo-EM data, allowing us to probe several key aspects that remained unexplored in the recent community analysis of the CASP submissions \cite{kretsch_blind_2026}. We find that limited sampling of inner-binding dynamics has a critical impact on ion distribution but moderate effect on cross-correlation with experimental maps. Simulations that fail to adequately sample inner-shell binding consistently show poorer agreement with experiment, highlighting the importance of capturing slow ion-exchange kinetics.
Simulations that omit divalent ions altogether yield substantially worse agreement, confirming that cryo-EM data are sensitive to their presence.
In contrast, differences between force fields are more difficult to resolve using cryo-EM data alone. Although the two tested force fields yield different fractions of inner- versus outer-bound ions, the experimental maps do not appear sufficiently sensitive to discriminate between them. This suggests that current cryo-EM data primarily constrain the overall ion distribution, while remaining only weakly sensitive to the detailed dynamical partitioning between binding modes. Notably, the total number of ions in the RNA region is comparable across models, suggesting that ion-counting experiments would likewise be unable to discriminate between them. Additional experimental observables would therefore be required to further discriminate between force-field models.
Finally, we assess the effect of restraints on RNA coordinates and observe that tighter restraints lead to measurably improved agreement with experiment, indicating that inaccuracies in RNA conformational dynamics remain a limiting factor. This suggests that further refinement of current RNA force fields could be beneficial, and that this system—combining structural heterogeneity with high-quality experimental data—provides a valuable benchmark for such developments. Notably, the restraints employed here are derived from a structure refined against a lower-resolution map, yet still improve agreement with a higher-resolution dataset. This indicates that independent cryo-EM experiments are sufficiently consistent to support force-field refinement without overfitting to a specific structure.

Importantly, our validation procedure does not merely rely on the solved PDB structures \cite{su_cryo-em_2021, kretsch_complex_2025} but rather includes the maps against which they were fitted, which is closer to the raw experimental data. It is relevant to note however that several steps are required to obtain such maps
from the EM images, most notably particle selection, alignment of the maps, and processing of the resulting map, all of which might indirectly affect the measured dynamics.
In fact, cryo-EM maps are routinely sharpened using a Gaussian filter or a more advanced technique in order to better resolve density peaks and ease fitting to a single native structure \cite{murshudov_refinement_2016, pintilie_validation_2021}. However, this is expected to attenuate the lower frequency features associated with static and dynamic disorder. Sharpening with large B-factor values may lead to ringing artifacts, i.e.\ non-physical negative values in the potential at the boundaries of sharp density peaks, which may decrease agreement with the forward model, by construction positive in our case. For that reason, we use the raw, unsharpened half maps as reference instead of the sharpened, main deposited map. This choice contrasts with the previous CASP16 assessment \cite{kretsch_blind_2026}, in which the sharpened map was used, leading to top-performing cross-correlation values much lower that those reported here ($0.42$ for \textit{bussilab\_replex}, compared with values exceeding $0.75$ in this work). We point out that, despite controlling for noise, automated sharpening procedures that rely on half-map cross-validation do \textit{not} improve signal accuracy, but merely accentuate features that may or may not be desirable for downstream interpretation \cite{murshudov_refinement_2016}. Hence, we believe that studies focused on dynamics, or machine learning approaches, should generally favour the raw, unsharpened half maps. Map sharpening is also of concern to the task of single-structure resolution and refinement, since dynamic disorder can spread electron density and result in local maxima whose positions violate stereochemical restraints. For example, it was recently shown that ensemble-based integrative modeling could resolve mismodeled RNA helices of a group II intron, a seemingly widespread problem in published RNA structures \cite{posani_ensemble_2025}.

Ion placement in biostructures is usually performed in a conservative way, only providing the best-resolved sites. Our simulations however depict highly dynamic binding patterns with a significant number of transient sites, and a smaller number of very stable keystone motifs of the overall RNA fold. Those transient sites might play a defining role in stabilizing alternative conformations at the secondary or tertiary structure level. This suggests a continuum between the nonspecific "electrostatic cloud" picture and a more site-specific picture whereby low-populated mono or bidentate sites become important tertiary interaction motifs through induce-fit/conformational-selection processes. That being said, quantitative assessment of transient inner-bound sites is extremely difficult experimentally and we cannot exclude overbinding artifacts due to force-field parameterization and/or lack of polarizability.

Because of the coupling of solvent, ion and RNA dynamics it is challenging to deconvolve which aspects of the modeling process contribute the most among system composition, RNA force field, ion parametrization, sampling, water model, conformational restraints, trajectory alignment, scattering and temperature factors. Nevertheless, our study constitutes a proof of principle that molecular imagery and MD simulation techniques have both matured enough to investigate the modeling parameter space in a systematic way.

Moving forward will bring forth additional challenges for both the experimental and the modeling aspects: artifacts related to sample preparation such as cryogenization or sample damage, biases introduced by processing such as particle image selection and 3D reconstruction algorithms, and post-processing procedures such as sharpening should be critically assessed when using cryo-EM data as a ground truth for investigating dynamics. The upcoming CASP17 challenge is set to establish a new milestone, offering two similar solvent-shell prediction targets at an unprecedented nominal map resolution.

  \section{Materials and Methods}

  \subsection{Simulation methods and force fields}

  All simulations were carried out using Gromacs 2023.5 patched with PLUMED 2.10-beta \cite{abraham_gromacs_2015,tribello_plumed_2014}.
  Isothermal-isobaric simulations were carried out with leap frog integrator at a 2-fs timestep, using the V-rescale thermostat (\qty{300}{\K}, $\tau_T = \qty{1}{\ps}$) and C-rescale barostat (\qty{1}{\bar}, $\tau_P = \qty{1}{\ps}$) \cite{bussi_canonical_2007, bernetti_pressure_2020}. Water geometry was fixed with SETTLE \cite{miyamoto_settle_1992} and bonds involving hydrogen atoms with LINCS \cite{hess_lincs_1997}. Van der Waals interactions were computed with a \qty{1}{\nm} cutoff and analytical long-range corrections to energy and pressure. Long-range electrostatics were treated with Particle Mesh Ewald \cite{essmann_smooth_1995} and cutoff $\qty{\geq 1}{\nm}$.

  Topologies were prepared with Gromacs using the parmBSC0\textchi OL3 force field for RNA \cite{perez_refinement_2007, zgarbova_refinement_2011} with TIP3P \cite{jorgensen_comparison_1983} water as included in the \textit{amber14sb\.ff} from the Gromacs User contributions webpage, modified to include the \textit{microMg} or \textit{nanoMg} non-bonded parameters from Grotz \cite{grotz_optimized_2021} and \ce{Na+}/\ce{Cl-} parameters from Mamatkulov \cite{mamatkulov_force_2018}.

  \subsection{Well-tempered Metadynamics}

  Diuridine (uridylyl-($3'$$\rightarrow$$5'$)-uridine) was solvated with one \ce{Mg^{2+}} in a rhombic dodecahedron box (\qtyproduct{41 x 41 x 29}{\angstrom}). Two coordination CVs $n_{\text{O}_\text{W}}$ and $n_{\text{O}_\text{P}}$ were defined between \ce{Mg^{2+}} and water oxygen atoms (OW) and non-bridging phosphate oxygen atoms (OP1, OP2) respectively, using a cosine switching function $f$

  \begin{equation*}
    f(r) =
    \begin{cases}
      1 \text{ if } r \le r_\text{min} \\
      0 \text{ if } r \ge r_\text{max} \\
      \frac{1}{2}\left[\cos\left(\frac{r - r_\text{min}}{r_\text{max} - r_\text{min}}\pi\right) + 1\right] \text{ else.}
    \end{cases}
  \end{equation*}

  \noindent between $r_\text{min} = \SI{1.8}{\angstrom}$ and $r_\text{max} = \SI{4.2}{\angstrom}$. The switching function was scaled by a factor $1/f(r_\text{Mg--O}) = 1.039661$ to obtain a value close to unity when Mg--O is at equilibrium distance $r_\text{Mg--O} = \SI{2.1}{\angstrom}$.
  Importantly, the employed PLUMED version included a backport of a bug fix correcting the behavior of the cosine switching function. The issue was identified during our preliminary simulations, and the corresponding fix was subsequently incorporated into the official PLUMED releases 2.9.5 and 2.10.1.
  Well-tempered Metadynamics \cite{barducci_well-tempered_2008} were performed for \qty{200}{\ns} on the 2-dimensional space formed by $n_{\text{O}_\text{P}}$ and $n_{\text{O}_\text{W}}$. Gaussian hills ($\sigma = \num{0.05}$) of unit height were deposited every \qty{1}{\ps} and a bias factor of \num{15} was used. The free-energy surface $G(n_{\text{O}_\text{P}}, n_{\text{O}_\text{W}})$ was estimated by weighted histogram ($0.05$ bin size), reweighting the WT-MetaD trajectory with the final accumulated bias \cite{branduardi_metadynamics_2012, schafer_data_2020}.

  \subsection{Barrier-flattening bias design}

  Previous approaches used the distance between the divalent ion and a phosphate oxygen as a collective variable, alone or in conjunction with the coordination number of the ion with water oxygen \cite{cunha_unraveling_2017, falkner_kinetic_2021, neumann_artificial_2022, kretsch_blind_2026}. We here preferred to use another coordination number because it naturally generalizes to the case of an ion bound to multiple phosphates. We design the barrier-flattening bias as follows. We define the following collective variables and the bias functional form as

  \begin{align*}
    &s = \frac{1}{2}(n_{\text{O}_\text{P}} - n_{\text{O}_\text{W}} + 6);
    &z = \frac{1}{2}(n_{\text{O}_\text{P}} + n_{\text{O}_\text{W}} - 6) \\
    &s' = s + \sigma_s (1 - \cos(2\pi s));
    &z' = z + \sigma_z (1 - \cos(2\pi s)) \\
  \end{align*}

  \noindent (note that $z'$ correctly depends on both $z$ and $s$)

  \begin{equation*}
    U_\text{flat}(s', z') =
    \begin{cases}
      -\epsilon (1 - \cos(2\pi s'))(1 - \cos(8\pi z')) \\
      \hspace{4em} \text{ if } 0 < s' \text{ and } 0 < z' < 1 \\
      0 \hspace{1em} \text{ else.} \\
    \end{cases} \\
  \end{equation*}

  The parameters $\theta = (\sigma_s, \sigma_z, \epsilon)$ are fitted by minimizing the loss

  \begin{equation}
    \mathcal{L}[\theta] = \max_s G^\mathcal{D}_{s}[\theta] - \min_s G^\mathcal{D}_s[\theta]
  \end{equation}

  where $G^\mathcal{D}_s[\theta]$ is the free-energy profile along the $s$ coordinate, integrated on the domain $\mathcal{D} = \{(s, z) \mid s \in [0, 1], z \in [0, 1]\}$. In other words we minimize the height of the free energy barrier projected on $s$.

  \subsection{Bias implementation}

  The bias is implemented efficiently for all \ce{Mg^{2+}} ion in a custom PLUMED collective variable named LIGANDEXCHANGE written in C++. It includes neighbor lists for cost-effective computation of $n_{\text{O}_\text{P}}$ and $n_{\text{O}_\text{W}}$. We benchmarked the neighbor list parameters to minimize violations (Table~\ref{stable:nlist}), then set the cutoff to \qty{1}{\nano\meter} and the stride to 100 timesteps (for the small UpU system) or 500 timesteps (for the group I intron HREx simulations).

  \subsection{Barrier-flattening bias validation}

  We performed a set of MD simulations of the UpU--Mg system biased with $\lambda U_\text{flat}$ for increasing scalings $\lambda \in \{0.6, 0.7, 0.8, 0.9, 1.0, 1.1, 1.2\}$. The distance between \ce{Mg^{2+}} and the center of mass of the phosphate oxygen atoms is restrained below \qty{0.8}{\nano\meter} with a semi-harmonic wall ($\kappa = \qty{50}{\kilo\joule\per\mol\per\nano\meter\squared}$). For comparison, similar simulations were done for unbiased μMg and nMg. Kinetic rates are estimated by detecting first passage times (FPTs) $t$ from $s \le 0$ to $s \ge 0.8$ and vice-versa. We computed the unbiased maximum likelihood estimate $\hat{k}$, accounting for right-censored data, and its exact 95\% confidence interval as:
  \begin{equation*}
    \hat{k} = \frac{r - 1}{T}, \quad \text{with } 95\% \text{ CI: } \left( \frac{\chi^2_{\alpha/2, 2r}}{2T}, \frac{\chi^2_{1-\alpha/2, 2r}}{2T} \right)
  \end{equation*}
  where $T$ is the total accumulated survival time, $r$ is the number of observed transitions, and $\chi^2_{p, 2r}$ is the $p$-th quantile of the Chi-Square distribution with $2r$ degrees of freedom ($\alpha = 0.05$). Rates for unbiased μMg could not be obtained directly due to the scarcity of observed events.
  To estimate unbiased rates $v^0_\text{on}$ and $v^0_\text{off}$ we first estimate the $\lambda$-dependent free-energy difference $\Delta G_\text{bind}(\lambda) = -RT\log [P_\text{bound}(\lambda)/P_\text{free}(\lambda)]$ between the free and bound states (excluding the barrier itself), $P_\text{free} = P(s \le 0)$ and $P_\text{bound} = P(s \ge 0.8)$, combining all biased simulations $\lambda \in \{0.6, 0.7, 0.8, 0.9, 1.0, 1.1, 1.2\}$ with weights computed with MBAR \cite{shirts_statistically_2008}. We then use the FPTs from the low-scaling/high-barrier regime ($\lambda = \{0.6, 0.7, 0.8\}$) to fit a simple kinetic model compatible with the observed $\Delta G_\text{bind}(\lambda)$:
  \begin{align*}
    v_\text{on}(\lambda) = v^0\exp(\beta\lambda\Delta E^\ddagger)\exp\left[-\frac{\beta}{2}\Delta G_\text{bind}(\lambda)\right], \\
    v_\text{off}(\lambda) = v^0\exp(\beta\lambda\Delta E^\ddagger)\exp\left[\frac{\beta}{2}\Delta G_\text{bind}(\lambda)\right], \\
  \end{align*}
  with $\beta$ the reciprocal thermodynamic temperature, and where $v^0$ stands for a rate and $\Delta E^\ddagger$ an activation free energy, using maximum likelihood (see SI). Error is estimated using bootstrap on blocks, where observed series of FPTs from each trajectory are divided in 4 blocks, yielding $4 \times 3 = 12$ sets of on and off FPTs, which are resampled with replacement for each bootstrap sample.

  \subsection{Affinity-corrected bias}

  Since the bias $U_\text{flat}$ is purely attractive, it results in an undesirable contribution to the binding free-energy. For the Hamiltonian Replica Exchange simulations, we account for it by adding a correction

  \begin{equation}
    U_\text{corr}(n_{\text{O}_\text{P}}, \lambda) = \alpha(\lambda)n_{\text{O}_\text{P}}
  \end{equation}

  where the slope $\alpha(\lambda)$ depends non-linearly on the bias scaling $\lambda$. The final scaled bias reads
  \begin{equation}
    \lambda U_\text{bias}(s', z',n_{\text{O}_\text{P}}, \lambda) = \lambda[U_\text{flat}(s', z') + U_\text{corr}(n_{\text{O}_\text{P}}, \lambda)]
  \end{equation}
  We optimize $\alpha(\lambda)$ on a grid $\lambda_i \in [0,1]$ with the loss
  \begin{equation}
    \mathcal{L}(\{\alpha_i\}) = \sum_i \left( -RT\ln\left\langle e^{-\beta U_\text{bias}(\lambda_i, \alpha_i)}\right\rangle_{0,\text{bound}} \right)^2
  \end{equation}
  where the ensemble average is taken with respect to the unbiased ensemble ($\lambda = 0$) restricted to the bound state ($n_{\text{O}_\text{P}} > 0$).

  In practice we reuse the set of UpU trajectories biased with $\lambda U_\text{flat}$ and reweight them using MBAR

  \subsection{\textit{Tetrahymena} ribozyme system preparation}

  Native structure of the \textit{Tetrahymena thermophila} group I self-
  splicing intron (Apo L-21 ScaI) was downloaded from the PDB (ID: \textsc{7ez0}) \cite{su_cryo-em_2021} and processed with PDBFixer \cite{eastman_openmm_2017}. The 27 \ce{Mg^{2+}} PDB ions were either kept or removed (\textit{novo}). The system’s principal axes were aligned to the xyz coordinates and it was solvated in a rectangular box with a \qty{2}{\nano\meter} margin, resulting in a \qtyproduct{17 x 12 x 13}{\nano\meter} cell. The total RNA charge (\num{-386}) was neutralized with a 3:1 ratio of \ce{Mg^{2+}} and \ce{Na+}, and extra \ce{Mg^{2+}}/\ce{Na+}/\ce{Cl-} ions were added to approximate a concentration of \qty{10}{mM} \ce{MgCl2} and \qty{50}{mM} \ce{NaCl} in a volume of \qty{86}{\percent} of the total water added, which was determined empirically from preliminary tests. The no-Mg system only used \ce{Na+} for neutralization and no added \ce{MgCl2}. Except from PDB \ce{Mg^{2+}}, all ions were added using \texttt{gmx genion}, using default minimum radius to non-solvent (\qty{0.6}{\nano\meter}). Different ion-placement seeds were used for each of the 16 replicas prepared for μMg, μMg-novo, nMg and nMg-novo, from which $2 \times 8$ were used for $2$ replicates of the replica exchange setup and $8$ for plain MD setups.

  Steepest descent energy minimization followed by \qty{1}{\nano\second} NPT equilibration were carried with default Gromacs position restraints on RNA heavy atoms and on the 27 PDB-informed ions, when included. As mentioned in the main text, some ions bound to the RNA during this process despite lying initially in the exclusion zone or outer-sphere.

  \subsection{\textit{Tetrahymena} ribozyme RNA restraints}

  During both MD and REx production runs, harmonic restraints were applied to the RMSD (center-of-mass aligned, without rotational fit) of a subset of RNA atoms (P, C1$'$, C2) to their native (\textsc{7ez0}) coordinates. This prevented unwanted rotational diffusion in the rectangular box in addition to limiting the solute’s internal degrees of freedom. A harmonic constant $\kappa = \qty{e5}{\kilo\joule\per\mol\per\nano\meter\squared}$ was used, except for the MD-soft setup ($\kappa = \qty{5e2}{\kilo\joule\per\mol\per\nano\meter\squared}$).

  \subsection{\textit{Tetrahymena} ribozyme plain MD simulations}

  NPT production runs were carried in 8 replicates for each setup using random ion-placement and velocity seeds, for \qty{600}{\nano\second} (μMg, nMg, μMg-novo, nMg-novo) or \qty{400}{\nano\second} (no-Mg).

  \subsection{Hamiltonian Replica Exchange simulations}

  Hamiltonian Replica Exchange was implemented as a custom PLUMED bias named REPLEX\_BIASVALUE. A set of 8 replicas with increasing $\lambda$ values between were used. We originally devised a $\lambda$-ladder of 12 replicas from $\lambda = 0$ to $\lambda = 1$, spaced to ensure homogeneous acceptance, as predicted from the UpU data, under the assumption that the Intron system behaves like $N$ independent sites. We only kept the 8 lower values since higher bias scaling offered diminishing returns in terms of kinetics versus phase-space overlap.

  Series of exchanges between neighboring replicas were made every 500 steps (\qty{1}{\pico\second}). To increase mixing in $\lambda$-space, a series consisted in 100 exchange attempts alternating between odd ($1 \leftrightarrow 2, 3 \leftrightarrow 4, ...$) and even ($2 \leftrightarrow 3, ...$) pairs. Average acceptance rate was \qtyrange{10}{30}{\%}.

  \subsection{Bulk concentration and preferential interaction coefficients}

  Bulk concentration was estimated by counting species at a distance $> \qty{4}{\nano\meter}$ from the RNA P atoms, and computing the molar concentration as $c_i^\text{bulk} = 55.5 \times n_i^\text{bulk} / n_\text{water}^\text{bulk}$~\unit{M}. Preferential interaction coefficients (Fig.~\ref{sfig:bulk_concentrations}), defined as the excess or deficit of ionic species due to the presence of the solute, were computed as the difference between the number of ions and the expected number given the bulk concentration, $\Gamma_i = n_i^\text{tot} - c_i^\text{bulk}n_\text{water}^\text{tot}/55.5$.

  \subsection{Motif analysis}

  Binding motifs were identified by listing non-water atoms around each metal cation with a \qty{2.8}{\angstrom} cutoff. Nomenclature for inner-shell binding motifs was inspired from the MgRNA study \cite{zheng_magnesium-binding_2015, naleem_cat_wiz_2026} and denotes the category of RNA ligand ($\text{O}_\text{ph}$: phosphate non-bridging oxygen; $\text{O}_\text{r}$: ribose oxygen or oxygen bridging phosphate and ribose; $\text{O}_\text{b}$: nucleobase oxygen; $\text{N}_\text{r}$: nucleobase nitrogen; $\text{X}$: other) and stereochemistry of their octahedral arrangement (\textit{cis}/\textit{trans} for bidentate, \textit{fac}/\textit{mer} for tridentate, \underline{Cis}/\underline{Trans} for tetradentate--emphasizing the relative position of the two remaining water molecules). Stereochemistry was automatically assigned based on the sum angles formed by pairs of ligands: \textit{cis} if $\alpha \le \qty{135}{\degree}$ in the bidentate case, \textit{fac} if $\alpha_{12} + \alpha_{13} + \alpha_{23} \le \qty{315}{\degree}$ (tridentate), \underline{Cis} if $\sum_{i,j>i} \alpha_{ij} \le \qty{675}{\degree}$ (tetradentate).

  \subsection{Cryo-EM validation}

  Forward modeling of the cryo-EM potential was performed as follows. The density map for a single frame was computed by aligning the coordinates to the reference PDB structure \textsc{9cbu} either globally or locally (flexible alignment), then accumulating 5-Gaussian contributions parameterized from scattering factors for all heavy atom types (C, N, O, Na, Mg, P, Cl) \cite{peng_robust_1996}, as described in \cite{hoff_accurate_2024}. For global alignment the system coordinates were rotated and translated to minimize the RMSD of the RNA heavy atoms. For flexible alignment, local neighborhoods of \qty{10}{\angstrom} around each residue were selected and aligned based on RNA heavy atoms found in that neighborhood; atoms appearing in multiple neighborhoods were assigned a fractional weight for each position corresponding to a different alignment, and the instantaneous map was computed as a weighted sum of all aligned positions. Average maps for each simulation was computed from a subsample of frames excluding the first \qty{25}{\percent} as equilibration, resulting in about 3000 frames per map.

  Pearson correlation with the \qty{2.2}{\angstrom} experimental map (EMD: 42499, unsharpened half maps downloaded from Das Lab’s GitHub page \cite{kretsch_water-cryoem-ribozymemaps_2026}) was computed for voxel selections based on the corresponding PDB structure: RNA with its solvation shell ($d_\text{RNA} < \qty{6}{\angstrom}$); solvation shell around phosphates only, excluding RNA itself ($d_\text{ph} < \qty{6}{\angstrom}$ and $d_\text{RNA} > \qty{1.8}{\angstrom}$). The generated maps were smoothed with uniform Gaussian filtering prior to the linear fit. Smoothing B-Factors were selected for each setup from a scan of values between 0 and \qty{100}{\angstrom^2} to maximize cross correlation on each voxel selection (Fig.~\ref{sfig:tempfactor_scan}). Confidence intervals and one-tailed p-values were quantified using Bayesian bootstrap on blocks of frames coming from continuous trajectories (8 replicates for MD, $2 \times 8$ demultiplexed replicas for REx): for each bootstrap sample, blocks were reweighted with weights sampled from a Dirichlet distribution (concentration parameter $\alpha = \vec{1}$) and the weighted average map computed. Experimental error was included similarly by reweighting the two reference half maps.

  \subsection{Integrative quantification of motifs}

  For each significant motif ($f \ge \qty{5}{\%}$ in at least one setup) a neighborhood was selected around the RNA atoms involved and a pair of maps was computed after local alignment of the neighborhood’s RNA heavy atoms by dividing frames where the motif was identified from those where it was not fully formed. Cross correlation was computed for mixtures of maps and its maximum yielded the optimal motif frequency. Error estimation used Bayesian bootstrap on blocks as described above.

  \section{Data availability}
  Code, input files and analysis scripts can be found at \url{https://github.com/ollyfutur/mg_sampling}.

  \section{Acknowledgements}
  This work has been funded by the Next Generation EU project PRIN 2022,
  Grant No.~2022Z4FZE9. O.L.-C.\ acknowledges the European Union's Horizon 2023 research and innovation program under Marie Skłodowska-Curie Grant Agreement No.~101152924.
  G.B.\ acknowledges support from the Italian Ministry of University and Research (MUR) through the Fondo Italiano per la Scienza (FIS 3), Grant No.~FIS-2024-01745, RNAScale.
  The authors acknowledge the CINECA award under the ISCRA initiative, Grant No.~HP10BE9GOQ, for the availability of high-performance computing resources and support.

\bibliographystyle{unsrt}
\bibliography{references}

\clearpage
\appendix

\setcounter{page}{1}
\setcounter{figure}{0}
\setcounter{table}{0}
\setcounter{equation}{0}

\renewcommand{\thefigure}{S\arabic{figure}}
\renewcommand{\thetable}{S\arabic{table}}
\renewcommand{\theequation}{S\arabic{equation}}

\begin{center}
   \Large \textbf{Supplementary Information:}

   \large \mytitle
   \vskip 0.5em

   \normalsize Olivier Languin-Cattoën, Elisa Posani, and Giovanni Bussi

\end{center}

\input{si.tex}

\end{document}

%% file: si.tex
\section*{Supplementary Figures}

\begin{figure}[htbp!]
  \centering
  \includegraphics[width=\linewidth]{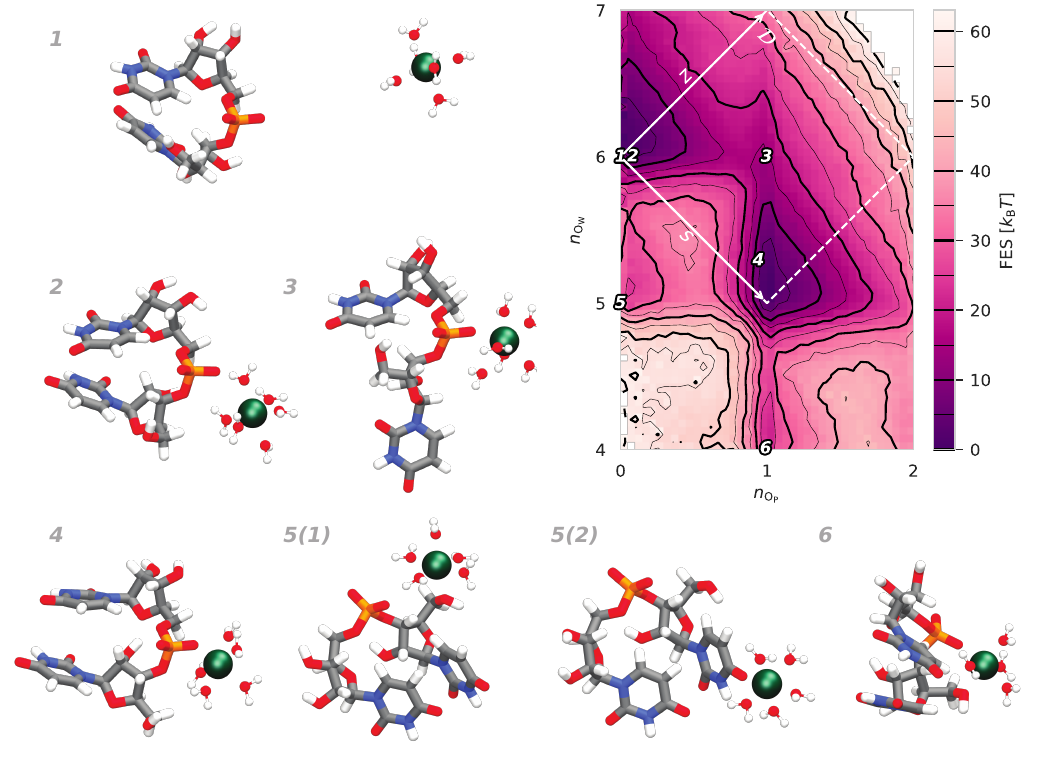}
  \caption[Free energy landscape of \ce{Mg^{2+}} binding to UpU phosphate oxygen atoms]{Free energy landscape of \ce{Mg^{2+}} binding to UpU phosphate oxygen atoms, obtained by performing well-tempered metadynamics on the two-dimensional space formed by the coordination number of \ce{Mg^{2+}} with water oxygen atoms ($n_{\text{O}_{\text{W}}}$) and phosphate OP1/OP2 ($n_{\text{O}_{\text{P}}}$). Snapshots of relevant conformational states: \textit{(\textbf{1})} Unbound. \textit{(\textbf{2})} Outer-bound. \textit{(\textbf{3})} Heptacoordinated binding intermediate. \textit{(\textbf{4})} Inner-bound. \textit{(\textbf{5-1})} Inner-bound to ribose oxygen atom O5$^\prime$. \textit{(\textbf{5-2})} Inner-bound to base oxygen atom O4. \textit{(\textbf{6})}. Clamp between phosphate and O5$^\prime$}
  \label{sfig:uu_snapshots}
\end{figure}

\begin{figure}[htbp!]
  \centering
  \includegraphics[width=\linewidth]{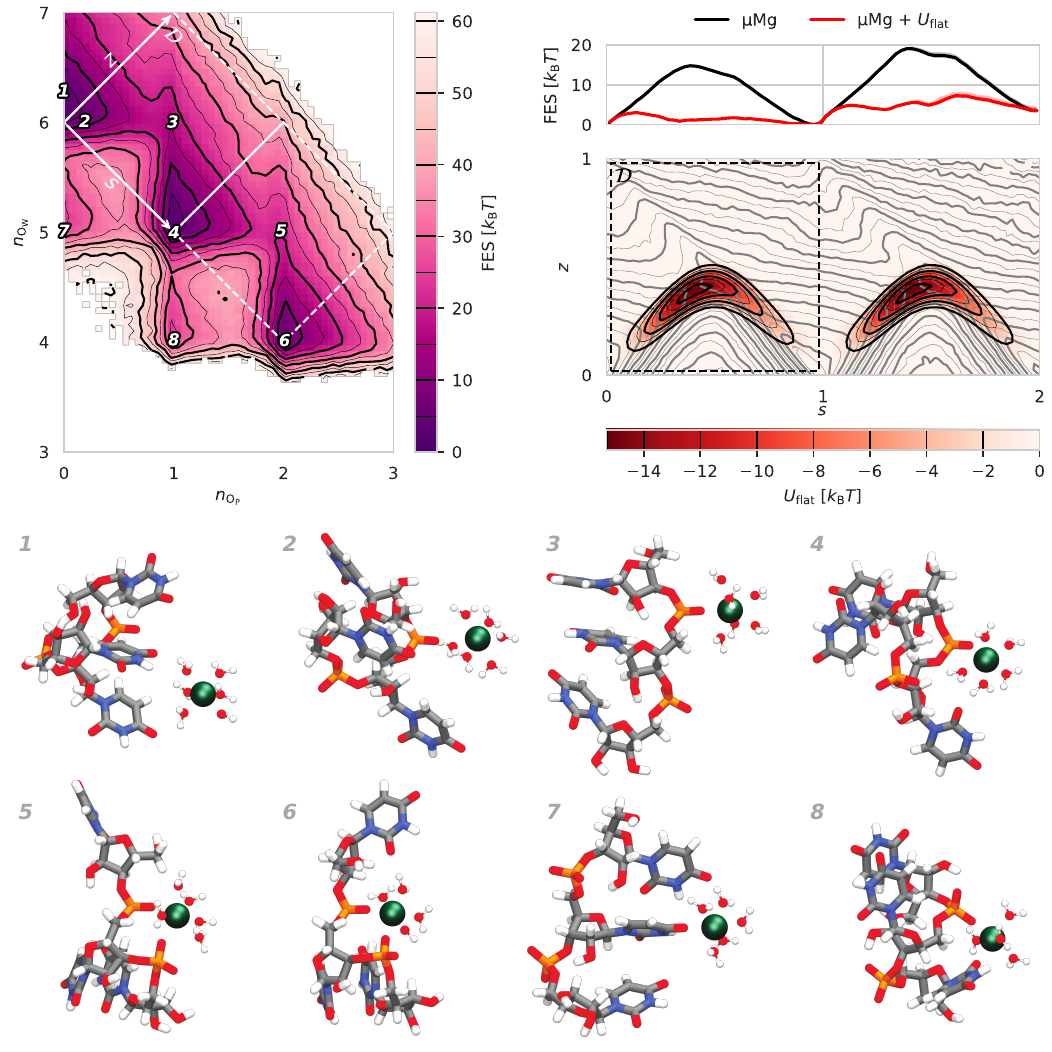}
  \caption[Well-tempered metadynamics free energy landscape of \ce{Mg^{2+}} binding to triuridine]{Well-tempered metadynamics free energy landscape of \ce{Mg^{2+}} binding to triuridine. Snapshots of relevant conformational states: \textit{(\textbf{1})} Unbound. \textit{(\textbf{2})} Outer-bound. \textit{(\textbf{3})} Heptacoordinated binding intermediate. \textit{(\textbf{4})} Inner-bound. \textit{(\textbf{5})} Heptacoordinated intermediate between mono- and bidentate states. \textit{(\textbf{6})} Bidentate phosphate clamp (10-membered ring). \textit{(\textbf{7})} Inner-bound to base oxygen atom O4. \textit{(\textbf{8})}. Bidentate clamp between phosphate and O2.}
  \label{sfig:uuu_snapshots}
\end{figure}

\begin{figure}[htbp!]
  \centering
  \begin{minipage}[b]{\linewidth}
    \includegraphics[width=0.49\linewidth]{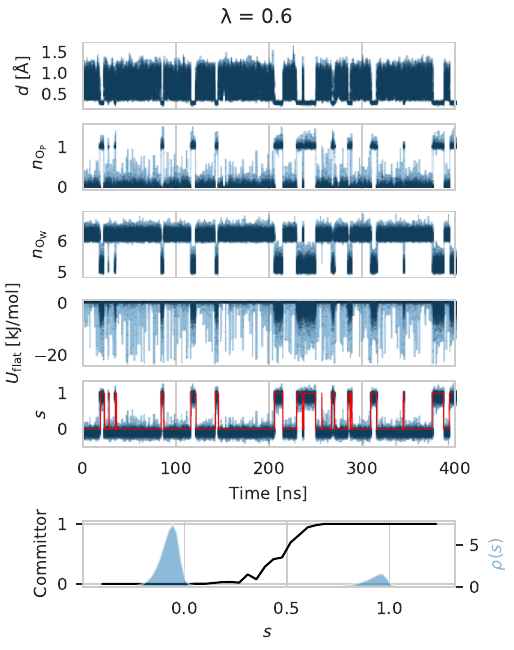}
    \includegraphics[width=0.49\linewidth]{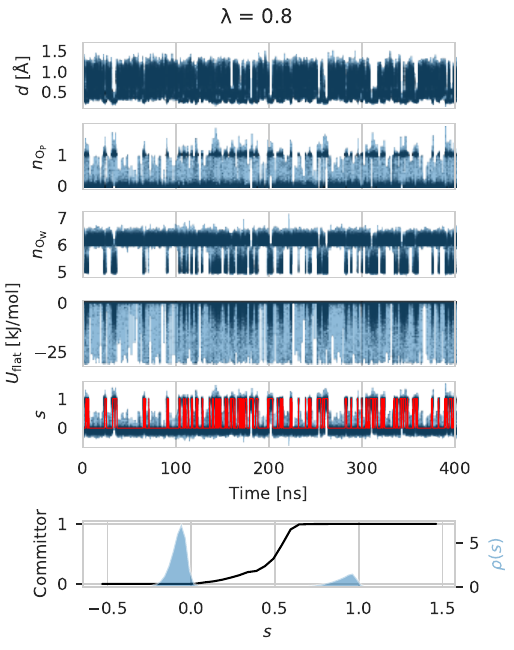}
  \end{minipage}
  \caption[MD simulations of UpU-\ce{Mg^{2+}} with varying barrier-flattening bias strength]{MD simulations of UpU-\ce{Mg^{2+}} with varying barrier-flattening bias strength. For each bias scaling factor $\lambda$ we show timeseries of the distance to the OP1/OP2 center-of-mass $d$, coordination number with phosphate oxygen ($n_{\text{O}_\text{P}}$) and water oxygen ($n_{\text{O}_\text{W}}$), bias energy $U_\text{flat}$ and binding coordinate $s$. The red curve corresponds to the state discretization used for rate estimations. We also show the committor along $s$, that is the probability for a trajectory starting at $s$ to end first in the bound state ($s > 0.8$) rather than the unbound state ($s < 0$), overlayed with the probability density along $s$. $\lambda = 0.6$ and $\lambda = 0.8$.}
  \label{sfig:bias_timeseries}
\end{figure}

\begin{figure}[htbp!]
  \ContinuedFloat
  \centering
  \begin{minipage}[b]{\linewidth}
    \includegraphics[width=0.49\linewidth]{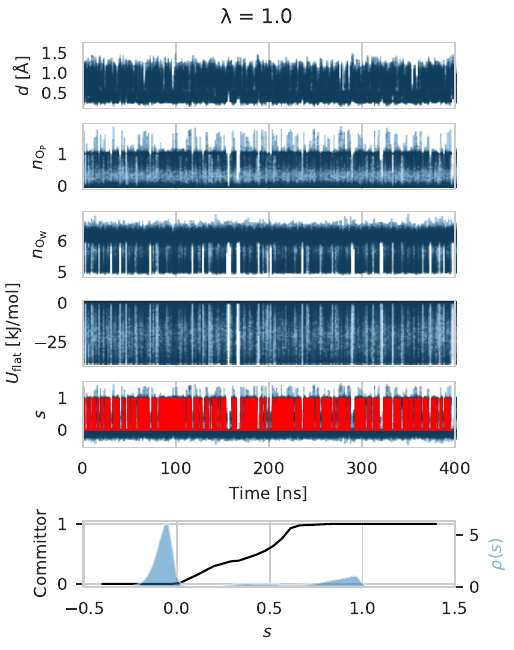}
    \includegraphics[width=0.49\linewidth]{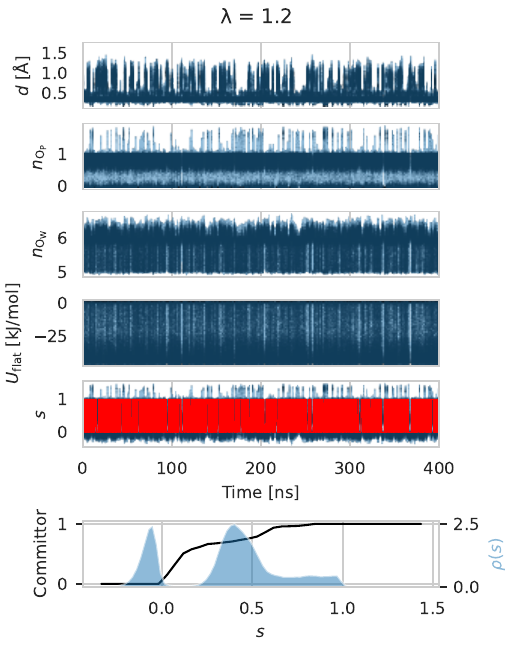}
  \end{minipage}
  \caption[]{(Continued) MD simulations of UpU-\ce{Mg^{2+}} with varying barrier-flattening bias strength. $\lambda = 1$ and $\lambda = 1.2$.}
\end{figure}

\begin{figure}[htbp!]
  \centering
  \begin{minipage}[b]{\linewidth}
    \stackinset{l}{1em}{t}{0.5em}{\textsf{\textbf{A}}}{
      \includegraphics[width=0.49\linewidth]{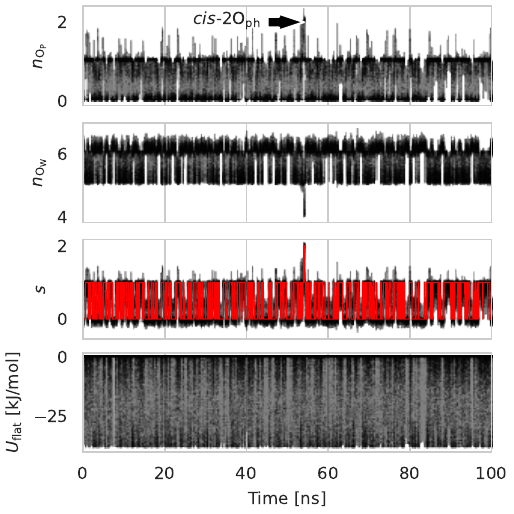}
    }
    \stackinset{l}{1em}{t}{0.5em}{\textsf{\textbf{B}}}{
      \includegraphics[width=0.49\linewidth]{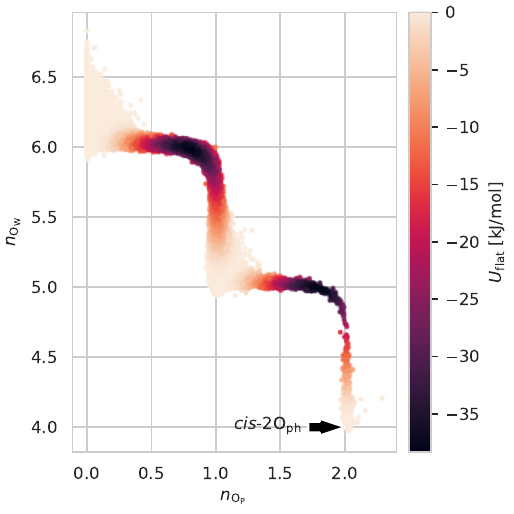}
    }
  \end{minipage}
  \caption[MD simulation of UpUpU--\ce{Mg^{2+}} biased with $U_\text{flat}$ ($\lambda = 1$)]{MD simulation of UpUpU--\ce{Mg^{2+}} biased with $U_\text{flat}$ ($\lambda = 1$). \textit{(A)} Timeseries of coordination number with phosphate oxygen ($n_{\text{O}_\text{P}}$) and water oxygen ($n_{\text{O}_\text{W}}$), binding coordinate $s$ and bias energy $U_\text{flat}$. The red curve corresponds to binding state discretization. \textit{(B)} Scatter plot of the trajectory projected onto the $n_{\text{O}_\text{P}}$/$n_{\text{O}_\text{W}}$ space, colored according to bias intensity. Annotations highlight the hard-to-sample phosphate clamp state (state \textbf{\textit{6}} on Fig.~\ref{sfig:uuu_snapshots}).}
  \label{sfig:uuu_bias_timeseries}
\end{figure}

\begin{figure}[htbp!]
  \centering
  \begin{minipage}[b]{\linewidth}
    \stackinset{l}{1em}{t}{0.5em}{\textsf{\textbf{A}}}{
      \includegraphics[width=0.49\linewidth]{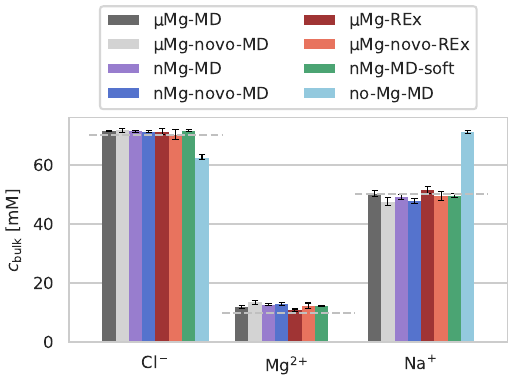}
    }
    \stackinset{l}{1em}{t}{0.5em}{\textsf{\textbf{B}}}{
      \includegraphics[width=0.49\linewidth]{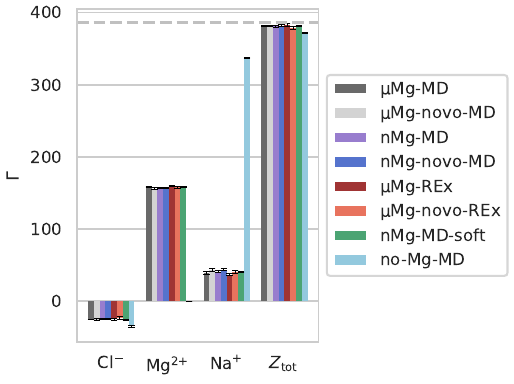}
    }
  \end{minipage}
  \caption[Ionic concentration in bulk and preferential interaction coefficients]{\textit{(A)} Ionic concentration in bulk, defined as the volume of solvent at radius $> \qty{4}{\nano\meter}$ from a representative subset of RNA atoms (P atoms). \textit{(B)} Preferential interaction coefficients $\Gamma$ for ionic species, defined as the excess number of ions recruited in a volume around the RNA with respect to what is expected from bulk concentration. Error bars indicate 95\% confidence intervals from Bayesian bootstrap on blocks.}
  \label{sfig:bulk_concentrations}
\end{figure}

\begin{figure}[htbp!]
  \centering
  \begin{minipage}[b]{\linewidth}
    \stackinset{l}{1em}{t}{0.5em}{\textsf{\textbf{A}}}{
      \includegraphics[width=0.49\linewidth]{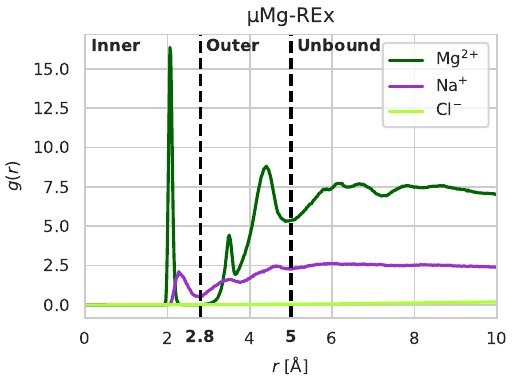}
    }
    \stackinset{l}{1em}{t}{0.5em}{\textsf{\textbf{B}}}{
      \includegraphics[width=0.49\linewidth]{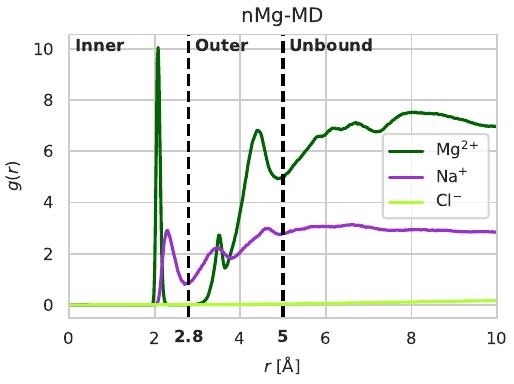}
    }
  \end{minipage}
  \caption[Radial distribution function $g(r)$ of RNA atoms around ions found in \textit{(A)} μMg-REx and \textit{(B)} nMg-MD simulations]{Radial distribution function $g(r)$ of RNA atoms around ions found in \textit{(A)} μMg-REx and \textit{(B)} nMg-MD simulations. The cutoffs used to categorize inner and outer binding are shown.}
  \label{sfig:rdfs}
\end{figure}

\begin{figure}[htbp!]
  \centering
  \begin{minipage}[b]{\linewidth}
    \centering
    \stackinset{l}{1em}{t}{0.5em}{\textsf{\textbf{A}}}{
      \includegraphics[width=0.49\linewidth]{figures/inner_bound_mg_timeseries}
    }
    \stackinset{l}{1em}{t}{0.5em}{\textsf{\textbf{B}}}{
      \includegraphics[width=0.49\linewidth]{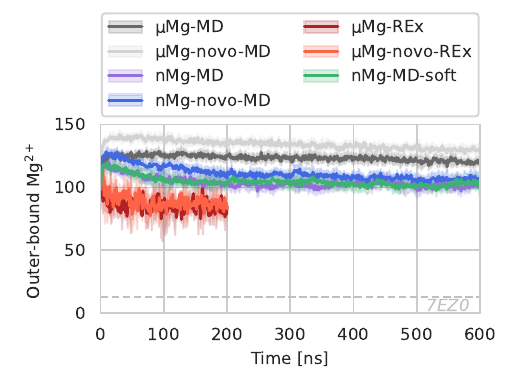}
    }
    \stackinset{l}{1em}{t}{0.5em}{\textsf{\textbf{C}}}{
      \includegraphics[width=0.49\linewidth]{figures/bound_mg_barplot}
    }
  \end{minipage}
  \caption[Magnesium binding to Apo L-21 ScaI]{Magnesium binding to Apo L-21 ScaI. \textit{(A)} Time evolution of the number of inner-bound ($d \le \qty{2.8}{\angstrom}$) \ce{Mg^{2+}} for the investigated setups. \textit{(B)} Time evolution of the number of outer-bound ($\qty{2.8}{\angstrom} < d \le \qty{5}{\angstrom}$) \ce{Mg^{2+}}. \textit{(C)} Number of \ce{Mg^{2+}} ions in the inner-sphere, outer-sphere and unbound, discarding the first \qty{25}{\%} of each trajectory. Error bars indicate 95\% CI estimated with Bayesian bootstrap on blocks.}
  \label{sfig:bound_mg}
\end{figure}

\begin{figure}[htbp!]
  \centering
  \begin{minipage}[b]{\linewidth}
    \centering
    \stackinset{l}{1em}{t}{0.5em}{\textsf{\textbf{A}}}{
      \includegraphics[width=0.49\linewidth]{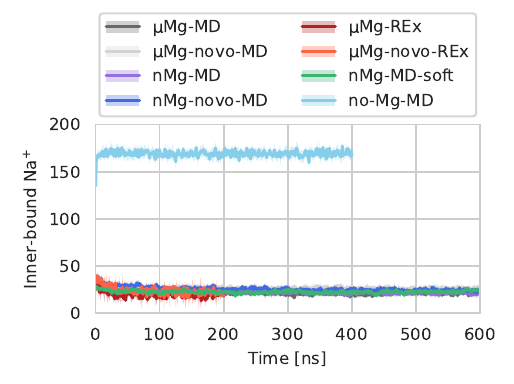}
    }
    \stackinset{l}{1em}{t}{0.5em}{\textsf{\textbf{B}}}{
      \includegraphics[width=0.49\linewidth]{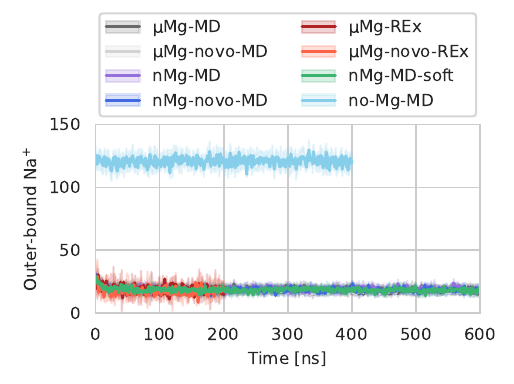}
    }
    \stackinset{l}{1em}{t}{0.5em}{\textsf{\textbf{C}}}{
      \includegraphics[width=0.49\linewidth]{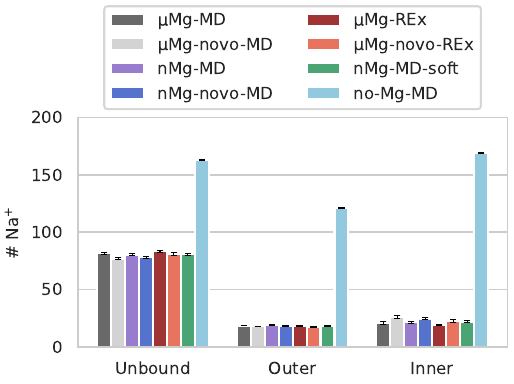}
    }
  \end{minipage}
  \caption[Sodium binding to Apo L-21 ScaI]{Sodium binding to Apo L-21 ScaI. \textit{(A)} Time evolution of the number of inner-bound ($d \le \qty{2.8}{\angstrom}$) \ce{Na+} for the investigated setups. \textit{(B)} Time evolution of the number of outer-bound ($\qty{2.8}{\angstrom} < d \le \qty{5}{\angstrom}$) \ce{Na+}. \textit{(C)} Number of \ce{Na+} ions in the inner-sphere, outer-sphere and unbound, discarding the first \qty{25}{\%} of each trajectory. Error bars indicate 95\% CI estimated with Bayesian bootstrap on blocks.}
  \label{sfig:bound_na}
\end{figure}

\begin{figure}[htbp!]
  \centering
  \stackinset{l}{1em}{t}{0.5em}{\textsf{\textbf{A}}}{
    \includegraphics[width=\linewidth]{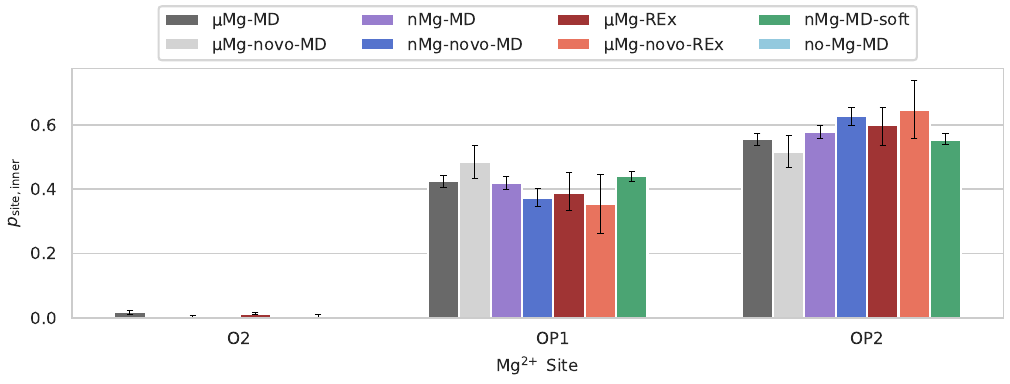}
  }
  \stackinset{l}{1em}{t}{0.5em}{\textsf{\textbf{B}}}{
    \includegraphics[width=\linewidth]{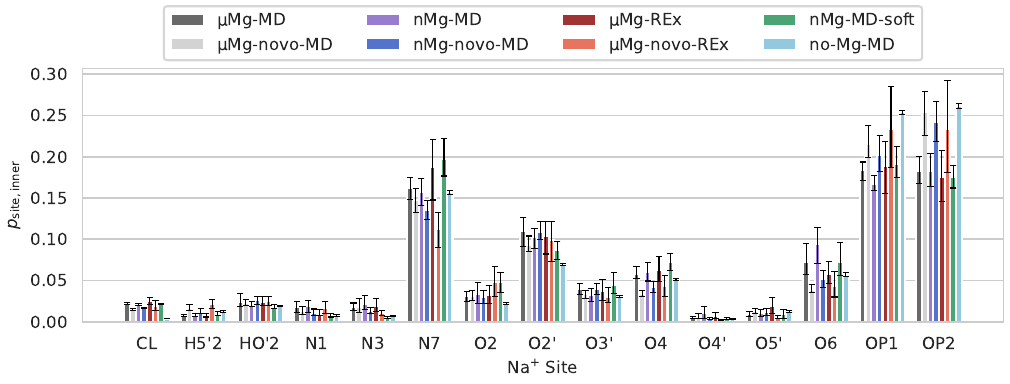}
  }
  \caption[Distribution of non-solvent atom types inner-bound to cations \textit{(A)} \ce{Mg^{2+}} and \textit{(B)} \ce{Na+}]{Distribution of non-solvent atom types inner-bound to cations \textit{(A)} \ce{Mg^{2+}} and \textit{(B)} \ce{Na+}. The frequency of each atom type is given relative to all observed inner-sphere interactions, and 95\% confidence intervals are obtained with Bayesian bootstrap on blocks. Atom types with frequency $< \qty{1}{\percent}$ in all setups are not shown.}
  \label{sfig:site_types}
\end{figure}

\begin{figure}[htbp!]
  \centering
  \includegraphics[width=\linewidth]{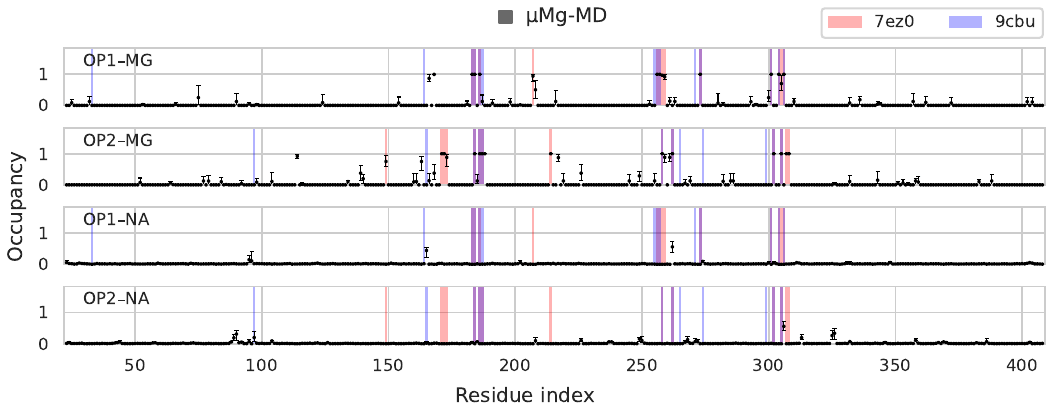}
  \includegraphics[width=\linewidth]{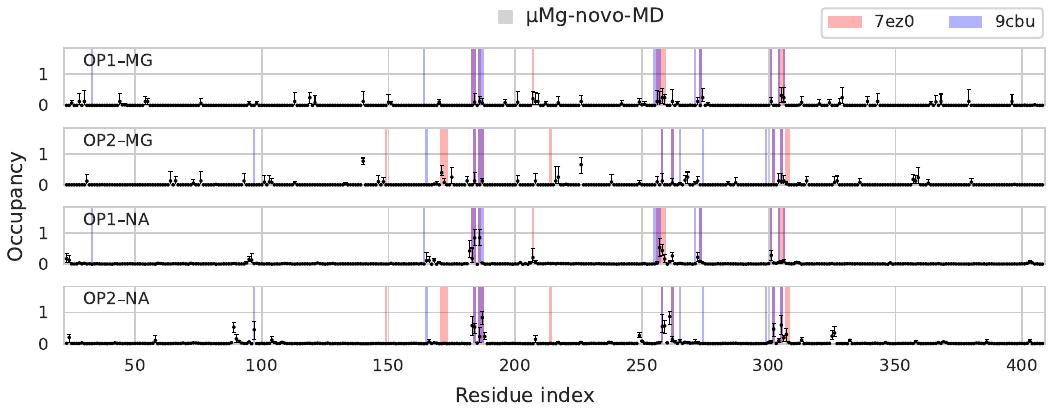}
  \caption[Occupancy of phosphate oxygen sites in simulations]{Occupancy of phosphate oxygen sites in simulations. The average number of \protect\ce{Mg^{2+}} and \protect\ce{Na+} inner-bound to each OP1 or OP2 site along the sequence is shown with error bars indicating 95\% CI estimated with Bayesian bootstrap on blocks. μMg-MD and μMg-novo-MD.}
  \label{sfig:residue_sites}
\end{figure}

\begin{figure}[htbp!]
  \ContinuedFloat
  \centering
  \includegraphics[width=\linewidth]{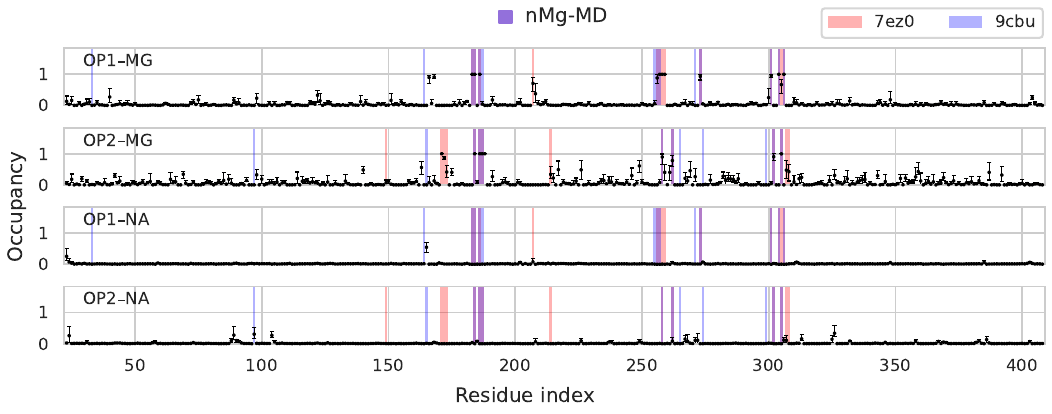}
  \includegraphics[width=\linewidth]{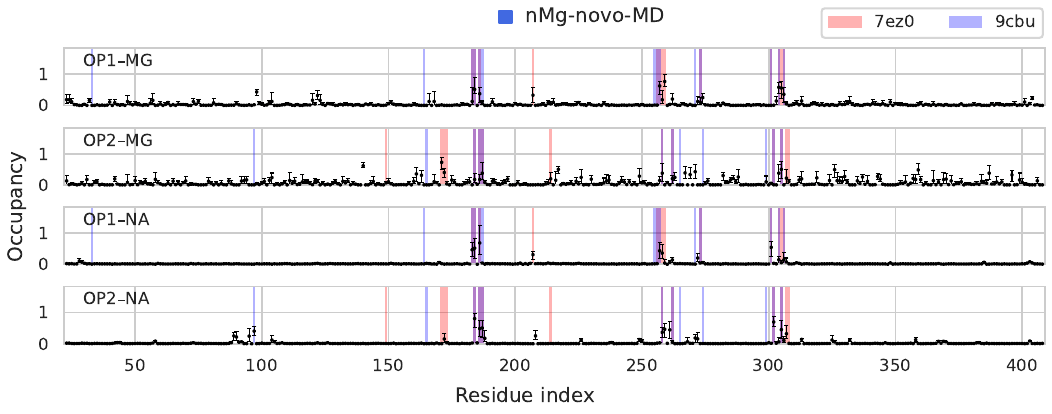}
  \caption[]{(Continued) Occupancy of phosphate oxygen sites in simulations. nMg-MD and nMg-novo-MD.}
\end{figure}

\begin{figure}[htbp!]
  \ContinuedFloat
  \centering
  \includegraphics[width=\linewidth]{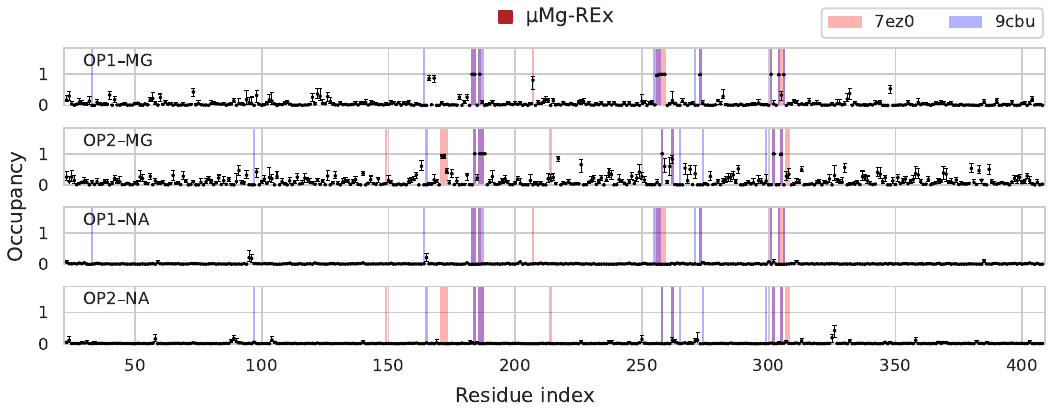}
  \includegraphics[width=\linewidth]{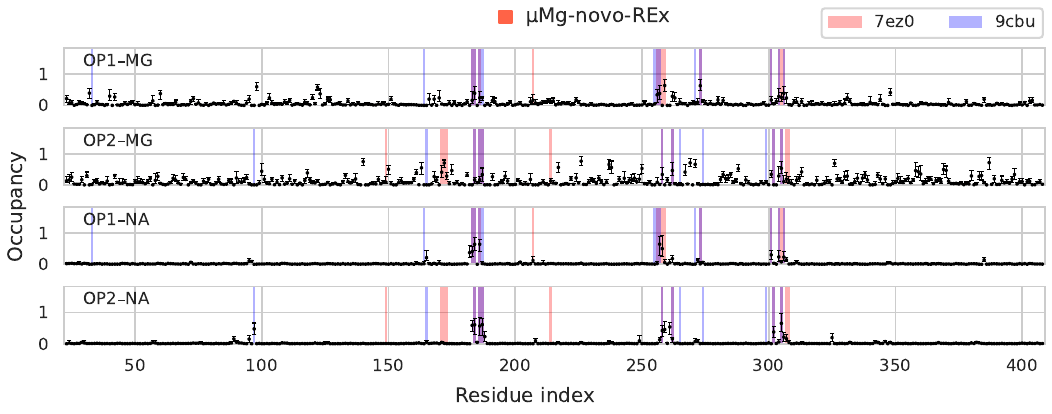}
  \caption[]{(Continued) Occupancy of phosphate oxygen sites in simulations. μMg-REx and μMg-novo-REx.}
\end{figure}

\begin{figure}[htbp!]
  \ContinuedFloat
  \centering
  \includegraphics[width=\linewidth]{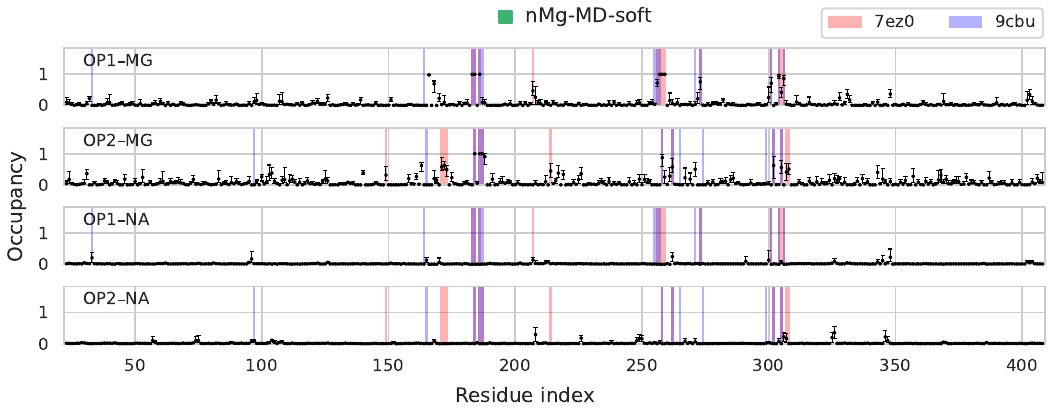}
  \includegraphics[width=\linewidth]{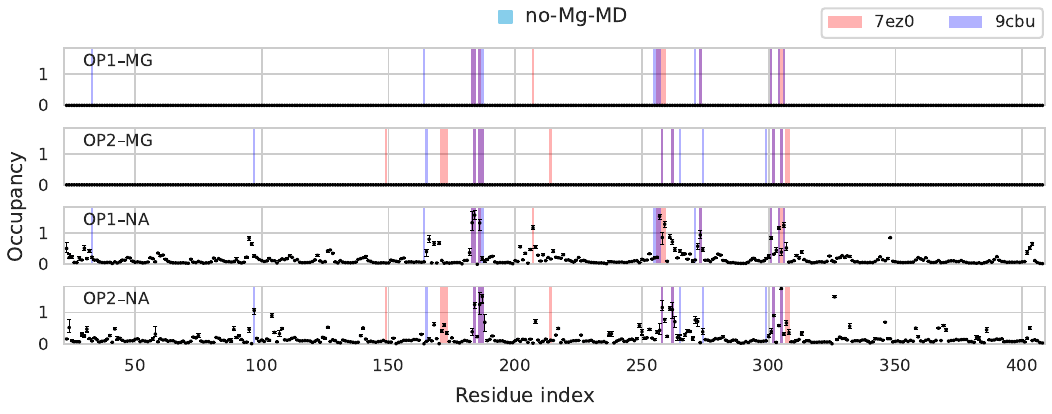}
  \caption[]{(Continued) Occupancy of phosphate oxygen sites in simulations. nMg-MD-soft and no-Mg-MD.}
\end{figure}

\begin{figure}[htbp!]
  \centering
  \begin{minipage}[b]{\linewidth}
    \stackinset{l}{1em}{t}{0.5em}{\textsf{\textbf{A}}}{
      \includegraphics[width=0.49\linewidth]{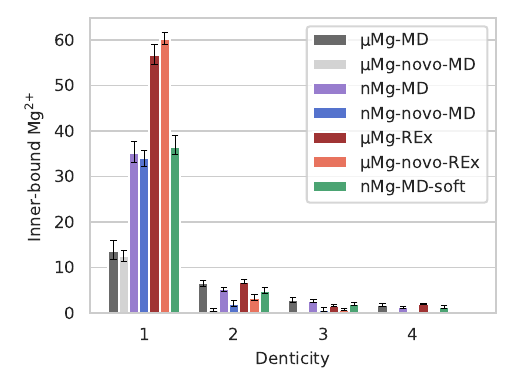}
    }
    \stackinset{l}{1em}{t}{0.5em}{\textsf{\textbf{B}}}{
      \includegraphics[width=0.49\linewidth]{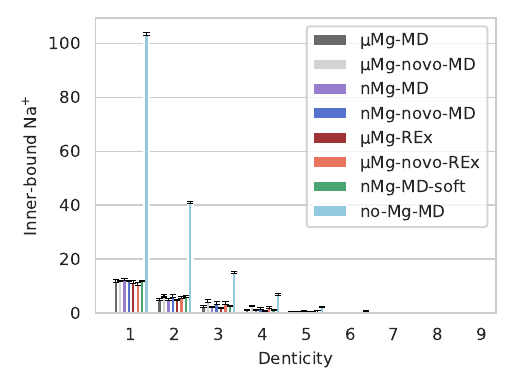}
    }
  \end{minipage}
  \caption[Number of observed motifs as a function of denticity (number of inner-bound, non-solvent ligands) for \textit{(A)} \ce{Mg^{2+}} and \textit{(B)} \ce{Na+}]{Number of observed motifs as a function of denticity (number of inner-bound, non-solvent ligands) for \textit{(A)} \ce{Mg^{2+}} and \textit{(B)} \ce{Na+}.}
  \label{sfig:denticities}
\end{figure}

\begin{figure}[htbp!]
  \centering
  \includegraphics[width=\linewidth]{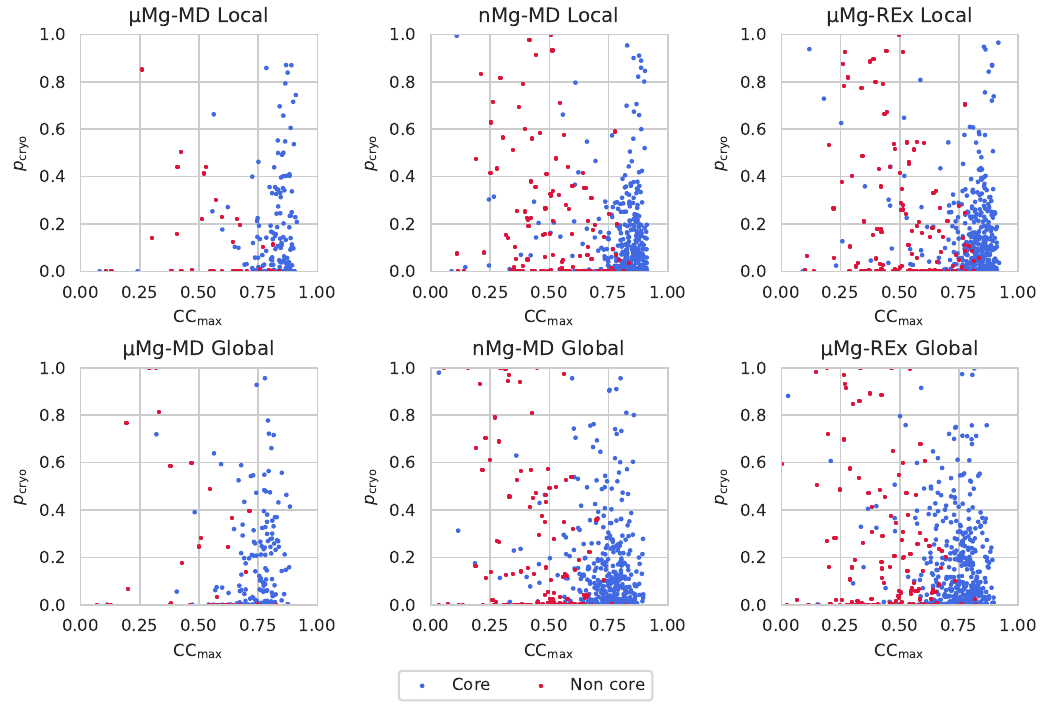}
  \caption[Cryo-EM-inferred occupancy $p_\text{cryo}$ of significant motifs as a function of the corresponding maximum cross-correlation, $\text{CC}_\text{max}$]{Cryo-EM-inferred occupancy $p_\text{cryo}$ of significant motifs as a function of the corresponding maximum cross-correlation, $\text{CC}_\text{max}$. Low $\text{CC}_\text{max}$ motifs correspond to unreliable predictions with erratic $p_\text{cryo}$ values and tend to lie outside of the core of the intron fold (red dots). $p_\text{cryo}$ values corresponding to $\text{CC}_\text{max} < 0.7$ were therefore not reported in Table~\ref{stable:motif_table}.}
  \label{sfig:ccmax_pcryo}
\end{figure}

\begin{figure}[htbp!]
  \centering
  \includegraphics[width=\linewidth]{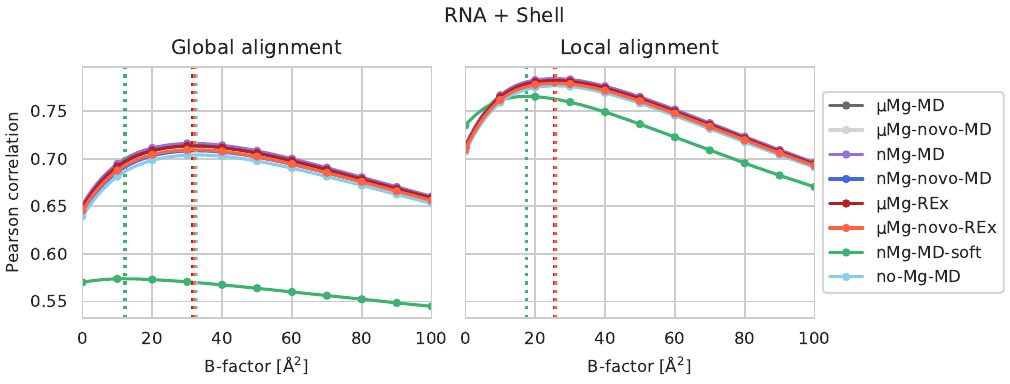}
  \includegraphics[width=\linewidth]{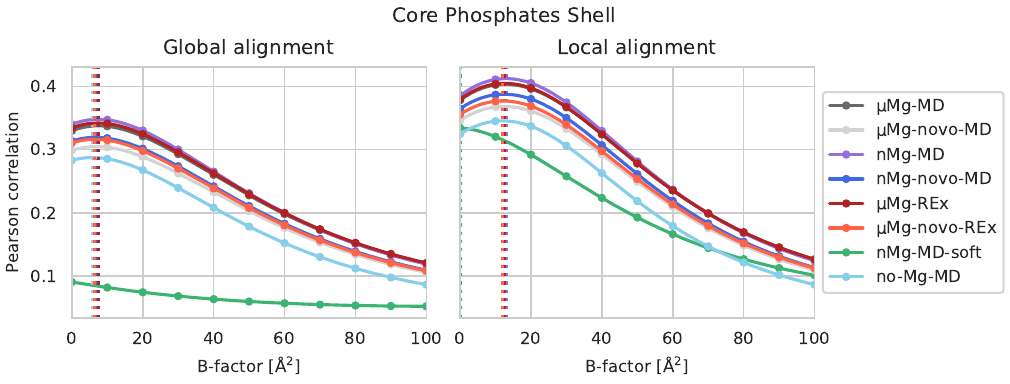}
  \caption[Selection of B-Factor smoothing for generated maps and unsharpened reference map]{Selection of B-Factor smoothing for generated maps and unsharpened reference map. The cross-correlation with experimental map (unsharpened, using the raw half maps, see Methods) on the given voxel selection is computed for different levels of applied B-factor (Gaussian) smoothing to determine the optimal value used in comparisons.}
  \label{sfig:tempfactor_scan}
\end{figure}

\begin{figure}[htbp!]
  \centering
  \includegraphics[width=\linewidth]{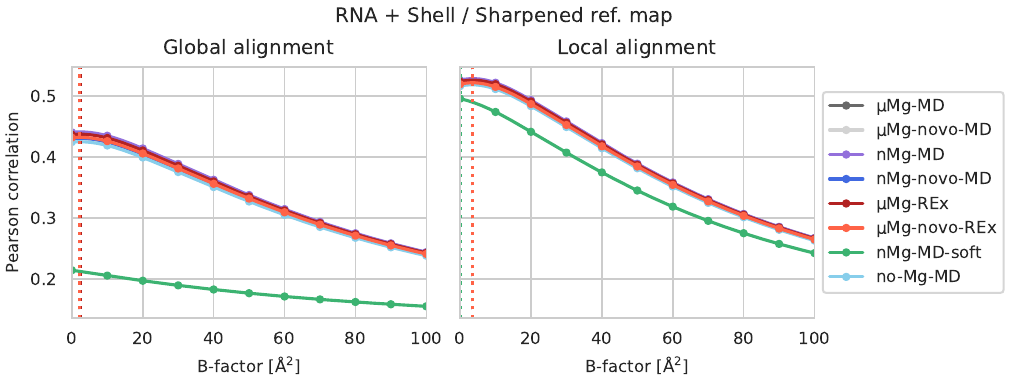}
  \includegraphics[width=\linewidth]{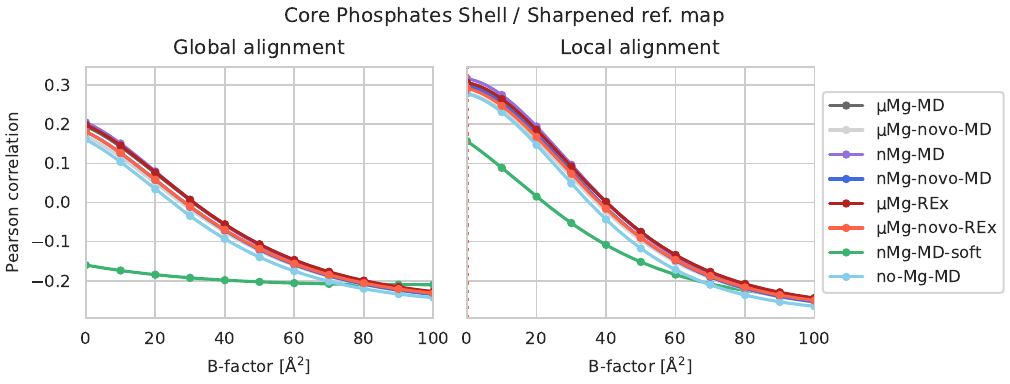}
  \caption[Selection of B-Factor smoothing for generated maps and sharpened reference map]{Selection of B-Factor smoothing for generated maps and sharpened reference map. The cross-correlation with experimental map (sharpened with B-factor \qty{79.23}{\angstrom\squared} \citesi{kretsch_complex_2025}--main deposited map in the EMD-42499 entry) on the given voxel selection is computed for different levels of applied B-factor (Gaussian) smoothing to determine the optimal value used in comparisons.}
  \label{sfig:tempfactor_scan_sharpened}
\end{figure}

\clearpage
\section*{Supplementary Tables}

\begin{table}[htbp!]
  \centering
  \caption[Summary of \textit{T.\ thermophila} intron simulations]{Summary of \textit{T.\ thermophila} intron simulations. For each setup we report the number of ions $n_i$ in the simulation box, the force constant associated with conformational restraints $\kappa$, the number of independent replicates and number of parallel replicas for REx simulations, simulation length, indicative performance and the corresponding total CPU time.}
  \label{stable:simulations}
  \fontfamily{phv}\fontseries{mc}\selectfont
  \begin{tabular}{lrrrrrrrrr}
    \toprule
    System & $n_\text{Mg}$ & $n_\text{Na}$ & $n_\text{Cl}$ & $\kappa$ & Replicates & Replicas & Duration & Perf.* & Time* \\
    &  & &  & [kJ/mol/nm$^2$] & & & [ns] & [h/ns] & [CPUh] \\
    \midrule
    {\color[HTML]{696969}$\blacksquare$} μMg-MD & 178 & 121 & 91 & \num{e5} & 8 & & 600 & 0.3 & 11520 \\
    {\color[HTML]{d3d3d3}$\blacksquare$} μMg-novo-MD & 178 & 121 & 91 & \num{e5} & 8 & & 600 & 0.3 & 11520 \\
    {\color[HTML]{9370db}$\blacksquare$} nMg-MD & 178 & 121 & 91 & \num{e5}  & 8 & & 600 & 0.3 & 11520 \\
    {\color[HTML]{4169e1}$\blacksquare$} nMg-novo-MD & 178 & 121 & 91 & \num{e5} & 8 & & 600 & 0.3 & 11520 \\
    {\color[HTML]{b22222}$\blacksquare$} μMg-REx & 178 & 121 & 91 & \num{e5} & 2 & 8 & 200 & 0.9 & 23040 \\
    {\color[HTML]{ff6347}$\blacksquare$} μMg-novo-REx & 178 & 121 & 91 & \num{e5} & 2 & 8 & 200 & 0.9 & 23040 \\
    {\color[HTML]{3cb371}$\blacksquare$} nMg-MD-soft & 178 & 121 & 91 & \num{5e2} & 8 & & 600 & 0.3 & 11520 \\
    {\color[HTML]{87ceeb}$\blacksquare$} no-Mg-MD & & 453 & 67 & \num{e5} & 8 & & 400 & 0.3 & 7680 \\
    \bottomrule
  \end{tabular}
    \begin{tablenotes} \item
    * Indicative only, benchmarked on 4-GPU, 32-core node architecture from CINECA's Leonardo HPC \citesi{turisini_leonardo_2024} with 1 GPU / 8 cores per replica. Actual setups and performances to produce the data presented in this work may have differed.
    \end{tablenotes}
\end{table}

\begin{table}[htbp!]\centering
  \caption[Benchmark for neighbor list parameters]{Benchmark for neighbor list parameters used for the accelerating bias in replica exchange simulations. We report the frequency of violations of the neighbor list during a \qty{5}{\nano\second} simulation of the UpU-\ce{Mg^{2+}} system, for increasing cutoffs and update strides.}
  \label{stable:nlist}
  \fontfamily{phv}\fontseries{mc}\selectfont
  \begin{tabular}{lrrrrr}
    \toprule
    & \multicolumn{5}{c}{$P_\text{violation}$} \\
    \midrule
    Cutoff [nm] & 0.7 & 0.8 & 0.9 & 1.0 & 1.1 \\
    Stride &  &  &  &  &  \\
    \midrule
    100 & 0 & 0 & 0 & 0 & 0 \\
    250 & \num{5e-5} & 0 & 0 & 0 & 0 \\
    500 & 0.001 & \num{6e-5} & \num{3e-5} & 0 & 0 \\
    1000 & 0.008 & 0.001 & 0.0003 & \num{6e-6} & 0 \\
    \bottomrule
  \end{tabular}
\end{table}

\begin{table}[htbp!]\centering
  \caption[Number of significant motifs]{Number of unique significant motifs ($p > \qty{5}{\percent}$) explored in different setups, in total and as a function of denticity. For comparison, number of observed motifs is also reported for the reference PDB structures (\textsc{7ez0} and \textsc{9cbu}) and for the post-equilibrated systems (Eq).}
  \label{stable:motif_count}
  \fontfamily{phv}\fontseries{mc}\selectfont
  \input{tables/motif_count}
\end{table}

\begin{table}[htbp!]\centering
  \caption[Number of reference motifs recovered]{Number of motifs from reference PDB structures recovered in simulations.}
  \label{stable:motif_recovery}
  \fontfamily{phv}\fontseries{mc}\selectfont
  \input{tables/motif_recovery}
\end{table}

\clearpage
\newgeometry{left=1.5cm, right=1.5cm, top=2cm, bottom=2cm, footskip=20pt}

\begin{landscape}
\scriptsize
\fontfamily{phv}\fontseries{mc}\selectfont
\setlength{\tabcolsep}{3pt}
\input{tables/motif_table}
\end{landscape}

\restoregeometry

\section*{Supplementary Methods}

\subsection*{Maximum Likelihood Estimate of $\lambda$-dependent Kinetic Parameters}

To extrapolate the on and off rates in the absence of barrier-flattening bias, we assume $\lambda$-dependent on ($+$) and off ($-$) rates have the form

\begin{align}
v^+(\lambda) &= v(\lambda)e^{-\beta\frac{\Delta G(\lambda)}{2}} = v_0e^{\beta\lambda\Delta E^\ddagger}e^{-\beta\frac{\Delta G\lambda)}{2}} \\
v^-(\lambda) &= v(\lambda)e^{+\beta\frac{\Delta G(\lambda)}{2}} = v_0e^{\beta\lambda\Delta E^\ddagger}e^{+\beta\frac{\Delta G(\lambda)}{2}}
\end{align}
where we assume the binding free energy difference $\Delta G(\lambda)$ is already known. The likelihood for the parameters $(v_0, \Delta E^\ddagger)$ from an observed first passage time $t$ at scaling $\lambda$ is given by
\begin{equation}
\mathcal{L}^\pm(v_0, \Delta E^\ddagger \mid t, \lambda) = v^\pm(\lambda)e^{-v^\pm(\lambda)t}
\end{equation}
hence the log-likelihood
\begin{equation}
\ell^\pm(v_0, \Delta E^\ddagger \mid t, \lambda) = \ln v_0 + \beta\lambda\Delta E^\ddagger \mp \beta\frac{\Delta G(\lambda)}{2} - v_0e^{\beta\lambda\Delta E^\ddagger}e^{\mp\beta\frac{\Delta G(\lambda)}{2}}t.
\end{equation}

For $K$ simulations (or blocks) indexed with $k$, each with a fixed bias scaling $\lambda_k$, a set of $n^+_k$ observed on-FPTs $\{t^+_{k,i}\}_i^{n^+_k}$ and $n^-_k$ off-FPTs $\{t^-_{k,j}\}_j^{n^-_k}$, the total log-likelihood reads

\begin{align}
&\ell\left(v_0, \Delta E^\ddagger \middle| X = \left\{ \left( \lambda_k, \{t^+_{k,i}\}_i^{n^+_k}, \{t^-_{k,j}\}_j^{n^-_k} \right) \right\}^{K}_k\right) \\
&= \sum_k^K\left[N_k(\ln v_0 + \beta\lambda_k\Delta E^\ddagger) - v_0e^{\beta\lambda_k\Delta E^\ddagger}\underbrace{\left(\sum_i^{n^+_k}e^{-\beta\frac{\Delta G(\lambda_k)}{2}}t^+_{k,i} + \sum_j^{n^-_k}e^{\beta\frac{\Delta G(\lambda_k)}{2}}t^-_{k,j} \right)}_{S_k}\right] + \text{const.}
\end{align}
where $N_k = n^+_k + n^-_k$. Taking the partial derivatives with respect to each parameter and searching for the extrema

\begin{align}
\frac{\partial\ell}{\partial v_0 }(v_0 , \Delta E^\ddagger; X) &= \sum_k^K \left[\frac{N_k}{v_0 } - e^{\beta\lambda_k\Delta E^\ddagger}S_k\right] \\
\frac{\partial\ell}{\partial \Delta E^\ddagger}(v_0 , \Delta E^\ddagger; X) &= \sum_k^K \left[\beta\lambda_kN_k - v_0 \beta\lambda_ke^{\beta\lambda_k\Delta E^\ddagger}S_k\right] \\
\frac{\partial\ell}{\partial v_0 }(v_0 , \Delta E^\ddagger; X) = 0 &\iff v_0  =  \frac{\sum_k^K N_k}{\sum_k^K e^{\beta\lambda_k\Delta E^\ddagger}S_k} \\
\frac{\partial\ell}{\partial \Delta E^\ddagger}(v_0 , \Delta E^\ddagger; X) = 0 &\iff v_0  = \frac{\sum^K_k \lambda_kN_k}{\sum^K_k \lambda_ke^{\beta\lambda_k\Delta E^\ddagger}S_k}
\end{align}
we find the maximum likelihood condition
\begin{equation}
\frac{\sum_k \lambda_kN_k}{\sum_k^K N_k} - \frac{\sum_k \lambda_ke^{\beta\lambda_k\Delta E^\ddagger}S_k}{\sum_k^K e^{\beta\lambda_k\Delta E^\ddagger}S_k} = 0\\
\end{equation}
that we solve numerically for $\Delta E^\ddagger$ by minimizing the squared difference.

\bibliographystylesi{unsrt}
\bibliographysi{references}

%% file: tables/motif_count.tex
\begin{tabular}{lrrrrrr}
\toprule
 & \multicolumn{5}{r}{Denticity} & Tot \\
 & 1 & 2 & 3 & 4 & 5 &  \\
\midrule
7ez0 & {\cellcolor[HTML]{FFF5F0}} \color[HTML]{000000} 6 & {\cellcolor[HTML]{FDC6B0}} \color[HTML]{000000} 7 & {\cellcolor[HTML]{FCAF93}} \color[HTML]{000000} 2 & {\cellcolor[HTML]{FCA082}} \color[HTML]{000000} 1 & {\cellcolor[HTML]{FFF5F0}} \color[HTML]{000000} 0 & {\cellcolor[HTML]{FFF5F0}} \color[HTML]{000000} 16 \\
9cbu & {\cellcolor[HTML]{FFF0E8}} \color[HTML]{000000} 17 & {\cellcolor[HTML]{FEE9DF}} \color[HTML]{000000} 5 & {\cellcolor[HTML]{FC8161}} \color[HTML]{F1F1F1} 3 & {\cellcolor[HTML]{FFF5F0}} \color[HTML]{000000} 0 & {\cellcolor[HTML]{FFF5F0}} \color[HTML]{000000} 0 & {\cellcolor[HTML]{FFF1EA}} \color[HTML]{000000} 25 \\
μMg-Eq & {\cellcolor[HTML]{FFEBE2}} \color[HTML]{000000} 26 & {\cellcolor[HTML]{EA362A}} \color[HTML]{F1F1F1} 13 & {\cellcolor[HTML]{F44F39}} \color[HTML]{F1F1F1} 4 & {\cellcolor[HTML]{67000D}} \color[HTML]{F1F1F1} 3 & {\cellcolor[HTML]{67000D}} \color[HTML]{F1F1F1} 1 & {\cellcolor[HTML]{FEE6DA}} \color[HTML]{000000} 47 \\
nMg-Eq & {\cellcolor[HTML]{FEE5D8}} \color[HTML]{000000} 38 & {\cellcolor[HTML]{F44F39}} \color[HTML]{F1F1F1} 12 & {\cellcolor[HTML]{F44F39}} \color[HTML]{F1F1F1} 4 & {\cellcolor[HTML]{67000D}} \color[HTML]{F1F1F1} 3 & {\cellcolor[HTML]{67000D}} \color[HTML]{F1F1F1} 1 & {\cellcolor[HTML]{FEE0D2}} \color[HTML]{000000} 58 \\
μMg-novo-Eq & {\cellcolor[HTML]{FEE2D5}} \color[HTML]{000000} 44 & {\cellcolor[HTML]{FFF5F0}} \color[HTML]{000000} 4 & {\cellcolor[HTML]{FEDBCC}} \color[HTML]{000000} 1 & {\cellcolor[HTML]{FFF5F0}} \color[HTML]{000000} 0 & {\cellcolor[HTML]{FFF5F0}} \color[HTML]{000000} 0 & {\cellcolor[HTML]{FEE5D8}} \color[HTML]{000000} 49 \\
nMg-novo-Eq & {\cellcolor[HTML]{FEE7DB}} \color[HTML]{000000} 35 & {\cellcolor[HTML]{FEE9DF}} \color[HTML]{000000} 5 & {\cellcolor[HTML]{FC8161}} \color[HTML]{F1F1F1} 3 & {\cellcolor[HTML]{FFF5F0}} \color[HTML]{000000} 0 & {\cellcolor[HTML]{FFF5F0}} \color[HTML]{000000} 0 & {\cellcolor[HTML]{FEE8DD}} \color[HTML]{000000} 43 \\
μMg-MD & {\cellcolor[HTML]{FDD5C4}} \color[HTML]{000000} 59 & {\cellcolor[HTML]{FB694A}} \color[HTML]{F1F1F1} 11 & {\cellcolor[HTML]{67000D}} \color[HTML]{F1F1F1} 7 & {\cellcolor[HTML]{E32F27}} \color[HTML]{F1F1F1} 2 & {\cellcolor[HTML]{FFF5F0}} \color[HTML]{000000} 0 & {\cellcolor[HTML]{FDCDB9}} \color[HTML]{000000} 79 \\
nMg-MD & {\cellcolor[HTML]{F96245}} \color[HTML]{F1F1F1} 176 & {\cellcolor[HTML]{D52221}} \color[HTML]{F1F1F1} 14 & {\cellcolor[HTML]{AA1016}} \color[HTML]{F1F1F1} 6 & {\cellcolor[HTML]{E32F27}} \color[HTML]{F1F1F1} 2 & {\cellcolor[HTML]{FFF5F0}} \color[HTML]{000000} 0 & {\cellcolor[HTML]{F7593F}} \color[HTML]{F1F1F1} 198 \\
nMg-MD-soft & {\cellcolor[HTML]{F14130}} \color[HTML]{F1F1F1} 203 & {\cellcolor[HTML]{67000D}} \color[HTML]{F1F1F1} 18 & {\cellcolor[HTML]{67000D}} \color[HTML]{F1F1F1} 7 & {\cellcolor[HTML]{E32F27}} \color[HTML]{F1F1F1} 2 & {\cellcolor[HTML]{FFF5F0}} \color[HTML]{000000} 0 & {\cellcolor[HTML]{EA362A}} \color[HTML]{F1F1F1} 230 \\
μMg-novo-MD & {\cellcolor[HTML]{FDC6B0}} \color[HTML]{000000} 75 & {\cellcolor[HTML]{FEE9DF}} \color[HTML]{000000} 5 & {\cellcolor[HTML]{FFF5F0}} \color[HTML]{000000} 0 & {\cellcolor[HTML]{FFF5F0}} \color[HTML]{000000} 0 & {\cellcolor[HTML]{FFF5F0}} \color[HTML]{000000} 0 & {\cellcolor[HTML]{FDCCB8}} \color[HTML]{000000} 80 \\
nMg-novo-MD & {\cellcolor[HTML]{FA6648}} \color[HTML]{F1F1F1} 172 & {\cellcolor[HTML]{FB694A}} \color[HTML]{F1F1F1} 11 & {\cellcolor[HTML]{FC8161}} \color[HTML]{F1F1F1} 3 & {\cellcolor[HTML]{FFF5F0}} \color[HTML]{000000} 0 & {\cellcolor[HTML]{FFF5F0}} \color[HTML]{000000} 0 & {\cellcolor[HTML]{FA6648}} \color[HTML]{F1F1F1} 186 \\
μMg-REx & {\cellcolor[HTML]{7E0610}} \color[HTML]{F1F1F1} 315 & {\cellcolor[HTML]{AA1016}} \color[HTML]{F1F1F1} 16 & {\cellcolor[HTML]{FCAF93}} \color[HTML]{000000} 2 & {\cellcolor[HTML]{E32F27}} \color[HTML]{F1F1F1} 2 & {\cellcolor[HTML]{FFF5F0}} \color[HTML]{000000} 0 & {\cellcolor[HTML]{7A0510}} \color[HTML]{F1F1F1} 335 \\
μMg-REx [1] & {\cellcolor[HTML]{960B13}} \color[HTML]{F1F1F1} 300 & {\cellcolor[HTML]{AA1016}} \color[HTML]{F1F1F1} 16 & {\cellcolor[HTML]{FCAF93}} \color[HTML]{000000} 2 & {\cellcolor[HTML]{E32F27}} \color[HTML]{F1F1F1} 2 & {\cellcolor[HTML]{FFF5F0}} \color[HTML]{000000} 0 & {\cellcolor[HTML]{920A13}} \color[HTML]{F1F1F1} 320 \\
μMg-REx [2] & {\cellcolor[HTML]{B11218}} \color[HTML]{F1F1F1} 277 & {\cellcolor[HTML]{D52221}} \color[HTML]{F1F1F1} 14 & {\cellcolor[HTML]{FCAF93}} \color[HTML]{000000} 2 & {\cellcolor[HTML]{E32F27}} \color[HTML]{F1F1F1} 2 & {\cellcolor[HTML]{FFF5F0}} \color[HTML]{000000} 0 & {\cellcolor[HTML]{B01217}} \color[HTML]{F1F1F1} 295 \\
μMg-novo-REx & {\cellcolor[HTML]{67000D}} \color[HTML]{F1F1F1} 331 & {\cellcolor[HTML]{D52221}} \color[HTML]{F1F1F1} 14 & {\cellcolor[HTML]{F44F39}} \color[HTML]{F1F1F1} 4 & {\cellcolor[HTML]{FFF5F0}} \color[HTML]{000000} 0 & {\cellcolor[HTML]{FFF5F0}} \color[HTML]{000000} 0 & {\cellcolor[HTML]{67000D}} \color[HTML]{F1F1F1} 349 \\
μMg-novo-REx [1] & {\cellcolor[HTML]{9A0C14}} \color[HTML]{F1F1F1} 297 & {\cellcolor[HTML]{D52221}} \color[HTML]{F1F1F1} 14 & {\cellcolor[HTML]{FCAF93}} \color[HTML]{000000} 2 & {\cellcolor[HTML]{FFF5F0}} \color[HTML]{000000} 0 & {\cellcolor[HTML]{FFF5F0}} \color[HTML]{000000} 0 & {\cellcolor[HTML]{9C0D14}} \color[HTML]{F1F1F1} 313 \\
μMg-novo-REx [2] & {\cellcolor[HTML]{820711}} \color[HTML]{F1F1F1} 313 & {\cellcolor[HTML]{FB694A}} \color[HTML]{F1F1F1} 11 & {\cellcolor[HTML]{FC8161}} \color[HTML]{F1F1F1} 3 & {\cellcolor[HTML]{FFF5F0}} \color[HTML]{000000} 0 & {\cellcolor[HTML]{FFF5F0}} \color[HTML]{000000} 0 & {\cellcolor[HTML]{860811}} \color[HTML]{F1F1F1} 327 \\
\bottomrule
\end{tabular}

%% file: tables/motif_recovery.tex
\begin{tabular}{lrrrrrrrrr}
\toprule
 & \multicolumn{5}{r}{7ez0} & \multicolumn{4}{r}{9cbu} \\
 & 1 & 2 & 3 & 4 & Tot & 1 & 2 & 3 & Tot \\
\midrule
7ez0 & {\cellcolor[HTML]{FCBEA5}} \color[HTML]{000000} 6/6 & {\cellcolor[HTML]{FCB296}} \color[HTML]{000000} 7/7 & {\cellcolor[HTML]{FEE8DD}} \color[HTML]{000000} 2/2 & {\cellcolor[HTML]{FFEEE7}} \color[HTML]{000000} 1/1 & {\cellcolor[HTML]{EB372A}} \color[HTML]{F1F1F1} 16/16 & {\cellcolor[HTML]{FFF5F0}} \color[HTML]{000000} 0/17 & {\cellcolor[HTML]{FEE1D4}} \color[HTML]{000000} 3/5 & {\cellcolor[HTML]{FEE8DD}} \color[HTML]{000000} 2/3 & {\cellcolor[HTML]{FDCAB5}} \color[HTML]{000000} 5/25 \\
9cbu & {\cellcolor[HTML]{FFF5F0}} \color[HTML]{000000} 0/6 & {\cellcolor[HTML]{FEE1D4}} \color[HTML]{000000} 3/7 & {\cellcolor[HTML]{FEE8DD}} \color[HTML]{000000} 2/2 & {\cellcolor[HTML]{FFF5F0}} \color[HTML]{000000} 0/1 & {\cellcolor[HTML]{FDCAB5}} \color[HTML]{000000} 5/16 & {\cellcolor[HTML]{DE2B25}} \color[HTML]{F1F1F1} 17/17 & {\cellcolor[HTML]{FDCAB5}} \color[HTML]{000000} 5/5 & {\cellcolor[HTML]{FEE1D4}} \color[HTML]{000000} 3/3 & {\cellcolor[HTML]{67000D}} \color[HTML]{F1F1F1} 25/25 \\
μMg-Eq & {\cellcolor[HTML]{FCBEA5}} \color[HTML]{000000} 6/6 & {\cellcolor[HTML]{FCBEA5}} \color[HTML]{000000} 6/7 & {\cellcolor[HTML]{FFEEE7}} \color[HTML]{000000} 1/2 & {\cellcolor[HTML]{FFEEE7}} \color[HTML]{000000} 1/1 & {\cellcolor[HTML]{F5533B}} \color[HTML]{F1F1F1} 14/16 & {\cellcolor[HTML]{FFF5F0}} \color[HTML]{000000} 0/17 & {\cellcolor[HTML]{FEE1D4}} \color[HTML]{000000} 3/5 & {\cellcolor[HTML]{FFEEE7}} \color[HTML]{000000} 1/3 & {\cellcolor[HTML]{FDD7C6}} \color[HTML]{000000} 4/25 \\
nMg-Eq & {\cellcolor[HTML]{FCBEA5}} \color[HTML]{000000} 6/6 & {\cellcolor[HTML]{FCBEA5}} \color[HTML]{000000} 6/7 & {\cellcolor[HTML]{FFEEE7}} \color[HTML]{000000} 1/2 & {\cellcolor[HTML]{FFEEE7}} \color[HTML]{000000} 1/1 & {\cellcolor[HTML]{F5533B}} \color[HTML]{F1F1F1} 14/16 & {\cellcolor[HTML]{FFF5F0}} \color[HTML]{000000} 0/17 & {\cellcolor[HTML]{FEE1D4}} \color[HTML]{000000} 3/5 & {\cellcolor[HTML]{FFEEE7}} \color[HTML]{000000} 1/3 & {\cellcolor[HTML]{FDD7C6}} \color[HTML]{000000} 4/25 \\
μMg-novo-Eq & {\cellcolor[HTML]{FEE1D4}} \color[HTML]{000000} 3/6 & {\cellcolor[HTML]{FFEEE7}} \color[HTML]{000000} 1/7 & {\cellcolor[HTML]{FFF5F0}} \color[HTML]{000000} 0/2 & {\cellcolor[HTML]{FFF5F0}} \color[HTML]{000000} 0/1 & {\cellcolor[HTML]{FDD7C6}} \color[HTML]{000000} 4/16 & {\cellcolor[HTML]{FFF5F0}} \color[HTML]{000000} 0/17 & {\cellcolor[HTML]{FFEEE7}} \color[HTML]{000000} 1/5 & {\cellcolor[HTML]{FFF5F0}} \color[HTML]{000000} 0/3 & {\cellcolor[HTML]{FFEEE7}} \color[HTML]{000000} 1/25 \\
nMg-novo-Eq & {\cellcolor[HTML]{FEE8DD}} \color[HTML]{000000} 2/6 & {\cellcolor[HTML]{FFEEE7}} \color[HTML]{000000} 1/7 & {\cellcolor[HTML]{FFEEE7}} \color[HTML]{000000} 1/2 & {\cellcolor[HTML]{FFF5F0}} \color[HTML]{000000} 0/1 & {\cellcolor[HTML]{FDD7C6}} \color[HTML]{000000} 4/16 & {\cellcolor[HTML]{FFEEE7}} \color[HTML]{000000} 1/17 & {\cellcolor[HTML]{FFF5F0}} \color[HTML]{000000} 0/5 & {\cellcolor[HTML]{FEE8DD}} \color[HTML]{000000} 2/3 & {\cellcolor[HTML]{FEE1D4}} \color[HTML]{000000} 3/25 \\
μMg-MD & {\cellcolor[HTML]{FCBEA5}} \color[HTML]{000000} 6/6 & {\cellcolor[HTML]{FCBEA5}} \color[HTML]{000000} 6/7 & {\cellcolor[HTML]{FFEEE7}} \color[HTML]{000000} 1/2 & {\cellcolor[HTML]{FFEEE7}} \color[HTML]{000000} 1/1 & {\cellcolor[HTML]{F5533B}} \color[HTML]{F1F1F1} 14/16 & {\cellcolor[HTML]{FEE8DD}} \color[HTML]{000000} 2/17 & {\cellcolor[HTML]{FEE1D4}} \color[HTML]{000000} 3/5 & {\cellcolor[HTML]{FEE8DD}} \color[HTML]{000000} 2/3 & {\cellcolor[HTML]{FCB296}} \color[HTML]{000000} 7/25 \\
nMg-MD & {\cellcolor[HTML]{FCBEA5}} \color[HTML]{000000} 6/6 & {\cellcolor[HTML]{FCBEA5}} \color[HTML]{000000} 6/7 & {\cellcolor[HTML]{FFEEE7}} \color[HTML]{000000} 1/2 & {\cellcolor[HTML]{FFEEE7}} \color[HTML]{000000} 1/1 & {\cellcolor[HTML]{F5533B}} \color[HTML]{F1F1F1} 14/16 & {\cellcolor[HTML]{FFEEE7}} \color[HTML]{000000} 1/17 & {\cellcolor[HTML]{FEE1D4}} \color[HTML]{000000} 3/5 & {\cellcolor[HTML]{FEE8DD}} \color[HTML]{000000} 2/3 & {\cellcolor[HTML]{FCBEA5}} \color[HTML]{000000} 6/25 \\
nMg-MD-soft & {\cellcolor[HTML]{FCBEA5}} \color[HTML]{000000} 6/6 & {\cellcolor[HTML]{FCBEA5}} \color[HTML]{000000} 6/7 & {\cellcolor[HTML]{FEE8DD}} \color[HTML]{000000} 2/2 & {\cellcolor[HTML]{FFEEE7}} \color[HTML]{000000} 1/1 & {\cellcolor[HTML]{F14432}} \color[HTML]{F1F1F1} 15/16 & {\cellcolor[HTML]{FEE8DD}} \color[HTML]{000000} 2/17 & {\cellcolor[HTML]{FEE1D4}} \color[HTML]{000000} 3/5 & {\cellcolor[HTML]{FEE1D4}} \color[HTML]{000000} 3/3 & {\cellcolor[HTML]{FCA588}} \color[HTML]{000000} 8/25 \\
μMg-novo-MD & {\cellcolor[HTML]{FDD7C6}} \color[HTML]{000000} 4/6 & {\cellcolor[HTML]{FFF5F0}} \color[HTML]{000000} 0/7 & {\cellcolor[HTML]{FFF5F0}} \color[HTML]{000000} 0/2 & {\cellcolor[HTML]{FFF5F0}} \color[HTML]{000000} 0/1 & {\cellcolor[HTML]{FDD7C6}} \color[HTML]{000000} 4/16 & {\cellcolor[HTML]{FFEEE7}} \color[HTML]{000000} 1/17 & {\cellcolor[HTML]{FFF5F0}} \color[HTML]{000000} 0/5 & {\cellcolor[HTML]{FFF5F0}} \color[HTML]{000000} 0/3 & {\cellcolor[HTML]{FFEEE7}} \color[HTML]{000000} 1/25 \\
nMg-novo-MD & {\cellcolor[HTML]{FDCAB5}} \color[HTML]{000000} 5/6 & {\cellcolor[HTML]{FEE1D4}} \color[HTML]{000000} 3/7 & {\cellcolor[HTML]{FEE8DD}} \color[HTML]{000000} 2/2 & {\cellcolor[HTML]{FFF5F0}} \color[HTML]{000000} 0/1 & {\cellcolor[HTML]{FC8A6A}} \color[HTML]{F1F1F1} 10/16 & {\cellcolor[HTML]{FEE1D4}} \color[HTML]{000000} 3/17 & {\cellcolor[HTML]{FFEEE7}} \color[HTML]{000000} 1/5 & {\cellcolor[HTML]{FEE1D4}} \color[HTML]{000000} 3/3 & {\cellcolor[HTML]{FCB296}} \color[HTML]{000000} 7/25 \\
μMg-REx & {\cellcolor[HTML]{FCBEA5}} \color[HTML]{000000} 6/6 & {\cellcolor[HTML]{FCBEA5}} \color[HTML]{000000} 6/7 & {\cellcolor[HTML]{FFEEE7}} \color[HTML]{000000} 1/2 & {\cellcolor[HTML]{FFEEE7}} \color[HTML]{000000} 1/1 & {\cellcolor[HTML]{F5533B}} \color[HTML]{F1F1F1} 14/16 & {\cellcolor[HTML]{FEE8DD}} \color[HTML]{000000} 2/17 & {\cellcolor[HTML]{FEE1D4}} \color[HTML]{000000} 3/5 & {\cellcolor[HTML]{FFEEE7}} \color[HTML]{000000} 1/3 & {\cellcolor[HTML]{FCBEA5}} \color[HTML]{000000} 6/25 \\
μMg-REx [1] & {\cellcolor[HTML]{FCBEA5}} \color[HTML]{000000} 6/6 & {\cellcolor[HTML]{FCBEA5}} \color[HTML]{000000} 6/7 & {\cellcolor[HTML]{FFEEE7}} \color[HTML]{000000} 1/2 & {\cellcolor[HTML]{FFEEE7}} \color[HTML]{000000} 1/1 & {\cellcolor[HTML]{F5533B}} \color[HTML]{F1F1F1} 14/16 & {\cellcolor[HTML]{FFEEE7}} \color[HTML]{000000} 1/17 & {\cellcolor[HTML]{FEE1D4}} \color[HTML]{000000} 3/5 & {\cellcolor[HTML]{FFEEE7}} \color[HTML]{000000} 1/3 & {\cellcolor[HTML]{FDCAB5}} \color[HTML]{000000} 5/25 \\
μMg-REx [2] & {\cellcolor[HTML]{FCBEA5}} \color[HTML]{000000} 6/6 & {\cellcolor[HTML]{FCBEA5}} \color[HTML]{000000} 6/7 & {\cellcolor[HTML]{FFEEE7}} \color[HTML]{000000} 1/2 & {\cellcolor[HTML]{FFEEE7}} \color[HTML]{000000} 1/1 & {\cellcolor[HTML]{F5533B}} \color[HTML]{F1F1F1} 14/16 & {\cellcolor[HTML]{FFEEE7}} \color[HTML]{000000} 1/17 & {\cellcolor[HTML]{FEE1D4}} \color[HTML]{000000} 3/5 & {\cellcolor[HTML]{FFEEE7}} \color[HTML]{000000} 1/3 & {\cellcolor[HTML]{FDCAB5}} \color[HTML]{000000} 5/25 \\
μMg-novo-REx & {\cellcolor[HTML]{FDCAB5}} \color[HTML]{000000} 5/6 & {\cellcolor[HTML]{FDCAB5}} \color[HTML]{000000} 5/7 & {\cellcolor[HTML]{FEE8DD}} \color[HTML]{000000} 2/2 & {\cellcolor[HTML]{FFF5F0}} \color[HTML]{000000} 0/1 & {\cellcolor[HTML]{FB7151}} \color[HTML]{F1F1F1} 12/16 & {\cellcolor[HTML]{FEE1D4}} \color[HTML]{000000} 3/17 & {\cellcolor[HTML]{FEE8DD}} \color[HTML]{000000} 2/5 & {\cellcolor[HTML]{FEE1D4}} \color[HTML]{000000} 3/3 & {\cellcolor[HTML]{FCA588}} \color[HTML]{000000} 8/25 \\
μMg-novo-REx [1] & {\cellcolor[HTML]{FDD7C6}} \color[HTML]{000000} 4/6 & {\cellcolor[HTML]{FDCAB5}} \color[HTML]{000000} 5/7 & {\cellcolor[HTML]{FFEEE7}} \color[HTML]{000000} 1/2 & {\cellcolor[HTML]{FFF5F0}} \color[HTML]{000000} 0/1 & {\cellcolor[HTML]{FC8A6A}} \color[HTML]{F1F1F1} 10/16 & {\cellcolor[HTML]{FDD7C6}} \color[HTML]{000000} 4/17 & {\cellcolor[HTML]{FEE8DD}} \color[HTML]{000000} 2/5 & {\cellcolor[HTML]{FEE8DD}} \color[HTML]{000000} 2/3 & {\cellcolor[HTML]{FCA588}} \color[HTML]{000000} 8/25 \\
μMg-novo-REx [2] & {\cellcolor[HTML]{FDCAB5}} \color[HTML]{000000} 5/6 & {\cellcolor[HTML]{FDCAB5}} \color[HTML]{000000} 5/7 & {\cellcolor[HTML]{FEE8DD}} \color[HTML]{000000} 2/2 & {\cellcolor[HTML]{FFF5F0}} \color[HTML]{000000} 0/1 & {\cellcolor[HTML]{FB7151}} \color[HTML]{F1F1F1} 12/16 & {\cellcolor[HTML]{FEE1D4}} \color[HTML]{000000} 3/17 & {\cellcolor[HTML]{FEE8DD}} \color[HTML]{000000} 2/5 & {\cellcolor[HTML]{FEE8DD}} \color[HTML]{000000} 2/3 & {\cellcolor[HTML]{FCB296}} \color[HTML]{000000} 7/25 \\
\bottomrule
\end{tabular}

%% file: tables/motif_table.tex
\begin{xltabular}{\linewidth}{b{0.3cm}b{3.5cm}b{0.6cm}b{0.6cm}b{0.5cm}b{0.5cm}b{0.5cm}b{0.5cm} *{11}{b{1.3cm}}}
\caption{Table of significant motifs and their populations observed in different simulation setups, just after system equilibration (-Eq) and during production phase, as well as populations obtained by integrating cryo-EM maps (+~Cryo). 95\% confidence intervals are computed with boostrap on blocks. White cells indicate absence of detection. Gray cells indicate that the population could not be inferred because of the lack of corresponding samples, or because the cross-correlation was below threshold (0.7).} \label{stable:motif_table} \\
\toprule
Id & Motif & 7ez0 & 9cbu & μMg-Eq & μMg-novo-Eq & nMg-Eq & nMg-novo-Eq & μMg-MD & nMg-MD & nMg-MD-soft & μMg-REx & μMg-novo-MD & nMg-novo-MD & μMg-novo-REx & μMg-MD +~Cryo & nMg-MD +~Cryo & μMg-REx +~Cryo \\
\midrule
\endfirsthead
\caption[]{Table of significant motifs and their populations observed in different simulation setups (continued).} \\
\toprule
Id & Motif & 7ez0 & 9cbu & μMg-Eq & μMg-novo-Eq & nMg-Eq & nMg-novo-Eq & μMg-MD & nMg-MD & nMg-MD-soft & μMg-REx & μMg-novo-MD & nMg-novo-MD & μMg-novo-REx & μMg-MD +~Cryo & nMg-MD +~Cryo & μMg-REx +~Cryo \\
\midrule
\endhead
{\cellcolor[HTML]{FEFEFE}} \color[HTML]{000000} 1 & {\cellcolor[HTML]{FEFEFE}} \color[HTML]{000000} $\text{A}^\text{183}_\text{OP1}$,$\text{A}^\text{184}_\text{OP2}$,$\text{A}^\text{186}_\text{OP2}$-\textit{fac}-3$\text{O}_\text{ph}$ & {\cellcolor[HTML]{FEFEFE}} \color[HTML]{000000} $\text{MG}^\text{505}$ & {\cellcolor[HTML]{FEFEFE}} \color[HTML]{000000} $\text{MG}^\text{502}$ & {\cellcolor[HTML]{FEFEFE}} \color[HTML]{000000} {\cellcolor[HTML]{67000D}} \color[HTML]{F1F1F1} 1.0 & {\cellcolor[HTML]{FEFEFE}} \color[HTML]{000000} {\cellcolor[HTML]{000000}} \color[HTML]{F1F1F1} {\cellcolor{white}}  & {\cellcolor[HTML]{FEFEFE}} \color[HTML]{000000} {\cellcolor[HTML]{67000D}} \color[HTML]{F1F1F1} 1.0 & {\cellcolor[HTML]{FEFEFE}} \color[HTML]{000000} {\cellcolor[HTML]{000000}} \color[HTML]{F1F1F1} {\cellcolor{white}}  & {\cellcolor[HTML]{FEFEFE}} \color[HTML]{000000} {\cellcolor[HTML]{67000D}} \color[HTML]{F1F1F1} 1.0 (1.0-1.0) & {\cellcolor[HTML]{FEFEFE}} \color[HTML]{000000} {\cellcolor[HTML]{67000D}} \color[HTML]{F1F1F1} 1.0 (1.0-1.0) & {\cellcolor[HTML]{FEFEFE}} \color[HTML]{000000} {\cellcolor[HTML]{69000D}} \color[HTML]{F1F1F1} 1.0 (1.0-1.0) & {\cellcolor[HTML]{FEFEFE}} \color[HTML]{000000} {\cellcolor[HTML]{67000D}} \color[HTML]{F1F1F1} 1.0 (1.0-1.0) & {\cellcolor[HTML]{FEFEFE}} \color[HTML]{000000} {\cellcolor[HTML]{000000}} \color[HTML]{F1F1F1} {\cellcolor{white}}  & {\cellcolor[HTML]{FEFEFE}} \color[HTML]{000000} {\cellcolor[HTML]{FEE2D5}} \color[HTML]{000000} 0.1 (0.0-0.4) & {\cellcolor[HTML]{FEFEFE}} \color[HTML]{000000} {\cellcolor[HTML]{FEE3D7}} \color[HTML]{000000} 0.1 (0.0-0.3) & {\cellcolor[HTML]{FEFEFE}} \color[HTML]{000000} {\cellcolor[HTML]{D92523}} \color[HTML]{F1F1F1} 0.7 (0.7-0.8) & {\cellcolor[HTML]{FEFEFE}} \color[HTML]{000000} {\cellcolor[HTML]{F6583E}} \color[HTML]{F1F1F1} 0.6 (0.5-0.7) & {\cellcolor[HTML]{FEFEFE}} \color[HTML]{000000} {\cellcolor[HTML]{D92523}} \color[HTML]{F1F1F1} 0.7 (0.6-0.8) \\
{\cellcolor[HTML]{FEFEFE}} \color[HTML]{000000} 2 & {\cellcolor[HTML]{FEFEFE}} \color[HTML]{000000} $\text{A}^\text{256}_\text{OP1}$,$\text{U}^\text{273}_\text{OP1}$-\textit{cis}-2$\text{O}_\text{ph}$ & {\cellcolor[HTML]{FEFEFE}} \color[HTML]{000000} $\text{MG}^\text{511}$ & {\cellcolor[HTML]{FEFEFE}} \color[HTML]{000000} $\text{MG}^\text{507}$ & {\cellcolor[HTML]{FEFEFE}} \color[HTML]{000000} {\cellcolor[HTML]{67000D}} \color[HTML]{F1F1F1} 1.0 & {\cellcolor[HTML]{FEFEFE}} \color[HTML]{000000} {\cellcolor[HTML]{000000}} \color[HTML]{F1F1F1} {\cellcolor{white}}  & {\cellcolor[HTML]{FEFEFE}} \color[HTML]{000000} {\cellcolor[HTML]{67000D}} \color[HTML]{F1F1F1} 1.0 & {\cellcolor[HTML]{FEFEFE}} \color[HTML]{000000} {\cellcolor[HTML]{000000}} \color[HTML]{F1F1F1} {\cellcolor{white}}  & {\cellcolor[HTML]{FEFEFE}} \color[HTML]{000000} {\cellcolor[HTML]{67000D}} \color[HTML]{F1F1F1} 1.0 (1.0-1.0) & {\cellcolor[HTML]{FEFEFE}} \color[HTML]{000000} {\cellcolor[HTML]{9D0D14}} \color[HTML]{F1F1F1} 0.9 (0.6-1.0) & {\cellcolor[HTML]{FEFEFE}} \color[HTML]{000000} {\cellcolor[HTML]{DA2723}} \color[HTML]{F1F1F1} 0.7 (0.5-0.9) & {\cellcolor[HTML]{FEFEFE}} \color[HTML]{000000} {\cellcolor[HTML]{71020E}} \color[HTML]{F1F1F1} 1.0 (1.0-1.0) & {\cellcolor[HTML]{FEFEFE}} \color[HTML]{000000} {\cellcolor[HTML]{000000}} \color[HTML]{F1F1F1} {\cellcolor{white}}  & {\cellcolor[HTML]{FEFEFE}} \color[HTML]{000000} {\cellcolor[HTML]{FFF4EF}} \color[HTML]{000000} 0.0 (0.0-0.0) & {\cellcolor[HTML]{FEFEFE}} \color[HTML]{000000} {\cellcolor[HTML]{FCA98C}} \color[HTML]{000000} 0.3 (0.1-0.5) & {\cellcolor[HTML]{FEFEFE}} \color[HTML]{000000} {\cellcolor[HTML]{F6583E}} \color[HTML]{F1F1F1} 0.6 (0.4-0.7) & {\cellcolor[HTML]{FEFEFE}} \color[HTML]{000000} {\cellcolor[HTML]{AC1117}} \color[HTML]{F1F1F1} 0.9 (0.6-1.0) & {\cellcolor[HTML]{FEFEFE}} \color[HTML]{000000} {\cellcolor[HTML]{CA181D}} \color[HTML]{F1F1F1} 0.8 (0.7-0.9) \\
{\cellcolor[HTML]{FEFEFE}} \color[HTML]{000000} 3 & {\cellcolor[HTML]{FEFEFE}} \color[HTML]{000000} $\text{A}^\text{304}_\text{OP1}$,$\text{A}^\text{306}_\text{OP1}$-\textit{cis}-2$\text{O}_\text{ph}$ & {\cellcolor[HTML]{FEFEFE}} \color[HTML]{000000} $\text{MG}^\text{526}$ & {\cellcolor[HTML]{FEFEFE}} \color[HTML]{000000} $\text{MG}^\text{508}$ & {\cellcolor[HTML]{FEFEFE}} \color[HTML]{000000} {\cellcolor[HTML]{B71319}} \color[HTML]{F1F1F1} 0.8 & {\cellcolor[HTML]{FEFEFE}} \color[HTML]{000000} {\cellcolor[HTML]{FEEAE1}} \color[HTML]{000000} 0.1 & {\cellcolor[HTML]{FEFEFE}} \color[HTML]{000000} {\cellcolor[HTML]{A30F15}} \color[HTML]{F1F1F1} 0.9 & {\cellcolor[HTML]{FEFEFE}} \color[HTML]{000000} {\cellcolor[HTML]{000000}} \color[HTML]{F1F1F1} {\cellcolor{white}}  & {\cellcolor[HTML]{FEFEFE}} \color[HTML]{000000} {\cellcolor[HTML]{FB6B4B}} \color[HTML]{F1F1F1} 0.5 (0.2-0.8) & {\cellcolor[HTML]{FEFEFE}} \color[HTML]{000000} {\cellcolor[HTML]{EF3C2C}} \color[HTML]{F1F1F1} 0.6 (0.3-0.9) & {\cellcolor[HTML]{FEFEFE}} \color[HTML]{000000} {\cellcolor[HTML]{F0402F}} \color[HTML]{F1F1F1} 0.6 (0.3-0.9) & {\cellcolor[HTML]{FEFEFE}} \color[HTML]{000000} {\cellcolor[HTML]{7C0510}} \color[HTML]{F1F1F1} 1.0 (0.9-1.0) & {\cellcolor[HTML]{FEFEFE}} \color[HTML]{000000} {\cellcolor[HTML]{000000}} \color[HTML]{F1F1F1} {\cellcolor{white}}  & {\cellcolor[HTML]{FEFEFE}} \color[HTML]{000000} {\cellcolor[HTML]{FCAF93}} \color[HTML]{000000} 0.3 (0.1-0.5) & {\cellcolor[HTML]{FEFEFE}} \color[HTML]{000000} {\cellcolor[HTML]{FCBBA1}} \color[HTML]{000000} 0.3 (0.1-0.5) & {\cellcolor[HTML]{FEFEFE}} \color[HTML]{000000} {\cellcolor[HTML]{BC141A}} \color[HTML]{F1F1F1} 0.8 (0.5-1.0) & {\cellcolor[HTML]{FEFEFE}} \color[HTML]{000000} {\cellcolor[HTML]{67000D}} \color[HTML]{F1F1F1} 1.0 (0.7-1.0) & {\cellcolor[HTML]{FEFEFE}} \color[HTML]{000000} {\cellcolor[HTML]{67000D}} \color[HTML]{F1F1F1} 1.0 (1.0-1.0) \\
{\cellcolor[HTML]{FEFEFE}} \color[HTML]{000000} 4 & {\cellcolor[HTML]{FEFEFE}} \color[HTML]{000000} $\text{A}^\text{301}_\text{OP1}$,$\text{A}^\text{302}_\text{OP2}$-\textit{cis}-2$\text{O}_\text{ph}$ & {\cellcolor[HTML]{FEFEFE}} \color[HTML]{000000} $\text{MG}^\text{519}$ & {\cellcolor[HTML]{FEFEFE}} \color[HTML]{000000} $\text{MG}^\text{504}$ & {\cellcolor[HTML]{FEFEFE}} \color[HTML]{000000} {\cellcolor[HTML]{67000D}} \color[HTML]{F1F1F1} 1.0 & {\cellcolor[HTML]{FEFEFE}} \color[HTML]{000000} {\cellcolor[HTML]{000000}} \color[HTML]{F1F1F1} {\cellcolor{white}}  & {\cellcolor[HTML]{FEFEFE}} \color[HTML]{000000} {\cellcolor[HTML]{67000D}} \color[HTML]{F1F1F1} 1.0 & {\cellcolor[HTML]{FEFEFE}} \color[HTML]{000000} {\cellcolor[HTML]{000000}} \color[HTML]{F1F1F1} {\cellcolor{white}}  & {\cellcolor[HTML]{FEFEFE}} \color[HTML]{000000} {\cellcolor[HTML]{CB181D}} \color[HTML]{F1F1F1} 0.7 (0.5-1.0) & {\cellcolor[HTML]{FEFEFE}} \color[HTML]{000000} {\cellcolor[HTML]{E22E27}} \color[HTML]{F1F1F1} 0.7 (0.4-0.9) & {\cellcolor[HTML]{FEFEFE}} \color[HTML]{000000} {\cellcolor[HTML]{FC8060}} \color[HTML]{F1F1F1} 0.4 (0.2-0.7) & {\cellcolor[HTML]{FEFEFE}} \color[HTML]{000000} {\cellcolor[HTML]{6D010E}} \color[HTML]{F1F1F1} 1.0 (0.9-1.0) & {\cellcolor[HTML]{FEFEFE}} \color[HTML]{000000} {\cellcolor[HTML]{000000}} \color[HTML]{F1F1F1} {\cellcolor{white}}  & {\cellcolor[HTML]{FEFEFE}} \color[HTML]{000000} {\cellcolor[HTML]{000000}} \color[HTML]{F1F1F1} {\cellcolor{white}}  & {\cellcolor[HTML]{FEFEFE}} \color[HTML]{000000} {\cellcolor[HTML]{000000}} \color[HTML]{F1F1F1} {\cellcolor{white}}  & {\cellcolor[HTML]{FEFEFE}} \color[HTML]{000000} {\cellcolor[HTML]{FB694A}} \color[HTML]{F1F1F1} 0.5 (0.5-0.7) & {\cellcolor[HTML]{FEFEFE}} \color[HTML]{000000} {\cellcolor[HTML]{F6583E}} \color[HTML]{F1F1F1} 0.6 (0.4-0.8) & {\cellcolor[HTML]{FEFEFE}} \color[HTML]{000000} {\cellcolor[HTML]{F14432}} \color[HTML]{F1F1F1} 0.6 (0.5-0.8) \\
{\cellcolor[HTML]{FEFEFE}} \color[HTML]{000000} 5 & {\cellcolor[HTML]{FEFEFE}} \color[HTML]{000000} $\text{A}^\text{184}_\text{OP1}$,$\text{A}^\text{186}_\text{OP1}$,$\text{A}^\text{187}_\text{OP2}$-\textit{mer}-3$\text{O}_\text{ph}$ & {\cellcolor[HTML]{FEFEFE}} \color[HTML]{000000} $\text{MG}^\text{504}$ & {\cellcolor[HTML]{FEFEFE}} \color[HTML]{000000} $\text{MG}^\text{503}$ & {\cellcolor[HTML]{FEFEFE}} \color[HTML]{000000} {\cellcolor[HTML]{000000}} \color[HTML]{F1F1F1} {\cellcolor{white}}  & {\cellcolor[HTML]{FEFEFE}} \color[HTML]{000000} {\cellcolor[HTML]{000000}} \color[HTML]{F1F1F1} {\cellcolor{white}}  & {\cellcolor[HTML]{FEFEFE}} \color[HTML]{000000} {\cellcolor[HTML]{000000}} \color[HTML]{F1F1F1} {\cellcolor{white}}  & {\cellcolor[HTML]{FEFEFE}} \color[HTML]{000000} {\cellcolor[HTML]{FC9272}} \color[HTML]{000000} 0.4 & {\cellcolor[HTML]{FEFEFE}} \color[HTML]{000000} {\cellcolor[HTML]{000000}} \color[HTML]{F1F1F1} {\cellcolor{white}}  & {\cellcolor[HTML]{FEFEFE}} \color[HTML]{000000} {\cellcolor[HTML]{000000}} \color[HTML]{F1F1F1} {\cellcolor{white}}  & {\cellcolor[HTML]{FEFEFE}} \color[HTML]{000000} {\cellcolor[HTML]{FEE9DF}} \color[HTML]{000000} 0.1 (0.0-0.2) & {\cellcolor[HTML]{FEFEFE}} \color[HTML]{000000} {\cellcolor[HTML]{000000}} \color[HTML]{F1F1F1} {\cellcolor{white}}  & {\cellcolor[HTML]{FEFEFE}} \color[HTML]{000000} {\cellcolor[HTML]{000000}} \color[HTML]{F1F1F1} {\cellcolor{white}}  & {\cellcolor[HTML]{FEFEFE}} \color[HTML]{000000} {\cellcolor[HTML]{FC9373}} \color[HTML]{000000} 0.4 (0.1-0.7) & {\cellcolor[HTML]{FEFEFE}} \color[HTML]{000000} {\cellcolor[HTML]{FCC1A8}} \color[HTML]{000000} 0.2 (0.1-0.4) & {\cellcolor[HTML]{FEFEFE}} \color[HTML]{000000} {\cellcolor[HTML]{000000}} \color[HTML]{F1F1F1} {\cellcolor{white}} {\cellcolor[HTML]{AAAAAA}}  & {\cellcolor[HTML]{FEFEFE}} \color[HTML]{000000} {\cellcolor[HTML]{FFF5F0}} \color[HTML]{000000} 0.0 (0.0-0.2) & {\cellcolor[HTML]{FEFEFE}} \color[HTML]{000000} {\cellcolor[HTML]{FFF5F0}} \color[HTML]{000000} 0.0 (0.0-0.2) \\
{\cellcolor[HTML]{FEFEFE}} \color[HTML]{000000} 6 & {\cellcolor[HTML]{FEFEFE}} \color[HTML]{000000} $\text{U}^\text{259}_\text{OP1}$-$\text{O}_\text{ph}$ & {\cellcolor[HTML]{FEFEFE}} \color[HTML]{000000} $\text{MG}^\text{517}$ & {\cellcolor[HTML]{FEFEFE}} \color[HTML]{000000}  & {\cellcolor[HTML]{FEFEFE}} \color[HTML]{000000} {\cellcolor[HTML]{FCBBA1}} \color[HTML]{000000} 0.2 & {\cellcolor[HTML]{FEFEFE}} \color[HTML]{000000} {\cellcolor[HTML]{FCBBA1}} \color[HTML]{000000} 0.2 & {\cellcolor[HTML]{FEFEFE}} \color[HTML]{000000} {\cellcolor[HTML]{CA181D}} \color[HTML]{F1F1F1} 0.8 & {\cellcolor[HTML]{FEFEFE}} \color[HTML]{000000} {\cellcolor[HTML]{FB694A}} \color[HTML]{F1F1F1} 0.5 & {\cellcolor[HTML]{FEFEFE}} \color[HTML]{000000} {\cellcolor[HTML]{D92523}} \color[HTML]{F1F1F1} 0.7 (0.5-0.8) & {\cellcolor[HTML]{FEFEFE}} \color[HTML]{000000} {\cellcolor[HTML]{7A0510}} \color[HTML]{F1F1F1} 1.0 (0.9-1.0) & {\cellcolor[HTML]{FEFEFE}} \color[HTML]{000000} {\cellcolor[HTML]{B01217}} \color[HTML]{F1F1F1} 0.8 (0.6-1.0) & {\cellcolor[HTML]{FEFEFE}} \color[HTML]{000000} {\cellcolor[HTML]{840711}} \color[HTML]{F1F1F1} 0.9 (0.9-1.0) & {\cellcolor[HTML]{FEFEFE}} \color[HTML]{000000} {\cellcolor[HTML]{FDC6B0}} \color[HTML]{000000} 0.2 (0.0-0.5) & {\cellcolor[HTML]{FEFEFE}} \color[HTML]{000000} {\cellcolor[HTML]{C5171C}} \color[HTML]{F1F1F1} 0.8 (0.5-0.9) & {\cellcolor[HTML]{FEFEFE}} \color[HTML]{000000} {\cellcolor[HTML]{F14331}} \color[HTML]{F1F1F1} 0.6 (0.3-0.8) & {\cellcolor[HTML]{FEFEFE}} \color[HTML]{000000} {\cellcolor[HTML]{FB7A5A}} \color[HTML]{F1F1F1} 0.5 (0.2-0.7) & {\cellcolor[HTML]{FEFEFE}} \color[HTML]{000000} {\cellcolor[HTML]{CA181D}} \color[HTML]{F1F1F1} 0.8 (0.5-0.9) & {\cellcolor[HTML]{FEFEFE}} \color[HTML]{000000} {\cellcolor[HTML]{CA181D}} \color[HTML]{F1F1F1} 0.8 (0.6-1.0) \\
{\cellcolor[HTML]{FEFEFE}} \color[HTML]{000000} 7 & {\cellcolor[HTML]{FEFEFE}} \color[HTML]{000000} $\text{U}^\text{305}_\text{OP1}$-$\text{O}_\text{ph}$ & {\cellcolor[HTML]{FEFEFE}} \color[HTML]{000000} $\text{MG}^\text{518}$ & {\cellcolor[HTML]{FEFEFE}} \color[HTML]{000000}  & {\cellcolor[HTML]{FEFEFE}} \color[HTML]{000000} {\cellcolor[HTML]{67000D}} \color[HTML]{F1F1F1} 1.0 & {\cellcolor[HTML]{FEFEFE}} \color[HTML]{000000} {\cellcolor[HTML]{FEEAE1}} \color[HTML]{000000} 0.1 & {\cellcolor[HTML]{FEFEFE}} \color[HTML]{000000} {\cellcolor[HTML]{67000D}} \color[HTML]{F1F1F1} 1.0 & {\cellcolor[HTML]{FEFEFE}} \color[HTML]{000000} {\cellcolor[HTML]{FEE0D2}} \color[HTML]{000000} 0.1 & {\cellcolor[HTML]{FEFEFE}} \color[HTML]{000000} {\cellcolor[HTML]{C9181D}} \color[HTML]{F1F1F1} 0.8 (0.5-0.9) & {\cellcolor[HTML]{FEFEFE}} \color[HTML]{000000} {\cellcolor[HTML]{E53228}} \color[HTML]{F1F1F1} 0.7 (0.4-0.9) & {\cellcolor[HTML]{FEFEFE}} \color[HTML]{000000} {\cellcolor[HTML]{FA6648}} \color[HTML]{F1F1F1} 0.5 (0.3-0.7) & {\cellcolor[HTML]{FEFEFE}} \color[HTML]{000000} {\cellcolor[HTML]{FC8565}} \color[HTML]{F1F1F1} 0.4 (0.3-0.6) & {\cellcolor[HTML]{FEFEFE}} \color[HTML]{000000} {\cellcolor[HTML]{FCAA8D}} \color[HTML]{000000} 0.3 (0.1-0.6) & {\cellcolor[HTML]{FEFEFE}} \color[HTML]{000000} {\cellcolor[HTML]{FC8767}} \color[HTML]{F1F1F1} 0.4 (0.2-0.7) & {\cellcolor[HTML]{FEFEFE}} \color[HTML]{000000} {\cellcolor[HTML]{FCB99F}} \color[HTML]{000000} 0.3 (0.1-0.4) & {\cellcolor[HTML]{FEFEFE}} \color[HTML]{000000} {\cellcolor[HTML]{67000D}} \color[HTML]{F1F1F1} 1.0 (0.7-1.0) & {\cellcolor[HTML]{FEFEFE}} \color[HTML]{000000} {\cellcolor[HTML]{67000D}} \color[HTML]{F1F1F1} 1.0 (0.5-1.0) & {\cellcolor[HTML]{FEFEFE}} \color[HTML]{000000} {\cellcolor[HTML]{E83429}} \color[HTML]{F1F1F1} 0.7 (0.3-0.9) \\
{\cellcolor[HTML]{FEFEFE}} \color[HTML]{000000} 8 & {\cellcolor[HTML]{FEFEFE}} \color[HTML]{000000} $\text{U}^\text{307}_\text{OP2}$,$\text{A}^\text{308}_\text{OP2}$-\textit{cis}-2$\text{O}_\text{ph}$ & {\cellcolor[HTML]{FEFEFE}} \color[HTML]{000000} $\text{MG}^\text{527}$ & {\cellcolor[HTML]{FEFEFE}} \color[HTML]{000000}  & {\cellcolor[HTML]{FEFEFE}} \color[HTML]{000000} {\cellcolor[HTML]{67000D}} \color[HTML]{F1F1F1} 1.0 & {\cellcolor[HTML]{FEFEFE}} \color[HTML]{000000} {\cellcolor[HTML]{000000}} \color[HTML]{F1F1F1} {\cellcolor{white}}  & {\cellcolor[HTML]{FEFEFE}} \color[HTML]{000000} {\cellcolor[HTML]{67000D}} \color[HTML]{F1F1F1} 1.0 & {\cellcolor[HTML]{FEFEFE}} \color[HTML]{000000} {\cellcolor[HTML]{FEE0D2}} \color[HTML]{000000} 0.1 & {\cellcolor[HTML]{FEFEFE}} \color[HTML]{000000} {\cellcolor[HTML]{67000D}} \color[HTML]{F1F1F1} 1.0 (1.0-1.0) & {\cellcolor[HTML]{FEFEFE}} \color[HTML]{000000} {\cellcolor[HTML]{FB7C5C}} \color[HTML]{F1F1F1} 0.4 (0.2-0.7) & {\cellcolor[HTML]{FEFEFE}} \color[HTML]{000000} {\cellcolor[HTML]{FB7C5C}} \color[HTML]{F1F1F1} 0.4 (0.2-0.7) & {\cellcolor[HTML]{FEFEFE}} \color[HTML]{000000} {\cellcolor[HTML]{FCAD90}} \color[HTML]{000000} 0.3 (0.1-0.5) & {\cellcolor[HTML]{FEFEFE}} \color[HTML]{000000} {\cellcolor[HTML]{000000}} \color[HTML]{F1F1F1} {\cellcolor{white}}  & {\cellcolor[HTML]{FEFEFE}} \color[HTML]{000000} {\cellcolor[HTML]{FEE3D6}} \color[HTML]{000000} 0.1 (0.0-0.3) & {\cellcolor[HTML]{FEFEFE}} \color[HTML]{000000} {\cellcolor[HTML]{FEE3D6}} \color[HTML]{000000} 0.1 (0.0-0.3) & {\cellcolor[HTML]{FEFEFE}} \color[HTML]{000000} {\cellcolor[HTML]{67000D}} \color[HTML]{F1F1F1} 1.0 (0.8-1.0) & {\cellcolor[HTML]{FEFEFE}} \color[HTML]{000000} {\cellcolor[HTML]{67000D}} \color[HTML]{F1F1F1} 1.0 (0.6-1.0) & {\cellcolor[HTML]{FEFEFE}} \color[HTML]{000000} {\cellcolor[HTML]{67000D}} \color[HTML]{F1F1F1} 1.0 (0.7-1.0) \\
{\cellcolor[HTML]{FEFEFE}} \color[HTML]{000000} 9 & {\cellcolor[HTML]{FEFEFE}} \color[HTML]{000000} $\text{A}^\text{172}_\text{OP2}$,$\text{A}^\text{173}_\text{OP2}$-\textit{cis}-2$\text{O}_\text{ph}$ & {\cellcolor[HTML]{FEFEFE}} \color[HTML]{000000} $\text{MG}^\text{523}$ & {\cellcolor[HTML]{FEFEFE}} \color[HTML]{000000}  & {\cellcolor[HTML]{FEFEFE}} \color[HTML]{000000} {\cellcolor[HTML]{67000D}} \color[HTML]{F1F1F1} 1.0 & {\cellcolor[HTML]{FEFEFE}} \color[HTML]{000000} {\cellcolor[HTML]{000000}} \color[HTML]{F1F1F1} {\cellcolor{white}}  & {\cellcolor[HTML]{FEFEFE}} \color[HTML]{000000} {\cellcolor[HTML]{67000D}} \color[HTML]{F1F1F1} 1.0 & {\cellcolor[HTML]{FEFEFE}} \color[HTML]{000000} {\cellcolor[HTML]{000000}} \color[HTML]{F1F1F1} {\cellcolor{white}}  & {\cellcolor[HTML]{FEFEFE}} \color[HTML]{000000} {\cellcolor[HTML]{920A13}} \color[HTML]{F1F1F1} 0.9 (0.7-1.0) & {\cellcolor[HTML]{FEFEFE}} \color[HTML]{000000} {\cellcolor[HTML]{FB7A5A}} \color[HTML]{F1F1F1} 0.5 (0.2-0.6) & {\cellcolor[HTML]{FEFEFE}} \color[HTML]{000000} {\cellcolor[HTML]{FB7555}} \color[HTML]{F1F1F1} 0.5 (0.3-0.6) & {\cellcolor[HTML]{FEFEFE}} \color[HTML]{000000} {\cellcolor[HTML]{FB7A5A}} \color[HTML]{F1F1F1} 0.4 (0.4-0.5) & {\cellcolor[HTML]{FEFEFE}} \color[HTML]{000000} {\cellcolor[HTML]{000000}} \color[HTML]{F1F1F1} {\cellcolor{white}}  & {\cellcolor[HTML]{FEFEFE}} \color[HTML]{000000} {\cellcolor[HTML]{FFF0E8}} \color[HTML]{000000} 0.0 (0.0-0.1) & {\cellcolor[HTML]{FEFEFE}} \color[HTML]{000000} {\cellcolor[HTML]{FDCAB5}} \color[HTML]{000000} 0.2 (0.1-0.3) & {\cellcolor[HTML]{FEFEFE}} \color[HTML]{000000} {\cellcolor[HTML]{E83429}} \color[HTML]{F1F1F1} 0.7 (0.6-0.8) & {\cellcolor[HTML]{FEFEFE}} \color[HTML]{000000} {\cellcolor[HTML]{67000D}} \color[HTML]{F1F1F1} 1.0 (0.8-1.0) & {\cellcolor[HTML]{FEFEFE}} \color[HTML]{000000} {\cellcolor[HTML]{67000D}} \color[HTML]{F1F1F1} 1.0 (0.9-1.0) \\
{\cellcolor[HTML]{FEFEFE}} \color[HTML]{000000} 10 & {\cellcolor[HTML]{FEFEFE}} \color[HTML]{000000} $\text{G}^\text{257}_\text{OP1}$,$\text{U}^\text{258}_\text{OP2}$,$\text{C}^\text{260}_\text{O2}$,$\text{U}^\text{305}_\text{OP2}$-\underline{Cis}-3$\text{O}_\text{ph}$$\text{O}_\text{b}$ & {\cellcolor[HTML]{FEFEFE}} \color[HTML]{000000} $\text{MG}^\text{515}$ & {\cellcolor[HTML]{FEFEFE}} \color[HTML]{000000}  & {\cellcolor[HTML]{FEFEFE}} \color[HTML]{000000} {\cellcolor[HTML]{67000D}} \color[HTML]{F1F1F1} 1.0 & {\cellcolor[HTML]{FEFEFE}} \color[HTML]{000000} {\cellcolor[HTML]{000000}} \color[HTML]{F1F1F1} {\cellcolor{white}}  & {\cellcolor[HTML]{FEFEFE}} \color[HTML]{000000} {\cellcolor[HTML]{840711}} \color[HTML]{F1F1F1} 0.9 & {\cellcolor[HTML]{FEFEFE}} \color[HTML]{000000} {\cellcolor[HTML]{000000}} \color[HTML]{F1F1F1} {\cellcolor{white}}  & {\cellcolor[HTML]{FEFEFE}} \color[HTML]{000000} {\cellcolor[HTML]{BC141A}} \color[HTML]{F1F1F1} 0.8 (0.5-1.0) & {\cellcolor[HTML]{FEFEFE}} \color[HTML]{000000} {\cellcolor[HTML]{FCA183}} \color[HTML]{000000} 0.3 (0.2-0.5) & {\cellcolor[HTML]{FEFEFE}} \color[HTML]{000000} {\cellcolor[HTML]{FC8161}} \color[HTML]{F1F1F1} 0.4 (0.2-0.7) & {\cellcolor[HTML]{FEFEFE}} \color[HTML]{000000} {\cellcolor[HTML]{6D010E}} \color[HTML]{F1F1F1} 1.0 (0.9-1.0) & {\cellcolor[HTML]{FEFEFE}} \color[HTML]{000000} {\cellcolor[HTML]{000000}} \color[HTML]{F1F1F1} {\cellcolor{white}}  & {\cellcolor[HTML]{FEFEFE}} \color[HTML]{000000} {\cellcolor[HTML]{000000}} \color[HTML]{F1F1F1} {\cellcolor{white}}  & {\cellcolor[HTML]{FEFEFE}} \color[HTML]{000000} {\cellcolor[HTML]{000000}} \color[HTML]{F1F1F1} {\cellcolor{white}}  & {\cellcolor[HTML]{FEFEFE}} \color[HTML]{000000} {\cellcolor[HTML]{FC9B7C}} \color[HTML]{000000} 0.4 (0.2-0.5) & {\cellcolor[HTML]{FEFEFE}} \color[HTML]{000000} {\cellcolor[HTML]{FC9B7C}} \color[HTML]{000000} 0.4 (0.2-0.5) & {\cellcolor[HTML]{FEFEFE}} \color[HTML]{000000} {\cellcolor[HTML]{FB694A}} \color[HTML]{F1F1F1} 0.5 (0.4-0.7) \\
{\cellcolor[HTML]{FEFEFE}} \color[HTML]{000000} 11 & {\cellcolor[HTML]{FEFEFE}} \color[HTML]{000000} $\text{A}^\text{171}_\text{OP2}$-$\text{O}_\text{ph}$ & {\cellcolor[HTML]{FEFEFE}} \color[HTML]{000000} $\text{MG}^\text{522}$ & {\cellcolor[HTML]{FEFEFE}} \color[HTML]{000000}  & {\cellcolor[HTML]{FEFEFE}} \color[HTML]{000000} {\cellcolor[HTML]{FB694A}} \color[HTML]{F1F1F1} 0.5 & {\cellcolor[HTML]{FEFEFE}} \color[HTML]{000000} {\cellcolor[HTML]{000000}} \color[HTML]{F1F1F1} {\cellcolor{white}}  & {\cellcolor[HTML]{FEFEFE}} \color[HTML]{000000} {\cellcolor[HTML]{A30F15}} \color[HTML]{F1F1F1} 0.9 & {\cellcolor[HTML]{FEFEFE}} \color[HTML]{000000} {\cellcolor[HTML]{000000}} \color[HTML]{F1F1F1} {\cellcolor{white}}  & {\cellcolor[HTML]{FEFEFE}} \color[HTML]{000000} {\cellcolor[HTML]{EE3A2C}} \color[HTML]{F1F1F1} 0.6 (0.3-0.9) & {\cellcolor[HTML]{FEFEFE}} \color[HTML]{000000} {\cellcolor[HTML]{A50F15}} \color[HTML]{F1F1F1} 0.9 (0.6-1.0) & {\cellcolor[HTML]{FEFEFE}} \color[HTML]{000000} {\cellcolor[HTML]{FB7252}} \color[HTML]{F1F1F1} 0.5 (0.2-0.8) & {\cellcolor[HTML]{FEFEFE}} \color[HTML]{000000} {\cellcolor[HTML]{B81419}} \color[HTML]{F1F1F1} 0.8 (0.7-0.9) & {\cellcolor[HTML]{FEFEFE}} \color[HTML]{000000} {\cellcolor[HTML]{FC9C7D}} \color[HTML]{000000} 0.3 (0.1-0.7) & {\cellcolor[HTML]{FEFEFE}} \color[HTML]{000000} {\cellcolor[HTML]{F14432}} \color[HTML]{F1F1F1} 0.6 (0.4-0.8) & {\cellcolor[HTML]{FEFEFE}} \color[HTML]{000000} {\cellcolor[HTML]{FC997A}} \color[HTML]{000000} 0.4 (0.2-0.6) & {\cellcolor[HTML]{FEFEFE}} \color[HTML]{000000} {\cellcolor[HTML]{FFF5F0}} \color[HTML]{000000} 0.0 (0.0-0.4) & {\cellcolor[HTML]{FEFEFE}} \color[HTML]{000000} {\cellcolor[HTML]{FFF5F0}} \color[HTML]{000000} 0.0 (0.0-0.1) & {\cellcolor[HTML]{FEFEFE}} \color[HTML]{000000} {\cellcolor[HTML]{FFF5F0}} \color[HTML]{000000} 0.0 (0.0-0.2) \\
{\cellcolor[HTML]{FEFEFE}} \color[HTML]{000000} 12 & {\cellcolor[HTML]{FEFEFE}} \color[HTML]{000000} $\text{A}^\text{214}_\text{OP2}$-$\text{O}_\text{ph}$ & {\cellcolor[HTML]{FEFEFE}} \color[HTML]{000000} $\text{MG}^\text{514}$ & {\cellcolor[HTML]{FEFEFE}} \color[HTML]{000000}  & {\cellcolor[HTML]{FEFEFE}} \color[HTML]{000000} {\cellcolor[HTML]{67000D}} \color[HTML]{F1F1F1} 1.0 & {\cellcolor[HTML]{FEFEFE}} \color[HTML]{000000} {\cellcolor[HTML]{000000}} \color[HTML]{F1F1F1} {\cellcolor{white}}  & {\cellcolor[HTML]{FEFEFE}} \color[HTML]{000000} {\cellcolor[HTML]{67000D}} \color[HTML]{F1F1F1} 1.0 & {\cellcolor[HTML]{FEFEFE}} \color[HTML]{000000} {\cellcolor[HTML]{000000}} \color[HTML]{F1F1F1} {\cellcolor{white}}  & {\cellcolor[HTML]{FEFEFE}} \color[HTML]{000000} {\cellcolor[HTML]{67000D}} \color[HTML]{F1F1F1} 1.0 (1.0-1.0) & {\cellcolor[HTML]{FEFEFE}} \color[HTML]{000000} {\cellcolor[HTML]{FB7858}} \color[HTML]{F1F1F1} 0.5 (0.2-0.7) & {\cellcolor[HTML]{FEFEFE}} \color[HTML]{000000} {\cellcolor[HTML]{FA6547}} \color[HTML]{F1F1F1} 0.5 (0.3-0.8) & {\cellcolor[HTML]{FEFEFE}} \color[HTML]{000000} {\cellcolor[HTML]{FCC2AA}} \color[HTML]{000000} 0.2 (0.1-0.4) & {\cellcolor[HTML]{FEFEFE}} \color[HTML]{000000} {\cellcolor[HTML]{000000}} \color[HTML]{F1F1F1} {\cellcolor{white}}  & {\cellcolor[HTML]{FEFEFE}} \color[HTML]{000000} {\cellcolor[HTML]{FED9C9}} \color[HTML]{000000} 0.1 (0.0-0.3) & {\cellcolor[HTML]{FEFEFE}} \color[HTML]{000000} {\cellcolor[HTML]{FEE2D5}} \color[HTML]{000000} 0.1 (0.1-0.2) & {\cellcolor[HTML]{FEFEFE}} \color[HTML]{000000} {\cellcolor[HTML]{FC8A6A}} \color[HTML]{F1F1F1} 0.4 (0.4-0.5) & {\cellcolor[HTML]{FEFEFE}} \color[HTML]{000000} {\cellcolor[HTML]{FCBBA1}} \color[HTML]{000000} 0.2 (0.1-0.5) & {\cellcolor[HTML]{FEFEFE}} \color[HTML]{000000} {\cellcolor[HTML]{FED9C9}} \color[HTML]{000000} 0.2 (0.1-0.4) \\
{\cellcolor[HTML]{FEFEFE}} \color[HTML]{000000} 13 & {\cellcolor[HTML]{FEFEFE}} \color[HTML]{000000} $\text{G}^\text{149}_\text{OP2}$-$\text{O}_\text{ph}$ & {\cellcolor[HTML]{FEFEFE}} \color[HTML]{000000} $\text{MG}^\text{510}$ & {\cellcolor[HTML]{FEFEFE}} \color[HTML]{000000}  & {\cellcolor[HTML]{FEFEFE}} \color[HTML]{000000} {\cellcolor[HTML]{67000D}} \color[HTML]{F1F1F1} 1.0 & {\cellcolor[HTML]{FEFEFE}} \color[HTML]{000000} {\cellcolor[HTML]{000000}} \color[HTML]{F1F1F1} {\cellcolor{white}}  & {\cellcolor[HTML]{FEFEFE}} \color[HTML]{000000} {\cellcolor[HTML]{67000D}} \color[HTML]{F1F1F1} 1.0 & {\cellcolor[HTML]{FEFEFE}} \color[HTML]{000000} {\cellcolor[HTML]{000000}} \color[HTML]{F1F1F1} {\cellcolor{white}}  & {\cellcolor[HTML]{FEFEFE}} \color[HTML]{000000} {\cellcolor[HTML]{BC141A}} \color[HTML]{F1F1F1} 0.8 (0.5-1.0) & {\cellcolor[HTML]{FEFEFE}} \color[HTML]{000000} {\cellcolor[HTML]{FDCCB8}} \color[HTML]{000000} 0.2 (0.1-0.3) & {\cellcolor[HTML]{FEFEFE}} \color[HTML]{000000} {\cellcolor[HTML]{FC7F5F}} \color[HTML]{F1F1F1} 0.4 (0.2-0.6) & {\cellcolor[HTML]{FEFEFE}} \color[HTML]{000000} {\cellcolor[HTML]{FEE5D9}} \color[HTML]{000000} 0.1 (0.1-0.1) & {\cellcolor[HTML]{FEFEFE}} \color[HTML]{000000} {\cellcolor[HTML]{000000}} \color[HTML]{F1F1F1} {\cellcolor{white}}  & {\cellcolor[HTML]{FEFEFE}} \color[HTML]{000000} {\cellcolor[HTML]{FFEDE5}} \color[HTML]{000000} 0.0 (0.0-0.1) & {\cellcolor[HTML]{FEFEFE}} \color[HTML]{000000} {\cellcolor[HTML]{FEE3D7}} \color[HTML]{000000} 0.1 (0.0-0.2) & {\cellcolor[HTML]{FEFEFE}} \color[HTML]{000000} {\cellcolor[HTML]{F6583E}} \color[HTML]{F1F1F1} 0.6 (0.4-0.7) & {\cellcolor[HTML]{FEFEFE}} \color[HTML]{000000} {\cellcolor[HTML]{FC8A6A}} \color[HTML]{F1F1F1} 0.4 (0.3-0.5) & {\cellcolor[HTML]{FEFEFE}} \color[HTML]{000000} {\cellcolor[HTML]{FC8A6A}} \color[HTML]{F1F1F1} 0.4 (0.3-0.5) \\
{\cellcolor[HTML]{FEFEFE}} \color[HTML]{000000} 14 & {\cellcolor[HTML]{FEFEFE}} \color[HTML]{000000} $\text{A}^\text{207}_\text{OP1}$,$\text{C}^\text{262}_\text{OP2}$-\textit{trans}-2$\text{O}_\text{ph}$ & {\cellcolor[HTML]{FEFEFE}} \color[HTML]{000000} $\text{MG}^\text{502}$ & {\cellcolor[HTML]{FEFEFE}} \color[HTML]{000000}  & {\cellcolor[HTML]{FEFEFE}} \color[HTML]{000000} {\cellcolor[HTML]{FCBBA1}} \color[HTML]{000000} 0.2 & {\cellcolor[HTML]{FEFEFE}} \color[HTML]{000000} {\cellcolor[HTML]{000000}} \color[HTML]{F1F1F1} {\cellcolor{white}}  & {\cellcolor[HTML]{FEFEFE}} \color[HTML]{000000} {\cellcolor[HTML]{B71319}} \color[HTML]{F1F1F1} 0.8 & {\cellcolor[HTML]{FEFEFE}} \color[HTML]{000000} {\cellcolor[HTML]{000000}} \color[HTML]{F1F1F1} {\cellcolor{white}}  & {\cellcolor[HTML]{FEFEFE}} \color[HTML]{000000} {\cellcolor[HTML]{FA6547}} \color[HTML]{F1F1F1} 0.5 (0.2-0.8) & {\cellcolor[HTML]{FEFEFE}} \color[HTML]{000000} {\cellcolor[HTML]{D92523}} \color[HTML]{F1F1F1} 0.7 (0.5-0.9) & {\cellcolor[HTML]{FEFEFE}} \color[HTML]{000000} {\cellcolor[HTML]{FEE6DA}} \color[HTML]{000000} 0.1 (0.0-0.2) & {\cellcolor[HTML]{FEFEFE}} \color[HTML]{000000} {\cellcolor[HTML]{B91419}} \color[HTML]{F1F1F1} 0.8 (0.6-0.9) & {\cellcolor[HTML]{FEFEFE}} \color[HTML]{000000} {\cellcolor[HTML]{000000}} \color[HTML]{F1F1F1} {\cellcolor{white}}  & {\cellcolor[HTML]{FEFEFE}} \color[HTML]{000000} {\cellcolor[HTML]{FEEAE0}} \color[HTML]{000000} 0.1 (0.0-0.2) & {\cellcolor[HTML]{FEFEFE}} \color[HTML]{000000} {\cellcolor[HTML]{FDD0BC}} \color[HTML]{000000} 0.2 (0.1-0.4) & {\cellcolor[HTML]{FEFEFE}} \color[HTML]{000000} {\cellcolor[HTML]{FB694A}} \color[HTML]{F1F1F1} 0.5 (0.4-0.7) & {\cellcolor[HTML]{FEFEFE}} \color[HTML]{000000} {\cellcolor[HTML]{F6583E}} \color[HTML]{F1F1F1} 0.6 (0.4-0.8) & {\cellcolor[HTML]{FEFEFE}} \color[HTML]{000000} {\cellcolor[HTML]{FB694A}} \color[HTML]{F1F1F1} 0.5 (0.4-0.8) \\
{\cellcolor[HTML]{FEFEFE}} \color[HTML]{000000} 15 & {\cellcolor[HTML]{FEFEFE}} \color[HTML]{000000} $\text{U}^\text{258}_\text{OP1}$-$\text{O}_\text{ph}$ & {\cellcolor[HTML]{FEFEFE}} \color[HTML]{000000} $\text{MG}^\text{516}$ & {\cellcolor[HTML]{FEFEFE}} \color[HTML]{000000}  & {\cellcolor[HTML]{FEFEFE}} \color[HTML]{000000} {\cellcolor[HTML]{FEEAE1}} \color[HTML]{000000} 0.1 & {\cellcolor[HTML]{FEFEFE}} \color[HTML]{000000} {\cellcolor[HTML]{FEE0D2}} \color[HTML]{000000} 0.1 & {\cellcolor[HTML]{FEFEFE}} \color[HTML]{000000} {\cellcolor[HTML]{FCBBA1}} \color[HTML]{000000} 0.2 & {\cellcolor[HTML]{FEFEFE}} \color[HTML]{000000} {\cellcolor[HTML]{000000}} \color[HTML]{F1F1F1} {\cellcolor{white}}  & {\cellcolor[HTML]{FEFEFE}} \color[HTML]{000000} {\cellcolor[HTML]{FEE7DB}} \color[HTML]{000000} 0.1 (0.0-0.3) & {\cellcolor[HTML]{FEFEFE}} \color[HTML]{000000} {\cellcolor[HTML]{F44F39}} \color[HTML]{F1F1F1} 0.6 (0.3-0.9) & {\cellcolor[HTML]{FEFEFE}} \color[HTML]{000000} {\cellcolor[HTML]{D42121}} \color[HTML]{F1F1F1} 0.7 (0.4-0.9) & {\cellcolor[HTML]{FEFEFE}} \color[HTML]{000000} {\cellcolor[HTML]{FC8D6D}} \color[HTML]{F1F1F1} 0.4 (0.2-0.7) & {\cellcolor[HTML]{FEFEFE}} \color[HTML]{000000} {\cellcolor[HTML]{FCBBA1}} \color[HTML]{000000} 0.2 (0.0-0.5) & {\cellcolor[HTML]{FEFEFE}} \color[HTML]{000000} {\cellcolor[HTML]{FED9C9}} \color[HTML]{000000} 0.2 (0.0-0.4) & {\cellcolor[HTML]{FEFEFE}} \color[HTML]{000000} {\cellcolor[HTML]{FFF4EE}} \color[HTML]{000000} 0.0 (0.0-0.1) & {\cellcolor[HTML]{FEFEFE}} \color[HTML]{000000} {\cellcolor[HTML]{FB694A}} \color[HTML]{F1F1F1} 0.5 (0.2-0.6) & {\cellcolor[HTML]{FEFEFE}} \color[HTML]{000000} {\cellcolor[HTML]{BC141A}} \color[HTML]{F1F1F1} 0.8 (0.7-1.0) & {\cellcolor[HTML]{FEFEFE}} \color[HTML]{000000} {\cellcolor[HTML]{FB694A}} \color[HTML]{F1F1F1} 0.5 (0.4-0.7) \\
{\cellcolor[HTML]{FEFEFE}} \color[HTML]{000000} 16 & {\cellcolor[HTML]{FEFEFE}} \color[HTML]{000000} $\text{G}^\text{163}_\text{N7}$,$\text{G}^\text{164}_\text{O6}$-\textit{cis}-$\text{O}_\text{b}$$\text{N}_\text{b}$ & {\cellcolor[HTML]{FEFEFE}} \color[HTML]{000000} $\text{MG}^\text{507}$ & {\cellcolor[HTML]{FEFEFE}} \color[HTML]{000000}  & {\cellcolor[HTML]{FEFEFE}} \color[HTML]{000000} {\cellcolor[HTML]{000000}} \color[HTML]{F1F1F1} {\cellcolor{white}}  & {\cellcolor[HTML]{FEFEFE}} \color[HTML]{000000} {\cellcolor[HTML]{000000}} \color[HTML]{F1F1F1} {\cellcolor{white}}  & {\cellcolor[HTML]{FEFEFE}} \color[HTML]{000000} {\cellcolor[HTML]{000000}} \color[HTML]{F1F1F1} {\cellcolor{white}}  & {\cellcolor[HTML]{FEFEFE}} \color[HTML]{000000} {\cellcolor[HTML]{000000}} \color[HTML]{F1F1F1} {\cellcolor{white}}  & {\cellcolor[HTML]{FEFEFE}} \color[HTML]{000000} {\cellcolor[HTML]{000000}} \color[HTML]{F1F1F1} {\cellcolor{white}}  & {\cellcolor[HTML]{FEFEFE}} \color[HTML]{000000} {\cellcolor[HTML]{000000}} \color[HTML]{F1F1F1} {\cellcolor{white}}  & {\cellcolor[HTML]{FEFEFE}} \color[HTML]{000000} {\cellcolor[HTML]{000000}} \color[HTML]{F1F1F1} {\cellcolor{white}}  & {\cellcolor[HTML]{FEFEFE}} \color[HTML]{000000} {\cellcolor[HTML]{000000}} \color[HTML]{F1F1F1} {\cellcolor{white}}  & {\cellcolor[HTML]{FEFEFE}} \color[HTML]{000000} {\cellcolor[HTML]{000000}} \color[HTML]{F1F1F1} {\cellcolor{white}}  & {\cellcolor[HTML]{FEFEFE}} \color[HTML]{000000} {\cellcolor[HTML]{000000}} \color[HTML]{F1F1F1} {\cellcolor{white}}  & {\cellcolor[HTML]{FEFEFE}} \color[HTML]{000000} {\cellcolor[HTML]{000000}} \color[HTML]{F1F1F1} {\cellcolor{white}}  & {\cellcolor[HTML]{FEFEFE}} \color[HTML]{000000} {\cellcolor[HTML]{000000}} \color[HTML]{F1F1F1} {\cellcolor{white}} {\cellcolor[HTML]{AAAAAA}}  & {\cellcolor[HTML]{FEFEFE}} \color[HTML]{000000} {\cellcolor[HTML]{000000}} \color[HTML]{F1F1F1} {\cellcolor{white}} {\cellcolor[HTML]{AAAAAA}}  & {\cellcolor[HTML]{FEFEFE}} \color[HTML]{000000} {\cellcolor[HTML]{000000}} \color[HTML]{F1F1F1} {\cellcolor{white}} {\cellcolor[HTML]{AAAAAA}}  \\
{\cellcolor[HTML]{FEFEFE}} \color[HTML]{000000} 17 & {\cellcolor[HTML]{FEFEFE}} \color[HTML]{000000} $\text{G}^\text{257}_\text{OP1}$,$\text{U}^\text{258}_\text{OP2}$,$\text{U}^\text{305}_\text{OP2}$-\textit{mer}-3$\text{O}_\text{ph}$ & {\cellcolor[HTML]{FEFEFE}} \color[HTML]{000000} {\cellcolor{white}}  & {\cellcolor[HTML]{FEFEFE}} \color[HTML]{000000} $\text{MG}^\text{501}$ & {\cellcolor[HTML]{FEFEFE}} \color[HTML]{000000} {\cellcolor[HTML]{000000}} \color[HTML]{F1F1F1} {\cellcolor{white}}  & {\cellcolor[HTML]{FEFEFE}} \color[HTML]{000000} {\cellcolor[HTML]{000000}} \color[HTML]{F1F1F1} {\cellcolor{white}}  & {\cellcolor[HTML]{FEFEFE}} \color[HTML]{000000} {\cellcolor[HTML]{000000}} \color[HTML]{F1F1F1} {\cellcolor{white}}  & {\cellcolor[HTML]{FEFEFE}} \color[HTML]{000000} {\cellcolor[HTML]{FC9272}} \color[HTML]{000000} 0.4 & {\cellcolor[HTML]{FEFEFE}} \color[HTML]{000000} {\cellcolor[HTML]{FEE5D9}} \color[HTML]{000000} 0.1 (0.0-0.3) & {\cellcolor[HTML]{FEFEFE}} \color[HTML]{000000} {\cellcolor[HTML]{FED9C9}} \color[HTML]{000000} 0.1 (0.1-0.3) & {\cellcolor[HTML]{FEFEFE}} \color[HTML]{000000} {\cellcolor[HTML]{FED8C7}} \color[HTML]{000000} 0.2 (0.0-0.3) & {\cellcolor[HTML]{FEFEFE}} \color[HTML]{000000} {\cellcolor[HTML]{FFF5F0}} \color[HTML]{000000} 0.0 (0.0-0.0) & {\cellcolor[HTML]{FEFEFE}} \color[HTML]{000000} {\cellcolor[HTML]{000000}} \color[HTML]{F1F1F1} {\cellcolor{white}}  & {\cellcolor[HTML]{FEFEFE}} \color[HTML]{000000} {\cellcolor[HTML]{FCB99F}} \color[HTML]{000000} 0.3 (0.1-0.6) & {\cellcolor[HTML]{FEFEFE}} \color[HTML]{000000} {\cellcolor[HTML]{FCC2AA}} \color[HTML]{000000} 0.2 (0.1-0.4) & {\cellcolor[HTML]{FEFEFE}} \color[HTML]{000000} {\cellcolor[HTML]{FCBBA1}} \color[HTML]{000000} 0.2 (0.2-0.4) & {\cellcolor[HTML]{FEFEFE}} \color[HTML]{000000} {\cellcolor[HTML]{FDCAB5}} \color[HTML]{000000} 0.2 (0.1-0.5) & {\cellcolor[HTML]{FEFEFE}} \color[HTML]{000000} {\cellcolor[HTML]{FFEDE5}} \color[HTML]{000000} 0.1 (0.0-0.2) \\
{\cellcolor[HTML]{FEFEFE}} \color[HTML]{000000} 18 & {\cellcolor[HTML]{FEFEFE}} \color[HTML]{000000} $\text{A}^\text{187}_\text{OP1}$-$\text{O}_\text{ph}$ & {\cellcolor[HTML]{FEFEFE}} \color[HTML]{000000} {\cellcolor{white}}  & {\cellcolor[HTML]{FEFEFE}} \color[HTML]{000000} $\text{MG}^\text{517}$ & {\cellcolor[HTML]{FEFEFE}} \color[HTML]{000000} {\cellcolor[HTML]{000000}} \color[HTML]{F1F1F1} {\cellcolor{white}}  & {\cellcolor[HTML]{FEFEFE}} \color[HTML]{000000} {\cellcolor[HTML]{000000}} \color[HTML]{F1F1F1} {\cellcolor{white}}  & {\cellcolor[HTML]{FEFEFE}} \color[HTML]{000000} {\cellcolor[HTML]{000000}} \color[HTML]{F1F1F1} {\cellcolor{white}}  & {\cellcolor[HTML]{FEFEFE}} \color[HTML]{000000} {\cellcolor[HTML]{000000}} \color[HTML]{F1F1F1} {\cellcolor{white}}  & {\cellcolor[HTML]{FEFEFE}} \color[HTML]{000000} {\cellcolor[HTML]{FEE6DA}} \color[HTML]{000000} 0.1 (0.0-0.3) & {\cellcolor[HTML]{FEFEFE}} \color[HTML]{000000} {\cellcolor[HTML]{FFF0E8}} \color[HTML]{000000} 0.0 (0.0-0.1) & {\cellcolor[HTML]{FEFEFE}} \color[HTML]{000000} {\cellcolor[HTML]{FEE4D8}} \color[HTML]{000000} 0.1 (0.0-0.2) & {\cellcolor[HTML]{FEFEFE}} \color[HTML]{000000} {\cellcolor[HTML]{FEE7DB}} \color[HTML]{000000} 0.1 (0.0-0.2) & {\cellcolor[HTML]{FEFEFE}} \color[HTML]{000000} {\cellcolor[HTML]{FFECE3}} \color[HTML]{000000} 0.1 (0.0-0.2) & {\cellcolor[HTML]{FEFEFE}} \color[HTML]{000000} {\cellcolor[HTML]{FFEBE2}} \color[HTML]{000000} 0.1 (0.0-0.1) & {\cellcolor[HTML]{FEFEFE}} \color[HTML]{000000} {\cellcolor[HTML]{FDC5AE}} \color[HTML]{000000} 0.2 (0.1-0.3) & {\cellcolor[HTML]{FEFEFE}} \color[HTML]{000000} {\cellcolor[HTML]{FFF5F0}} \color[HTML]{000000} 0.0 (0.0-0.2) & {\cellcolor[HTML]{FEFEFE}} \color[HTML]{000000} {\cellcolor[HTML]{FCAB8F}} \color[HTML]{000000} 0.3 (0.0-0.5) & {\cellcolor[HTML]{FEFEFE}} \color[HTML]{000000} {\cellcolor[HTML]{FED9C9}} \color[HTML]{000000} 0.2 (0.0-0.4) \\
{\cellcolor[HTML]{FEFEFE}} \color[HTML]{000000} 19 & {\cellcolor[HTML]{FEFEFE}} \color[HTML]{000000} $\text{C}^\text{262}_\text{OP2}$-$\text{O}_\text{ph}$ & {\cellcolor[HTML]{FEFEFE}} \color[HTML]{000000} {\cellcolor{white}}  & {\cellcolor[HTML]{FEFEFE}} \color[HTML]{000000} $\text{MG}^\text{521}$ & {\cellcolor[HTML]{FEFEFE}} \color[HTML]{000000} {\cellcolor[HTML]{000000}} \color[HTML]{F1F1F1} {\cellcolor{white}}  & {\cellcolor[HTML]{FEFEFE}} \color[HTML]{000000} {\cellcolor[HTML]{000000}} \color[HTML]{F1F1F1} {\cellcolor{white}}  & {\cellcolor[HTML]{FEFEFE}} \color[HTML]{000000} {\cellcolor[HTML]{000000}} \color[HTML]{F1F1F1} {\cellcolor{white}}  & {\cellcolor[HTML]{FEFEFE}} \color[HTML]{000000} {\cellcolor[HTML]{000000}} \color[HTML]{F1F1F1} {\cellcolor{white}}  & {\cellcolor[HTML]{FEFEFE}} \color[HTML]{000000} {\cellcolor[HTML]{FEE9DF}} \color[HTML]{000000} 0.1 (0.0-0.3) & {\cellcolor[HTML]{FEFEFE}} \color[HTML]{000000} {\cellcolor[HTML]{FEE6DA}} \color[HTML]{000000} 0.1 (0.0-0.2) & {\cellcolor[HTML]{FEFEFE}} \color[HTML]{000000} {\cellcolor[HTML]{FCBBA1}} \color[HTML]{000000} 0.3 (0.1-0.4) & {\cellcolor[HTML]{FEFEFE}} \color[HTML]{000000} {\cellcolor[HTML]{FFF1EA}} \color[HTML]{000000} 0.0 (0.0-0.1) & {\cellcolor[HTML]{FEFEFE}} \color[HTML]{000000} {\cellcolor[HTML]{FFEEE6}} \color[HTML]{000000} 0.0 (0.0-0.2) & {\cellcolor[HTML]{FEFEFE}} \color[HTML]{000000} {\cellcolor[HTML]{FEE7DB}} \color[HTML]{000000} 0.1 (0.0-0.2) & {\cellcolor[HTML]{FEFEFE}} \color[HTML]{000000} {\cellcolor[HTML]{FCB89E}} \color[HTML]{000000} 0.3 (0.1-0.4) & {\cellcolor[HTML]{FEFEFE}} \color[HTML]{000000} {\cellcolor[HTML]{FFF5F0}} \color[HTML]{000000} 0.0 (0.0-0.2) & {\cellcolor[HTML]{FEFEFE}} \color[HTML]{000000} {\cellcolor[HTML]{FFF5F0}} \color[HTML]{000000} 0.0 (0.0-0.2) & {\cellcolor[HTML]{FEFEFE}} \color[HTML]{000000} {\cellcolor[HTML]{FCBBA1}} \color[HTML]{000000} 0.2 (0.1-0.5) \\
{\cellcolor[HTML]{FEFEFE}} \color[HTML]{000000} 20 & {\cellcolor[HTML]{FEFEFE}} \color[HTML]{000000} $\text{G}^\text{164}_\text{OP1}$-$\text{O}_\text{ph}$ & {\cellcolor[HTML]{FEFEFE}} \color[HTML]{000000} {\cellcolor{white}}  & {\cellcolor[HTML]{FEFEFE}} \color[HTML]{000000} $\text{MG}^\text{525}$ & {\cellcolor[HTML]{FEFEFE}} \color[HTML]{000000} {\cellcolor[HTML]{000000}} \color[HTML]{F1F1F1} {\cellcolor{white}}  & {\cellcolor[HTML]{FEFEFE}} \color[HTML]{000000} {\cellcolor[HTML]{000000}} \color[HTML]{F1F1F1} {\cellcolor{white}}  & {\cellcolor[HTML]{FEFEFE}} \color[HTML]{000000} {\cellcolor[HTML]{000000}} \color[HTML]{F1F1F1} {\cellcolor{white}}  & {\cellcolor[HTML]{FEFEFE}} \color[HTML]{000000} {\cellcolor[HTML]{FCBBA1}} \color[HTML]{000000} 0.2 & {\cellcolor[HTML]{FEFEFE}} \color[HTML]{000000} {\cellcolor[HTML]{000000}} \color[HTML]{F1F1F1} {\cellcolor{white}}  & {\cellcolor[HTML]{FEFEFE}} \color[HTML]{000000} {\cellcolor[HTML]{000000}} \color[HTML]{F1F1F1} {\cellcolor{white}}  & {\cellcolor[HTML]{FEFEFE}} \color[HTML]{000000} {\cellcolor[HTML]{000000}} \color[HTML]{F1F1F1} {\cellcolor{white}}  & {\cellcolor[HTML]{FEFEFE}} \color[HTML]{000000} {\cellcolor[HTML]{FFF5F0}} \color[HTML]{000000} 0.0 (0.0-0.0) & {\cellcolor[HTML]{FEFEFE}} \color[HTML]{000000} {\cellcolor[HTML]{000000}} \color[HTML]{F1F1F1} {\cellcolor{white}}  & {\cellcolor[HTML]{FEFEFE}} \color[HTML]{000000} {\cellcolor[HTML]{FFF2EB}} \color[HTML]{000000} 0.0 (0.0-0.0) & {\cellcolor[HTML]{FEFEFE}} \color[HTML]{000000} {\cellcolor[HTML]{FFF5F0}} \color[HTML]{000000} 0.0 (0.0-0.0) & {\cellcolor[HTML]{FEFEFE}} \color[HTML]{000000} {\cellcolor[HTML]{000000}} \color[HTML]{F1F1F1} {\cellcolor{white}} {\cellcolor[HTML]{AAAAAA}}  & {\cellcolor[HTML]{FEFEFE}} \color[HTML]{000000} {\cellcolor[HTML]{CA181D}} \color[HTML]{F1F1F1} 0.8 (0.6-0.8) & {\cellcolor[HTML]{FEFEFE}} \color[HTML]{000000} {\cellcolor[HTML]{000000}} \color[HTML]{F1F1F1} {\cellcolor{white}} {\cellcolor[HTML]{AAAAAA}}  \\
{\cellcolor[HTML]{FEFEFE}} \color[HTML]{000000} 21 & {\cellcolor[HTML]{FEFEFE}} \color[HTML]{000000} $\text{C}^\text{255}_\text{OP1}$-$\text{O}_\text{ph}$ & {\cellcolor[HTML]{FEFEFE}} \color[HTML]{000000} {\cellcolor{white}}  & {\cellcolor[HTML]{FEFEFE}} \color[HTML]{000000} $\text{MG}^\text{514}$ & {\cellcolor[HTML]{FEFEFE}} \color[HTML]{000000} {\cellcolor[HTML]{000000}} \color[HTML]{F1F1F1} {\cellcolor{white}}  & {\cellcolor[HTML]{FEFEFE}} \color[HTML]{000000} {\cellcolor[HTML]{000000}} \color[HTML]{F1F1F1} {\cellcolor{white}}  & {\cellcolor[HTML]{FEFEFE}} \color[HTML]{000000} {\cellcolor[HTML]{000000}} \color[HTML]{F1F1F1} {\cellcolor{white}}  & {\cellcolor[HTML]{FEFEFE}} \color[HTML]{000000} {\cellcolor[HTML]{000000}} \color[HTML]{F1F1F1} {\cellcolor{white}}  & {\cellcolor[HTML]{FEFEFE}} \color[HTML]{000000} {\cellcolor[HTML]{000000}} \color[HTML]{F1F1F1} {\cellcolor{white}}  & {\cellcolor[HTML]{FEFEFE}} \color[HTML]{000000} {\cellcolor[HTML]{FFEEE7}} \color[HTML]{000000} 0.0 (0.0-0.1) & {\cellcolor[HTML]{FEFEFE}} \color[HTML]{000000} {\cellcolor[HTML]{FFF1EA}} \color[HTML]{000000} 0.0 (0.0-0.1) & {\cellcolor[HTML]{FEFEFE}} \color[HTML]{000000} {\cellcolor[HTML]{FFF3ED}} \color[HTML]{000000} 0.0 (0.0-0.0) & {\cellcolor[HTML]{FEFEFE}} \color[HTML]{000000} {\cellcolor[HTML]{000000}} \color[HTML]{F1F1F1} {\cellcolor{white}}  & {\cellcolor[HTML]{FEFEFE}} \color[HTML]{000000} {\cellcolor[HTML]{FFF4EE}} \color[HTML]{000000} 0.0 (0.0-0.0) & {\cellcolor[HTML]{FEFEFE}} \color[HTML]{000000} {\cellcolor[HTML]{FFEEE7}} \color[HTML]{000000} 0.0 (0.0-0.1) & {\cellcolor[HTML]{FEFEFE}} \color[HTML]{000000} {\cellcolor[HTML]{000000}} \color[HTML]{F1F1F1} {\cellcolor{white}} {\cellcolor[HTML]{AAAAAA}}  & {\cellcolor[HTML]{FEFEFE}} \color[HTML]{000000} {\cellcolor[HTML]{FFF5F0}} \color[HTML]{000000} 0.0 (0.0-0.1) & {\cellcolor[HTML]{FEFEFE}} \color[HTML]{000000} {\cellcolor[HTML]{FC8A6A}} \color[HTML]{F1F1F1} 0.4 (0.0-0.6) \\
{\cellcolor[HTML]{FEFEFE}} \color[HTML]{000000} 22 & {\cellcolor[HTML]{FEFEFE}} \color[HTML]{000000} $\text{A}^\text{299}_\text{OP2}$-$\text{O}_\text{ph}$ & {\cellcolor[HTML]{FEFEFE}} \color[HTML]{000000} {\cellcolor{white}}  & {\cellcolor[HTML]{FEFEFE}} \color[HTML]{000000} $\text{MG}^\text{513}$ & {\cellcolor[HTML]{FEFEFE}} \color[HTML]{000000} {\cellcolor[HTML]{000000}} \color[HTML]{F1F1F1} {\cellcolor{white}}  & {\cellcolor[HTML]{FEFEFE}} \color[HTML]{000000} {\cellcolor[HTML]{000000}} \color[HTML]{F1F1F1} {\cellcolor{white}}  & {\cellcolor[HTML]{FEFEFE}} \color[HTML]{000000} {\cellcolor[HTML]{000000}} \color[HTML]{F1F1F1} {\cellcolor{white}}  & {\cellcolor[HTML]{FEFEFE}} \color[HTML]{000000} {\cellcolor[HTML]{000000}} \color[HTML]{F1F1F1} {\cellcolor{white}}  & {\cellcolor[HTML]{FEFEFE}} \color[HTML]{000000} {\cellcolor[HTML]{000000}} \color[HTML]{F1F1F1} {\cellcolor{white}}  & {\cellcolor[HTML]{FEFEFE}} \color[HTML]{000000} {\cellcolor[HTML]{FFF4EF}} \color[HTML]{000000} 0.0 (0.0-0.0) & {\cellcolor[HTML]{FEFEFE}} \color[HTML]{000000} {\cellcolor[HTML]{FFF1EA}} \color[HTML]{000000} 0.0 (0.0-0.1) & {\cellcolor[HTML]{FEFEFE}} \color[HTML]{000000} {\cellcolor[HTML]{FEE5D8}} \color[HTML]{000000} 0.1 (0.0-0.2) & {\cellcolor[HTML]{FEFEFE}} \color[HTML]{000000} {\cellcolor[HTML]{000000}} \color[HTML]{F1F1F1} {\cellcolor{white}}  & {\cellcolor[HTML]{FEFEFE}} \color[HTML]{000000} {\cellcolor[HTML]{FEE9DF}} \color[HTML]{000000} 0.1 (0.0-0.2) & {\cellcolor[HTML]{FEFEFE}} \color[HTML]{000000} {\cellcolor[HTML]{FFECE3}} \color[HTML]{000000} 0.1 (0.0-0.1) & {\cellcolor[HTML]{FEFEFE}} \color[HTML]{000000} {\cellcolor[HTML]{000000}} \color[HTML]{F1F1F1} {\cellcolor{white}} {\cellcolor[HTML]{AAAAAA}}  & {\cellcolor[HTML]{FEFEFE}} \color[HTML]{000000} {\cellcolor[HTML]{FFF5F0}} \color[HTML]{000000} 0.0 (0.0-0.1) & {\cellcolor[HTML]{FEFEFE}} \color[HTML]{000000} {\cellcolor[HTML]{FFF5F0}} \color[HTML]{000000} 0.0 (0.0-0.0) \\
{\cellcolor[HTML]{FEFEFE}} \color[HTML]{000000} 23 & {\cellcolor[HTML]{FEFEFE}} \color[HTML]{000000} $\text{U}^\text{271}_\text{OP1}$-$\text{O}_\text{ph}$ & {\cellcolor[HTML]{FEFEFE}} \color[HTML]{000000} {\cellcolor{white}}  & {\cellcolor[HTML]{FEFEFE}} \color[HTML]{000000} $\text{MG}^\text{520}$ & {\cellcolor[HTML]{FEFEFE}} \color[HTML]{000000} {\cellcolor[HTML]{000000}} \color[HTML]{F1F1F1} {\cellcolor{white}}  & {\cellcolor[HTML]{FEFEFE}} \color[HTML]{000000} {\cellcolor[HTML]{000000}} \color[HTML]{F1F1F1} {\cellcolor{white}}  & {\cellcolor[HTML]{FEFEFE}} \color[HTML]{000000} {\cellcolor[HTML]{000000}} \color[HTML]{F1F1F1} {\cellcolor{white}}  & {\cellcolor[HTML]{FEFEFE}} \color[HTML]{000000} {\cellcolor[HTML]{000000}} \color[HTML]{F1F1F1} {\cellcolor{white}}  & {\cellcolor[HTML]{FEFEFE}} \color[HTML]{000000} {\cellcolor[HTML]{000000}} \color[HTML]{F1F1F1} {\cellcolor{white}}  & {\cellcolor[HTML]{FEFEFE}} \color[HTML]{000000} {\cellcolor[HTML]{000000}} \color[HTML]{F1F1F1} {\cellcolor{white}}  & {\cellcolor[HTML]{FEFEFE}} \color[HTML]{000000} {\cellcolor[HTML]{FFF1EA}} \color[HTML]{000000} 0.0 (0.0-0.1) & {\cellcolor[HTML]{FEFEFE}} \color[HTML]{000000} {\cellcolor[HTML]{FFF4EF}} \color[HTML]{000000} 0.0 (0.0-0.0) & {\cellcolor[HTML]{FEFEFE}} \color[HTML]{000000} {\cellcolor[HTML]{000000}} \color[HTML]{F1F1F1} {\cellcolor{white}}  & {\cellcolor[HTML]{FEFEFE}} \color[HTML]{000000} {\cellcolor[HTML]{000000}} \color[HTML]{F1F1F1} {\cellcolor{white}}  & {\cellcolor[HTML]{FEFEFE}} \color[HTML]{000000} {\cellcolor[HTML]{FFF0E9}} \color[HTML]{000000} 0.0 (0.0-0.1) & {\cellcolor[HTML]{FEFEFE}} \color[HTML]{000000} {\cellcolor[HTML]{000000}} \color[HTML]{F1F1F1} {\cellcolor{white}} {\cellcolor[HTML]{AAAAAA}}  & {\cellcolor[HTML]{FEFEFE}} \color[HTML]{000000} {\cellcolor[HTML]{000000}} \color[HTML]{F1F1F1} {\cellcolor{white}} {\cellcolor[HTML]{AAAAAA}}  & {\cellcolor[HTML]{FEFEFE}} \color[HTML]{000000} {\cellcolor[HTML]{FFF5F0}} \color[HTML]{000000} 0.0 (0.0-0.0) \\
{\cellcolor[HTML]{FEFEFE}} \color[HTML]{000000} 24 & {\cellcolor[HTML]{FEFEFE}} \color[HTML]{000000} $\text{C}^\text{274}_\text{OP2}$-$\text{O}_\text{ph}$ & {\cellcolor[HTML]{FEFEFE}} \color[HTML]{000000} {\cellcolor{white}}  & {\cellcolor[HTML]{FEFEFE}} \color[HTML]{000000} $\text{MG}^\text{512}$ & {\cellcolor[HTML]{FEFEFE}} \color[HTML]{000000} {\cellcolor[HTML]{000000}} \color[HTML]{F1F1F1} {\cellcolor{white}}  & {\cellcolor[HTML]{FEFEFE}} \color[HTML]{000000} {\cellcolor[HTML]{000000}} \color[HTML]{F1F1F1} {\cellcolor{white}}  & {\cellcolor[HTML]{FEFEFE}} \color[HTML]{000000} {\cellcolor[HTML]{000000}} \color[HTML]{F1F1F1} {\cellcolor{white}}  & {\cellcolor[HTML]{FEFEFE}} \color[HTML]{000000} {\cellcolor[HTML]{000000}} \color[HTML]{F1F1F1} {\cellcolor{white}}  & {\cellcolor[HTML]{FEFEFE}} \color[HTML]{000000} {\cellcolor[HTML]{000000}} \color[HTML]{F1F1F1} {\cellcolor{white}}  & {\cellcolor[HTML]{FEFEFE}} \color[HTML]{000000} {\cellcolor[HTML]{000000}} \color[HTML]{F1F1F1} {\cellcolor{white}}  & {\cellcolor[HTML]{FEFEFE}} \color[HTML]{000000} {\cellcolor[HTML]{000000}} \color[HTML]{F1F1F1} {\cellcolor{white}}  & {\cellcolor[HTML]{FEFEFE}} \color[HTML]{000000} {\cellcolor[HTML]{000000}} \color[HTML]{F1F1F1} {\cellcolor{white}}  & {\cellcolor[HTML]{FEFEFE}} \color[HTML]{000000} {\cellcolor[HTML]{000000}} \color[HTML]{F1F1F1} {\cellcolor{white}}  & {\cellcolor[HTML]{FEFEFE}} \color[HTML]{000000} {\cellcolor[HTML]{000000}} \color[HTML]{F1F1F1} {\cellcolor{white}}  & {\cellcolor[HTML]{FEFEFE}} \color[HTML]{000000} {\cellcolor[HTML]{FFF4EF}} \color[HTML]{000000} 0.0 (0.0-0.0) & {\cellcolor[HTML]{FEFEFE}} \color[HTML]{000000} {\cellcolor[HTML]{000000}} \color[HTML]{F1F1F1} {\cellcolor{white}} {\cellcolor[HTML]{AAAAAA}}  & {\cellcolor[HTML]{FEFEFE}} \color[HTML]{000000} {\cellcolor[HTML]{000000}} \color[HTML]{F1F1F1} {\cellcolor{white}} {\cellcolor[HTML]{AAAAAA}}  & {\cellcolor[HTML]{FEFEFE}} \color[HTML]{000000} {\cellcolor[HTML]{FFF5F0}} \color[HTML]{000000} 0.0 (0.0-0.0) \\
{\cellcolor[HTML]{FEFEFE}} \color[HTML]{000000} 25 & {\cellcolor[HTML]{FEFEFE}} \color[HTML]{000000} $\text{A}^\text{178}_\text{N1}$-$\text{N}_\text{b}$ & {\cellcolor[HTML]{FEFEFE}} \color[HTML]{000000} {\cellcolor{white}}  & {\cellcolor[HTML]{FEFEFE}} \color[HTML]{000000} $\text{MG}^\text{519}$ & {\cellcolor[HTML]{FEFEFE}} \color[HTML]{000000} {\cellcolor[HTML]{000000}} \color[HTML]{F1F1F1} {\cellcolor{white}}  & {\cellcolor[HTML]{FEFEFE}} \color[HTML]{000000} {\cellcolor[HTML]{000000}} \color[HTML]{F1F1F1} {\cellcolor{white}}  & {\cellcolor[HTML]{FEFEFE}} \color[HTML]{000000} {\cellcolor[HTML]{000000}} \color[HTML]{F1F1F1} {\cellcolor{white}}  & {\cellcolor[HTML]{FEFEFE}} \color[HTML]{000000} {\cellcolor[HTML]{000000}} \color[HTML]{F1F1F1} {\cellcolor{white}}  & {\cellcolor[HTML]{FEFEFE}} \color[HTML]{000000} {\cellcolor[HTML]{000000}} \color[HTML]{F1F1F1} {\cellcolor{white}}  & {\cellcolor[HTML]{FEFEFE}} \color[HTML]{000000} {\cellcolor[HTML]{000000}} \color[HTML]{F1F1F1} {\cellcolor{white}}  & {\cellcolor[HTML]{FEFEFE}} \color[HTML]{000000} {\cellcolor[HTML]{000000}} \color[HTML]{F1F1F1} {\cellcolor{white}}  & {\cellcolor[HTML]{FEFEFE}} \color[HTML]{000000} {\cellcolor[HTML]{000000}} \color[HTML]{F1F1F1} {\cellcolor{white}}  & {\cellcolor[HTML]{FEFEFE}} \color[HTML]{000000} {\cellcolor[HTML]{000000}} \color[HTML]{F1F1F1} {\cellcolor{white}}  & {\cellcolor[HTML]{FEFEFE}} \color[HTML]{000000} {\cellcolor[HTML]{000000}} \color[HTML]{F1F1F1} {\cellcolor{white}}  & {\cellcolor[HTML]{FEFEFE}} \color[HTML]{000000} {\cellcolor[HTML]{000000}} \color[HTML]{F1F1F1} {\cellcolor{white}}  & {\cellcolor[HTML]{FEFEFE}} \color[HTML]{000000} {\cellcolor[HTML]{000000}} \color[HTML]{F1F1F1} {\cellcolor{white}} {\cellcolor[HTML]{AAAAAA}}  & {\cellcolor[HTML]{FEFEFE}} \color[HTML]{000000} {\cellcolor[HTML]{000000}} \color[HTML]{F1F1F1} {\cellcolor{white}} {\cellcolor[HTML]{AAAAAA}}  & {\cellcolor[HTML]{FEFEFE}} \color[HTML]{000000} {\cellcolor[HTML]{000000}} \color[HTML]{F1F1F1} {\cellcolor{white}} {\cellcolor[HTML]{AAAAAA}}  \\
{\cellcolor[HTML]{FEFEFE}} \color[HTML]{000000} 26 & {\cellcolor[HTML]{FEFEFE}} \color[HTML]{000000} $\text{A}^\text{261}_\text{N7}$,$\text{A}^\text{265}_\text{OP2}$-\textit{cis}-$\text{O}_\text{ph}$$\text{N}_\text{b}$ & {\cellcolor[HTML]{FEFEFE}} \color[HTML]{000000} {\cellcolor{white}}  & {\cellcolor[HTML]{FEFEFE}} \color[HTML]{000000} $\text{MG}^\text{509}$ & {\cellcolor[HTML]{FEFEFE}} \color[HTML]{000000} {\cellcolor[HTML]{000000}} \color[HTML]{F1F1F1} {\cellcolor{white}}  & {\cellcolor[HTML]{FEFEFE}} \color[HTML]{000000} {\cellcolor[HTML]{000000}} \color[HTML]{F1F1F1} {\cellcolor{white}}  & {\cellcolor[HTML]{FEFEFE}} \color[HTML]{000000} {\cellcolor[HTML]{000000}} \color[HTML]{F1F1F1} {\cellcolor{white}}  & {\cellcolor[HTML]{FEFEFE}} \color[HTML]{000000} {\cellcolor[HTML]{000000}} \color[HTML]{F1F1F1} {\cellcolor{white}}  & {\cellcolor[HTML]{FEFEFE}} \color[HTML]{000000} {\cellcolor[HTML]{000000}} \color[HTML]{F1F1F1} {\cellcolor{white}}  & {\cellcolor[HTML]{FEFEFE}} \color[HTML]{000000} {\cellcolor[HTML]{000000}} \color[HTML]{F1F1F1} {\cellcolor{white}}  & {\cellcolor[HTML]{FEFEFE}} \color[HTML]{000000} {\cellcolor[HTML]{000000}} \color[HTML]{F1F1F1} {\cellcolor{white}}  & {\cellcolor[HTML]{FEFEFE}} \color[HTML]{000000} {\cellcolor[HTML]{000000}} \color[HTML]{F1F1F1} {\cellcolor{white}}  & {\cellcolor[HTML]{FEFEFE}} \color[HTML]{000000} {\cellcolor[HTML]{000000}} \color[HTML]{F1F1F1} {\cellcolor{white}}  & {\cellcolor[HTML]{FEFEFE}} \color[HTML]{000000} {\cellcolor[HTML]{000000}} \color[HTML]{F1F1F1} {\cellcolor{white}}  & {\cellcolor[HTML]{FEFEFE}} \color[HTML]{000000} {\cellcolor[HTML]{000000}} \color[HTML]{F1F1F1} {\cellcolor{white}}  & {\cellcolor[HTML]{FEFEFE}} \color[HTML]{000000} {\cellcolor[HTML]{000000}} \color[HTML]{F1F1F1} {\cellcolor{white}} {\cellcolor[HTML]{AAAAAA}}  & {\cellcolor[HTML]{FEFEFE}} \color[HTML]{000000} {\cellcolor[HTML]{000000}} \color[HTML]{F1F1F1} {\cellcolor{white}} {\cellcolor[HTML]{AAAAAA}}  & {\cellcolor[HTML]{FEFEFE}} \color[HTML]{000000} {\cellcolor[HTML]{000000}} \color[HTML]{F1F1F1} {\cellcolor{white}} {\cellcolor[HTML]{AAAAAA}}  \\
{\cellcolor[HTML]{FEFEFE}} \color[HTML]{000000} 27 & {\cellcolor[HTML]{FEFEFE}} \color[HTML]{000000} $\text{A}^\text{308}_\text{N7}$-$\text{N}_\text{b}$ & {\cellcolor[HTML]{FEFEFE}} \color[HTML]{000000} {\cellcolor{white}}  & {\cellcolor[HTML]{FEFEFE}} \color[HTML]{000000} $\text{MG}^\text{506}$ & {\cellcolor[HTML]{FEFEFE}} \color[HTML]{000000} {\cellcolor[HTML]{000000}} \color[HTML]{F1F1F1} {\cellcolor{white}}  & {\cellcolor[HTML]{FEFEFE}} \color[HTML]{000000} {\cellcolor[HTML]{000000}} \color[HTML]{F1F1F1} {\cellcolor{white}}  & {\cellcolor[HTML]{FEFEFE}} \color[HTML]{000000} {\cellcolor[HTML]{000000}} \color[HTML]{F1F1F1} {\cellcolor{white}}  & {\cellcolor[HTML]{FEFEFE}} \color[HTML]{000000} {\cellcolor[HTML]{000000}} \color[HTML]{F1F1F1} {\cellcolor{white}}  & {\cellcolor[HTML]{FEFEFE}} \color[HTML]{000000} {\cellcolor[HTML]{000000}} \color[HTML]{F1F1F1} {\cellcolor{white}}  & {\cellcolor[HTML]{FEFEFE}} \color[HTML]{000000} {\cellcolor[HTML]{000000}} \color[HTML]{F1F1F1} {\cellcolor{white}}  & {\cellcolor[HTML]{FEFEFE}} \color[HTML]{000000} {\cellcolor[HTML]{000000}} \color[HTML]{F1F1F1} {\cellcolor{white}}  & {\cellcolor[HTML]{FEFEFE}} \color[HTML]{000000} {\cellcolor[HTML]{000000}} \color[HTML]{F1F1F1} {\cellcolor{white}}  & {\cellcolor[HTML]{FEFEFE}} \color[HTML]{000000} {\cellcolor[HTML]{000000}} \color[HTML]{F1F1F1} {\cellcolor{white}}  & {\cellcolor[HTML]{FEFEFE}} \color[HTML]{000000} {\cellcolor[HTML]{000000}} \color[HTML]{F1F1F1} {\cellcolor{white}}  & {\cellcolor[HTML]{FEFEFE}} \color[HTML]{000000} {\cellcolor[HTML]{000000}} \color[HTML]{F1F1F1} {\cellcolor{white}}  & {\cellcolor[HTML]{FEFEFE}} \color[HTML]{000000} {\cellcolor[HTML]{000000}} \color[HTML]{F1F1F1} {\cellcolor{white}} {\cellcolor[HTML]{AAAAAA}}  & {\cellcolor[HTML]{FEFEFE}} \color[HTML]{000000} {\cellcolor[HTML]{000000}} \color[HTML]{F1F1F1} {\cellcolor{white}} {\cellcolor[HTML]{AAAAAA}}  & {\cellcolor[HTML]{FEFEFE}} \color[HTML]{000000} {\cellcolor[HTML]{000000}} \color[HTML]{F1F1F1} {\cellcolor{white}} {\cellcolor[HTML]{AAAAAA}}  \\
{\cellcolor[HTML]{FEFEFE}} \color[HTML]{000000} 28 & {\cellcolor[HTML]{FEFEFE}} \color[HTML]{000000} $\text{A}^\text{97}_\text{OP2}$-$\text{O}_\text{ph}$ & {\cellcolor[HTML]{FEFEFE}} \color[HTML]{000000} {\cellcolor{white}}  & {\cellcolor[HTML]{FEFEFE}} \color[HTML]{000000} $\text{MG}^\text{511}$ & {\cellcolor[HTML]{FEFEFE}} \color[HTML]{000000} {\cellcolor[HTML]{000000}} \color[HTML]{F1F1F1} {\cellcolor{white}}  & {\cellcolor[HTML]{FEFEFE}} \color[HTML]{000000} {\cellcolor[HTML]{000000}} \color[HTML]{F1F1F1} {\cellcolor{white}}  & {\cellcolor[HTML]{FEFEFE}} \color[HTML]{000000} {\cellcolor[HTML]{000000}} \color[HTML]{F1F1F1} {\cellcolor{white}}  & {\cellcolor[HTML]{FEFEFE}} \color[HTML]{000000} {\cellcolor[HTML]{000000}} \color[HTML]{F1F1F1} {\cellcolor{white}}  & {\cellcolor[HTML]{FEFEFE}} \color[HTML]{000000} {\cellcolor[HTML]{000000}} \color[HTML]{F1F1F1} {\cellcolor{white}}  & {\cellcolor[HTML]{FEFEFE}} \color[HTML]{000000} {\cellcolor[HTML]{000000}} \color[HTML]{F1F1F1} {\cellcolor{white}}  & {\cellcolor[HTML]{FEFEFE}} \color[HTML]{000000} {\cellcolor[HTML]{000000}} \color[HTML]{F1F1F1} {\cellcolor{white}}  & {\cellcolor[HTML]{FEFEFE}} \color[HTML]{000000} {\cellcolor[HTML]{000000}} \color[HTML]{F1F1F1} {\cellcolor{white}}  & {\cellcolor[HTML]{FEFEFE}} \color[HTML]{000000} {\cellcolor[HTML]{000000}} \color[HTML]{F1F1F1} {\cellcolor{white}}  & {\cellcolor[HTML]{FEFEFE}} \color[HTML]{000000} {\cellcolor[HTML]{000000}} \color[HTML]{F1F1F1} {\cellcolor{white}}  & {\cellcolor[HTML]{FEFEFE}} \color[HTML]{000000} {\cellcolor[HTML]{000000}} \color[HTML]{F1F1F1} {\cellcolor{white}}  & {\cellcolor[HTML]{FEFEFE}} \color[HTML]{000000} {\cellcolor[HTML]{000000}} \color[HTML]{F1F1F1} {\cellcolor{white}} {\cellcolor[HTML]{AAAAAA}}  & {\cellcolor[HTML]{FEFEFE}} \color[HTML]{000000} {\cellcolor[HTML]{000000}} \color[HTML]{F1F1F1} {\cellcolor{white}} {\cellcolor[HTML]{AAAAAA}}  & {\cellcolor[HTML]{FEFEFE}} \color[HTML]{000000} {\cellcolor[HTML]{000000}} \color[HTML]{F1F1F1} {\cellcolor{white}} {\cellcolor[HTML]{AAAAAA}}  \\
{\cellcolor[HTML]{FEFEFE}} \color[HTML]{000000} 29 & {\cellcolor[HTML]{FEFEFE}} \color[HTML]{000000} $\text{C}^\text{165}_\text{OP2}$,$\text{U}^\text{167}_\text{O2'}$-\textit{cis}-$\text{O}_\text{ph}$$\text{O}_\text{r}$ & {\cellcolor[HTML]{FEFEFE}} \color[HTML]{000000} {\cellcolor{white}}  & {\cellcolor[HTML]{FEFEFE}} \color[HTML]{000000} $\text{MG}^\text{516}$ & {\cellcolor[HTML]{FEFEFE}} \color[HTML]{000000} {\cellcolor[HTML]{000000}} \color[HTML]{F1F1F1} {\cellcolor{white}}  & {\cellcolor[HTML]{FEFEFE}} \color[HTML]{000000} {\cellcolor[HTML]{000000}} \color[HTML]{F1F1F1} {\cellcolor{white}}  & {\cellcolor[HTML]{FEFEFE}} \color[HTML]{000000} {\cellcolor[HTML]{000000}} \color[HTML]{F1F1F1} {\cellcolor{white}}  & {\cellcolor[HTML]{FEFEFE}} \color[HTML]{000000} {\cellcolor[HTML]{000000}} \color[HTML]{F1F1F1} {\cellcolor{white}}  & {\cellcolor[HTML]{FEFEFE}} \color[HTML]{000000} {\cellcolor[HTML]{000000}} \color[HTML]{F1F1F1} {\cellcolor{white}}  & {\cellcolor[HTML]{FEFEFE}} \color[HTML]{000000} {\cellcolor[HTML]{000000}} \color[HTML]{F1F1F1} {\cellcolor{white}}  & {\cellcolor[HTML]{FEFEFE}} \color[HTML]{000000} {\cellcolor[HTML]{000000}} \color[HTML]{F1F1F1} {\cellcolor{white}}  & {\cellcolor[HTML]{FEFEFE}} \color[HTML]{000000} {\cellcolor[HTML]{000000}} \color[HTML]{F1F1F1} {\cellcolor{white}}  & {\cellcolor[HTML]{FEFEFE}} \color[HTML]{000000} {\cellcolor[HTML]{000000}} \color[HTML]{F1F1F1} {\cellcolor{white}}  & {\cellcolor[HTML]{FEFEFE}} \color[HTML]{000000} {\cellcolor[HTML]{000000}} \color[HTML]{F1F1F1} {\cellcolor{white}}  & {\cellcolor[HTML]{FEFEFE}} \color[HTML]{000000} {\cellcolor[HTML]{000000}} \color[HTML]{F1F1F1} {\cellcolor{white}}  & {\cellcolor[HTML]{FEFEFE}} \color[HTML]{000000} {\cellcolor[HTML]{000000}} \color[HTML]{F1F1F1} {\cellcolor{white}} {\cellcolor[HTML]{AAAAAA}}  & {\cellcolor[HTML]{FEFEFE}} \color[HTML]{000000} {\cellcolor[HTML]{000000}} \color[HTML]{F1F1F1} {\cellcolor{white}} {\cellcolor[HTML]{AAAAAA}}  & {\cellcolor[HTML]{FEFEFE}} \color[HTML]{000000} {\cellcolor[HTML]{000000}} \color[HTML]{F1F1F1} {\cellcolor{white}} {\cellcolor[HTML]{AAAAAA}}  \\
{\cellcolor[HTML]{FEFEFE}} \color[HTML]{000000} 30 & {\cellcolor[HTML]{FEFEFE}} \color[HTML]{000000} $\text{G}^\text{272}_\text{O6}$-$\text{O}_\text{b}$ & {\cellcolor[HTML]{FEFEFE}} \color[HTML]{000000} {\cellcolor{white}}  & {\cellcolor[HTML]{FEFEFE}} \color[HTML]{000000} $\text{MG}^\text{515}$ & {\cellcolor[HTML]{FEFEFE}} \color[HTML]{000000} {\cellcolor[HTML]{000000}} \color[HTML]{F1F1F1} {\cellcolor{white}}  & {\cellcolor[HTML]{FEFEFE}} \color[HTML]{000000} {\cellcolor[HTML]{000000}} \color[HTML]{F1F1F1} {\cellcolor{white}}  & {\cellcolor[HTML]{FEFEFE}} \color[HTML]{000000} {\cellcolor[HTML]{000000}} \color[HTML]{F1F1F1} {\cellcolor{white}}  & {\cellcolor[HTML]{FEFEFE}} \color[HTML]{000000} {\cellcolor[HTML]{000000}} \color[HTML]{F1F1F1} {\cellcolor{white}}  & {\cellcolor[HTML]{FEFEFE}} \color[HTML]{000000} {\cellcolor[HTML]{000000}} \color[HTML]{F1F1F1} {\cellcolor{white}}  & {\cellcolor[HTML]{FEFEFE}} \color[HTML]{000000} {\cellcolor[HTML]{000000}} \color[HTML]{F1F1F1} {\cellcolor{white}}  & {\cellcolor[HTML]{FEFEFE}} \color[HTML]{000000} {\cellcolor[HTML]{000000}} \color[HTML]{F1F1F1} {\cellcolor{white}}  & {\cellcolor[HTML]{FEFEFE}} \color[HTML]{000000} {\cellcolor[HTML]{000000}} \color[HTML]{F1F1F1} {\cellcolor{white}}  & {\cellcolor[HTML]{FEFEFE}} \color[HTML]{000000} {\cellcolor[HTML]{000000}} \color[HTML]{F1F1F1} {\cellcolor{white}}  & {\cellcolor[HTML]{FEFEFE}} \color[HTML]{000000} {\cellcolor[HTML]{000000}} \color[HTML]{F1F1F1} {\cellcolor{white}}  & {\cellcolor[HTML]{FEFEFE}} \color[HTML]{000000} {\cellcolor[HTML]{000000}} \color[HTML]{F1F1F1} {\cellcolor{white}}  & {\cellcolor[HTML]{FEFEFE}} \color[HTML]{000000} {\cellcolor[HTML]{000000}} \color[HTML]{F1F1F1} {\cellcolor{white}} {\cellcolor[HTML]{AAAAAA}}  & {\cellcolor[HTML]{FEFEFE}} \color[HTML]{000000} {\cellcolor[HTML]{000000}} \color[HTML]{F1F1F1} {\cellcolor{white}} {\cellcolor[HTML]{AAAAAA}}  & {\cellcolor[HTML]{FEFEFE}} \color[HTML]{000000} {\cellcolor[HTML]{000000}} \color[HTML]{F1F1F1} {\cellcolor{white}} {\cellcolor[HTML]{AAAAAA}}  \\
{\cellcolor[HTML]{FEFEFE}} \color[HTML]{000000} 31 & {\cellcolor[HTML]{FEFEFE}} \color[HTML]{000000} $\text{G}^\text{275}_\text{O2'}$-$\text{O}_\text{r}$ & {\cellcolor[HTML]{FEFEFE}} \color[HTML]{000000} {\cellcolor{white}}  & {\cellcolor[HTML]{FEFEFE}} \color[HTML]{000000} $\text{MG}^\text{523}$ & {\cellcolor[HTML]{FEFEFE}} \color[HTML]{000000} {\cellcolor[HTML]{000000}} \color[HTML]{F1F1F1} {\cellcolor{white}}  & {\cellcolor[HTML]{FEFEFE}} \color[HTML]{000000} {\cellcolor[HTML]{000000}} \color[HTML]{F1F1F1} {\cellcolor{white}}  & {\cellcolor[HTML]{FEFEFE}} \color[HTML]{000000} {\cellcolor[HTML]{000000}} \color[HTML]{F1F1F1} {\cellcolor{white}}  & {\cellcolor[HTML]{FEFEFE}} \color[HTML]{000000} {\cellcolor[HTML]{000000}} \color[HTML]{F1F1F1} {\cellcolor{white}}  & {\cellcolor[HTML]{FEFEFE}} \color[HTML]{000000} {\cellcolor[HTML]{000000}} \color[HTML]{F1F1F1} {\cellcolor{white}}  & {\cellcolor[HTML]{FEFEFE}} \color[HTML]{000000} {\cellcolor[HTML]{000000}} \color[HTML]{F1F1F1} {\cellcolor{white}}  & {\cellcolor[HTML]{FEFEFE}} \color[HTML]{000000} {\cellcolor[HTML]{000000}} \color[HTML]{F1F1F1} {\cellcolor{white}}  & {\cellcolor[HTML]{FEFEFE}} \color[HTML]{000000} {\cellcolor[HTML]{000000}} \color[HTML]{F1F1F1} {\cellcolor{white}}  & {\cellcolor[HTML]{FEFEFE}} \color[HTML]{000000} {\cellcolor[HTML]{000000}} \color[HTML]{F1F1F1} {\cellcolor{white}}  & {\cellcolor[HTML]{FEFEFE}} \color[HTML]{000000} {\cellcolor[HTML]{000000}} \color[HTML]{F1F1F1} {\cellcolor{white}}  & {\cellcolor[HTML]{FEFEFE}} \color[HTML]{000000} {\cellcolor[HTML]{000000}} \color[HTML]{F1F1F1} {\cellcolor{white}}  & {\cellcolor[HTML]{FEFEFE}} \color[HTML]{000000} {\cellcolor[HTML]{000000}} \color[HTML]{F1F1F1} {\cellcolor{white}} {\cellcolor[HTML]{AAAAAA}}  & {\cellcolor[HTML]{FEFEFE}} \color[HTML]{000000} {\cellcolor[HTML]{000000}} \color[HTML]{F1F1F1} {\cellcolor{white}} {\cellcolor[HTML]{AAAAAA}}  & {\cellcolor[HTML]{FEFEFE}} \color[HTML]{000000} {\cellcolor[HTML]{000000}} \color[HTML]{F1F1F1} {\cellcolor{white}} {\cellcolor[HTML]{AAAAAA}}  \\
{\cellcolor[HTML]{FEFEFE}} \color[HTML]{000000} 32 & {\cellcolor[HTML]{FEFEFE}} \color[HTML]{000000} $\text{G}^\text{313}_\text{O2'}$-$\text{O}_\text{r}$ & {\cellcolor[HTML]{FEFEFE}} \color[HTML]{000000} {\cellcolor{white}}  & {\cellcolor[HTML]{FEFEFE}} \color[HTML]{000000} $\text{MG}^\text{522}$ & {\cellcolor[HTML]{FEFEFE}} \color[HTML]{000000} {\cellcolor[HTML]{000000}} \color[HTML]{F1F1F1} {\cellcolor{white}}  & {\cellcolor[HTML]{FEFEFE}} \color[HTML]{000000} {\cellcolor[HTML]{000000}} \color[HTML]{F1F1F1} {\cellcolor{white}}  & {\cellcolor[HTML]{FEFEFE}} \color[HTML]{000000} {\cellcolor[HTML]{000000}} \color[HTML]{F1F1F1} {\cellcolor{white}}  & {\cellcolor[HTML]{FEFEFE}} \color[HTML]{000000} {\cellcolor[HTML]{000000}} \color[HTML]{F1F1F1} {\cellcolor{white}}  & {\cellcolor[HTML]{FEFEFE}} \color[HTML]{000000} {\cellcolor[HTML]{000000}} \color[HTML]{F1F1F1} {\cellcolor{white}}  & {\cellcolor[HTML]{FEFEFE}} \color[HTML]{000000} {\cellcolor[HTML]{000000}} \color[HTML]{F1F1F1} {\cellcolor{white}}  & {\cellcolor[HTML]{FEFEFE}} \color[HTML]{000000} {\cellcolor[HTML]{000000}} \color[HTML]{F1F1F1} {\cellcolor{white}}  & {\cellcolor[HTML]{FEFEFE}} \color[HTML]{000000} {\cellcolor[HTML]{000000}} \color[HTML]{F1F1F1} {\cellcolor{white}}  & {\cellcolor[HTML]{FEFEFE}} \color[HTML]{000000} {\cellcolor[HTML]{000000}} \color[HTML]{F1F1F1} {\cellcolor{white}}  & {\cellcolor[HTML]{FEFEFE}} \color[HTML]{000000} {\cellcolor[HTML]{000000}} \color[HTML]{F1F1F1} {\cellcolor{white}}  & {\cellcolor[HTML]{FEFEFE}} \color[HTML]{000000} {\cellcolor[HTML]{000000}} \color[HTML]{F1F1F1} {\cellcolor{white}}  & {\cellcolor[HTML]{FEFEFE}} \color[HTML]{000000} {\cellcolor[HTML]{000000}} \color[HTML]{F1F1F1} {\cellcolor{white}} {\cellcolor[HTML]{AAAAAA}}  & {\cellcolor[HTML]{FEFEFE}} \color[HTML]{000000} {\cellcolor[HTML]{000000}} \color[HTML]{F1F1F1} {\cellcolor{white}} {\cellcolor[HTML]{AAAAAA}}  & {\cellcolor[HTML]{FEFEFE}} \color[HTML]{000000} {\cellcolor[HTML]{000000}} \color[HTML]{F1F1F1} {\cellcolor{white}} {\cellcolor[HTML]{AAAAAA}}  \\
{\cellcolor[HTML]{FEFEFE}} \color[HTML]{000000} 33 & {\cellcolor[HTML]{FEFEFE}} \color[HTML]{000000} $\text{G}^\text{96}_\text{N7}$-$\text{N}_\text{b}$ & {\cellcolor[HTML]{FEFEFE}} \color[HTML]{000000} {\cellcolor{white}}  & {\cellcolor[HTML]{FEFEFE}} \color[HTML]{000000} $\text{MG}^\text{505}$ & {\cellcolor[HTML]{FEFEFE}} \color[HTML]{000000} {\cellcolor[HTML]{000000}} \color[HTML]{F1F1F1} {\cellcolor{white}}  & {\cellcolor[HTML]{FEFEFE}} \color[HTML]{000000} {\cellcolor[HTML]{000000}} \color[HTML]{F1F1F1} {\cellcolor{white}}  & {\cellcolor[HTML]{FEFEFE}} \color[HTML]{000000} {\cellcolor[HTML]{000000}} \color[HTML]{F1F1F1} {\cellcolor{white}}  & {\cellcolor[HTML]{FEFEFE}} \color[HTML]{000000} {\cellcolor[HTML]{000000}} \color[HTML]{F1F1F1} {\cellcolor{white}}  & {\cellcolor[HTML]{FEFEFE}} \color[HTML]{000000} {\cellcolor[HTML]{000000}} \color[HTML]{F1F1F1} {\cellcolor{white}}  & {\cellcolor[HTML]{FEFEFE}} \color[HTML]{000000} {\cellcolor[HTML]{000000}} \color[HTML]{F1F1F1} {\cellcolor{white}}  & {\cellcolor[HTML]{FEFEFE}} \color[HTML]{000000} {\cellcolor[HTML]{000000}} \color[HTML]{F1F1F1} {\cellcolor{white}}  & {\cellcolor[HTML]{FEFEFE}} \color[HTML]{000000} {\cellcolor[HTML]{000000}} \color[HTML]{F1F1F1} {\cellcolor{white}}  & {\cellcolor[HTML]{FEFEFE}} \color[HTML]{000000} {\cellcolor[HTML]{000000}} \color[HTML]{F1F1F1} {\cellcolor{white}}  & {\cellcolor[HTML]{FEFEFE}} \color[HTML]{000000} {\cellcolor[HTML]{000000}} \color[HTML]{F1F1F1} {\cellcolor{white}}  & {\cellcolor[HTML]{FEFEFE}} \color[HTML]{000000} {\cellcolor[HTML]{000000}} \color[HTML]{F1F1F1} {\cellcolor{white}}  & {\cellcolor[HTML]{FEFEFE}} \color[HTML]{000000} {\cellcolor[HTML]{000000}} \color[HTML]{F1F1F1} {\cellcolor{white}} {\cellcolor[HTML]{AAAAAA}}  & {\cellcolor[HTML]{FEFEFE}} \color[HTML]{000000} {\cellcolor[HTML]{000000}} \color[HTML]{F1F1F1} {\cellcolor{white}} {\cellcolor[HTML]{AAAAAA}}  & {\cellcolor[HTML]{FEFEFE}} \color[HTML]{000000} {\cellcolor[HTML]{000000}} \color[HTML]{F1F1F1} {\cellcolor{white}} {\cellcolor[HTML]{AAAAAA}}  \\
{\cellcolor[HTML]{FEFEFE}} \color[HTML]{000000} 34 & {\cellcolor[HTML]{FEFEFE}} \color[HTML]{000000} $\text{U}^\text{277}_\text{O4}$-$\text{O}_\text{b}$ & {\cellcolor[HTML]{FEFEFE}} \color[HTML]{000000} {\cellcolor{white}}  & {\cellcolor[HTML]{FEFEFE}} \color[HTML]{000000} $\text{MG}^\text{510}$ & {\cellcolor[HTML]{FEFEFE}} \color[HTML]{000000} {\cellcolor[HTML]{000000}} \color[HTML]{F1F1F1} {\cellcolor{white}}  & {\cellcolor[HTML]{FEFEFE}} \color[HTML]{000000} {\cellcolor[HTML]{000000}} \color[HTML]{F1F1F1} {\cellcolor{white}}  & {\cellcolor[HTML]{FEFEFE}} \color[HTML]{000000} {\cellcolor[HTML]{000000}} \color[HTML]{F1F1F1} {\cellcolor{white}}  & {\cellcolor[HTML]{FEFEFE}} \color[HTML]{000000} {\cellcolor[HTML]{000000}} \color[HTML]{F1F1F1} {\cellcolor{white}}  & {\cellcolor[HTML]{FEFEFE}} \color[HTML]{000000} {\cellcolor[HTML]{000000}} \color[HTML]{F1F1F1} {\cellcolor{white}}  & {\cellcolor[HTML]{FEFEFE}} \color[HTML]{000000} {\cellcolor[HTML]{000000}} \color[HTML]{F1F1F1} {\cellcolor{white}}  & {\cellcolor[HTML]{FEFEFE}} \color[HTML]{000000} {\cellcolor[HTML]{000000}} \color[HTML]{F1F1F1} {\cellcolor{white}}  & {\cellcolor[HTML]{FEFEFE}} \color[HTML]{000000} {\cellcolor[HTML]{000000}} \color[HTML]{F1F1F1} {\cellcolor{white}}  & {\cellcolor[HTML]{FEFEFE}} \color[HTML]{000000} {\cellcolor[HTML]{000000}} \color[HTML]{F1F1F1} {\cellcolor{white}}  & {\cellcolor[HTML]{FEFEFE}} \color[HTML]{000000} {\cellcolor[HTML]{000000}} \color[HTML]{F1F1F1} {\cellcolor{white}}  & {\cellcolor[HTML]{FEFEFE}} \color[HTML]{000000} {\cellcolor[HTML]{000000}} \color[HTML]{F1F1F1} {\cellcolor{white}}  & {\cellcolor[HTML]{FEFEFE}} \color[HTML]{000000} {\cellcolor[HTML]{000000}} \color[HTML]{F1F1F1} {\cellcolor{white}} {\cellcolor[HTML]{AAAAAA}}  & {\cellcolor[HTML]{FEFEFE}} \color[HTML]{000000} {\cellcolor[HTML]{000000}} \color[HTML]{F1F1F1} {\cellcolor{white}} {\cellcolor[HTML]{AAAAAA}}  & {\cellcolor[HTML]{FEFEFE}} \color[HTML]{000000} {\cellcolor[HTML]{000000}} \color[HTML]{F1F1F1} {\cellcolor{white}} {\cellcolor[HTML]{AAAAAA}}  \\
{\cellcolor[HTML]{FEFEFE}} \color[HTML]{000000} 35 & {\cellcolor[HTML]{FEFEFE}} \color[HTML]{000000} $\text{U}^\text{326}_\text{O4}$-$\text{O}_\text{b}$ & {\cellcolor[HTML]{FEFEFE}} \color[HTML]{000000} {\cellcolor{white}}  & {\cellcolor[HTML]{FEFEFE}} \color[HTML]{000000} $\text{MG}^\text{524}$ & {\cellcolor[HTML]{FEFEFE}} \color[HTML]{000000} {\cellcolor[HTML]{000000}} \color[HTML]{F1F1F1} {\cellcolor{white}}  & {\cellcolor[HTML]{FEFEFE}} \color[HTML]{000000} {\cellcolor[HTML]{000000}} \color[HTML]{F1F1F1} {\cellcolor{white}}  & {\cellcolor[HTML]{FEFEFE}} \color[HTML]{000000} {\cellcolor[HTML]{000000}} \color[HTML]{F1F1F1} {\cellcolor{white}}  & {\cellcolor[HTML]{FEFEFE}} \color[HTML]{000000} {\cellcolor[HTML]{000000}} \color[HTML]{F1F1F1} {\cellcolor{white}}  & {\cellcolor[HTML]{FEFEFE}} \color[HTML]{000000} {\cellcolor[HTML]{000000}} \color[HTML]{F1F1F1} {\cellcolor{white}}  & {\cellcolor[HTML]{FEFEFE}} \color[HTML]{000000} {\cellcolor[HTML]{000000}} \color[HTML]{F1F1F1} {\cellcolor{white}}  & {\cellcolor[HTML]{FEFEFE}} \color[HTML]{000000} {\cellcolor[HTML]{000000}} \color[HTML]{F1F1F1} {\cellcolor{white}}  & {\cellcolor[HTML]{FEFEFE}} \color[HTML]{000000} {\cellcolor[HTML]{000000}} \color[HTML]{F1F1F1} {\cellcolor{white}}  & {\cellcolor[HTML]{FEFEFE}} \color[HTML]{000000} {\cellcolor[HTML]{000000}} \color[HTML]{F1F1F1} {\cellcolor{white}}  & {\cellcolor[HTML]{FEFEFE}} \color[HTML]{000000} {\cellcolor[HTML]{000000}} \color[HTML]{F1F1F1} {\cellcolor{white}}  & {\cellcolor[HTML]{FEFEFE}} \color[HTML]{000000} {\cellcolor[HTML]{000000}} \color[HTML]{F1F1F1} {\cellcolor{white}}  & {\cellcolor[HTML]{FEFEFE}} \color[HTML]{000000} {\cellcolor[HTML]{000000}} \color[HTML]{F1F1F1} {\cellcolor{white}} {\cellcolor[HTML]{AAAAAA}}  & {\cellcolor[HTML]{FEFEFE}} \color[HTML]{000000} {\cellcolor[HTML]{000000}} \color[HTML]{F1F1F1} {\cellcolor{white}} {\cellcolor[HTML]{AAAAAA}}  & {\cellcolor[HTML]{FEFEFE}} \color[HTML]{000000} {\cellcolor[HTML]{000000}} \color[HTML]{F1F1F1} {\cellcolor{white}} {\cellcolor[HTML]{AAAAAA}}  \\
{\cellcolor[HTML]{FEFEFE}} \color[HTML]{000000} 36 & {\cellcolor[HTML]{FEFEFE}} \color[HTML]{000000} $\text{U}^\text{33}_\text{OP1}$-$\text{O}_\text{ph}$ & {\cellcolor[HTML]{FEFEFE}} \color[HTML]{000000} {\cellcolor{white}}  & {\cellcolor[HTML]{FEFEFE}} \color[HTML]{000000} $\text{MG}^\text{518}$ & {\cellcolor[HTML]{FEFEFE}} \color[HTML]{000000} {\cellcolor[HTML]{000000}} \color[HTML]{F1F1F1} {\cellcolor{white}}  & {\cellcolor[HTML]{FEFEFE}} \color[HTML]{000000} {\cellcolor[HTML]{000000}} \color[HTML]{F1F1F1} {\cellcolor{white}}  & {\cellcolor[HTML]{FEFEFE}} \color[HTML]{000000} {\cellcolor[HTML]{000000}} \color[HTML]{F1F1F1} {\cellcolor{white}}  & {\cellcolor[HTML]{FEFEFE}} \color[HTML]{000000} {\cellcolor[HTML]{000000}} \color[HTML]{F1F1F1} {\cellcolor{white}}  & {\cellcolor[HTML]{FEFEFE}} \color[HTML]{000000} {\cellcolor[HTML]{000000}} \color[HTML]{F1F1F1} {\cellcolor{white}}  & {\cellcolor[HTML]{FEFEFE}} \color[HTML]{000000} {\cellcolor[HTML]{000000}} \color[HTML]{F1F1F1} {\cellcolor{white}}  & {\cellcolor[HTML]{FEFEFE}} \color[HTML]{000000} {\cellcolor[HTML]{000000}} \color[HTML]{F1F1F1} {\cellcolor{white}}  & {\cellcolor[HTML]{FEFEFE}} \color[HTML]{000000} {\cellcolor[HTML]{000000}} \color[HTML]{F1F1F1} {\cellcolor{white}}  & {\cellcolor[HTML]{FEFEFE}} \color[HTML]{000000} {\cellcolor[HTML]{000000}} \color[HTML]{F1F1F1} {\cellcolor{white}}  & {\cellcolor[HTML]{FEFEFE}} \color[HTML]{000000} {\cellcolor[HTML]{000000}} \color[HTML]{F1F1F1} {\cellcolor{white}}  & {\cellcolor[HTML]{FEFEFE}} \color[HTML]{000000} {\cellcolor[HTML]{000000}} \color[HTML]{F1F1F1} {\cellcolor{white}}  & {\cellcolor[HTML]{FEFEFE}} \color[HTML]{000000} {\cellcolor[HTML]{000000}} \color[HTML]{F1F1F1} {\cellcolor{white}} {\cellcolor[HTML]{AAAAAA}}  & {\cellcolor[HTML]{FEFEFE}} \color[HTML]{000000} {\cellcolor[HTML]{000000}} \color[HTML]{F1F1F1} {\cellcolor{white}} {\cellcolor[HTML]{AAAAAA}}  & {\cellcolor[HTML]{FEFEFE}} \color[HTML]{000000} {\cellcolor[HTML]{000000}} \color[HTML]{F1F1F1} {\cellcolor{white}} {\cellcolor[HTML]{AAAAAA}}  \\
{\cellcolor[HTML]{FEFEFE}} \color[HTML]{000000} 37 & {\cellcolor[HTML]{FEFEFE}} \color[HTML]{000000} $\text{A}^\text{184}_\text{OP1}$,$\text{A}^\text{186}_\text{OP1}$,$\text{A}^\text{187}_\text{OP2}$,$\text{G}^\text{188}_\text{OP2}$-\underline{Trans}-4$\text{O}_\text{ph}$ & {\cellcolor[HTML]{FEFEFE}} \color[HTML]{000000} {\cellcolor{white}}  & {\cellcolor[HTML]{FEFEFE}} \color[HTML]{000000}  & {\cellcolor[HTML]{FEFEFE}} \color[HTML]{000000} {\cellcolor[HTML]{840711}} \color[HTML]{F1F1F1} 0.9 & {\cellcolor[HTML]{FEFEFE}} \color[HTML]{000000} {\cellcolor[HTML]{000000}} \color[HTML]{F1F1F1} {\cellcolor{white}}  & {\cellcolor[HTML]{FEFEFE}} \color[HTML]{000000} {\cellcolor[HTML]{840711}} \color[HTML]{F1F1F1} 0.9 & {\cellcolor[HTML]{FEFEFE}} \color[HTML]{000000} {\cellcolor[HTML]{000000}} \color[HTML]{F1F1F1} {\cellcolor{white}}  & {\cellcolor[HTML]{FEFEFE}} \color[HTML]{000000} {\cellcolor[HTML]{67000D}} \color[HTML]{F1F1F1} 1.0 (1.0-1.0) & {\cellcolor[HTML]{FEFEFE}} \color[HTML]{000000} {\cellcolor[HTML]{67000D}} \color[HTML]{F1F1F1} 1.0 (1.0-1.0) & {\cellcolor[HTML]{FEFEFE}} \color[HTML]{000000} {\cellcolor[HTML]{8A0812}} \color[HTML]{F1F1F1} 0.9 (0.8-1.0) & {\cellcolor[HTML]{FEFEFE}} \color[HTML]{000000} {\cellcolor[HTML]{67000D}} \color[HTML]{F1F1F1} 1.0 (1.0-1.0) & {\cellcolor[HTML]{FEFEFE}} \color[HTML]{000000} {\cellcolor[HTML]{000000}} \color[HTML]{F1F1F1} {\cellcolor{white}}  & {\cellcolor[HTML]{FEFEFE}} \color[HTML]{000000} {\cellcolor[HTML]{000000}} \color[HTML]{F1F1F1} {\cellcolor{white}}  & {\cellcolor[HTML]{FEFEFE}} \color[HTML]{000000} {\cellcolor[HTML]{000000}} \color[HTML]{F1F1F1} {\cellcolor{white}}  & {\cellcolor[HTML]{FEFEFE}} \color[HTML]{000000} {\cellcolor[HTML]{D92523}} \color[HTML]{F1F1F1} 0.7 (0.7-0.8) & {\cellcolor[HTML]{FEFEFE}} \color[HTML]{000000} {\cellcolor[HTML]{F14432}} \color[HTML]{F1F1F1} 0.6 (0.5-0.7) & {\cellcolor[HTML]{FEFEFE}} \color[HTML]{000000} {\cellcolor[HTML]{E83429}} \color[HTML]{F1F1F1} 0.7 (0.6-0.8) \\
{\cellcolor[HTML]{FEFEFE}} \color[HTML]{000000} 38 & {\cellcolor[HTML]{FEFEFE}} \color[HTML]{000000} $\text{C}^\text{217}_\text{OP2}$-$\text{O}_\text{ph}$ & {\cellcolor[HTML]{FEFEFE}} \color[HTML]{000000} {\cellcolor{white}}  & {\cellcolor[HTML]{FEFEFE}} \color[HTML]{000000}  & {\cellcolor[HTML]{FEFEFE}} \color[HTML]{000000} {\cellcolor[HTML]{67000D}} \color[HTML]{F1F1F1} 1.0 & {\cellcolor[HTML]{FEFEFE}} \color[HTML]{000000} {\cellcolor[HTML]{FEE0D2}} \color[HTML]{000000} 0.1 & {\cellcolor[HTML]{FEFEFE}} \color[HTML]{000000} {\cellcolor[HTML]{67000D}} \color[HTML]{F1F1F1} 1.0 & {\cellcolor[HTML]{FEFEFE}} \color[HTML]{000000} {\cellcolor[HTML]{FEE0D2}} \color[HTML]{000000} 0.1 & {\cellcolor[HTML]{FEFEFE}} \color[HTML]{000000} {\cellcolor[HTML]{9C0D14}} \color[HTML]{F1F1F1} 0.9 (0.6-1.0) & {\cellcolor[HTML]{FEFEFE}} \color[HTML]{000000} {\cellcolor[HTML]{F14432}} \color[HTML]{F1F1F1} 0.6 (0.4-0.8) & {\cellcolor[HTML]{FEFEFE}} \color[HTML]{000000} {\cellcolor[HTML]{F85F43}} \color[HTML]{F1F1F1} 0.5 (0.4-0.7) & {\cellcolor[HTML]{FEFEFE}} \color[HTML]{000000} {\cellcolor[HTML]{A10E15}} \color[HTML]{F1F1F1} 0.9 (0.8-0.9) & {\cellcolor[HTML]{FEFEFE}} \color[HTML]{000000} {\cellcolor[HTML]{FCC1A8}} \color[HTML]{000000} 0.2 (0.0-0.5) & {\cellcolor[HTML]{FEFEFE}} \color[HTML]{000000} {\cellcolor[HTML]{FB7A5A}} \color[HTML]{F1F1F1} 0.5 (0.3-0.7) & {\cellcolor[HTML]{FEFEFE}} \color[HTML]{000000} {\cellcolor[HTML]{F6563D}} \color[HTML]{F1F1F1} 0.6 (0.4-0.7) & {\cellcolor[HTML]{FEFEFE}} \color[HTML]{000000} {\cellcolor[HTML]{FDCAB5}} \color[HTML]{000000} 0.2 (0.1-0.4) & {\cellcolor[HTML]{FEFEFE}} \color[HTML]{000000} {\cellcolor[HTML]{FC8A6A}} \color[HTML]{F1F1F1} 0.4 (0.3-0.5) & {\cellcolor[HTML]{FEFEFE}} \color[HTML]{000000} {\cellcolor[HTML]{FC9B7C}} \color[HTML]{000000} 0.4 (0.2-0.4) \\
{\cellcolor[HTML]{FEFEFE}} \color[HTML]{000000} 39 & {\cellcolor[HTML]{FEFEFE}} \color[HTML]{000000} $\text{C}^\text{166}_\text{OP1}$,$\text{U}^\text{168}_\text{OP1}$-\textit{cis}-2$\text{O}_\text{ph}$ & {\cellcolor[HTML]{FEFEFE}} \color[HTML]{000000} {\cellcolor{white}}  & {\cellcolor[HTML]{FEFEFE}} \color[HTML]{000000}  & {\cellcolor[HTML]{FEFEFE}} \color[HTML]{000000} {\cellcolor[HTML]{EE3A2C}} \color[HTML]{F1F1F1} 0.6 & {\cellcolor[HTML]{FEFEFE}} \color[HTML]{000000} {\cellcolor[HTML]{000000}} \color[HTML]{F1F1F1} {\cellcolor{white}}  & {\cellcolor[HTML]{FEFEFE}} \color[HTML]{000000} {\cellcolor[HTML]{840711}} \color[HTML]{F1F1F1} 0.9 & {\cellcolor[HTML]{FEFEFE}} \color[HTML]{000000} {\cellcolor[HTML]{000000}} \color[HTML]{F1F1F1} {\cellcolor{white}}  & {\cellcolor[HTML]{FEFEFE}} \color[HTML]{000000} {\cellcolor[HTML]{B81419}} \color[HTML]{F1F1F1} 0.8 (0.6-0.9) & {\cellcolor[HTML]{FEFEFE}} \color[HTML]{000000} {\cellcolor[HTML]{B71319}} \color[HTML]{F1F1F1} 0.8 (0.7-0.9) & {\cellcolor[HTML]{FEFEFE}} \color[HTML]{000000} {\cellcolor[HTML]{D82422}} \color[HTML]{F1F1F1} 0.7 (0.5-0.9) & {\cellcolor[HTML]{FEFEFE}} \color[HTML]{000000} {\cellcolor[HTML]{B81419}} \color[HTML]{F1F1F1} 0.8 (0.7-0.9) & {\cellcolor[HTML]{FEFEFE}} \color[HTML]{000000} {\cellcolor[HTML]{000000}} \color[HTML]{F1F1F1} {\cellcolor{white}}  & {\cellcolor[HTML]{FEFEFE}} \color[HTML]{000000} {\cellcolor[HTML]{FEE3D7}} \color[HTML]{000000} 0.1 (0.0-0.3) & {\cellcolor[HTML]{FEFEFE}} \color[HTML]{000000} {\cellcolor[HTML]{FDD4C2}} \color[HTML]{000000} 0.2 (0.0-0.4) & {\cellcolor[HTML]{FEFEFE}} \color[HTML]{000000} {\cellcolor[HTML]{FDCAB5}} \color[HTML]{000000} 0.2 (0.1-0.4) & {\cellcolor[HTML]{FEFEFE}} \color[HTML]{000000} {\cellcolor[HTML]{FEE5D8}} \color[HTML]{000000} 0.1 (0.0-0.2) & {\cellcolor[HTML]{FEFEFE}} \color[HTML]{000000} {\cellcolor[HTML]{FC9B7C}} \color[HTML]{000000} 0.4 (0.2-0.6) \\
{\cellcolor[HTML]{FEFEFE}} \color[HTML]{000000} 40 & {\cellcolor[HTML]{FEFEFE}} \color[HTML]{000000} $\text{A}^\text{140}_\text{OP2}$-$\text{O}_\text{ph}$ & {\cellcolor[HTML]{FEFEFE}} \color[HTML]{000000} {\cellcolor{white}}  & {\cellcolor[HTML]{FEFEFE}} \color[HTML]{000000}  & {\cellcolor[HTML]{FEFEFE}} \color[HTML]{000000} {\cellcolor[HTML]{000000}} \color[HTML]{F1F1F1} {\cellcolor{white}}  & {\cellcolor[HTML]{FEFEFE}} \color[HTML]{000000} {\cellcolor[HTML]{000000}} \color[HTML]{F1F1F1} {\cellcolor{white}}  & {\cellcolor[HTML]{FEFEFE}} \color[HTML]{000000} {\cellcolor[HTML]{000000}} \color[HTML]{F1F1F1} {\cellcolor{white}}  & {\cellcolor[HTML]{FEFEFE}} \color[HTML]{000000} {\cellcolor[HTML]{FCBBA1}} \color[HTML]{000000} 0.2 & {\cellcolor[HTML]{FEFEFE}} \color[HTML]{000000} {\cellcolor[HTML]{FDD1BE}} \color[HTML]{000000} 0.2 (0.0-0.4) & {\cellcolor[HTML]{FEFEFE}} \color[HTML]{000000} {\cellcolor[HTML]{FB7C5C}} \color[HTML]{F1F1F1} 0.4 (0.3-0.5) & {\cellcolor[HTML]{FEFEFE}} \color[HTML]{000000} {\cellcolor[HTML]{FC8F6F}} \color[HTML]{000000} 0.4 (0.3-0.5) & {\cellcolor[HTML]{FEFEFE}} \color[HTML]{000000} {\cellcolor[HTML]{FCA486}} \color[HTML]{000000} 0.3 (0.2-0.4) & {\cellcolor[HTML]{FEFEFE}} \color[HTML]{000000} {\cellcolor[HTML]{E53228}} \color[HTML]{F1F1F1} 0.7 (0.5-0.8) & {\cellcolor[HTML]{FEFEFE}} \color[HTML]{000000} {\cellcolor[HTML]{F24633}} \color[HTML]{F1F1F1} 0.6 (0.4-0.7) & {\cellcolor[HTML]{FEFEFE}} \color[HTML]{000000} {\cellcolor[HTML]{CB181D}} \color[HTML]{F1F1F1} 0.7 (0.6-0.9) & {\cellcolor[HTML]{FEFEFE}} \color[HTML]{000000} {\cellcolor[HTML]{FCBBA1}} \color[HTML]{000000} 0.2 (0.2-0.4) & {\cellcolor[HTML]{FEFEFE}} \color[HTML]{000000} {\cellcolor[HTML]{FB7A5A}} \color[HTML]{F1F1F1} 0.5 (0.4-0.6) & {\cellcolor[HTML]{FEFEFE}} \color[HTML]{000000} {\cellcolor[HTML]{FCAB8F}} \color[HTML]{000000} 0.3 (0.2-0.4) \\
{\cellcolor[HTML]{FEFEFE}} \color[HTML]{000000} 41 & {\cellcolor[HTML]{FEFEFE}} \color[HTML]{000000} $\text{U}^\text{258}_\text{OP1}$,$\text{U}^\text{259}_\text{OP2}$,$\text{A}^\text{261}_\text{OP2}$-\textit{mer}-3$\text{O}_\text{ph}$ & {\cellcolor[HTML]{FEFEFE}} \color[HTML]{000000} {\cellcolor{white}}  & {\cellcolor[HTML]{FEFEFE}} \color[HTML]{000000}  & {\cellcolor[HTML]{FEFEFE}} \color[HTML]{000000} {\cellcolor[HTML]{A30F15}} \color[HTML]{F1F1F1} 0.9 & {\cellcolor[HTML]{FEFEFE}} \color[HTML]{000000} {\cellcolor[HTML]{000000}} \color[HTML]{F1F1F1} {\cellcolor{white}}  & {\cellcolor[HTML]{FEFEFE}} \color[HTML]{000000} {\cellcolor[HTML]{DC2924}} \color[HTML]{F1F1F1} 0.7 & {\cellcolor[HTML]{FEFEFE}} \color[HTML]{000000} {\cellcolor[HTML]{000000}} \color[HTML]{F1F1F1} {\cellcolor{white}}  & {\cellcolor[HTML]{FEFEFE}} \color[HTML]{000000} {\cellcolor[HTML]{B51318}} \color[HTML]{F1F1F1} 0.8 (0.6-0.9) & {\cellcolor[HTML]{FEFEFE}} \color[HTML]{000000} {\cellcolor[HTML]{FC8D6D}} \color[HTML]{F1F1F1} 0.4 (0.1-0.7) & {\cellcolor[HTML]{FEFEFE}} \color[HTML]{000000} {\cellcolor[HTML]{FCC1A8}} \color[HTML]{000000} 0.2 (0.1-0.4) & {\cellcolor[HTML]{FEFEFE}} \color[HTML]{000000} {\cellcolor[HTML]{F34C37}} \color[HTML]{F1F1F1} 0.6 (0.4-0.8) & {\cellcolor[HTML]{FEFEFE}} \color[HTML]{000000} {\cellcolor[HTML]{000000}} \color[HTML]{F1F1F1} {\cellcolor{white}}  & {\cellcolor[HTML]{FEFEFE}} \color[HTML]{000000} {\cellcolor[HTML]{000000}} \color[HTML]{F1F1F1} {\cellcolor{white}}  & {\cellcolor[HTML]{FEFEFE}} \color[HTML]{000000} {\cellcolor[HTML]{000000}} \color[HTML]{F1F1F1} {\cellcolor{white}}  & {\cellcolor[HTML]{FEFEFE}} \color[HTML]{000000} {\cellcolor[HTML]{FC9B7C}} \color[HTML]{000000} 0.4 (0.2-0.5) & {\cellcolor[HTML]{FEFEFE}} \color[HTML]{000000} {\cellcolor[HTML]{FCAB8F}} \color[HTML]{000000} 0.3 (0.1-0.5) & {\cellcolor[HTML]{FEFEFE}} \color[HTML]{000000} {\cellcolor[HTML]{FDCAB5}} \color[HTML]{000000} 0.2 (0.1-0.4) \\
{\cellcolor[HTML]{FEFEFE}} \color[HTML]{000000} 42 & {\cellcolor[HTML]{FEFEFE}} \color[HTML]{000000} $\text{G}^\text{163}_\text{OP2}$-$\text{O}_\text{ph}$ & {\cellcolor[HTML]{FEFEFE}} \color[HTML]{000000} {\cellcolor{white}}  & {\cellcolor[HTML]{FEFEFE}} \color[HTML]{000000}  & {\cellcolor[HTML]{FEFEFE}} \color[HTML]{000000} {\cellcolor[HTML]{EE3A2C}} \color[HTML]{F1F1F1} 0.6 & {\cellcolor[HTML]{FEFEFE}} \color[HTML]{000000} {\cellcolor[HTML]{000000}} \color[HTML]{F1F1F1} {\cellcolor{white}}  & {\cellcolor[HTML]{FEFEFE}} \color[HTML]{000000} {\cellcolor[HTML]{FC9272}} \color[HTML]{000000} 0.4 & {\cellcolor[HTML]{FEFEFE}} \color[HTML]{000000} {\cellcolor[HTML]{FEE0D2}} \color[HTML]{000000} 0.1 & {\cellcolor[HTML]{FEFEFE}} \color[HTML]{000000} {\cellcolor[HTML]{CB181D}} \color[HTML]{F1F1F1} 0.7 (0.4-0.9) & {\cellcolor[HTML]{FEFEFE}} \color[HTML]{000000} {\cellcolor[HTML]{FA6547}} \color[HTML]{F1F1F1} 0.5 (0.3-0.7) & {\cellcolor[HTML]{FEFEFE}} \color[HTML]{000000} {\cellcolor[HTML]{F14432}} \color[HTML]{F1F1F1} 0.6 (0.5-0.7) & {\cellcolor[HTML]{FEFEFE}} \color[HTML]{000000} {\cellcolor[HTML]{F24734}} \color[HTML]{F1F1F1} 0.6 (0.4-0.7) & {\cellcolor[HTML]{FEFEFE}} \color[HTML]{000000} {\cellcolor[HTML]{000000}} \color[HTML]{F1F1F1} {\cellcolor{white}}  & {\cellcolor[HTML]{FEFEFE}} \color[HTML]{000000} {\cellcolor[HTML]{FCB99F}} \color[HTML]{000000} 0.3 (0.1-0.5) & {\cellcolor[HTML]{FEFEFE}} \color[HTML]{000000} {\cellcolor[HTML]{FC7F5F}} \color[HTML]{F1F1F1} 0.4 (0.3-0.6) & {\cellcolor[HTML]{FEFEFE}} \color[HTML]{000000} {\cellcolor[HTML]{FFF5F0}} \color[HTML]{000000} 0.0 (0.0-0.0) & {\cellcolor[HTML]{FEFEFE}} \color[HTML]{000000} {\cellcolor[HTML]{FFF5F0}} \color[HTML]{000000} 0.0 (0.0-0.1) & {\cellcolor[HTML]{FEFEFE}} \color[HTML]{000000} {\cellcolor[HTML]{FFF5F0}} \color[HTML]{000000} 0.0 (0.0-0.0) \\
{\cellcolor[HTML]{FEFEFE}} \color[HTML]{000000} 43 & {\cellcolor[HTML]{FEFEFE}} \color[HTML]{000000} $\text{A}^\text{226}_\text{OP2}$-$\text{O}_\text{ph}$ & {\cellcolor[HTML]{FEFEFE}} \color[HTML]{000000} {\cellcolor{white}}  & {\cellcolor[HTML]{FEFEFE}} \color[HTML]{000000}  & {\cellcolor[HTML]{FEFEFE}} \color[HTML]{000000} {\cellcolor[HTML]{000000}} \color[HTML]{F1F1F1} {\cellcolor{white}}  & {\cellcolor[HTML]{FEFEFE}} \color[HTML]{000000} {\cellcolor[HTML]{000000}} \color[HTML]{F1F1F1} {\cellcolor{white}}  & {\cellcolor[HTML]{FEFEFE}} \color[HTML]{000000} {\cellcolor[HTML]{000000}} \color[HTML]{F1F1F1} {\cellcolor{white}}  & {\cellcolor[HTML]{FEFEFE}} \color[HTML]{000000} {\cellcolor[HTML]{000000}} \color[HTML]{F1F1F1} {\cellcolor{white}}  & {\cellcolor[HTML]{FEFEFE}} \color[HTML]{000000} {\cellcolor[HTML]{FCA285}} \color[HTML]{000000} 0.3 (0.1-0.6) & {\cellcolor[HTML]{FEFEFE}} \color[HTML]{000000} {\cellcolor[HTML]{FB7050}} \color[HTML]{F1F1F1} 0.5 (0.2-0.8) & {\cellcolor[HTML]{FEFEFE}} \color[HTML]{000000} {\cellcolor[HTML]{FC9777}} \color[HTML]{000000} 0.4 (0.1-0.6) & {\cellcolor[HTML]{FEFEFE}} \color[HTML]{000000} {\cellcolor[HTML]{EE3A2C}} \color[HTML]{F1F1F1} 0.6 (0.4-0.8) & {\cellcolor[HTML]{FEFEFE}} \color[HTML]{000000} {\cellcolor[HTML]{F14130}} \color[HTML]{F1F1F1} 0.6 (0.3-0.8) & {\cellcolor[HTML]{FEFEFE}} \color[HTML]{000000} {\cellcolor[HTML]{FCBFA7}} \color[HTML]{000000} 0.2 (0.0-0.5) & {\cellcolor[HTML]{FEFEFE}} \color[HTML]{000000} {\cellcolor[HTML]{DA2723}} \color[HTML]{F1F1F1} 0.7 (0.5-0.9) & {\cellcolor[HTML]{FEFEFE}} \color[HTML]{000000} {\cellcolor[HTML]{FED9C9}} \color[HTML]{000000} 0.2 (0.1-0.3) & {\cellcolor[HTML]{FEFEFE}} \color[HTML]{000000} {\cellcolor[HTML]{FC9B7C}} \color[HTML]{000000} 0.4 (0.2-0.4) & {\cellcolor[HTML]{FEFEFE}} \color[HTML]{000000} {\cellcolor[HTML]{FDCAB5}} \color[HTML]{000000} 0.2 (0.1-0.3) \\
{\cellcolor[HTML]{FEFEFE}} \color[HTML]{000000} 44 & {\cellcolor[HTML]{FEFEFE}} \color[HTML]{000000} $\text{A}^\text{359}_\text{OP2}$-$\text{O}_\text{ph}$ & {\cellcolor[HTML]{FEFEFE}} \color[HTML]{000000} {\cellcolor{white}}  & {\cellcolor[HTML]{FEFEFE}} \color[HTML]{000000}  & {\cellcolor[HTML]{FEFEFE}} \color[HTML]{000000} {\cellcolor[HTML]{000000}} \color[HTML]{F1F1F1} {\cellcolor{white}}  & {\cellcolor[HTML]{FEFEFE}} \color[HTML]{000000} {\cellcolor[HTML]{FEEAE1}} \color[HTML]{000000} 0.1 & {\cellcolor[HTML]{FEFEFE}} \color[HTML]{000000} {\cellcolor[HTML]{000000}} \color[HTML]{F1F1F1} {\cellcolor{white}}  & {\cellcolor[HTML]{FEFEFE}} \color[HTML]{000000} {\cellcolor[HTML]{000000}} \color[HTML]{F1F1F1} {\cellcolor{white}}  & {\cellcolor[HTML]{FEFEFE}} \color[HTML]{000000} {\cellcolor[HTML]{FEEAE0}} \color[HTML]{000000} 0.1 (0.0-0.3) & {\cellcolor[HTML]{FEFEFE}} \color[HTML]{000000} {\cellcolor[HTML]{FC997A}} \color[HTML]{000000} 0.4 (0.2-0.6) & {\cellcolor[HTML]{FEFEFE}} \color[HTML]{000000} {\cellcolor[HTML]{FEE1D4}} \color[HTML]{000000} 0.1 (0.0-0.2) & {\cellcolor[HTML]{FEFEFE}} \color[HTML]{000000} {\cellcolor[HTML]{FCB095}} \color[HTML]{000000} 0.3 (0.1-0.5) & {\cellcolor[HTML]{FEFEFE}} \color[HTML]{000000} {\cellcolor[HTML]{FDC9B3}} \color[HTML]{000000} 0.2 (0.0-0.5) & {\cellcolor[HTML]{FEFEFE}} \color[HTML]{000000} {\cellcolor[HTML]{FB7A5A}} \color[HTML]{F1F1F1} 0.5 (0.3-0.7) & {\cellcolor[HTML]{FEFEFE}} \color[HTML]{000000} {\cellcolor[HTML]{FB7A5A}} \color[HTML]{F1F1F1} 0.5 (0.3-0.6) & {\cellcolor[HTML]{FEFEFE}} \color[HTML]{000000} {\cellcolor[HTML]{FB7A5A}} \color[HTML]{F1F1F1} 0.5 (0.3-0.6) & {\cellcolor[HTML]{FEFEFE}} \color[HTML]{000000} {\cellcolor[HTML]{FC9B7C}} \color[HTML]{000000} 0.4 (0.3-0.5) & {\cellcolor[HTML]{FEFEFE}} \color[HTML]{000000} {\cellcolor[HTML]{F6583E}} \color[HTML]{F1F1F1} 0.6 (0.5-0.7) \\
{\cellcolor[HTML]{FEFEFE}} \color[HTML]{000000} 45 & {\cellcolor[HTML]{FEFEFE}} \color[HTML]{000000} $\text{A}^\text{114}_\text{OP2}$-$\text{O}_\text{ph}$ & {\cellcolor[HTML]{FEFEFE}} \color[HTML]{000000} {\cellcolor{white}}  & {\cellcolor[HTML]{FEFEFE}} \color[HTML]{000000}  & {\cellcolor[HTML]{FEFEFE}} \color[HTML]{000000} {\cellcolor[HTML]{840711}} \color[HTML]{F1F1F1} 0.9 & {\cellcolor[HTML]{FEFEFE}} \color[HTML]{000000} {\cellcolor[HTML]{000000}} \color[HTML]{F1F1F1} {\cellcolor{white}}  & {\cellcolor[HTML]{FEFEFE}} \color[HTML]{000000} {\cellcolor[HTML]{FDCDB9}} \color[HTML]{000000} 0.2 & {\cellcolor[HTML]{FEFEFE}} \color[HTML]{000000} {\cellcolor[HTML]{000000}} \color[HTML]{F1F1F1} {\cellcolor{white}}  & {\cellcolor[HTML]{FEFEFE}} \color[HTML]{000000} {\cellcolor[HTML]{840711}} \color[HTML]{F1F1F1} 0.9 (0.8-1.0) & {\cellcolor[HTML]{FEFEFE}} \color[HTML]{000000} {\cellcolor[HTML]{FDC5AE}} \color[HTML]{000000} 0.2 (0.0-0.5) & {\cellcolor[HTML]{FEFEFE}} \color[HTML]{000000} {\cellcolor[HTML]{FFF0E9}} \color[HTML]{000000} 0.0 (0.0-0.1) & {\cellcolor[HTML]{FEFEFE}} \color[HTML]{000000} {\cellcolor[HTML]{FCA588}} \color[HTML]{000000} 0.3 (0.2-0.5) & {\cellcolor[HTML]{FEFEFE}} \color[HTML]{000000} {\cellcolor[HTML]{000000}} \color[HTML]{F1F1F1} {\cellcolor{white}}  & {\cellcolor[HTML]{FEFEFE}} \color[HTML]{000000} {\cellcolor[HTML]{FFF4EE}} \color[HTML]{000000} 0.0 (0.0-0.0) & {\cellcolor[HTML]{FEFEFE}} \color[HTML]{000000} {\cellcolor[HTML]{000000}} \color[HTML]{F1F1F1} {\cellcolor{white}}  & {\cellcolor[HTML]{FEFEFE}} \color[HTML]{000000} {\cellcolor[HTML]{FC9B7C}} \color[HTML]{000000} 0.4 (0.3-0.5) & {\cellcolor[HTML]{FEFEFE}} \color[HTML]{000000} {\cellcolor[HTML]{FEE5D8}} \color[HTML]{000000} 0.1 (0.0-0.2) & {\cellcolor[HTML]{FEFEFE}} \color[HTML]{000000} {\cellcolor[HTML]{FCBBA1}} \color[HTML]{000000} 0.2 (0.2-0.4) \\
{\cellcolor[HTML]{FEFEFE}} \color[HTML]{000000} 46 & {\cellcolor[HTML]{FEFEFE}} \color[HTML]{000000} $\text{A}^\text{104}_\text{OP2}$-$\text{O}_\text{ph}$ & {\cellcolor[HTML]{FEFEFE}} \color[HTML]{000000} {\cellcolor{white}}  & {\cellcolor[HTML]{FEFEFE}} \color[HTML]{000000}  & {\cellcolor[HTML]{FEFEFE}} \color[HTML]{000000} {\cellcolor[HTML]{000000}} \color[HTML]{F1F1F1} {\cellcolor{white}}  & {\cellcolor[HTML]{FEFEFE}} \color[HTML]{000000} {\cellcolor[HTML]{000000}} \color[HTML]{F1F1F1} {\cellcolor{white}}  & {\cellcolor[HTML]{FEFEFE}} \color[HTML]{000000} {\cellcolor[HTML]{FEE0D2}} \color[HTML]{000000} 0.1 & {\cellcolor[HTML]{FEFEFE}} \color[HTML]{000000} {\cellcolor[HTML]{FEE0D2}} \color[HTML]{000000} 0.1 & {\cellcolor[HTML]{FEFEFE}} \color[HTML]{000000} {\cellcolor[HTML]{FEE8DD}} \color[HTML]{000000} 0.1 (0.0-0.3) & {\cellcolor[HTML]{FEFEFE}} \color[HTML]{000000} {\cellcolor[HTML]{FEE7DC}} \color[HTML]{000000} 0.1 (0.0-0.1) & {\cellcolor[HTML]{FEFEFE}} \color[HTML]{000000} {\cellcolor[HTML]{FCAD90}} \color[HTML]{000000} 0.3 (0.1-0.5) & {\cellcolor[HTML]{FEFEFE}} \color[HTML]{000000} {\cellcolor[HTML]{FCBBA1}} \color[HTML]{000000} 0.3 (0.1-0.4) & {\cellcolor[HTML]{FEFEFE}} \color[HTML]{000000} {\cellcolor[HTML]{FFECE3}} \color[HTML]{000000} 0.1 (0.0-0.2) & {\cellcolor[HTML]{FEFEFE}} \color[HTML]{000000} {\cellcolor[HTML]{FDCAB5}} \color[HTML]{000000} 0.2 (0.0-0.4) & {\cellcolor[HTML]{FEFEFE}} \color[HTML]{000000} {\cellcolor[HTML]{FDD2BF}} \color[HTML]{000000} 0.2 (0.1-0.3) & {\cellcolor[HTML]{FEFEFE}} \color[HTML]{000000} {\cellcolor[HTML]{FCAB8F}} \color[HTML]{000000} 0.3 (0.1-0.6) & {\cellcolor[HTML]{FEFEFE}} \color[HTML]{000000} {\cellcolor[HTML]{CA181D}} \color[HTML]{F1F1F1} 0.8 (0.3-1.0) & {\cellcolor[HTML]{FEFEFE}} \color[HTML]{000000} {\cellcolor[HTML]{D92523}} \color[HTML]{F1F1F1} 0.7 (0.4-1.0) \\
{\cellcolor[HTML]{FEFEFE}} \color[HTML]{000000} 47 & {\cellcolor[HTML]{FEFEFE}} \color[HTML]{000000} $\text{U}^\text{267}_\text{OP2}$-$\text{O}_\text{ph}$ & {\cellcolor[HTML]{FEFEFE}} \color[HTML]{000000} {\cellcolor{white}}  & {\cellcolor[HTML]{FEFEFE}} \color[HTML]{000000}  & {\cellcolor[HTML]{FEFEFE}} \color[HTML]{000000} {\cellcolor[HTML]{000000}} \color[HTML]{F1F1F1} {\cellcolor{white}}  & {\cellcolor[HTML]{FEFEFE}} \color[HTML]{000000} {\cellcolor[HTML]{FEEAE1}} \color[HTML]{000000} 0.1 & {\cellcolor[HTML]{FEFEFE}} \color[HTML]{000000} {\cellcolor[HTML]{FEEAE1}} \color[HTML]{000000} 0.1 & {\cellcolor[HTML]{FEFEFE}} \color[HTML]{000000} {\cellcolor[HTML]{FB694A}} \color[HTML]{F1F1F1} 0.5 & {\cellcolor[HTML]{FEFEFE}} \color[HTML]{000000} {\cellcolor[HTML]{FFEFE8}} \color[HTML]{000000} 0.0 (0.0-0.1) & {\cellcolor[HTML]{FEFEFE}} \color[HTML]{000000} {\cellcolor[HTML]{FDC5AE}} \color[HTML]{000000} 0.2 (0.0-0.5) & {\cellcolor[HTML]{FEFEFE}} \color[HTML]{000000} {\cellcolor[HTML]{FDCAB5}} \color[HTML]{000000} 0.2 (0.0-0.5) & {\cellcolor[HTML]{FEFEFE}} \color[HTML]{000000} {\cellcolor[HTML]{FB7C5C}} \color[HTML]{F1F1F1} 0.4 (0.3-0.6) & {\cellcolor[HTML]{FEFEFE}} \color[HTML]{000000} {\cellcolor[HTML]{FEE3D6}} \color[HTML]{000000} 0.1 (0.0-0.3) & {\cellcolor[HTML]{FEFEFE}} \color[HTML]{000000} {\cellcolor[HTML]{FC8F6F}} \color[HTML]{000000} 0.4 (0.2-0.6) & {\cellcolor[HTML]{FEFEFE}} \color[HTML]{000000} {\cellcolor[HTML]{FC8969}} \color[HTML]{F1F1F1} 0.4 (0.2-0.6) & {\cellcolor[HTML]{FEFEFE}} \color[HTML]{000000} {\cellcolor[HTML]{FCBBA1}} \color[HTML]{000000} 0.2 (0.1-0.4) & {\cellcolor[HTML]{FEFEFE}} \color[HTML]{000000} {\cellcolor[HTML]{FEE5D8}} \color[HTML]{000000} 0.1 (0.0-0.2) & {\cellcolor[HTML]{FEFEFE}} \color[HTML]{000000} {\cellcolor[HTML]{FCAB8F}} \color[HTML]{000000} 0.3 (0.2-0.5) \\
{\cellcolor[HTML]{FEFEFE}} \color[HTML]{000000} 48 & {\cellcolor[HTML]{FEFEFE}} \color[HTML]{000000} $\text{A}^\text{207}_\text{OP1}$,$\text{C}^\text{262}_\text{OP2}$-\textit{cis}-2$\text{O}_\text{ph}$ & {\cellcolor[HTML]{FEFEFE}} \color[HTML]{000000} {\cellcolor{white}}  & {\cellcolor[HTML]{FEFEFE}} \color[HTML]{000000}  & {\cellcolor[HTML]{FEFEFE}} \color[HTML]{000000} {\cellcolor[HTML]{CA181D}} \color[HTML]{F1F1F1} 0.8 & {\cellcolor[HTML]{FEFEFE}} \color[HTML]{000000} {\cellcolor[HTML]{000000}} \color[HTML]{F1F1F1} {\cellcolor{white}}  & {\cellcolor[HTML]{FEFEFE}} \color[HTML]{000000} {\cellcolor[HTML]{FDCDB9}} \color[HTML]{000000} 0.2 & {\cellcolor[HTML]{FEFEFE}} \color[HTML]{000000} {\cellcolor[HTML]{000000}} \color[HTML]{F1F1F1} {\cellcolor{white}}  & {\cellcolor[HTML]{FEFEFE}} \color[HTML]{000000} {\cellcolor[HTML]{FC8666}} \color[HTML]{F1F1F1} 0.4 (0.1-0.7) & {\cellcolor[HTML]{FEFEFE}} \color[HTML]{000000} {\cellcolor[HTML]{FFEEE7}} \color[HTML]{000000} 0.0 (0.0-0.1) & {\cellcolor[HTML]{FEFEFE}} \color[HTML]{000000} {\cellcolor[HTML]{FCA486}} \color[HTML]{000000} 0.3 (0.1-0.6) & {\cellcolor[HTML]{FEFEFE}} \color[HTML]{000000} {\cellcolor[HTML]{FFF2EC}} \color[HTML]{000000} 0.0 (0.0-0.0) & {\cellcolor[HTML]{FEFEFE}} \color[HTML]{000000} {\cellcolor[HTML]{000000}} \color[HTML]{F1F1F1} {\cellcolor{white}}  & {\cellcolor[HTML]{FEFEFE}} \color[HTML]{000000} {\cellcolor[HTML]{FFF5F0}} \color[HTML]{000000} 0.0 (0.0-0.0) & {\cellcolor[HTML]{FEFEFE}} \color[HTML]{000000} {\cellcolor[HTML]{FFF4EF}} \color[HTML]{000000} 0.0 (0.0-0.0) & {\cellcolor[HTML]{FEFEFE}} \color[HTML]{000000} {\cellcolor[HTML]{FCBBA1}} \color[HTML]{000000} 0.2 (0.2-0.4) & {\cellcolor[HTML]{FEFEFE}} \color[HTML]{000000} {\cellcolor[HTML]{FCBBA1}} \color[HTML]{000000} 0.2 (0.2-0.4) & {\cellcolor[HTML]{FEFEFE}} \color[HTML]{000000} {\cellcolor[HTML]{FED9C9}} \color[HTML]{000000} 0.2 (0.0-0.4) \\
{\cellcolor[HTML]{FEFEFE}} \color[HTML]{000000} 49 & {\cellcolor[HTML]{FEFEFE}} \color[HTML]{000000} $\text{C}^\text{208}_\text{OP1}$,$\text{A}^\text{304}_\text{OP1}$,$\text{A}^\text{306}_\text{OP1}$-\textit{fac}-3$\text{O}_\text{ph}$ & {\cellcolor[HTML]{FEFEFE}} \color[HTML]{000000} {\cellcolor{white}}  & {\cellcolor[HTML]{FEFEFE}} \color[HTML]{000000}  & {\cellcolor[HTML]{FEFEFE}} \color[HTML]{000000} {\cellcolor[HTML]{FEE0D2}} \color[HTML]{000000} 0.1 & {\cellcolor[HTML]{FEFEFE}} \color[HTML]{000000} {\cellcolor[HTML]{000000}} \color[HTML]{F1F1F1} {\cellcolor{white}}  & {\cellcolor[HTML]{FEFEFE}} \color[HTML]{000000} {\cellcolor[HTML]{FEE0D2}} \color[HTML]{000000} 0.1 & {\cellcolor[HTML]{FEFEFE}} \color[HTML]{000000} {\cellcolor[HTML]{000000}} \color[HTML]{F1F1F1} {\cellcolor{white}}  & {\cellcolor[HTML]{FEFEFE}} \color[HTML]{000000} {\cellcolor[HTML]{FC9272}} \color[HTML]{000000} 0.4 (0.1-0.7) & {\cellcolor[HTML]{FEFEFE}} \color[HTML]{000000} {\cellcolor[HTML]{FC9474}} \color[HTML]{000000} 0.4 (0.1-0.7) & {\cellcolor[HTML]{FEFEFE}} \color[HTML]{000000} {\cellcolor[HTML]{FDD5C4}} \color[HTML]{000000} 0.2 (0.0-0.4) & {\cellcolor[HTML]{FEFEFE}} \color[HTML]{000000} {\cellcolor[HTML]{FFF5F0}} \color[HTML]{000000} 0.0 (0.0-0.0) & {\cellcolor[HTML]{FEFEFE}} \color[HTML]{000000} {\cellcolor[HTML]{000000}} \color[HTML]{F1F1F1} {\cellcolor{white}}  & {\cellcolor[HTML]{FEFEFE}} \color[HTML]{000000} {\cellcolor[HTML]{000000}} \color[HTML]{F1F1F1} {\cellcolor{white}}  & {\cellcolor[HTML]{FEFEFE}} \color[HTML]{000000} {\cellcolor[HTML]{000000}} \color[HTML]{F1F1F1} {\cellcolor{white}}  & {\cellcolor[HTML]{FEFEFE}} \color[HTML]{000000} {\cellcolor[HTML]{FB694A}} \color[HTML]{F1F1F1} 0.5 (0.3-0.6) & {\cellcolor[HTML]{FEFEFE}} \color[HTML]{000000} {\cellcolor[HTML]{FC8A6A}} \color[HTML]{F1F1F1} 0.4 (0.3-0.5) & {\cellcolor[HTML]{FEFEFE}} \color[HTML]{000000} {\cellcolor[HTML]{FCAB8F}} \color[HTML]{000000} 0.3 (0.2-0.4) \\
{\cellcolor[HTML]{FEFEFE}} \color[HTML]{000000} 50 & {\cellcolor[HTML]{FEFEFE}} \color[HTML]{000000} $\text{A}^\text{343}_\text{OP2}$-$\text{O}_\text{ph}$ & {\cellcolor[HTML]{FEFEFE}} \color[HTML]{000000} {\cellcolor{white}}  & {\cellcolor[HTML]{FEFEFE}} \color[HTML]{000000}  & {\cellcolor[HTML]{FEFEFE}} \color[HTML]{000000} {\cellcolor[HTML]{000000}} \color[HTML]{F1F1F1} {\cellcolor{white}}  & {\cellcolor[HTML]{FEFEFE}} \color[HTML]{000000} {\cellcolor[HTML]{000000}} \color[HTML]{F1F1F1} {\cellcolor{white}}  & {\cellcolor[HTML]{FEFEFE}} \color[HTML]{000000} {\cellcolor[HTML]{000000}} \color[HTML]{F1F1F1} {\cellcolor{white}}  & {\cellcolor[HTML]{FEFEFE}} \color[HTML]{000000} {\cellcolor[HTML]{000000}} \color[HTML]{F1F1F1} {\cellcolor{white}}  & {\cellcolor[HTML]{FEFEFE}} \color[HTML]{000000} {\cellcolor[HTML]{FEE3D6}} \color[HTML]{000000} 0.1 (0.0-0.3) & {\cellcolor[HTML]{FEFEFE}} \color[HTML]{000000} {\cellcolor[HTML]{FCB296}} \color[HTML]{000000} 0.3 (0.1-0.4) & {\cellcolor[HTML]{FEFEFE}} \color[HTML]{000000} {\cellcolor[HTML]{FCA486}} \color[HTML]{000000} 0.3 (0.2-0.6) & {\cellcolor[HTML]{FEFEFE}} \color[HTML]{000000} {\cellcolor[HTML]{FC8D6D}} \color[HTML]{F1F1F1} 0.4 (0.2-0.5) & {\cellcolor[HTML]{FEFEFE}} \color[HTML]{000000} {\cellcolor[HTML]{000000}} \color[HTML]{F1F1F1} {\cellcolor{white}}  & {\cellcolor[HTML]{FEFEFE}} \color[HTML]{000000} {\cellcolor[HTML]{FCBEA5}} \color[HTML]{000000} 0.2 (0.1-0.4) & {\cellcolor[HTML]{FEFEFE}} \color[HTML]{000000} {\cellcolor[HTML]{FC9777}} \color[HTML]{000000} 0.4 (0.3-0.5) & {\cellcolor[HTML]{FEFEFE}} \color[HTML]{000000} {\cellcolor[HTML]{FFF5F0}} \color[HTML]{000000} 0.0 (0.0-0.1) & {\cellcolor[HTML]{FEFEFE}} \color[HTML]{000000} {\cellcolor[HTML]{FCBBA1}} \color[HTML]{000000} 0.2 (0.1-0.5) & {\cellcolor[HTML]{FEFEFE}} \color[HTML]{000000} {\cellcolor[HTML]{FCAB8F}} \color[HTML]{000000} 0.3 (0.2-0.4) \\
{\cellcolor[HTML]{FEFEFE}} \color[HTML]{000000} 51 & {\cellcolor[HTML]{FEFEFE}} \color[HTML]{000000} $\text{G}^\text{175}_\text{OP2}$-$\text{O}_\text{ph}$ & {\cellcolor[HTML]{FEFEFE}} \color[HTML]{000000} {\cellcolor{white}}  & {\cellcolor[HTML]{FEFEFE}} \color[HTML]{000000}  & {\cellcolor[HTML]{FEFEFE}} \color[HTML]{000000} {\cellcolor[HTML]{000000}} \color[HTML]{F1F1F1} {\cellcolor{white}}  & {\cellcolor[HTML]{FEFEFE}} \color[HTML]{000000} {\cellcolor[HTML]{000000}} \color[HTML]{F1F1F1} {\cellcolor{white}}  & {\cellcolor[HTML]{FEFEFE}} \color[HTML]{000000} {\cellcolor[HTML]{000000}} \color[HTML]{F1F1F1} {\cellcolor{white}}  & {\cellcolor[HTML]{FEFEFE}} \color[HTML]{000000} {\cellcolor[HTML]{000000}} \color[HTML]{F1F1F1} {\cellcolor{white}}  & {\cellcolor[HTML]{FEFEFE}} \color[HTML]{000000} {\cellcolor[HTML]{000000}} \color[HTML]{F1F1F1} {\cellcolor{white}}  & {\cellcolor[HTML]{FEFEFE}} \color[HTML]{000000} {\cellcolor[HTML]{FC9777}} \color[HTML]{000000} 0.4 (0.3-0.5) & {\cellcolor[HTML]{FEFEFE}} \color[HTML]{000000} {\cellcolor[HTML]{FDCBB6}} \color[HTML]{000000} 0.2 (0.1-0.3) & {\cellcolor[HTML]{FEFEFE}} \color[HTML]{000000} {\cellcolor[HTML]{FCA78B}} \color[HTML]{000000} 0.3 (0.2-0.4) & {\cellcolor[HTML]{FEFEFE}} \color[HTML]{000000} {\cellcolor[HTML]{FDCDB9}} \color[HTML]{000000} 0.2 (0.0-0.5) & {\cellcolor[HTML]{FEFEFE}} \color[HTML]{000000} {\cellcolor[HTML]{FEDCCD}} \color[HTML]{000000} 0.1 (0.1-0.3) & {\cellcolor[HTML]{FEFEFE}} \color[HTML]{000000} {\cellcolor[HTML]{FC8262}} \color[HTML]{F1F1F1} 0.4 (0.3-0.6) & {\cellcolor[HTML]{FEFEFE}} \color[HTML]{000000} {\cellcolor[HTML]{FCBBA1}} \color[HTML]{000000} 0.2 (0.0-0.5) & {\cellcolor[HTML]{FEFEFE}} \color[HTML]{000000} {\cellcolor[HTML]{FFEDE5}} \color[HTML]{000000} 0.1 (0.0-0.3) & {\cellcolor[HTML]{FEFEFE}} \color[HTML]{000000} {\cellcolor[HTML]{FCAB8F}} \color[HTML]{000000} 0.3 (0.2-0.5) \\
{\cellcolor[HTML]{FEFEFE}} \color[HTML]{000000} 52 & {\cellcolor[HTML]{FEFEFE}} \color[HTML]{000000} $\text{A}^\text{172}_\text{OP2}$-$\text{O}_\text{ph}$ & {\cellcolor[HTML]{FEFEFE}} \color[HTML]{000000} {\cellcolor{white}}  & {\cellcolor[HTML]{FEFEFE}} \color[HTML]{000000}  & {\cellcolor[HTML]{FEFEFE}} \color[HTML]{000000} {\cellcolor[HTML]{000000}} \color[HTML]{F1F1F1} {\cellcolor{white}}  & {\cellcolor[HTML]{FEFEFE}} \color[HTML]{000000} {\cellcolor[HTML]{FEE0D2}} \color[HTML]{000000} 0.1 & {\cellcolor[HTML]{FEFEFE}} \color[HTML]{000000} {\cellcolor[HTML]{000000}} \color[HTML]{F1F1F1} {\cellcolor{white}}  & {\cellcolor[HTML]{FEFEFE}} \color[HTML]{000000} {\cellcolor[HTML]{000000}} \color[HTML]{F1F1F1} {\cellcolor{white}}  & {\cellcolor[HTML]{FEFEFE}} \color[HTML]{000000} {\cellcolor[HTML]{FEE7DB}} \color[HTML]{000000} 0.1 (0.0-0.3) & {\cellcolor[HTML]{FEFEFE}} \color[HTML]{000000} {\cellcolor[HTML]{FC8969}} \color[HTML]{F1F1F1} 0.4 (0.2-0.6) & {\cellcolor[HTML]{FEFEFE}} \color[HTML]{000000} {\cellcolor[HTML]{FDC6B0}} \color[HTML]{000000} 0.2 (0.1-0.3) & {\cellcolor[HTML]{FEFEFE}} \color[HTML]{000000} {\cellcolor[HTML]{FB7252}} \color[HTML]{F1F1F1} 0.5 (0.4-0.6) & {\cellcolor[HTML]{FEFEFE}} \color[HTML]{000000} {\cellcolor[HTML]{FFECE3}} \color[HTML]{000000} 0.1 (0.0-0.2) & {\cellcolor[HTML]{FEFEFE}} \color[HTML]{000000} {\cellcolor[HTML]{FCAB8F}} \color[HTML]{000000} 0.3 (0.2-0.4) & {\cellcolor[HTML]{FEFEFE}} \color[HTML]{000000} {\cellcolor[HTML]{FB6C4C}} \color[HTML]{F1F1F1} 0.5 (0.4-0.6) & {\cellcolor[HTML]{FEFEFE}} \color[HTML]{000000} {\cellcolor[HTML]{FFF5F0}} \color[HTML]{000000} 0.0 (0.0-0.0) & {\cellcolor[HTML]{FEFEFE}} \color[HTML]{000000} {\cellcolor[HTML]{FFF5F0}} \color[HTML]{000000} 0.0 (0.0-0.0) & {\cellcolor[HTML]{FEFEFE}} \color[HTML]{000000} {\cellcolor[HTML]{FFF5F0}} \color[HTML]{000000} 0.0 (0.0-0.0) \\
{\cellcolor[HTML]{FEFEFE}} \color[HTML]{000000} 53 & {\cellcolor[HTML]{FEFEFE}} \color[HTML]{000000} $\text{A}^\text{139}_\text{OP2}$-$\text{O}_\text{ph}$ & {\cellcolor[HTML]{FEFEFE}} \color[HTML]{000000} {\cellcolor{white}}  & {\cellcolor[HTML]{FEFEFE}} \color[HTML]{000000}  & {\cellcolor[HTML]{FEFEFE}} \color[HTML]{000000} {\cellcolor[HTML]{FDCDB9}} \color[HTML]{000000} 0.2 & {\cellcolor[HTML]{FEFEFE}} \color[HTML]{000000} {\cellcolor[HTML]{000000}} \color[HTML]{F1F1F1} {\cellcolor{white}}  & {\cellcolor[HTML]{FEFEFE}} \color[HTML]{000000} {\cellcolor[HTML]{FEE0D2}} \color[HTML]{000000} 0.1 & {\cellcolor[HTML]{FEFEFE}} \color[HTML]{000000} {\cellcolor[HTML]{000000}} \color[HTML]{F1F1F1} {\cellcolor{white}}  & {\cellcolor[HTML]{FEFEFE}} \color[HTML]{000000} {\cellcolor[HTML]{FC8A6A}} \color[HTML]{F1F1F1} 0.4 (0.2-0.7) & {\cellcolor[HTML]{FEFEFE}} \color[HTML]{000000} {\cellcolor[HTML]{FFF3ED}} \color[HTML]{000000} 0.0 (0.0-0.0) & {\cellcolor[HTML]{FEFEFE}} \color[HTML]{000000} {\cellcolor[HTML]{FFF5F0}} \color[HTML]{000000} 0.0 (0.0-0.0) & {\cellcolor[HTML]{FEFEFE}} \color[HTML]{000000} {\cellcolor[HTML]{FFF2EC}} \color[HTML]{000000} 0.0 (0.0-0.0) & {\cellcolor[HTML]{FEFEFE}} \color[HTML]{000000} {\cellcolor[HTML]{000000}} \color[HTML]{F1F1F1} {\cellcolor{white}}  & {\cellcolor[HTML]{FEFEFE}} \color[HTML]{000000} {\cellcolor[HTML]{000000}} \color[HTML]{F1F1F1} {\cellcolor{white}}  & {\cellcolor[HTML]{FEFEFE}} \color[HTML]{000000} {\cellcolor[HTML]{000000}} \color[HTML]{F1F1F1} {\cellcolor{white}}  & {\cellcolor[HTML]{FEFEFE}} \color[HTML]{000000} {\cellcolor[HTML]{F6583E}} \color[HTML]{F1F1F1} 0.6 (0.4-0.7) & {\cellcolor[HTML]{FEFEFE}} \color[HTML]{000000} {\cellcolor[HTML]{FC9B7C}} \color[HTML]{000000} 0.4 (0.2-0.5) & {\cellcolor[HTML]{FEFEFE}} \color[HTML]{000000} {\cellcolor[HTML]{FB694A}} \color[HTML]{F1F1F1} 0.5 (0.4-0.6) \\
{\cellcolor[HTML]{FEFEFE}} \color[HTML]{000000} 54 & {\cellcolor[HTML]{FEFEFE}} \color[HTML]{000000} $\text{A}^\text{178}_\text{OP1}$-$\text{O}_\text{ph}$ & {\cellcolor[HTML]{FEFEFE}} \color[HTML]{000000} {\cellcolor{white}}  & {\cellcolor[HTML]{FEFEFE}} \color[HTML]{000000}  & {\cellcolor[HTML]{FEFEFE}} \color[HTML]{000000} {\cellcolor[HTML]{000000}} \color[HTML]{F1F1F1} {\cellcolor{white}}  & {\cellcolor[HTML]{FEFEFE}} \color[HTML]{000000} {\cellcolor[HTML]{000000}} \color[HTML]{F1F1F1} {\cellcolor{white}}  & {\cellcolor[HTML]{FEFEFE}} \color[HTML]{000000} {\cellcolor[HTML]{000000}} \color[HTML]{F1F1F1} {\cellcolor{white}}  & {\cellcolor[HTML]{FEFEFE}} \color[HTML]{000000} {\cellcolor[HTML]{000000}} \color[HTML]{F1F1F1} {\cellcolor{white}}  & {\cellcolor[HTML]{FEFEFE}} \color[HTML]{000000} {\cellcolor[HTML]{000000}} \color[HTML]{F1F1F1} {\cellcolor{white}}  & {\cellcolor[HTML]{FEFEFE}} \color[HTML]{000000} {\cellcolor[HTML]{FFECE3}} \color[HTML]{000000} 0.1 (0.0-0.1) & {\cellcolor[HTML]{FEFEFE}} \color[HTML]{000000} {\cellcolor[HTML]{FFEDE5}} \color[HTML]{000000} 0.0 (0.0-0.1) & {\cellcolor[HTML]{FEFEFE}} \color[HTML]{000000} {\cellcolor[HTML]{FCBEA5}} \color[HTML]{000000} 0.2 (0.2-0.3) & {\cellcolor[HTML]{FEFEFE}} \color[HTML]{000000} {\cellcolor[HTML]{000000}} \color[HTML]{F1F1F1} {\cellcolor{white}}  & {\cellcolor[HTML]{FEFEFE}} \color[HTML]{000000} {\cellcolor[HTML]{FFF5F0}} \color[HTML]{000000} 0.0 (0.0-0.0) & {\cellcolor[HTML]{FEFEFE}} \color[HTML]{000000} {\cellcolor[HTML]{FFECE3}} \color[HTML]{000000} 0.1 (0.0-0.1) & {\cellcolor[HTML]{FEFEFE}} \color[HTML]{000000} {\cellcolor[HTML]{000000}} \color[HTML]{F1F1F1} {\cellcolor{white}} {\cellcolor[HTML]{AAAAAA}}  & {\cellcolor[HTML]{FEFEFE}} \color[HTML]{000000} {\cellcolor[HTML]{BC141A}} \color[HTML]{F1F1F1} 0.8 (0.1-0.8) & {\cellcolor[HTML]{FEFEFE}} \color[HTML]{000000} {\cellcolor[HTML]{980C13}} \color[HTML]{F1F1F1} 0.9 (0.0-1.0) \\
{\cellcolor[HTML]{FEFEFE}} \color[HTML]{000000} 55 & {\cellcolor[HTML]{FEFEFE}} \color[HTML]{000000} $\text{A}^\text{269}_\text{OP2}$-$\text{O}_\text{ph}$ & {\cellcolor[HTML]{FEFEFE}} \color[HTML]{000000} {\cellcolor{white}}  & {\cellcolor[HTML]{FEFEFE}} \color[HTML]{000000}  & {\cellcolor[HTML]{FEFEFE}} \color[HTML]{000000} {\cellcolor[HTML]{000000}} \color[HTML]{F1F1F1} {\cellcolor{white}}  & {\cellcolor[HTML]{FEFEFE}} \color[HTML]{000000} {\cellcolor[HTML]{000000}} \color[HTML]{F1F1F1} {\cellcolor{white}}  & {\cellcolor[HTML]{FEFEFE}} \color[HTML]{000000} {\cellcolor[HTML]{FEE0D2}} \color[HTML]{000000} 0.1 & {\cellcolor[HTML]{FEFEFE}} \color[HTML]{000000} {\cellcolor[HTML]{000000}} \color[HTML]{F1F1F1} {\cellcolor{white}}  & {\cellcolor[HTML]{FEFEFE}} \color[HTML]{000000} {\cellcolor[HTML]{FEE3D6}} \color[HTML]{000000} 0.1 (0.0-0.4) & {\cellcolor[HTML]{FEFEFE}} \color[HTML]{000000} {\cellcolor[HTML]{FC8D6D}} \color[HTML]{F1F1F1} 0.4 (0.2-0.6) & {\cellcolor[HTML]{FEFEFE}} \color[HTML]{000000} {\cellcolor[HTML]{FDD2BF}} \color[HTML]{000000} 0.2 (0.0-0.4) & {\cellcolor[HTML]{FEFEFE}} \color[HTML]{000000} {\cellcolor[HTML]{FC7F5F}} \color[HTML]{F1F1F1} 0.4 (0.3-0.6) & {\cellcolor[HTML]{FEFEFE}} \color[HTML]{000000} {\cellcolor[HTML]{000000}} \color[HTML]{F1F1F1} {\cellcolor{white}}  & {\cellcolor[HTML]{FEFEFE}} \color[HTML]{000000} {\cellcolor[HTML]{FCB69B}} \color[HTML]{000000} 0.3 (0.1-0.4) & {\cellcolor[HTML]{FEFEFE}} \color[HTML]{000000} {\cellcolor[HTML]{F24734}} \color[HTML]{F1F1F1} 0.6 (0.5-0.7) & {\cellcolor[HTML]{FEFEFE}} \color[HTML]{000000} {\cellcolor[HTML]{000000}} \color[HTML]{F1F1F1} {\cellcolor{white}} {\cellcolor[HTML]{AAAAAA}}  & {\cellcolor[HTML]{FEFEFE}} \color[HTML]{000000} {\cellcolor[HTML]{000000}} \color[HTML]{F1F1F1} {\cellcolor{white}} {\cellcolor[HTML]{AAAAAA}}  & {\cellcolor[HTML]{FEFEFE}} \color[HTML]{000000} {\cellcolor[HTML]{000000}} \color[HTML]{F1F1F1} {\cellcolor{white}} {\cellcolor[HTML]{AAAAAA}}  \\
{\cellcolor[HTML]{FEFEFE}} \color[HTML]{000000} 56 & {\cellcolor[HTML]{FEFEFE}} \color[HTML]{000000} $\text{G}^\text{169}_\text{OP2}$-$\text{O}_\text{ph}$ & {\cellcolor[HTML]{FEFEFE}} \color[HTML]{000000} {\cellcolor{white}}  & {\cellcolor[HTML]{FEFEFE}} \color[HTML]{000000}  & {\cellcolor[HTML]{FEFEFE}} \color[HTML]{000000} {\cellcolor[HTML]{000000}} \color[HTML]{F1F1F1} {\cellcolor{white}}  & {\cellcolor[HTML]{FEFEFE}} \color[HTML]{000000} {\cellcolor[HTML]{000000}} \color[HTML]{F1F1F1} {\cellcolor{white}}  & {\cellcolor[HTML]{FEFEFE}} \color[HTML]{000000} {\cellcolor[HTML]{000000}} \color[HTML]{F1F1F1} {\cellcolor{white}}  & {\cellcolor[HTML]{FEFEFE}} \color[HTML]{000000} {\cellcolor[HTML]{000000}} \color[HTML]{F1F1F1} {\cellcolor{white}}  & {\cellcolor[HTML]{FEFEFE}} \color[HTML]{000000} {\cellcolor[HTML]{000000}} \color[HTML]{F1F1F1} {\cellcolor{white}}  & {\cellcolor[HTML]{FEFEFE}} \color[HTML]{000000} {\cellcolor[HTML]{FEEAE1}} \color[HTML]{000000} 0.1 (0.0-0.1) & {\cellcolor[HTML]{FEFEFE}} \color[HTML]{000000} {\cellcolor[HTML]{FEDECF}} \color[HTML]{000000} 0.1 (0.1-0.3) & {\cellcolor[HTML]{FEFEFE}} \color[HTML]{000000} {\cellcolor[HTML]{FEE2D5}} \color[HTML]{000000} 0.1 (0.1-0.2) & {\cellcolor[HTML]{FEFEFE}} \color[HTML]{000000} {\cellcolor[HTML]{FFF1EA}} \color[HTML]{000000} 0.0 (0.0-0.1) & {\cellcolor[HTML]{FEFEFE}} \color[HTML]{000000} {\cellcolor[HTML]{FFECE3}} \color[HTML]{000000} 0.1 (0.0-0.1) & {\cellcolor[HTML]{FEFEFE}} \color[HTML]{000000} {\cellcolor[HTML]{FEDECF}} \color[HTML]{000000} 0.1 (0.1-0.2) & {\cellcolor[HTML]{FEFEFE}} \color[HTML]{000000} {\cellcolor[HTML]{FB7A5A}} \color[HTML]{F1F1F1} 0.5 (0.4-0.6) & {\cellcolor[HTML]{FEFEFE}} \color[HTML]{000000} {\cellcolor[HTML]{D92523}} \color[HTML]{F1F1F1} 0.7 (0.3-0.9) & {\cellcolor[HTML]{FEFEFE}} \color[HTML]{000000} {\cellcolor[HTML]{FC8A6A}} \color[HTML]{F1F1F1} 0.4 (0.0-0.7) \\
{\cellcolor[HTML]{FEFEFE}} \color[HTML]{000000} 57 & {\cellcolor[HTML]{FEFEFE}} \color[HTML]{000000} $\text{A}^\text{387}_\text{OP2}$-$\text{O}_\text{ph}$ & {\cellcolor[HTML]{FEFEFE}} \color[HTML]{000000} {\cellcolor{white}}  & {\cellcolor[HTML]{FEFEFE}} \color[HTML]{000000}  & {\cellcolor[HTML]{FEFEFE}} \color[HTML]{000000} {\cellcolor[HTML]{000000}} \color[HTML]{F1F1F1} {\cellcolor{white}}  & {\cellcolor[HTML]{FEFEFE}} \color[HTML]{000000} {\cellcolor[HTML]{000000}} \color[HTML]{F1F1F1} {\cellcolor{white}}  & {\cellcolor[HTML]{FEFEFE}} \color[HTML]{000000} {\cellcolor[HTML]{000000}} \color[HTML]{F1F1F1} {\cellcolor{white}}  & {\cellcolor[HTML]{FEFEFE}} \color[HTML]{000000} {\cellcolor[HTML]{000000}} \color[HTML]{F1F1F1} {\cellcolor{white}}  & {\cellcolor[HTML]{FEFEFE}} \color[HTML]{000000} {\cellcolor[HTML]{000000}} \color[HTML]{F1F1F1} {\cellcolor{white}}  & {\cellcolor[HTML]{FEFEFE}} \color[HTML]{000000} {\cellcolor[HTML]{FC9576}} \color[HTML]{000000} 0.4 (0.2-0.7) & {\cellcolor[HTML]{FEFEFE}} \color[HTML]{000000} {\cellcolor[HTML]{FDCDB9}} \color[HTML]{000000} 0.2 (0.0-0.5) & {\cellcolor[HTML]{FEFEFE}} \color[HTML]{000000} {\cellcolor[HTML]{F96245}} \color[HTML]{F1F1F1} 0.5 (0.3-0.8) & {\cellcolor[HTML]{FEFEFE}} \color[HTML]{000000} {\cellcolor[HTML]{000000}} \color[HTML]{F1F1F1} {\cellcolor{white}}  & {\cellcolor[HTML]{FEFEFE}} \color[HTML]{000000} {\cellcolor[HTML]{FCAB8F}} \color[HTML]{000000} 0.3 (0.1-0.6) & {\cellcolor[HTML]{FEFEFE}} \color[HTML]{000000} {\cellcolor[HTML]{EC382B}} \color[HTML]{F1F1F1} 0.6 (0.4-0.8) & {\cellcolor[HTML]{FEFEFE}} \color[HTML]{000000} {\cellcolor[HTML]{000000}} \color[HTML]{F1F1F1} {\cellcolor{white}} {\cellcolor[HTML]{AAAAAA}}  & {\cellcolor[HTML]{FEFEFE}} \color[HTML]{000000} {\cellcolor[HTML]{000000}} \color[HTML]{F1F1F1} {\cellcolor{white}} {\cellcolor[HTML]{AAAAAA}}  & {\cellcolor[HTML]{FEFEFE}} \color[HTML]{000000} {\cellcolor[HTML]{000000}} \color[HTML]{F1F1F1} {\cellcolor{white}} {\cellcolor[HTML]{AAAAAA}}  \\
{\cellcolor[HTML]{FEFEFE}} \color[HTML]{000000} 58 & {\cellcolor[HTML]{FEFEFE}} \color[HTML]{000000} $\text{G}^\text{150}_\text{OP2}$-$\text{O}_\text{ph}$ & {\cellcolor[HTML]{FEFEFE}} \color[HTML]{000000} {\cellcolor{white}}  & {\cellcolor[HTML]{FEFEFE}} \color[HTML]{000000}  & {\cellcolor[HTML]{FEFEFE}} \color[HTML]{000000} {\cellcolor[HTML]{000000}} \color[HTML]{F1F1F1} {\cellcolor{white}}  & {\cellcolor[HTML]{FEFEFE}} \color[HTML]{000000} {\cellcolor[HTML]{000000}} \color[HTML]{F1F1F1} {\cellcolor{white}}  & {\cellcolor[HTML]{FEFEFE}} \color[HTML]{000000} {\cellcolor[HTML]{000000}} \color[HTML]{F1F1F1} {\cellcolor{white}}  & {\cellcolor[HTML]{FEFEFE}} \color[HTML]{000000} {\cellcolor[HTML]{000000}} \color[HTML]{F1F1F1} {\cellcolor{white}}  & {\cellcolor[HTML]{FEFEFE}} \color[HTML]{000000} {\cellcolor[HTML]{000000}} \color[HTML]{F1F1F1} {\cellcolor{white}}  & {\cellcolor[HTML]{FEFEFE}} \color[HTML]{000000} {\cellcolor[HTML]{FEE4D8}} \color[HTML]{000000} 0.1 (0.0-0.2) & {\cellcolor[HTML]{FEFEFE}} \color[HTML]{000000} {\cellcolor[HTML]{000000}} \color[HTML]{F1F1F1} {\cellcolor{white}}  & {\cellcolor[HTML]{FEFEFE}} \color[HTML]{000000} {\cellcolor[HTML]{FEE0D2}} \color[HTML]{000000} 0.1 (0.1-0.2) & {\cellcolor[HTML]{FEFEFE}} \color[HTML]{000000} {\cellcolor[HTML]{000000}} \color[HTML]{F1F1F1} {\cellcolor{white}}  & {\cellcolor[HTML]{FEFEFE}} \color[HTML]{000000} {\cellcolor[HTML]{FEE7DC}} \color[HTML]{000000} 0.1 (0.0-0.2) & {\cellcolor[HTML]{FEFEFE}} \color[HTML]{000000} {\cellcolor[HTML]{FC8969}} \color[HTML]{F1F1F1} 0.4 (0.3-0.6) & {\cellcolor[HTML]{FEFEFE}} \color[HTML]{000000} {\cellcolor[HTML]{000000}} \color[HTML]{F1F1F1} {\cellcolor{white}} {\cellcolor[HTML]{AAAAAA}}  & {\cellcolor[HTML]{FEFEFE}} \color[HTML]{000000} {\cellcolor[HTML]{E83429}} \color[HTML]{F1F1F1} 0.7 (0.5-0.7) & {\cellcolor[HTML]{FEFEFE}} \color[HTML]{000000} {\cellcolor[HTML]{F6583E}} \color[HTML]{F1F1F1} 0.6 (0.5-0.6) \\
{\cellcolor[HTML]{FEFEFE}} \color[HTML]{000000} 59 & {\cellcolor[HTML]{FEFEFE}} \color[HTML]{000000} $\text{C}^\text{170}_\text{OP1}$-$\text{O}_\text{ph}$ & {\cellcolor[HTML]{FEFEFE}} \color[HTML]{000000} {\cellcolor{white}}  & {\cellcolor[HTML]{FEFEFE}} \color[HTML]{000000}  & {\cellcolor[HTML]{FEFEFE}} \color[HTML]{000000} {\cellcolor[HTML]{000000}} \color[HTML]{F1F1F1} {\cellcolor{white}}  & {\cellcolor[HTML]{FEFEFE}} \color[HTML]{000000} {\cellcolor[HTML]{000000}} \color[HTML]{F1F1F1} {\cellcolor{white}}  & {\cellcolor[HTML]{FEFEFE}} \color[HTML]{000000} {\cellcolor[HTML]{000000}} \color[HTML]{F1F1F1} {\cellcolor{white}}  & {\cellcolor[HTML]{FEFEFE}} \color[HTML]{000000} {\cellcolor[HTML]{000000}} \color[HTML]{F1F1F1} {\cellcolor{white}}  & {\cellcolor[HTML]{FEFEFE}} \color[HTML]{000000} {\cellcolor[HTML]{000000}} \color[HTML]{F1F1F1} {\cellcolor{white}}  & {\cellcolor[HTML]{FEFEFE}} \color[HTML]{000000} {\cellcolor[HTML]{000000}} \color[HTML]{F1F1F1} {\cellcolor{white}}  & {\cellcolor[HTML]{FEFEFE}} \color[HTML]{000000} {\cellcolor[HTML]{FDD3C1}} \color[HTML]{000000} 0.2 (0.0-0.3) & {\cellcolor[HTML]{FEFEFE}} \color[HTML]{000000} {\cellcolor[HTML]{FFF5F0}} \color[HTML]{000000} 0.0 (0.0-0.0) & {\cellcolor[HTML]{FEFEFE}} \color[HTML]{000000} {\cellcolor[HTML]{FEE8DD}} \color[HTML]{000000} 0.1 (0.0-0.2) & {\cellcolor[HTML]{FEFEFE}} \color[HTML]{000000} {\cellcolor[HTML]{FFECE4}} \color[HTML]{000000} 0.1 (0.0-0.1) & {\cellcolor[HTML]{FEFEFE}} \color[HTML]{000000} {\cellcolor[HTML]{FDD0BC}} \color[HTML]{000000} 0.2 (0.1-0.4) & {\cellcolor[HTML]{FEFEFE}} \color[HTML]{000000} {\cellcolor[HTML]{FC8A6A}} \color[HTML]{F1F1F1} 0.4 (0.2-0.5) & {\cellcolor[HTML]{FEFEFE}} \color[HTML]{000000} {\cellcolor[HTML]{FB694A}} \color[HTML]{F1F1F1} 0.5 (0.4-0.6) & {\cellcolor[HTML]{FEFEFE}} \color[HTML]{000000} {\cellcolor[HTML]{FB694A}} \color[HTML]{F1F1F1} 0.5 (0.5-0.6) \\
{\cellcolor[HTML]{FEFEFE}} \color[HTML]{000000} 60 & {\cellcolor[HTML]{FEFEFE}} \color[HTML]{000000} $\text{U}^\text{249}_\text{OP2}$-$\text{O}_\text{ph}$ & {\cellcolor[HTML]{FEFEFE}} \color[HTML]{000000} {\cellcolor{white}}  & {\cellcolor[HTML]{FEFEFE}} \color[HTML]{000000}  & {\cellcolor[HTML]{FEFEFE}} \color[HTML]{000000} {\cellcolor[HTML]{000000}} \color[HTML]{F1F1F1} {\cellcolor{white}}  & {\cellcolor[HTML]{FEFEFE}} \color[HTML]{000000} {\cellcolor[HTML]{000000}} \color[HTML]{F1F1F1} {\cellcolor{white}}  & {\cellcolor[HTML]{FEFEFE}} \color[HTML]{000000} {\cellcolor[HTML]{000000}} \color[HTML]{F1F1F1} {\cellcolor{white}}  & {\cellcolor[HTML]{FEFEFE}} \color[HTML]{000000} {\cellcolor[HTML]{000000}} \color[HTML]{F1F1F1} {\cellcolor{white}}  & {\cellcolor[HTML]{FEFEFE}} \color[HTML]{000000} {\cellcolor[HTML]{FCBFA7}} \color[HTML]{000000} 0.2 (0.1-0.4) & {\cellcolor[HTML]{FEFEFE}} \color[HTML]{000000} {\cellcolor[HTML]{F5533B}} \color[HTML]{F1F1F1} 0.6 (0.4-0.7) & {\cellcolor[HTML]{FEFEFE}} \color[HTML]{000000} {\cellcolor[HTML]{FED9C9}} \color[HTML]{000000} 0.2 (0.0-0.3) & {\cellcolor[HTML]{FEFEFE}} \color[HTML]{000000} {\cellcolor[HTML]{FC9E80}} \color[HTML]{000000} 0.3 (0.2-0.5) & {\cellcolor[HTML]{FEFEFE}} \color[HTML]{000000} {\cellcolor[HTML]{FFEFE8}} \color[HTML]{000000} 0.0 (0.0-0.1) & {\cellcolor[HTML]{FEFEFE}} \color[HTML]{000000} {\cellcolor[HTML]{FCB79C}} \color[HTML]{000000} 0.3 (0.0-0.6) & {\cellcolor[HTML]{FEFEFE}} \color[HTML]{000000} {\cellcolor[HTML]{FDCBB6}} \color[HTML]{000000} 0.2 (0.1-0.4) & {\cellcolor[HTML]{FEFEFE}} \color[HTML]{000000} {\cellcolor[HTML]{FFF5F0}} \color[HTML]{000000} 0.0 (0.0-0.1) & {\cellcolor[HTML]{FEFEFE}} \color[HTML]{000000} {\cellcolor[HTML]{FFF5F0}} \color[HTML]{000000} 0.0 (0.0-0.1) & {\cellcolor[HTML]{FEFEFE}} \color[HTML]{000000} {\cellcolor[HTML]{FFEDE5}} \color[HTML]{000000} 0.1 (0.0-0.3) \\
{\cellcolor[HTML]{FEFEFE}} \color[HTML]{000000} 61 & {\cellcolor[HTML]{FEFEFE}} \color[HTML]{000000} $\text{U}^\text{271}_\text{OP2}$-$\text{O}_\text{ph}$ & {\cellcolor[HTML]{FEFEFE}} \color[HTML]{000000} {\cellcolor{white}}  & {\cellcolor[HTML]{FEFEFE}} \color[HTML]{000000}  & {\cellcolor[HTML]{FEFEFE}} \color[HTML]{000000} {\cellcolor[HTML]{000000}} \color[HTML]{F1F1F1} {\cellcolor{white}}  & {\cellcolor[HTML]{FEFEFE}} \color[HTML]{000000} {\cellcolor[HTML]{000000}} \color[HTML]{F1F1F1} {\cellcolor{white}}  & {\cellcolor[HTML]{FEFEFE}} \color[HTML]{000000} {\cellcolor[HTML]{000000}} \color[HTML]{F1F1F1} {\cellcolor{white}}  & {\cellcolor[HTML]{FEFEFE}} \color[HTML]{000000} {\cellcolor[HTML]{000000}} \color[HTML]{F1F1F1} {\cellcolor{white}}  & {\cellcolor[HTML]{FEFEFE}} \color[HTML]{000000} {\cellcolor[HTML]{000000}} \color[HTML]{F1F1F1} {\cellcolor{white}}  & {\cellcolor[HTML]{FEFEFE}} \color[HTML]{000000} {\cellcolor[HTML]{FDCBB6}} \color[HTML]{000000} 0.2 (0.0-0.4) & {\cellcolor[HTML]{FEFEFE}} \color[HTML]{000000} {\cellcolor[HTML]{FC8B6B}} \color[HTML]{F1F1F1} 0.4 (0.2-0.6) & {\cellcolor[HTML]{FEFEFE}} \color[HTML]{000000} {\cellcolor[HTML]{FCB296}} \color[HTML]{000000} 0.3 (0.1-0.5) & {\cellcolor[HTML]{FEFEFE}} \color[HTML]{000000} {\cellcolor[HTML]{FFEEE6}} \color[HTML]{000000} 0.0 (0.0-0.1) & {\cellcolor[HTML]{FEFEFE}} \color[HTML]{000000} {\cellcolor[HTML]{FC9E80}} \color[HTML]{000000} 0.3 (0.1-0.6) & {\cellcolor[HTML]{FEFEFE}} \color[HTML]{000000} {\cellcolor[HTML]{F5523A}} \color[HTML]{F1F1F1} 0.6 (0.4-0.7) & {\cellcolor[HTML]{FEFEFE}} \color[HTML]{000000} {\cellcolor[HTML]{FFF5F0}} \color[HTML]{000000} 0.0 (0.0-0.1) & {\cellcolor[HTML]{FEFEFE}} \color[HTML]{000000} {\cellcolor[HTML]{FFF5F0}} \color[HTML]{000000} 0.0 (0.0-0.0) & {\cellcolor[HTML]{FEFEFE}} \color[HTML]{000000} {\cellcolor[HTML]{FFF5F0}} \color[HTML]{000000} 0.0 (0.0-0.0) \\
{\cellcolor[HTML]{FEFEFE}} \color[HTML]{000000} 62 & {\cellcolor[HTML]{FEFEFE}} \color[HTML]{000000} $\text{A}^\text{139}_\text{OP1}$-$\text{O}_\text{ph}$ & {\cellcolor[HTML]{FEFEFE}} \color[HTML]{000000} {\cellcolor{white}}  & {\cellcolor[HTML]{FEFEFE}} \color[HTML]{000000}  & {\cellcolor[HTML]{FEFEFE}} \color[HTML]{000000} {\cellcolor[HTML]{FDCDB9}} \color[HTML]{000000} 0.2 & {\cellcolor[HTML]{FEFEFE}} \color[HTML]{000000} {\cellcolor[HTML]{000000}} \color[HTML]{F1F1F1} {\cellcolor{white}}  & {\cellcolor[HTML]{FEFEFE}} \color[HTML]{000000} {\cellcolor[HTML]{FCBBA1}} \color[HTML]{000000} 0.2 & {\cellcolor[HTML]{FEFEFE}} \color[HTML]{000000} {\cellcolor[HTML]{000000}} \color[HTML]{F1F1F1} {\cellcolor{white}}  & {\cellcolor[HTML]{FEFEFE}} \color[HTML]{000000} {\cellcolor[HTML]{FFF5F0}} \color[HTML]{000000} 0.0 (0.0-0.0) & {\cellcolor[HTML]{FEFEFE}} \color[HTML]{000000} {\cellcolor[HTML]{FEE1D3}} \color[HTML]{000000} 0.1 (0.0-0.4) & {\cellcolor[HTML]{FEFEFE}} \color[HTML]{000000} {\cellcolor[HTML]{FDCCB8}} \color[HTML]{000000} 0.2 (0.0-0.4) & {\cellcolor[HTML]{FEFEFE}} \color[HTML]{000000} {\cellcolor[HTML]{FFECE3}} \color[HTML]{000000} 0.1 (0.0-0.1) & {\cellcolor[HTML]{FEFEFE}} \color[HTML]{000000} {\cellcolor[HTML]{000000}} \color[HTML]{F1F1F1} {\cellcolor{white}}  & {\cellcolor[HTML]{FEFEFE}} \color[HTML]{000000} {\cellcolor[HTML]{000000}} \color[HTML]{F1F1F1} {\cellcolor{white}}  & {\cellcolor[HTML]{FEFEFE}} \color[HTML]{000000} {\cellcolor[HTML]{FFF2EC}} \color[HTML]{000000} 0.0 (0.0-0.0) & {\cellcolor[HTML]{FEFEFE}} \color[HTML]{000000} {\cellcolor[HTML]{FCAB8F}} \color[HTML]{000000} 0.3 (0.2-0.3) & {\cellcolor[HTML]{FEFEFE}} \color[HTML]{000000} {\cellcolor[HTML]{FCBBA1}} \color[HTML]{000000} 0.2 (0.1-0.4) & {\cellcolor[HTML]{FEFEFE}} \color[HTML]{000000} {\cellcolor[HTML]{FC8A6A}} \color[HTML]{F1F1F1} 0.4 (0.1-0.6) \\
{\cellcolor[HTML]{FEFEFE}} \color[HTML]{000000} 63 & {\cellcolor[HTML]{FEFEFE}} \color[HTML]{000000} $\text{U}^\text{259}_\text{OP1}$,$\text{A}^\text{261}_\text{OP1}$-\textit{cis}-2$\text{O}_\text{ph}$ & {\cellcolor[HTML]{FEFEFE}} \color[HTML]{000000} {\cellcolor{white}}  & {\cellcolor[HTML]{FEFEFE}} \color[HTML]{000000}  & {\cellcolor[HTML]{FEFEFE}} \color[HTML]{000000} {\cellcolor[HTML]{CA181D}} \color[HTML]{F1F1F1} 0.8 & {\cellcolor[HTML]{FEFEFE}} \color[HTML]{000000} {\cellcolor[HTML]{000000}} \color[HTML]{F1F1F1} {\cellcolor{white}}  & {\cellcolor[HTML]{FEFEFE}} \color[HTML]{000000} {\cellcolor[HTML]{FCBBA1}} \color[HTML]{000000} 0.2 & {\cellcolor[HTML]{FEFEFE}} \color[HTML]{000000} {\cellcolor[HTML]{000000}} \color[HTML]{F1F1F1} {\cellcolor{white}}  & {\cellcolor[HTML]{FEFEFE}} \color[HTML]{000000} {\cellcolor[HTML]{FCBFA7}} \color[HTML]{000000} 0.2 (0.1-0.5) & {\cellcolor[HTML]{FEFEFE}} \color[HTML]{000000} {\cellcolor[HTML]{FFEEE7}} \color[HTML]{000000} 0.0 (0.0-0.1) & {\cellcolor[HTML]{FEFEFE}} \color[HTML]{000000} {\cellcolor[HTML]{FDD5C4}} \color[HTML]{000000} 0.2 (0.0-0.5) & {\cellcolor[HTML]{FEFEFE}} \color[HTML]{000000} {\cellcolor[HTML]{FFECE3}} \color[HTML]{000000} 0.1 (0.0-0.1) & {\cellcolor[HTML]{FEFEFE}} \color[HTML]{000000} {\cellcolor[HTML]{000000}} \color[HTML]{F1F1F1} {\cellcolor{white}}  & {\cellcolor[HTML]{FEFEFE}} \color[HTML]{000000} {\cellcolor[HTML]{FFF1EA}} \color[HTML]{000000} 0.0 (0.0-0.1) & {\cellcolor[HTML]{FEFEFE}} \color[HTML]{000000} {\cellcolor[HTML]{FFF4EF}} \color[HTML]{000000} 0.0 (0.0-0.0) & {\cellcolor[HTML]{FEFEFE}} \color[HTML]{000000} {\cellcolor[HTML]{FFF5F0}} \color[HTML]{000000} 0.0 (0.0-0.2) & {\cellcolor[HTML]{FEFEFE}} \color[HTML]{000000} {\cellcolor[HTML]{FCBBA1}} \color[HTML]{000000} 0.2 (0.1-0.4) & {\cellcolor[HTML]{FEFEFE}} \color[HTML]{000000} {\cellcolor[HTML]{FFF5F0}} \color[HTML]{000000} 0.0 (0.0-0.2) \\
{\cellcolor[HTML]{FEFEFE}} \color[HTML]{000000} 64 & {\cellcolor[HTML]{FEFEFE}} \color[HTML]{000000} $\text{U}^\text{326}_\text{OP2}$-$\text{O}_\text{ph}$ & {\cellcolor[HTML]{FEFEFE}} \color[HTML]{000000} {\cellcolor{white}}  & {\cellcolor[HTML]{FEFEFE}} \color[HTML]{000000}  & {\cellcolor[HTML]{FEFEFE}} \color[HTML]{000000} {\cellcolor[HTML]{000000}} \color[HTML]{F1F1F1} {\cellcolor{white}}  & {\cellcolor[HTML]{FEFEFE}} \color[HTML]{000000} {\cellcolor[HTML]{000000}} \color[HTML]{F1F1F1} {\cellcolor{white}}  & {\cellcolor[HTML]{FEFEFE}} \color[HTML]{000000} {\cellcolor[HTML]{000000}} \color[HTML]{F1F1F1} {\cellcolor{white}}  & {\cellcolor[HTML]{FEFEFE}} \color[HTML]{000000} {\cellcolor[HTML]{000000}} \color[HTML]{F1F1F1} {\cellcolor{white}}  & {\cellcolor[HTML]{FEFEFE}} \color[HTML]{000000} {\cellcolor[HTML]{FFF1EA}} \color[HTML]{000000} 0.0 (0.0-0.1) & {\cellcolor[HTML]{FEFEFE}} \color[HTML]{000000} {\cellcolor[HTML]{FCB398}} \color[HTML]{000000} 0.3 (0.1-0.5) & {\cellcolor[HTML]{FEFEFE}} \color[HTML]{000000} {\cellcolor[HTML]{FDCCB8}} \color[HTML]{000000} 0.2 (0.0-0.5) & {\cellcolor[HTML]{FEFEFE}} \color[HTML]{000000} {\cellcolor[HTML]{FCB89E}} \color[HTML]{000000} 0.3 (0.1-0.4) & {\cellcolor[HTML]{FEFEFE}} \color[HTML]{000000} {\cellcolor[HTML]{FFECE3}} \color[HTML]{000000} 0.1 (0.0-0.2) & {\cellcolor[HTML]{FEFEFE}} \color[HTML]{000000} {\cellcolor[HTML]{FC8F6F}} \color[HTML]{000000} 0.4 (0.2-0.6) & {\cellcolor[HTML]{FEFEFE}} \color[HTML]{000000} {\cellcolor[HTML]{F44D38}} \color[HTML]{F1F1F1} 0.6 (0.4-0.7) & {\cellcolor[HTML]{FEFEFE}} \color[HTML]{000000} {\cellcolor[HTML]{FFF5F0}} \color[HTML]{000000} 0.0 (0.0-0.1) & {\cellcolor[HTML]{FEFEFE}} \color[HTML]{000000} {\cellcolor[HTML]{FFF5F0}} \color[HTML]{000000} 0.0 (0.0-0.0) & {\cellcolor[HTML]{FEFEFE}} \color[HTML]{000000} {\cellcolor[HTML]{FFF5F0}} \color[HTML]{000000} 0.0 (0.0-0.0) \\
{\cellcolor[HTML]{FEFEFE}} \color[HTML]{000000} 65 & {\cellcolor[HTML]{FEFEFE}} \color[HTML]{000000} $\text{G}^\text{309}_\text{OP1}$-$\text{O}_\text{ph}$ & {\cellcolor[HTML]{FEFEFE}} \color[HTML]{000000} {\cellcolor{white}}  & {\cellcolor[HTML]{FEFEFE}} \color[HTML]{000000}  & {\cellcolor[HTML]{FEFEFE}} \color[HTML]{000000} {\cellcolor[HTML]{000000}} \color[HTML]{F1F1F1} {\cellcolor{white}}  & {\cellcolor[HTML]{FEFEFE}} \color[HTML]{000000} {\cellcolor[HTML]{000000}} \color[HTML]{F1F1F1} {\cellcolor{white}}  & {\cellcolor[HTML]{FEFEFE}} \color[HTML]{000000} {\cellcolor[HTML]{000000}} \color[HTML]{F1F1F1} {\cellcolor{white}}  & {\cellcolor[HTML]{FEFEFE}} \color[HTML]{000000} {\cellcolor[HTML]{000000}} \color[HTML]{F1F1F1} {\cellcolor{white}}  & {\cellcolor[HTML]{FEFEFE}} \color[HTML]{000000} {\cellcolor[HTML]{000000}} \color[HTML]{F1F1F1} {\cellcolor{white}}  & {\cellcolor[HTML]{FEFEFE}} \color[HTML]{000000} {\cellcolor[HTML]{FFF4EF}} \color[HTML]{000000} 0.0 (0.0-0.0) & {\cellcolor[HTML]{FEFEFE}} \color[HTML]{000000} {\cellcolor[HTML]{FFF4EF}} \color[HTML]{000000} 0.0 (0.0-0.0) & {\cellcolor[HTML]{FEFEFE}} \color[HTML]{000000} {\cellcolor[HTML]{FFEEE7}} \color[HTML]{000000} 0.0 (0.0-0.1) & {\cellcolor[HTML]{FEFEFE}} \color[HTML]{000000} {\cellcolor[HTML]{000000}} \color[HTML]{F1F1F1} {\cellcolor{white}}  & {\cellcolor[HTML]{FEFEFE}} \color[HTML]{000000} {\cellcolor[HTML]{FFF4EE}} \color[HTML]{000000} 0.0 (0.0-0.0) & {\cellcolor[HTML]{FEFEFE}} \color[HTML]{000000} {\cellcolor[HTML]{FFF4EE}} \color[HTML]{000000} 0.0 (0.0-0.0) & {\cellcolor[HTML]{FEFEFE}} \color[HTML]{000000} {\cellcolor[HTML]{000000}} \color[HTML]{F1F1F1} {\cellcolor{white}} {\cellcolor[HTML]{AAAAAA}}  & {\cellcolor[HTML]{FEFEFE}} \color[HTML]{000000} {\cellcolor[HTML]{E83429}} \color[HTML]{F1F1F1} 0.7 (0.0-0.8) & {\cellcolor[HTML]{FEFEFE}} \color[HTML]{000000} {\cellcolor[HTML]{67000D}} \color[HTML]{F1F1F1} 1.0 (0.0-1.0) \\
{\cellcolor[HTML]{FEFEFE}} \color[HTML]{000000} 66 & {\cellcolor[HTML]{FEFEFE}} \color[HTML]{000000} $\text{G}^\text{358}_\text{OP2}$-$\text{O}_\text{ph}$ & {\cellcolor[HTML]{FEFEFE}} \color[HTML]{000000} {\cellcolor{white}}  & {\cellcolor[HTML]{FEFEFE}} \color[HTML]{000000}  & {\cellcolor[HTML]{FEFEFE}} \color[HTML]{000000} {\cellcolor[HTML]{000000}} \color[HTML]{F1F1F1} {\cellcolor{white}}  & {\cellcolor[HTML]{FEFEFE}} \color[HTML]{000000} {\cellcolor[HTML]{000000}} \color[HTML]{F1F1F1} {\cellcolor{white}}  & {\cellcolor[HTML]{FEFEFE}} \color[HTML]{000000} {\cellcolor[HTML]{FEEAE1}} \color[HTML]{000000} 0.1 & {\cellcolor[HTML]{FEFEFE}} \color[HTML]{000000} {\cellcolor[HTML]{000000}} \color[HTML]{F1F1F1} {\cellcolor{white}}  & {\cellcolor[HTML]{FEFEFE}} \color[HTML]{000000} {\cellcolor[HTML]{FEE4D8}} \color[HTML]{000000} 0.1 (0.0-0.3) & {\cellcolor[HTML]{FEFEFE}} \color[HTML]{000000} {\cellcolor[HTML]{FCC4AD}} \color[HTML]{000000} 0.2 (0.1-0.4) & {\cellcolor[HTML]{FEFEFE}} \color[HTML]{000000} {\cellcolor[HTML]{000000}} \color[HTML]{F1F1F1} {\cellcolor{white}}  & {\cellcolor[HTML]{FEFEFE}} \color[HTML]{000000} {\cellcolor[HTML]{FCAB8F}} \color[HTML]{000000} 0.3 (0.2-0.5) & {\cellcolor[HTML]{FEFEFE}} \color[HTML]{000000} {\cellcolor[HTML]{FEE6DA}} \color[HTML]{000000} 0.1 (0.0-0.3) & {\cellcolor[HTML]{FEFEFE}} \color[HTML]{000000} {\cellcolor[HTML]{FEDACA}} \color[HTML]{000000} 0.1 (0.0-0.4) & {\cellcolor[HTML]{FEFEFE}} \color[HTML]{000000} {\cellcolor[HTML]{FC8565}} \color[HTML]{F1F1F1} 0.4 (0.2-0.6) & {\cellcolor[HTML]{FEFEFE}} \color[HTML]{000000} {\cellcolor[HTML]{FCBBA1}} \color[HTML]{000000} 0.2 (0.2-0.4) & {\cellcolor[HTML]{FEFEFE}} \color[HTML]{000000} {\cellcolor[HTML]{FEE5D8}} \color[HTML]{000000} 0.1 (0.0-0.2) & {\cellcolor[HTML]{FEFEFE}} \color[HTML]{000000} {\cellcolor[HTML]{FFF5F0}} \color[HTML]{000000} 0.0 (0.0-0.1) \\
{\cellcolor[HTML]{FEFEFE}} \color[HTML]{000000} 67 & {\cellcolor[HTML]{FEFEFE}} \color[HTML]{000000} $\text{U}^\text{168}_\text{OP2}$,$\text{A}^\text{171}_\text{OP2}$-\textit{trans}-2$\text{O}_\text{ph}$ & {\cellcolor[HTML]{FEFEFE}} \color[HTML]{000000} {\cellcolor{white}}  & {\cellcolor[HTML]{FEFEFE}} \color[HTML]{000000}  & {\cellcolor[HTML]{FEFEFE}} \color[HTML]{000000} {\cellcolor[HTML]{FB7D5D}} \color[HTML]{F1F1F1} 0.4 & {\cellcolor[HTML]{FEFEFE}} \color[HTML]{000000} {\cellcolor[HTML]{000000}} \color[HTML]{F1F1F1} {\cellcolor{white}}  & {\cellcolor[HTML]{FEFEFE}} \color[HTML]{000000} {\cellcolor[HTML]{FEEAE1}} \color[HTML]{000000} 0.1 & {\cellcolor[HTML]{FEFEFE}} \color[HTML]{000000} {\cellcolor[HTML]{000000}} \color[HTML]{F1F1F1} {\cellcolor{white}}  & {\cellcolor[HTML]{FEFEFE}} \color[HTML]{000000} {\cellcolor[HTML]{FC9373}} \color[HTML]{000000} 0.4 (0.1-0.7) & {\cellcolor[HTML]{FEFEFE}} \color[HTML]{000000} {\cellcolor[HTML]{FEE0D2}} \color[HTML]{000000} 0.1 (0.0-0.4) & {\cellcolor[HTML]{FEFEFE}} \color[HTML]{000000} {\cellcolor[HTML]{FEE0D2}} \color[HTML]{000000} 0.1 (0.0-0.4) & {\cellcolor[HTML]{FEFEFE}} \color[HTML]{000000} {\cellcolor[HTML]{FEDBCC}} \color[HTML]{000000} 0.1 (0.1-0.3) & {\cellcolor[HTML]{FEFEFE}} \color[HTML]{000000} {\cellcolor[HTML]{000000}} \color[HTML]{F1F1F1} {\cellcolor{white}}  & {\cellcolor[HTML]{FEFEFE}} \color[HTML]{000000} {\cellcolor[HTML]{000000}} \color[HTML]{F1F1F1} {\cellcolor{white}}  & {\cellcolor[HTML]{FEFEFE}} \color[HTML]{000000} {\cellcolor[HTML]{000000}} \color[HTML]{F1F1F1} {\cellcolor{white}}  & {\cellcolor[HTML]{FEFEFE}} \color[HTML]{000000} {\cellcolor[HTML]{FEE5D8}} \color[HTML]{000000} 0.1 (0.1-0.2) & {\cellcolor[HTML]{FEFEFE}} \color[HTML]{000000} {\cellcolor[HTML]{FEE5D8}} \color[HTML]{000000} 0.1 (0.1-0.2) & {\cellcolor[HTML]{FEFEFE}} \color[HTML]{000000} {\cellcolor[HTML]{FEE5D8}} \color[HTML]{000000} 0.1 (0.1-0.2) \\
{\cellcolor[HTML]{FEFEFE}} \color[HTML]{000000} 68 & {\cellcolor[HTML]{FEFEFE}} \color[HTML]{000000} $\text{C}^\text{98}_\text{OP1}$-$\text{O}_\text{ph}$ & {\cellcolor[HTML]{FEFEFE}} \color[HTML]{000000} {\cellcolor{white}}  & {\cellcolor[HTML]{FEFEFE}} \color[HTML]{000000}  & {\cellcolor[HTML]{FEFEFE}} \color[HTML]{000000} {\cellcolor[HTML]{000000}} \color[HTML]{F1F1F1} {\cellcolor{white}}  & {\cellcolor[HTML]{FEFEFE}} \color[HTML]{000000} {\cellcolor[HTML]{000000}} \color[HTML]{F1F1F1} {\cellcolor{white}}  & {\cellcolor[HTML]{FEFEFE}} \color[HTML]{000000} {\cellcolor[HTML]{000000}} \color[HTML]{F1F1F1} {\cellcolor{white}}  & {\cellcolor[HTML]{FEFEFE}} \color[HTML]{000000} {\cellcolor[HTML]{000000}} \color[HTML]{F1F1F1} {\cellcolor{white}}  & {\cellcolor[HTML]{FEFEFE}} \color[HTML]{000000} {\cellcolor[HTML]{FFF2EC}} \color[HTML]{000000} 0.0 (0.0-0.1) & {\cellcolor[HTML]{FEFEFE}} \color[HTML]{000000} {\cellcolor[HTML]{FDD2BF}} \color[HTML]{000000} 0.2 (0.1-0.3) & {\cellcolor[HTML]{FEFEFE}} \color[HTML]{000000} {\cellcolor[HTML]{FED8C7}} \color[HTML]{000000} 0.2 (0.0-0.4) & {\cellcolor[HTML]{FEFEFE}} \color[HTML]{000000} {\cellcolor[HTML]{FCB095}} \color[HTML]{000000} 0.3 (0.2-0.5) & {\cellcolor[HTML]{FEFEFE}} \color[HTML]{000000} {\cellcolor[HTML]{FFEDE5}} \color[HTML]{000000} 0.0 (0.0-0.1) & {\cellcolor[HTML]{FEFEFE}} \color[HTML]{000000} {\cellcolor[HTML]{FC8767}} \color[HTML]{F1F1F1} 0.4 (0.3-0.6) & {\cellcolor[HTML]{FEFEFE}} \color[HTML]{000000} {\cellcolor[HTML]{FB7050}} \color[HTML]{F1F1F1} 0.5 (0.3-0.6) & {\cellcolor[HTML]{FEFEFE}} \color[HTML]{000000} {\cellcolor[HTML]{000000}} \color[HTML]{F1F1F1} {\cellcolor{white}} {\cellcolor[HTML]{AAAAAA}}  & {\cellcolor[HTML]{FEFEFE}} \color[HTML]{000000} {\cellcolor[HTML]{000000}} \color[HTML]{F1F1F1} {\cellcolor{white}} {\cellcolor[HTML]{AAAAAA}}  & {\cellcolor[HTML]{FEFEFE}} \color[HTML]{000000} {\cellcolor[HTML]{FFF5F0}} \color[HTML]{000000} 0.0 (0.0-0.0) \\
{\cellcolor[HTML]{FEFEFE}} \color[HTML]{000000} 69 & {\cellcolor[HTML]{FEFEFE}} \color[HTML]{000000} $\text{G}^\text{257}_\text{OP1}$,$\text{U}^\text{258}_\text{OP2}$-\textit{cis}-2$\text{O}_\text{ph}$ & {\cellcolor[HTML]{FEFEFE}} \color[HTML]{000000} {\cellcolor{white}}  & {\cellcolor[HTML]{FEFEFE}} \color[HTML]{000000}  & {\cellcolor[HTML]{FEFEFE}} \color[HTML]{000000} {\cellcolor[HTML]{000000}} \color[HTML]{F1F1F1} {\cellcolor{white}}  & {\cellcolor[HTML]{FEFEFE}} \color[HTML]{000000} {\cellcolor[HTML]{000000}} \color[HTML]{F1F1F1} {\cellcolor{white}}  & {\cellcolor[HTML]{FEFEFE}} \color[HTML]{000000} {\cellcolor[HTML]{000000}} \color[HTML]{F1F1F1} {\cellcolor{white}}  & {\cellcolor[HTML]{FEFEFE}} \color[HTML]{000000} {\cellcolor[HTML]{000000}} \color[HTML]{F1F1F1} {\cellcolor{white}}  & {\cellcolor[HTML]{FEFEFE}} \color[HTML]{000000} {\cellcolor[HTML]{000000}} \color[HTML]{F1F1F1} {\cellcolor{white}}  & {\cellcolor[HTML]{FEFEFE}} \color[HTML]{000000} {\cellcolor[HTML]{FFF5F0}} \color[HTML]{000000} 0.0 (0.0-0.0) & {\cellcolor[HTML]{FEFEFE}} \color[HTML]{000000} {\cellcolor[HTML]{FCA78B}} \color[HTML]{000000} 0.3 (0.1-0.5) & {\cellcolor[HTML]{FEFEFE}} \color[HTML]{000000} {\cellcolor[HTML]{FFF5F0}} \color[HTML]{000000} 0.0 (0.0-0.0) & {\cellcolor[HTML]{FEFEFE}} \color[HTML]{000000} {\cellcolor[HTML]{FEE1D3}} \color[HTML]{000000} 0.1 (0.0-0.5) & {\cellcolor[HTML]{FEFEFE}} \color[HTML]{000000} {\cellcolor[HTML]{FEE1D3}} \color[HTML]{000000} 0.1 (0.0-0.4) & {\cellcolor[HTML]{FEFEFE}} \color[HTML]{000000} {\cellcolor[HTML]{FFEFE8}} \color[HTML]{000000} 0.0 (0.0-0.1) & {\cellcolor[HTML]{FEFEFE}} \color[HTML]{000000} {\cellcolor[HTML]{FCBBA1}} \color[HTML]{000000} 0.2 (0.1-0.5) & {\cellcolor[HTML]{FEFEFE}} \color[HTML]{000000} {\cellcolor[HTML]{FEE5D8}} \color[HTML]{000000} 0.1 (0.1-0.4) & {\cellcolor[HTML]{FEFEFE}} \color[HTML]{000000} {\cellcolor[HTML]{F6583E}} \color[HTML]{F1F1F1} 0.6 (0.4-0.7) \\
{\cellcolor[HTML]{FEFEFE}} \color[HTML]{000000} 70 & {\cellcolor[HTML]{FEFEFE}} \color[HTML]{000000} $\text{G}^\text{329}_\text{OP1}$-$\text{O}_\text{ph}$ & {\cellcolor[HTML]{FEFEFE}} \color[HTML]{000000} {\cellcolor{white}}  & {\cellcolor[HTML]{FEFEFE}} \color[HTML]{000000}  & {\cellcolor[HTML]{FEFEFE}} \color[HTML]{000000} {\cellcolor[HTML]{000000}} \color[HTML]{F1F1F1} {\cellcolor{white}}  & {\cellcolor[HTML]{FEFEFE}} \color[HTML]{000000} {\cellcolor[HTML]{FCBBA1}} \color[HTML]{000000} 0.2 & {\cellcolor[HTML]{FEFEFE}} \color[HTML]{000000} {\cellcolor[HTML]{000000}} \color[HTML]{F1F1F1} {\cellcolor{white}}  & {\cellcolor[HTML]{FEFEFE}} \color[HTML]{000000} {\cellcolor[HTML]{FEE0D2}} \color[HTML]{000000} 0.1 & {\cellcolor[HTML]{FEFEFE}} \color[HTML]{000000} {\cellcolor[HTML]{000000}} \color[HTML]{F1F1F1} {\cellcolor{white}}  & {\cellcolor[HTML]{FEFEFE}} \color[HTML]{000000} {\cellcolor[HTML]{FFF4EE}} \color[HTML]{000000} 0.0 (0.0-0.0) & {\cellcolor[HTML]{FEFEFE}} \color[HTML]{000000} {\cellcolor[HTML]{FFF3ED}} \color[HTML]{000000} 0.0 (0.0-0.0) & {\cellcolor[HTML]{FEFEFE}} \color[HTML]{000000} {\cellcolor[HTML]{FFF4EE}} \color[HTML]{000000} 0.0 (0.0-0.0) & {\cellcolor[HTML]{FEFEFE}} \color[HTML]{000000} {\cellcolor[HTML]{FCBBA1}} \color[HTML]{000000} 0.2 (0.0-0.6) & {\cellcolor[HTML]{FEFEFE}} \color[HTML]{000000} {\cellcolor[HTML]{FFF1EA}} \color[HTML]{000000} 0.0 (0.0-0.1) & {\cellcolor[HTML]{FEFEFE}} \color[HTML]{000000} {\cellcolor[HTML]{FFECE4}} \color[HTML]{000000} 0.1 (0.0-0.1) & {\cellcolor[HTML]{FEFEFE}} \color[HTML]{000000} {\cellcolor[HTML]{FCAB8F}} \color[HTML]{000000} 0.3 (0.2-0.4) & {\cellcolor[HTML]{FEFEFE}} \color[HTML]{000000} {\cellcolor[HTML]{FED9C9}} \color[HTML]{000000} 0.2 (0.0-0.2) & {\cellcolor[HTML]{FEFEFE}} \color[HTML]{000000} {\cellcolor[HTML]{FCAB8F}} \color[HTML]{000000} 0.3 (0.1-0.4) \\
{\cellcolor[HTML]{FEFEFE}} \color[HTML]{000000} 71 & {\cellcolor[HTML]{FEFEFE}} \color[HTML]{000000} $\text{G}^\text{285}_\text{OP2}$-$\text{O}_\text{ph}$ & {\cellcolor[HTML]{FEFEFE}} \color[HTML]{000000} {\cellcolor{white}}  & {\cellcolor[HTML]{FEFEFE}} \color[HTML]{000000}  & {\cellcolor[HTML]{FEFEFE}} \color[HTML]{000000} {\cellcolor[HTML]{000000}} \color[HTML]{F1F1F1} {\cellcolor{white}}  & {\cellcolor[HTML]{FEFEFE}} \color[HTML]{000000} {\cellcolor[HTML]{000000}} \color[HTML]{F1F1F1} {\cellcolor{white}}  & {\cellcolor[HTML]{FEFEFE}} \color[HTML]{000000} {\cellcolor[HTML]{FEEAE1}} \color[HTML]{000000} 0.1 & {\cellcolor[HTML]{FEFEFE}} \color[HTML]{000000} {\cellcolor[HTML]{000000}} \color[HTML]{F1F1F1} {\cellcolor{white}}  & {\cellcolor[HTML]{FEFEFE}} \color[HTML]{000000} {\cellcolor[HTML]{FEE3D6}} \color[HTML]{000000} 0.1 (0.0-0.4) & {\cellcolor[HTML]{FEFEFE}} \color[HTML]{000000} {\cellcolor[HTML]{FDCDB9}} \color[HTML]{000000} 0.2 (0.1-0.3) & {\cellcolor[HTML]{FEFEFE}} \color[HTML]{000000} {\cellcolor[HTML]{FEDCCD}} \color[HTML]{000000} 0.1 (0.1-0.2) & {\cellcolor[HTML]{FEFEFE}} \color[HTML]{000000} {\cellcolor[HTML]{FDCAB5}} \color[HTML]{000000} 0.2 (0.1-0.3) & {\cellcolor[HTML]{FEFEFE}} \color[HTML]{000000} {\cellcolor[HTML]{000000}} \color[HTML]{F1F1F1} {\cellcolor{white}}  & {\cellcolor[HTML]{FEFEFE}} \color[HTML]{000000} {\cellcolor[HTML]{FEE1D4}} \color[HTML]{000000} 0.1 (0.0-0.3) & {\cellcolor[HTML]{FEFEFE}} \color[HTML]{000000} {\cellcolor[HTML]{FCB99F}} \color[HTML]{000000} 0.3 (0.2-0.3) & {\cellcolor[HTML]{FEFEFE}} \color[HTML]{000000} {\cellcolor[HTML]{FC8A6A}} \color[HTML]{F1F1F1} 0.4 (0.3-0.5) & {\cellcolor[HTML]{FEFEFE}} \color[HTML]{000000} {\cellcolor[HTML]{000000}} \color[HTML]{F1F1F1} {\cellcolor{white}} {\cellcolor[HTML]{AAAAAA}}  & {\cellcolor[HTML]{FEFEFE}} \color[HTML]{000000} {\cellcolor[HTML]{000000}} \color[HTML]{F1F1F1} {\cellcolor{white}} {\cellcolor[HTML]{AAAAAA}}  \\
{\cellcolor[HTML]{FEFEFE}} \color[HTML]{000000} 72 & {\cellcolor[HTML]{FEFEFE}} \color[HTML]{000000} $\text{A}^\text{42}_\text{OP2}$-$\text{O}_\text{ph}$ & {\cellcolor[HTML]{FEFEFE}} \color[HTML]{000000} {\cellcolor{white}}  & {\cellcolor[HTML]{FEFEFE}} \color[HTML]{000000}  & {\cellcolor[HTML]{FEFEFE}} \color[HTML]{000000} {\cellcolor[HTML]{000000}} \color[HTML]{F1F1F1} {\cellcolor{white}}  & {\cellcolor[HTML]{FEFEFE}} \color[HTML]{000000} {\cellcolor[HTML]{000000}} \color[HTML]{F1F1F1} {\cellcolor{white}}  & {\cellcolor[HTML]{FEFEFE}} \color[HTML]{000000} {\cellcolor[HTML]{000000}} \color[HTML]{F1F1F1} {\cellcolor{white}}  & {\cellcolor[HTML]{FEFEFE}} \color[HTML]{000000} {\cellcolor[HTML]{000000}} \color[HTML]{F1F1F1} {\cellcolor{white}}  & {\cellcolor[HTML]{FEFEFE}} \color[HTML]{000000} {\cellcolor[HTML]{000000}} \color[HTML]{F1F1F1} {\cellcolor{white}}  & {\cellcolor[HTML]{FEFEFE}} \color[HTML]{000000} {\cellcolor[HTML]{FCB69B}} \color[HTML]{000000} 0.3 (0.2-0.4) & {\cellcolor[HTML]{FEFEFE}} \color[HTML]{000000} {\cellcolor[HTML]{FEE3D6}} \color[HTML]{000000} 0.1 (0.0-0.2) & {\cellcolor[HTML]{FEFEFE}} \color[HTML]{000000} {\cellcolor[HTML]{FDD0BC}} \color[HTML]{000000} 0.2 (0.1-0.3) & {\cellcolor[HTML]{FEFEFE}} \color[HTML]{000000} {\cellcolor[HTML]{000000}} \color[HTML]{F1F1F1} {\cellcolor{white}}  & {\cellcolor[HTML]{FEFEFE}} \color[HTML]{000000} {\cellcolor[HTML]{FDCBB6}} \color[HTML]{000000} 0.2 (0.1-0.3) & {\cellcolor[HTML]{FEFEFE}} \color[HTML]{000000} {\cellcolor[HTML]{FCB79C}} \color[HTML]{000000} 0.3 (0.2-0.4) & {\cellcolor[HTML]{FEFEFE}} \color[HTML]{000000} {\cellcolor[HTML]{000000}} \color[HTML]{F1F1F1} {\cellcolor{white}} {\cellcolor[HTML]{AAAAAA}}  & {\cellcolor[HTML]{FEFEFE}} \color[HTML]{000000} {\cellcolor[HTML]{FCAB8F}} \color[HTML]{000000} 0.3 (0.2-0.4) & {\cellcolor[HTML]{FEFEFE}} \color[HTML]{000000} {\cellcolor[HTML]{FED9C9}} \color[HTML]{000000} 0.2 (0.0-0.4) \\
{\cellcolor[HTML]{FEFEFE}} \color[HTML]{000000} 73 & {\cellcolor[HTML]{FEFEFE}} \color[HTML]{000000} $\text{C}^\text{166}_\text{OP2}$,$\text{U}^\text{168}_\text{OP1}$-\textit{cis}-2$\text{O}_\text{ph}$ & {\cellcolor[HTML]{FEFEFE}} \color[HTML]{000000} {\cellcolor{white}}  & {\cellcolor[HTML]{FEFEFE}} \color[HTML]{000000}  & {\cellcolor[HTML]{FEFEFE}} \color[HTML]{000000} {\cellcolor[HTML]{FC9272}} \color[HTML]{000000} 0.4 & {\cellcolor[HTML]{FEFEFE}} \color[HTML]{000000} {\cellcolor[HTML]{000000}} \color[HTML]{F1F1F1} {\cellcolor{white}}  & {\cellcolor[HTML]{FEFEFE}} \color[HTML]{000000} {\cellcolor[HTML]{FEEAE1}} \color[HTML]{000000} 0.1 & {\cellcolor[HTML]{FEFEFE}} \color[HTML]{000000} {\cellcolor[HTML]{000000}} \color[HTML]{F1F1F1} {\cellcolor{white}}  & {\cellcolor[HTML]{FEFEFE}} \color[HTML]{000000} {\cellcolor[HTML]{FDCEBB}} \color[HTML]{000000} 0.2 (0.0-0.4) & {\cellcolor[HTML]{FEFEFE}} \color[HTML]{000000} {\cellcolor[HTML]{FEE6DA}} \color[HTML]{000000} 0.1 (0.0-0.2) & {\cellcolor[HTML]{FEFEFE}} \color[HTML]{000000} {\cellcolor[HTML]{FFF1EA}} \color[HTML]{000000} 0.0 (0.0-0.1) & {\cellcolor[HTML]{FEFEFE}} \color[HTML]{000000} {\cellcolor[HTML]{FFF0E9}} \color[HTML]{000000} 0.0 (0.0-0.1) & {\cellcolor[HTML]{FEFEFE}} \color[HTML]{000000} {\cellcolor[HTML]{000000}} \color[HTML]{F1F1F1} {\cellcolor{white}}  & {\cellcolor[HTML]{FEFEFE}} \color[HTML]{000000} {\cellcolor[HTML]{000000}} \color[HTML]{F1F1F1} {\cellcolor{white}}  & {\cellcolor[HTML]{FEFEFE}} \color[HTML]{000000} {\cellcolor[HTML]{000000}} \color[HTML]{F1F1F1} {\cellcolor{white}}  & {\cellcolor[HTML]{FEFEFE}} \color[HTML]{000000} {\cellcolor[HTML]{FDCAB5}} \color[HTML]{000000} 0.2 (0.1-0.3) & {\cellcolor[HTML]{FEFEFE}} \color[HTML]{000000} {\cellcolor[HTML]{FCBBA1}} \color[HTML]{000000} 0.2 (0.2-0.4) & {\cellcolor[HTML]{FEFEFE}} \color[HTML]{000000} {\cellcolor[HTML]{FCBBA1}} \color[HTML]{000000} 0.2 (0.2-0.5) \\
{\cellcolor[HTML]{FEFEFE}} \color[HTML]{000000} 74 & {\cellcolor[HTML]{FEFEFE}} \color[HTML]{000000} $\text{A}^\text{113}_\text{OP2}$-$\text{O}_\text{ph}$ & {\cellcolor[HTML]{FEFEFE}} \color[HTML]{000000} {\cellcolor{white}}  & {\cellcolor[HTML]{FEFEFE}} \color[HTML]{000000}  & {\cellcolor[HTML]{FEFEFE}} \color[HTML]{000000} {\cellcolor[HTML]{000000}} \color[HTML]{F1F1F1} {\cellcolor{white}}  & {\cellcolor[HTML]{FEFEFE}} \color[HTML]{000000} {\cellcolor[HTML]{FEEAE1}} \color[HTML]{000000} 0.1 & {\cellcolor[HTML]{FEFEFE}} \color[HTML]{000000} {\cellcolor[HTML]{000000}} \color[HTML]{F1F1F1} {\cellcolor{white}}  & {\cellcolor[HTML]{FEFEFE}} \color[HTML]{000000} {\cellcolor[HTML]{FEE0D2}} \color[HTML]{000000} 0.1 & {\cellcolor[HTML]{FEFEFE}} \color[HTML]{000000} {\cellcolor[HTML]{000000}} \color[HTML]{F1F1F1} {\cellcolor{white}}  & {\cellcolor[HTML]{FEFEFE}} \color[HTML]{000000} {\cellcolor[HTML]{FFF3ED}} \color[HTML]{000000} 0.0 (0.0-0.0) & {\cellcolor[HTML]{FEFEFE}} \color[HTML]{000000} {\cellcolor[HTML]{FEEAE0}} \color[HTML]{000000} 0.1 (0.0-0.1) & {\cellcolor[HTML]{FEFEFE}} \color[HTML]{000000} {\cellcolor[HTML]{FEEAE0}} \color[HTML]{000000} 0.1 (0.0-0.2) & {\cellcolor[HTML]{FEFEFE}} \color[HTML]{000000} {\cellcolor[HTML]{FFEDE5}} \color[HTML]{000000} 0.0 (0.0-0.2) & {\cellcolor[HTML]{FEFEFE}} \color[HTML]{000000} {\cellcolor[HTML]{FEE2D5}} \color[HTML]{000000} 0.1 (0.0-0.2) & {\cellcolor[HTML]{FEFEFE}} \color[HTML]{000000} {\cellcolor[HTML]{FEE1D3}} \color[HTML]{000000} 0.1 (0.1-0.2) & {\cellcolor[HTML]{FEFEFE}} \color[HTML]{000000} {\cellcolor[HTML]{FDCAB5}} \color[HTML]{000000} 0.2 (0.1-0.4) & {\cellcolor[HTML]{FEFEFE}} \color[HTML]{000000} {\cellcolor[HTML]{FED9C9}} \color[HTML]{000000} 0.2 (0.0-0.6) & {\cellcolor[HTML]{FEFEFE}} \color[HTML]{000000} {\cellcolor[HTML]{FB7A5A}} \color[HTML]{F1F1F1} 0.5 (0.1-0.8) \\
{\cellcolor[HTML]{FEFEFE}} \color[HTML]{000000} 75 & {\cellcolor[HTML]{FEFEFE}} \color[HTML]{000000} $\text{G}^\text{32}_\text{OP1}$-$\text{O}_\text{ph}$ & {\cellcolor[HTML]{FEFEFE}} \color[HTML]{000000} {\cellcolor{white}}  & {\cellcolor[HTML]{FEFEFE}} \color[HTML]{000000}  & {\cellcolor[HTML]{FEFEFE}} \color[HTML]{000000} {\cellcolor[HTML]{000000}} \color[HTML]{F1F1F1} {\cellcolor{white}}  & {\cellcolor[HTML]{FEFEFE}} \color[HTML]{000000} {\cellcolor[HTML]{000000}} \color[HTML]{F1F1F1} {\cellcolor{white}}  & {\cellcolor[HTML]{FEFEFE}} \color[HTML]{000000} {\cellcolor[HTML]{000000}} \color[HTML]{F1F1F1} {\cellcolor{white}}  & {\cellcolor[HTML]{FEFEFE}} \color[HTML]{000000} {\cellcolor[HTML]{000000}} \color[HTML]{F1F1F1} {\cellcolor{white}}  & {\cellcolor[HTML]{FEFEFE}} \color[HTML]{000000} {\cellcolor[HTML]{FEE0D2}} \color[HTML]{000000} 0.1 (0.0-0.4) & {\cellcolor[HTML]{FEFEFE}} \color[HTML]{000000} {\cellcolor[HTML]{FEDCCD}} \color[HTML]{000000} 0.1 (0.0-0.3) & {\cellcolor[HTML]{FEFEFE}} \color[HTML]{000000} {\cellcolor[HTML]{FDD4C2}} \color[HTML]{000000} 0.2 (0.1-0.3) & {\cellcolor[HTML]{FEFEFE}} \color[HTML]{000000} {\cellcolor[HTML]{FDC6B0}} \color[HTML]{000000} 0.2 (0.1-0.4) & {\cellcolor[HTML]{FEFEFE}} \color[HTML]{000000} {\cellcolor[HTML]{000000}} \color[HTML]{F1F1F1} {\cellcolor{white}}  & {\cellcolor[HTML]{FEFEFE}} \color[HTML]{000000} {\cellcolor[HTML]{FEDBCC}} \color[HTML]{000000} 0.1 (0.1-0.2) & {\cellcolor[HTML]{FEFEFE}} \color[HTML]{000000} {\cellcolor[HTML]{FCA486}} \color[HTML]{000000} 0.3 (0.2-0.4) & {\cellcolor[HTML]{FEFEFE}} \color[HTML]{000000} {\cellcolor[HTML]{FCAB8F}} \color[HTML]{000000} 0.3 (0.2-0.4) & {\cellcolor[HTML]{FEFEFE}} \color[HTML]{000000} {\cellcolor[HTML]{FFF5F0}} \color[HTML]{000000} 0.0 (0.0-0.2) & {\cellcolor[HTML]{FEFEFE}} \color[HTML]{000000} {\cellcolor[HTML]{FFF5F0}} \color[HTML]{000000} 0.0 (0.0-0.0) \\
{\cellcolor[HTML]{FEFEFE}} \color[HTML]{000000} 76 & {\cellcolor[HTML]{FEFEFE}} \color[HTML]{000000} $\text{C}^\text{278}_\text{OP2}$-$\text{O}_\text{ph}$ & {\cellcolor[HTML]{FEFEFE}} \color[HTML]{000000} {\cellcolor{white}}  & {\cellcolor[HTML]{FEFEFE}} \color[HTML]{000000}  & {\cellcolor[HTML]{FEFEFE}} \color[HTML]{000000} {\cellcolor[HTML]{000000}} \color[HTML]{F1F1F1} {\cellcolor{white}}  & {\cellcolor[HTML]{FEFEFE}} \color[HTML]{000000} {\cellcolor[HTML]{000000}} \color[HTML]{F1F1F1} {\cellcolor{white}}  & {\cellcolor[HTML]{FEFEFE}} \color[HTML]{000000} {\cellcolor[HTML]{000000}} \color[HTML]{F1F1F1} {\cellcolor{white}}  & {\cellcolor[HTML]{FEFEFE}} \color[HTML]{000000} {\cellcolor[HTML]{000000}} \color[HTML]{F1F1F1} {\cellcolor{white}}  & {\cellcolor[HTML]{FEFEFE}} \color[HTML]{000000} {\cellcolor[HTML]{000000}} \color[HTML]{F1F1F1} {\cellcolor{white}}  & {\cellcolor[HTML]{FEFEFE}} \color[HTML]{000000} {\cellcolor[HTML]{FFF4EE}} \color[HTML]{000000} 0.0 (0.0-0.0) & {\cellcolor[HTML]{FEFEFE}} \color[HTML]{000000} {\cellcolor[HTML]{FEEAE0}} \color[HTML]{000000} 0.1 (0.0-0.1) & {\cellcolor[HTML]{FEFEFE}} \color[HTML]{000000} {\cellcolor[HTML]{FFF4EE}} \color[HTML]{000000} 0.0 (0.0-0.0) & {\cellcolor[HTML]{FEFEFE}} \color[HTML]{000000} {\cellcolor[HTML]{000000}} \color[HTML]{F1F1F1} {\cellcolor{white}}  & {\cellcolor[HTML]{FEFEFE}} \color[HTML]{000000} {\cellcolor[HTML]{000000}} \color[HTML]{F1F1F1} {\cellcolor{white}}  & {\cellcolor[HTML]{FEFEFE}} \color[HTML]{000000} {\cellcolor[HTML]{FFF5F0}} \color[HTML]{000000} 0.0 (0.0-0.0) & {\cellcolor[HTML]{FEFEFE}} \color[HTML]{000000} {\cellcolor[HTML]{000000}} \color[HTML]{F1F1F1} {\cellcolor{white}} {\cellcolor[HTML]{AAAAAA}}  & {\cellcolor[HTML]{FEFEFE}} \color[HTML]{000000} {\cellcolor[HTML]{FB694A}} \color[HTML]{F1F1F1} 0.5 (0.4-0.6) & {\cellcolor[HTML]{FEFEFE}} \color[HTML]{000000} {\cellcolor[HTML]{BC141A}} \color[HTML]{F1F1F1} 0.8 (0.4-0.9) \\
{\cellcolor[HTML]{FEFEFE}} \color[HTML]{000000} 77 & {\cellcolor[HTML]{FEFEFE}} \color[HTML]{000000} $\text{A}^\text{161}_\text{OP2}$-$\text{O}_\text{ph}$ & {\cellcolor[HTML]{FEFEFE}} \color[HTML]{000000} {\cellcolor{white}}  & {\cellcolor[HTML]{FEFEFE}} \color[HTML]{000000}  & {\cellcolor[HTML]{FEFEFE}} \color[HTML]{000000} {\cellcolor[HTML]{000000}} \color[HTML]{F1F1F1} {\cellcolor{white}}  & {\cellcolor[HTML]{FEFEFE}} \color[HTML]{000000} {\cellcolor[HTML]{000000}} \color[HTML]{F1F1F1} {\cellcolor{white}}  & {\cellcolor[HTML]{FEFEFE}} \color[HTML]{000000} {\cellcolor[HTML]{000000}} \color[HTML]{F1F1F1} {\cellcolor{white}}  & {\cellcolor[HTML]{FEFEFE}} \color[HTML]{000000} {\cellcolor[HTML]{000000}} \color[HTML]{F1F1F1} {\cellcolor{white}}  & {\cellcolor[HTML]{FEFEFE}} \color[HTML]{000000} {\cellcolor[HTML]{FEE8DD}} \color[HTML]{000000} 0.1 (0.0-0.3) & {\cellcolor[HTML]{FEFEFE}} \color[HTML]{000000} {\cellcolor[HTML]{FEE8DD}} \color[HTML]{000000} 0.1 (0.0-0.2) & {\cellcolor[HTML]{FEFEFE}} \color[HTML]{000000} {\cellcolor[HTML]{FDC6B0}} \color[HTML]{000000} 0.2 (0.1-0.3) & {\cellcolor[HTML]{FEFEFE}} \color[HTML]{000000} {\cellcolor[HTML]{FCA98C}} \color[HTML]{000000} 0.3 (0.2-0.4) & {\cellcolor[HTML]{FEFEFE}} \color[HTML]{000000} {\cellcolor[HTML]{000000}} \color[HTML]{F1F1F1} {\cellcolor{white}}  & {\cellcolor[HTML]{FEFEFE}} \color[HTML]{000000} {\cellcolor[HTML]{FCB095}} \color[HTML]{000000} 0.3 (0.1-0.5) & {\cellcolor[HTML]{FEFEFE}} \color[HTML]{000000} {\cellcolor[HTML]{FC8464}} \color[HTML]{F1F1F1} 0.4 (0.3-0.5) & {\cellcolor[HTML]{FEFEFE}} \color[HTML]{000000} {\cellcolor[HTML]{000000}} \color[HTML]{F1F1F1} {\cellcolor{white}} {\cellcolor[HTML]{AAAAAA}}  & {\cellcolor[HTML]{FEFEFE}} \color[HTML]{000000} {\cellcolor[HTML]{000000}} \color[HTML]{F1F1F1} {\cellcolor{white}} {\cellcolor[HTML]{AAAAAA}}  & {\cellcolor[HTML]{FEFEFE}} \color[HTML]{000000} {\cellcolor[HTML]{000000}} \color[HTML]{F1F1F1} {\cellcolor{white}} {\cellcolor[HTML]{AAAAAA}}  \\
{\cellcolor[HTML]{FEFEFE}} \color[HTML]{000000} 78 & {\cellcolor[HTML]{FEFEFE}} \color[HTML]{000000} $\text{U}^\text{404}_\text{OP1}$-$\text{O}_\text{ph}$ & {\cellcolor[HTML]{FEFEFE}} \color[HTML]{000000} {\cellcolor{white}}  & {\cellcolor[HTML]{FEFEFE}} \color[HTML]{000000}  & {\cellcolor[HTML]{FEFEFE}} \color[HTML]{000000} {\cellcolor[HTML]{000000}} \color[HTML]{F1F1F1} {\cellcolor{white}}  & {\cellcolor[HTML]{FEFEFE}} \color[HTML]{000000} {\cellcolor[HTML]{000000}} \color[HTML]{F1F1F1} {\cellcolor{white}}  & {\cellcolor[HTML]{FEFEFE}} \color[HTML]{000000} {\cellcolor[HTML]{000000}} \color[HTML]{F1F1F1} {\cellcolor{white}}  & {\cellcolor[HTML]{FEFEFE}} \color[HTML]{000000} {\cellcolor[HTML]{000000}} \color[HTML]{F1F1F1} {\cellcolor{white}}  & {\cellcolor[HTML]{FEFEFE}} \color[HTML]{000000} {\cellcolor[HTML]{FEE8DD}} \color[HTML]{000000} 0.1 (0.0-0.2) & {\cellcolor[HTML]{FEFEFE}} \color[HTML]{000000} {\cellcolor[HTML]{FCBDA4}} \color[HTML]{000000} 0.2 (0.1-0.4) & {\cellcolor[HTML]{FEFEFE}} \color[HTML]{000000} {\cellcolor[HTML]{FFF0E8}} \color[HTML]{000000} 0.0 (0.0-0.1) & {\cellcolor[HTML]{FEFEFE}} \color[HTML]{000000} {\cellcolor[HTML]{FDC9B3}} \color[HTML]{000000} 0.2 (0.1-0.3) & {\cellcolor[HTML]{FEFEFE}} \color[HTML]{000000} {\cellcolor[HTML]{FFF5F0}} \color[HTML]{000000} 0.0 (0.0-0.0) & {\cellcolor[HTML]{FEFEFE}} \color[HTML]{000000} {\cellcolor[HTML]{FCC4AD}} \color[HTML]{000000} 0.2 (0.1-0.3) & {\cellcolor[HTML]{FEFEFE}} \color[HTML]{000000} {\cellcolor[HTML]{FDD3C1}} \color[HTML]{000000} 0.2 (0.1-0.3) & {\cellcolor[HTML]{FEFEFE}} \color[HTML]{000000} {\cellcolor[HTML]{FCBBA1}} \color[HTML]{000000} 0.2 (0.1-0.5) & {\cellcolor[HTML]{FEFEFE}} \color[HTML]{000000} {\cellcolor[HTML]{FED9C9}} \color[HTML]{000000} 0.2 (0.0-0.5) & {\cellcolor[HTML]{FEFEFE}} \color[HTML]{000000} {\cellcolor[HTML]{FFF5F0}} \color[HTML]{000000} 0.0 (0.0-0.2) \\
{\cellcolor[HTML]{FEFEFE}} \color[HTML]{000000} 79 & {\cellcolor[HTML]{FEFEFE}} \color[HTML]{000000} $\text{G}^\text{112}_\text{OP2}$-$\text{O}_\text{ph}$ & {\cellcolor[HTML]{FEFEFE}} \color[HTML]{000000} {\cellcolor{white}}  & {\cellcolor[HTML]{FEFEFE}} \color[HTML]{000000}  & {\cellcolor[HTML]{FEFEFE}} \color[HTML]{000000} {\cellcolor[HTML]{000000}} \color[HTML]{F1F1F1} {\cellcolor{white}}  & {\cellcolor[HTML]{FEFEFE}} \color[HTML]{000000} {\cellcolor[HTML]{000000}} \color[HTML]{F1F1F1} {\cellcolor{white}}  & {\cellcolor[HTML]{FEFEFE}} \color[HTML]{000000} {\cellcolor[HTML]{000000}} \color[HTML]{F1F1F1} {\cellcolor{white}}  & {\cellcolor[HTML]{FEFEFE}} \color[HTML]{000000} {\cellcolor[HTML]{000000}} \color[HTML]{F1F1F1} {\cellcolor{white}}  & {\cellcolor[HTML]{FEFEFE}} \color[HTML]{000000} {\cellcolor[HTML]{000000}} \color[HTML]{F1F1F1} {\cellcolor{white}}  & {\cellcolor[HTML]{FEFEFE}} \color[HTML]{000000} {\cellcolor[HTML]{000000}} \color[HTML]{F1F1F1} {\cellcolor{white}}  & {\cellcolor[HTML]{FEFEFE}} \color[HTML]{000000} {\cellcolor[HTML]{FEDECF}} \color[HTML]{000000} 0.1 (0.0-0.3) & {\cellcolor[HTML]{FEFEFE}} \color[HTML]{000000} {\cellcolor[HTML]{FFF4EE}} \color[HTML]{000000} 0.0 (0.0-0.0) & {\cellcolor[HTML]{FEFEFE}} \color[HTML]{000000} {\cellcolor[HTML]{000000}} \color[HTML]{F1F1F1} {\cellcolor{white}}  & {\cellcolor[HTML]{FEFEFE}} \color[HTML]{000000} {\cellcolor[HTML]{FFEEE7}} \color[HTML]{000000} 0.0 (0.0-0.1) & {\cellcolor[HTML]{FEFEFE}} \color[HTML]{000000} {\cellcolor[HTML]{FFF2EC}} \color[HTML]{000000} 0.0 (0.0-0.0) & {\cellcolor[HTML]{FEFEFE}} \color[HTML]{000000} {\cellcolor[HTML]{000000}} \color[HTML]{F1F1F1} {\cellcolor{white}} {\cellcolor[HTML]{AAAAAA}}  & {\cellcolor[HTML]{FEFEFE}} \color[HTML]{000000} {\cellcolor[HTML]{FC9B7C}} \color[HTML]{000000} 0.4 (0.2-0.5) & {\cellcolor[HTML]{FEFEFE}} \color[HTML]{000000} {\cellcolor[HTML]{BC141A}} \color[HTML]{F1F1F1} 0.8 (0.4-0.9) \\
{\cellcolor[HTML]{FEFEFE}} \color[HTML]{000000} 80 & {\cellcolor[HTML]{FEFEFE}} \color[HTML]{000000} $\text{G}^\text{313}_\text{OP2}$-$\text{O}_\text{ph}$ & {\cellcolor[HTML]{FEFEFE}} \color[HTML]{000000} {\cellcolor{white}}  & {\cellcolor[HTML]{FEFEFE}} \color[HTML]{000000}  & {\cellcolor[HTML]{FEFEFE}} \color[HTML]{000000} {\cellcolor[HTML]{000000}} \color[HTML]{F1F1F1} {\cellcolor{white}}  & {\cellcolor[HTML]{FEFEFE}} \color[HTML]{000000} {\cellcolor[HTML]{000000}} \color[HTML]{F1F1F1} {\cellcolor{white}}  & {\cellcolor[HTML]{FEFEFE}} \color[HTML]{000000} {\cellcolor[HTML]{000000}} \color[HTML]{F1F1F1} {\cellcolor{white}}  & {\cellcolor[HTML]{FEFEFE}} \color[HTML]{000000} {\cellcolor[HTML]{000000}} \color[HTML]{F1F1F1} {\cellcolor{white}}  & {\cellcolor[HTML]{FEFEFE}} \color[HTML]{000000} {\cellcolor[HTML]{000000}} \color[HTML]{F1F1F1} {\cellcolor{white}}  & {\cellcolor[HTML]{FEFEFE}} \color[HTML]{000000} {\cellcolor[HTML]{FDCBB6}} \color[HTML]{000000} 0.2 (0.1-0.4) & {\cellcolor[HTML]{FEFEFE}} \color[HTML]{000000} {\cellcolor[HTML]{FFEEE7}} \color[HTML]{000000} 0.0 (0.0-0.1) & {\cellcolor[HTML]{FEFEFE}} \color[HTML]{000000} {\cellcolor[HTML]{FB7252}} \color[HTML]{F1F1F1} 0.5 (0.3-0.6) & {\cellcolor[HTML]{FEFEFE}} \color[HTML]{000000} {\cellcolor[HTML]{000000}} \color[HTML]{F1F1F1} {\cellcolor{white}}  & {\cellcolor[HTML]{FEFEFE}} \color[HTML]{000000} {\cellcolor[HTML]{FDCBB6}} \color[HTML]{000000} 0.2 (0.1-0.4) & {\cellcolor[HTML]{FEFEFE}} \color[HTML]{000000} {\cellcolor[HTML]{FB7C5C}} \color[HTML]{F1F1F1} 0.4 (0.3-0.6) & {\cellcolor[HTML]{FEFEFE}} \color[HTML]{000000} {\cellcolor[HTML]{000000}} \color[HTML]{F1F1F1} {\cellcolor{white}} {\cellcolor[HTML]{AAAAAA}}  & {\cellcolor[HTML]{FEFEFE}} \color[HTML]{000000} {\cellcolor[HTML]{000000}} \color[HTML]{F1F1F1} {\cellcolor{white}} {\cellcolor[HTML]{AAAAAA}}  & {\cellcolor[HTML]{FEFEFE}} \color[HTML]{000000} {\cellcolor[HTML]{000000}} \color[HTML]{F1F1F1} {\cellcolor{white}} {\cellcolor[HTML]{AAAAAA}}  \\
{\cellcolor[HTML]{FEFEFE}} \color[HTML]{000000} 81 & {\cellcolor[HTML]{FEFEFE}} \color[HTML]{000000} $\text{G}^\text{288}_\text{OP2}$-$\text{O}_\text{ph}$ & {\cellcolor[HTML]{FEFEFE}} \color[HTML]{000000} {\cellcolor{white}}  & {\cellcolor[HTML]{FEFEFE}} \color[HTML]{000000}  & {\cellcolor[HTML]{FEFEFE}} \color[HTML]{000000} {\cellcolor[HTML]{000000}} \color[HTML]{F1F1F1} {\cellcolor{white}}  & {\cellcolor[HTML]{FEFEFE}} \color[HTML]{000000} {\cellcolor[HTML]{000000}} \color[HTML]{F1F1F1} {\cellcolor{white}}  & {\cellcolor[HTML]{FEFEFE}} \color[HTML]{000000} {\cellcolor[HTML]{000000}} \color[HTML]{F1F1F1} {\cellcolor{white}}  & {\cellcolor[HTML]{FEFEFE}} \color[HTML]{000000} {\cellcolor[HTML]{000000}} \color[HTML]{F1F1F1} {\cellcolor{white}}  & {\cellcolor[HTML]{FEFEFE}} \color[HTML]{000000} {\cellcolor[HTML]{000000}} \color[HTML]{F1F1F1} {\cellcolor{white}}  & {\cellcolor[HTML]{FEFEFE}} \color[HTML]{000000} {\cellcolor[HTML]{FED9C9}} \color[HTML]{000000} 0.2 (0.1-0.3) & {\cellcolor[HTML]{FEFEFE}} \color[HTML]{000000} {\cellcolor[HTML]{FFECE4}} \color[HTML]{000000} 0.1 (0.0-0.1) & {\cellcolor[HTML]{FEFEFE}} \color[HTML]{000000} {\cellcolor[HTML]{FC8565}} \color[HTML]{F1F1F1} 0.4 (0.3-0.5) & {\cellcolor[HTML]{FEFEFE}} \color[HTML]{000000} {\cellcolor[HTML]{000000}} \color[HTML]{F1F1F1} {\cellcolor{white}}  & {\cellcolor[HTML]{FEFEFE}} \color[HTML]{000000} {\cellcolor[HTML]{FCBDA4}} \color[HTML]{000000} 0.2 (0.1-0.5) & {\cellcolor[HTML]{FEFEFE}} \color[HTML]{000000} {\cellcolor[HTML]{FB7757}} \color[HTML]{F1F1F1} 0.5 (0.4-0.5) & {\cellcolor[HTML]{FEFEFE}} \color[HTML]{000000} {\cellcolor[HTML]{000000}} \color[HTML]{F1F1F1} {\cellcolor{white}} {\cellcolor[HTML]{AAAAAA}}  & {\cellcolor[HTML]{FEFEFE}} \color[HTML]{000000} {\cellcolor[HTML]{000000}} \color[HTML]{F1F1F1} {\cellcolor{white}} {\cellcolor[HTML]{AAAAAA}}  & {\cellcolor[HTML]{FEFEFE}} \color[HTML]{000000} {\cellcolor[HTML]{000000}} \color[HTML]{F1F1F1} {\cellcolor{white}} {\cellcolor[HTML]{AAAAAA}}  \\
{\cellcolor[HTML]{FEFEFE}} \color[HTML]{000000} 82 & {\cellcolor[HTML]{FEFEFE}} \color[HTML]{000000} $\text{C}^\text{208}_\text{OP1}$,$\text{A}^\text{304}_\text{OP1}$,$\text{A}^\text{306}_\text{OP1}$-\textit{mer}-3$\text{O}_\text{ph}$ & {\cellcolor[HTML]{FEFEFE}} \color[HTML]{000000} {\cellcolor{white}}  & {\cellcolor[HTML]{FEFEFE}} \color[HTML]{000000}  & {\cellcolor[HTML]{FEFEFE}} \color[HTML]{000000} {\cellcolor[HTML]{FEEAE1}} \color[HTML]{000000} 0.1 & {\cellcolor[HTML]{FEFEFE}} \color[HTML]{000000} {\cellcolor[HTML]{000000}} \color[HTML]{F1F1F1} {\cellcolor{white}}  & {\cellcolor[HTML]{FEFEFE}} \color[HTML]{000000} {\cellcolor[HTML]{000000}} \color[HTML]{F1F1F1} {\cellcolor{white}}  & {\cellcolor[HTML]{FEFEFE}} \color[HTML]{000000} {\cellcolor[HTML]{FEE0D2}} \color[HTML]{000000} 0.1 & {\cellcolor[HTML]{FEFEFE}} \color[HTML]{000000} {\cellcolor[HTML]{FEE0D2}} \color[HTML]{000000} 0.1 (0.0-0.4) & {\cellcolor[HTML]{FEFEFE}} \color[HTML]{000000} {\cellcolor[HTML]{FFF4EF}} \color[HTML]{000000} 0.0 (0.0-0.0) & {\cellcolor[HTML]{FEFEFE}} \color[HTML]{000000} {\cellcolor[HTML]{FEE1D3}} \color[HTML]{000000} 0.1 (0.0-0.4) & {\cellcolor[HTML]{FEFEFE}} \color[HTML]{000000} {\cellcolor[HTML]{FFF5F0}} \color[HTML]{000000} 0.0 (0.0-0.0) & {\cellcolor[HTML]{FEFEFE}} \color[HTML]{000000} {\cellcolor[HTML]{000000}} \color[HTML]{F1F1F1} {\cellcolor{white}}  & {\cellcolor[HTML]{FEFEFE}} \color[HTML]{000000} {\cellcolor[HTML]{FFF4EF}} \color[HTML]{000000} 0.0 (0.0-0.0) & {\cellcolor[HTML]{FEFEFE}} \color[HTML]{000000} {\cellcolor[HTML]{000000}} \color[HTML]{F1F1F1} {\cellcolor{white}}  & {\cellcolor[HTML]{FEFEFE}} \color[HTML]{000000} {\cellcolor[HTML]{FCBBA1}} \color[HTML]{000000} 0.2 (0.1-0.5) & {\cellcolor[HTML]{FEFEFE}} \color[HTML]{000000} {\cellcolor[HTML]{FCAB8F}} \color[HTML]{000000} 0.3 (0.1-0.5) & {\cellcolor[HTML]{FEFEFE}} \color[HTML]{000000} {\cellcolor[HTML]{FCAB8F}} \color[HTML]{000000} 0.3 (0.2-0.4) \\
{\cellcolor[HTML]{FEFEFE}} \color[HTML]{000000} 83 & {\cellcolor[HTML]{FEFEFE}} \color[HTML]{000000} $\text{C}^\text{311}_\text{OP2}$-$\text{O}_\text{ph}$ & {\cellcolor[HTML]{FEFEFE}} \color[HTML]{000000} {\cellcolor{white}}  & {\cellcolor[HTML]{FEFEFE}} \color[HTML]{000000}  & {\cellcolor[HTML]{FEFEFE}} \color[HTML]{000000} {\cellcolor[HTML]{000000}} \color[HTML]{F1F1F1} {\cellcolor{white}}  & {\cellcolor[HTML]{FEFEFE}} \color[HTML]{000000} {\cellcolor[HTML]{000000}} \color[HTML]{F1F1F1} {\cellcolor{white}}  & {\cellcolor[HTML]{FEFEFE}} \color[HTML]{000000} {\cellcolor[HTML]{000000}} \color[HTML]{F1F1F1} {\cellcolor{white}}  & {\cellcolor[HTML]{FEFEFE}} \color[HTML]{000000} {\cellcolor[HTML]{000000}} \color[HTML]{F1F1F1} {\cellcolor{white}}  & {\cellcolor[HTML]{FEFEFE}} \color[HTML]{000000} {\cellcolor[HTML]{000000}} \color[HTML]{F1F1F1} {\cellcolor{white}}  & {\cellcolor[HTML]{FEFEFE}} \color[HTML]{000000} {\cellcolor[HTML]{FFF4EE}} \color[HTML]{000000} 0.0 (0.0-0.0) & {\cellcolor[HTML]{FEFEFE}} \color[HTML]{000000} {\cellcolor[HTML]{FFF4EF}} \color[HTML]{000000} 0.0 (0.0-0.0) & {\cellcolor[HTML]{FEFEFE}} \color[HTML]{000000} {\cellcolor[HTML]{FEE8DD}} \color[HTML]{000000} 0.1 (0.0-0.2) & {\cellcolor[HTML]{FEFEFE}} \color[HTML]{000000} {\cellcolor[HTML]{000000}} \color[HTML]{F1F1F1} {\cellcolor{white}}  & {\cellcolor[HTML]{FEFEFE}} \color[HTML]{000000} {\cellcolor[HTML]{FFF3ED}} \color[HTML]{000000} 0.0 (0.0-0.0) & {\cellcolor[HTML]{FEFEFE}} \color[HTML]{000000} {\cellcolor[HTML]{FEE1D3}} \color[HTML]{000000} 0.1 (0.1-0.2) & {\cellcolor[HTML]{FEFEFE}} \color[HTML]{000000} {\cellcolor[HTML]{000000}} \color[HTML]{F1F1F1} {\cellcolor{white}} {\cellcolor[HTML]{AAAAAA}}  & {\cellcolor[HTML]{FEFEFE}} \color[HTML]{000000} {\cellcolor[HTML]{FC8A6A}} \color[HTML]{F1F1F1} 0.4 (0.0-0.8) & {\cellcolor[HTML]{FEFEFE}} \color[HTML]{000000} {\cellcolor[HTML]{E83429}} \color[HTML]{F1F1F1} 0.7 (0.2-0.9) \\
{\cellcolor[HTML]{FEFEFE}} \color[HTML]{000000} 84 & {\cellcolor[HTML]{FEFEFE}} \color[HTML]{000000} $\text{G}^\text{257}_\text{OP1}$,$\text{U}^\text{258}_\text{OP2}$,$\text{U}^\text{305}_\text{OP2}$-\textit{fac}-3$\text{O}_\text{ph}$ & {\cellcolor[HTML]{FEFEFE}} \color[HTML]{000000} {\cellcolor{white}}  & {\cellcolor[HTML]{FEFEFE}} \color[HTML]{000000}  & {\cellcolor[HTML]{FEFEFE}} \color[HTML]{000000} {\cellcolor[HTML]{000000}} \color[HTML]{F1F1F1} {\cellcolor{white}}  & {\cellcolor[HTML]{FEFEFE}} \color[HTML]{000000} {\cellcolor[HTML]{000000}} \color[HTML]{F1F1F1} {\cellcolor{white}}  & {\cellcolor[HTML]{FEFEFE}} \color[HTML]{000000} {\cellcolor[HTML]{FEEAE1}} \color[HTML]{000000} 0.1 & {\cellcolor[HTML]{FEFEFE}} \color[HTML]{000000} {\cellcolor[HTML]{000000}} \color[HTML]{F1F1F1} {\cellcolor{white}}  & {\cellcolor[HTML]{FEFEFE}} \color[HTML]{000000} {\cellcolor[HTML]{FEE4D8}} \color[HTML]{000000} 0.1 (0.0-0.4) & {\cellcolor[HTML]{FEFEFE}} \color[HTML]{000000} {\cellcolor[HTML]{FB7D5D}} \color[HTML]{F1F1F1} 0.4 (0.2-0.6) & {\cellcolor[HTML]{FEFEFE}} \color[HTML]{000000} {\cellcolor[HTML]{FFF5F0}} \color[HTML]{000000} 0.0 (0.0-0.0) & {\cellcolor[HTML]{FEFEFE}} \color[HTML]{000000} {\cellcolor[HTML]{FFF4EE}} \color[HTML]{000000} 0.0 (0.0-0.1) & {\cellcolor[HTML]{FEFEFE}} \color[HTML]{000000} {\cellcolor[HTML]{000000}} \color[HTML]{F1F1F1} {\cellcolor{white}}  & {\cellcolor[HTML]{FEFEFE}} \color[HTML]{000000} {\cellcolor[HTML]{000000}} \color[HTML]{F1F1F1} {\cellcolor{white}}  & {\cellcolor[HTML]{FEFEFE}} \color[HTML]{000000} {\cellcolor[HTML]{FEE6DA}} \color[HTML]{000000} 0.1 (0.0-0.2) & {\cellcolor[HTML]{FEFEFE}} \color[HTML]{000000} {\cellcolor[HTML]{FEE5D8}} \color[HTML]{000000} 0.1 (0.0-0.2) & {\cellcolor[HTML]{FEFEFE}} \color[HTML]{000000} {\cellcolor[HTML]{FB7A5A}} \color[HTML]{F1F1F1} 0.5 (0.2-0.6) & {\cellcolor[HTML]{FEFEFE}} \color[HTML]{000000} {\cellcolor[HTML]{FFF5F0}} \color[HTML]{000000} 0.0 (0.0-0.2) \\
{\cellcolor[HTML]{FEFEFE}} \color[HTML]{000000} 85 & {\cellcolor[HTML]{FEFEFE}} \color[HTML]{000000} $\text{C}^\text{311}_\text{OP1}$-$\text{O}_\text{ph}$ & {\cellcolor[HTML]{FEFEFE}} \color[HTML]{000000} {\cellcolor{white}}  & {\cellcolor[HTML]{FEFEFE}} \color[HTML]{000000}  & {\cellcolor[HTML]{FEFEFE}} \color[HTML]{000000} {\cellcolor[HTML]{000000}} \color[HTML]{F1F1F1} {\cellcolor{white}}  & {\cellcolor[HTML]{FEFEFE}} \color[HTML]{000000} {\cellcolor[HTML]{000000}} \color[HTML]{F1F1F1} {\cellcolor{white}}  & {\cellcolor[HTML]{FEFEFE}} \color[HTML]{000000} {\cellcolor[HTML]{000000}} \color[HTML]{F1F1F1} {\cellcolor{white}}  & {\cellcolor[HTML]{FEFEFE}} \color[HTML]{000000} {\cellcolor[HTML]{000000}} \color[HTML]{F1F1F1} {\cellcolor{white}}  & {\cellcolor[HTML]{FEFEFE}} \color[HTML]{000000} {\cellcolor[HTML]{000000}} \color[HTML]{F1F1F1} {\cellcolor{white}}  & {\cellcolor[HTML]{FEFEFE}} \color[HTML]{000000} {\cellcolor[HTML]{FEEAE0}} \color[HTML]{000000} 0.1 (0.0-0.2) & {\cellcolor[HTML]{FEFEFE}} \color[HTML]{000000} {\cellcolor[HTML]{FEE7DB}} \color[HTML]{000000} 0.1 (0.0-0.2) & {\cellcolor[HTML]{FEFEFE}} \color[HTML]{000000} {\cellcolor[HTML]{FEDCCD}} \color[HTML]{000000} 0.1 (0.1-0.2) & {\cellcolor[HTML]{FEFEFE}} \color[HTML]{000000} {\cellcolor[HTML]{000000}} \color[HTML]{F1F1F1} {\cellcolor{white}}  & {\cellcolor[HTML]{FEFEFE}} \color[HTML]{000000} {\cellcolor[HTML]{FEE8DD}} \color[HTML]{000000} 0.1 (0.0-0.2) & {\cellcolor[HTML]{FEFEFE}} \color[HTML]{000000} {\cellcolor[HTML]{FEE5D9}} \color[HTML]{000000} 0.1 (0.1-0.2) & {\cellcolor[HTML]{FEFEFE}} \color[HTML]{000000} {\cellcolor[HTML]{000000}} \color[HTML]{F1F1F1} {\cellcolor{white}} {\cellcolor[HTML]{AAAAAA}}  & {\cellcolor[HTML]{FEFEFE}} \color[HTML]{000000} {\cellcolor[HTML]{FB694A}} \color[HTML]{F1F1F1} 0.5 (0.3-0.8) & {\cellcolor[HTML]{FEFEFE}} \color[HTML]{000000} {\cellcolor[HTML]{FCBBA1}} \color[HTML]{000000} 0.2 (0.0-0.6) \\
{\cellcolor[HTML]{FEFEFE}} \color[HTML]{000000} 86 & {\cellcolor[HTML]{FEFEFE}} \color[HTML]{000000} $\text{U}^\text{300}_\text{OP1}$,$\text{A}^\text{301}_\text{OP1}$,$\text{A}^\text{302}_\text{OP2}$-\textit{fac}-3$\text{O}_\text{ph}$ & {\cellcolor[HTML]{FEFEFE}} \color[HTML]{000000} {\cellcolor{white}}  & {\cellcolor[HTML]{FEFEFE}} \color[HTML]{000000}  & {\cellcolor[HTML]{FEFEFE}} \color[HTML]{000000} {\cellcolor[HTML]{000000}} \color[HTML]{F1F1F1} {\cellcolor{white}}  & {\cellcolor[HTML]{FEFEFE}} \color[HTML]{000000} {\cellcolor[HTML]{000000}} \color[HTML]{F1F1F1} {\cellcolor{white}}  & {\cellcolor[HTML]{FEFEFE}} \color[HTML]{000000} {\cellcolor[HTML]{000000}} \color[HTML]{F1F1F1} {\cellcolor{white}}  & {\cellcolor[HTML]{FEFEFE}} \color[HTML]{000000} {\cellcolor[HTML]{000000}} \color[HTML]{F1F1F1} {\cellcolor{white}}  & {\cellcolor[HTML]{FEFEFE}} \color[HTML]{000000} {\cellcolor[HTML]{FCBCA2}} \color[HTML]{000000} 0.2 (0.0-0.6) & {\cellcolor[HTML]{FEFEFE}} \color[HTML]{000000} {\cellcolor[HTML]{FCBCA2}} \color[HTML]{000000} 0.2 (0.0-0.6) & {\cellcolor[HTML]{FEFEFE}} \color[HTML]{000000} {\cellcolor[HTML]{FCBCA2}} \color[HTML]{000000} 0.2 (0.0-0.5) & {\cellcolor[HTML]{FEFEFE}} \color[HTML]{000000} {\cellcolor[HTML]{000000}} \color[HTML]{F1F1F1} {\cellcolor{white}}  & {\cellcolor[HTML]{FEFEFE}} \color[HTML]{000000} {\cellcolor[HTML]{000000}} \color[HTML]{F1F1F1} {\cellcolor{white}}  & {\cellcolor[HTML]{FEFEFE}} \color[HTML]{000000} {\cellcolor[HTML]{000000}} \color[HTML]{F1F1F1} {\cellcolor{white}}  & {\cellcolor[HTML]{FEFEFE}} \color[HTML]{000000} {\cellcolor[HTML]{000000}} \color[HTML]{F1F1F1} {\cellcolor{white}}  & {\cellcolor[HTML]{FEFEFE}} \color[HTML]{000000} {\cellcolor[HTML]{FCBBA1}} \color[HTML]{000000} 0.2 (0.2-0.3) & {\cellcolor[HTML]{FEFEFE}} \color[HTML]{000000} {\cellcolor[HTML]{FDCAB5}} \color[HTML]{000000} 0.2 (0.2-0.2) & {\cellcolor[HTML]{FEFEFE}} \color[HTML]{000000} {\cellcolor[HTML]{000000}} \color[HTML]{F1F1F1} {\cellcolor{white}} {\cellcolor[HTML]{AAAAAA}}  \\
{\cellcolor[HTML]{FEFEFE}} \color[HTML]{000000} 87 & {\cellcolor[HTML]{FEFEFE}} \color[HTML]{000000} $\text{U}^\text{33}_\text{OP2}$-$\text{O}_\text{ph}$ & {\cellcolor[HTML]{FEFEFE}} \color[HTML]{000000} {\cellcolor{white}}  & {\cellcolor[HTML]{FEFEFE}} \color[HTML]{000000}  & {\cellcolor[HTML]{FEFEFE}} \color[HTML]{000000} {\cellcolor[HTML]{000000}} \color[HTML]{F1F1F1} {\cellcolor{white}}  & {\cellcolor[HTML]{FEFEFE}} \color[HTML]{000000} {\cellcolor[HTML]{000000}} \color[HTML]{F1F1F1} {\cellcolor{white}}  & {\cellcolor[HTML]{FEFEFE}} \color[HTML]{000000} {\cellcolor[HTML]{000000}} \color[HTML]{F1F1F1} {\cellcolor{white}}  & {\cellcolor[HTML]{FEFEFE}} \color[HTML]{000000} {\cellcolor[HTML]{000000}} \color[HTML]{F1F1F1} {\cellcolor{white}}  & {\cellcolor[HTML]{FEFEFE}} \color[HTML]{000000} {\cellcolor[HTML]{000000}} \color[HTML]{F1F1F1} {\cellcolor{white}}  & {\cellcolor[HTML]{FEFEFE}} \color[HTML]{000000} {\cellcolor[HTML]{FFF2EC}} \color[HTML]{000000} 0.0 (0.0-0.0) & {\cellcolor[HTML]{FEFEFE}} \color[HTML]{000000} {\cellcolor[HTML]{000000}} \color[HTML]{F1F1F1} {\cellcolor{white}}  & {\cellcolor[HTML]{FEFEFE}} \color[HTML]{000000} {\cellcolor[HTML]{FEEAE1}} \color[HTML]{000000} 0.1 (0.0-0.1) & {\cellcolor[HTML]{FEFEFE}} \color[HTML]{000000} {\cellcolor[HTML]{000000}} \color[HTML]{F1F1F1} {\cellcolor{white}}  & {\cellcolor[HTML]{FEFEFE}} \color[HTML]{000000} {\cellcolor[HTML]{000000}} \color[HTML]{F1F1F1} {\cellcolor{white}}  & {\cellcolor[HTML]{FEFEFE}} \color[HTML]{000000} {\cellcolor[HTML]{FFF2EC}} \color[HTML]{000000} 0.0 (0.0-0.1) & {\cellcolor[HTML]{FEFEFE}} \color[HTML]{000000} {\cellcolor[HTML]{000000}} \color[HTML]{F1F1F1} {\cellcolor{white}} {\cellcolor[HTML]{AAAAAA}}  & {\cellcolor[HTML]{FEFEFE}} \color[HTML]{000000} {\cellcolor[HTML]{F6583E}} \color[HTML]{F1F1F1} 0.6 (0.2-0.7) & {\cellcolor[HTML]{FEFEFE}} \color[HTML]{000000} {\cellcolor[HTML]{F6583E}} \color[HTML]{F1F1F1} 0.6 (0.1-0.7) \\
{\cellcolor[HTML]{FEFEFE}} \color[HTML]{000000} 88 & {\cellcolor[HTML]{FEFEFE}} \color[HTML]{000000} $\text{G}^\text{250}_\text{OP2}$-$\text{O}_\text{ph}$ & {\cellcolor[HTML]{FEFEFE}} \color[HTML]{000000} {\cellcolor{white}}  & {\cellcolor[HTML]{FEFEFE}} \color[HTML]{000000}  & {\cellcolor[HTML]{FEFEFE}} \color[HTML]{000000} {\cellcolor[HTML]{000000}} \color[HTML]{F1F1F1} {\cellcolor{white}}  & {\cellcolor[HTML]{FEFEFE}} \color[HTML]{000000} {\cellcolor[HTML]{000000}} \color[HTML]{F1F1F1} {\cellcolor{white}}  & {\cellcolor[HTML]{FEFEFE}} \color[HTML]{000000} {\cellcolor[HTML]{000000}} \color[HTML]{F1F1F1} {\cellcolor{white}}  & {\cellcolor[HTML]{FEFEFE}} \color[HTML]{000000} {\cellcolor[HTML]{000000}} \color[HTML]{F1F1F1} {\cellcolor{white}}  & {\cellcolor[HTML]{FEFEFE}} \color[HTML]{000000} {\cellcolor[HTML]{FFF5F0}} \color[HTML]{000000} 0.0 (0.0-0.0) & {\cellcolor[HTML]{FEFEFE}} \color[HTML]{000000} {\cellcolor[HTML]{FEE5D8}} \color[HTML]{000000} 0.1 (0.0-0.2) & {\cellcolor[HTML]{FEFEFE}} \color[HTML]{000000} {\cellcolor[HTML]{FFF5F0}} \color[HTML]{000000} 0.0 (0.0-0.0) & {\cellcolor[HTML]{FEFEFE}} \color[HTML]{000000} {\cellcolor[HTML]{FCC1A8}} \color[HTML]{000000} 0.2 (0.1-0.4) & {\cellcolor[HTML]{FEFEFE}} \color[HTML]{000000} {\cellcolor[HTML]{000000}} \color[HTML]{F1F1F1} {\cellcolor{white}}  & {\cellcolor[HTML]{FEFEFE}} \color[HTML]{000000} {\cellcolor[HTML]{FEE8DD}} \color[HTML]{000000} 0.1 (0.0-0.1) & {\cellcolor[HTML]{FEFEFE}} \color[HTML]{000000} {\cellcolor[HTML]{FC8262}} \color[HTML]{F1F1F1} 0.4 (0.3-0.5) & {\cellcolor[HTML]{FEFEFE}} \color[HTML]{000000} {\cellcolor[HTML]{FFF5F0}} \color[HTML]{000000} 0.0 (0.0-0.0) & {\cellcolor[HTML]{FEFEFE}} \color[HTML]{000000} {\cellcolor[HTML]{FDCAB5}} \color[HTML]{000000} 0.2 (0.1-0.5) & {\cellcolor[HTML]{FEFEFE}} \color[HTML]{000000} {\cellcolor[HTML]{FED9C9}} \color[HTML]{000000} 0.2 (0.0-0.5) \\
{\cellcolor[HTML]{FEFEFE}} \color[HTML]{000000} 89 & {\cellcolor[HTML]{FEFEFE}} \color[HTML]{000000} $\text{A}^\text{330}_\text{OP2}$-$\text{O}_\text{ph}$ & {\cellcolor[HTML]{FEFEFE}} \color[HTML]{000000} {\cellcolor{white}}  & {\cellcolor[HTML]{FEFEFE}} \color[HTML]{000000}  & {\cellcolor[HTML]{FEFEFE}} \color[HTML]{000000} {\cellcolor[HTML]{000000}} \color[HTML]{F1F1F1} {\cellcolor{white}}  & {\cellcolor[HTML]{FEFEFE}} \color[HTML]{000000} {\cellcolor[HTML]{000000}} \color[HTML]{F1F1F1} {\cellcolor{white}}  & {\cellcolor[HTML]{FEFEFE}} \color[HTML]{000000} {\cellcolor[HTML]{000000}} \color[HTML]{F1F1F1} {\cellcolor{white}}  & {\cellcolor[HTML]{FEFEFE}} \color[HTML]{000000} {\cellcolor[HTML]{000000}} \color[HTML]{F1F1F1} {\cellcolor{white}}  & {\cellcolor[HTML]{FEFEFE}} \color[HTML]{000000} {\cellcolor[HTML]{000000}} \color[HTML]{F1F1F1} {\cellcolor{white}}  & {\cellcolor[HTML]{FEFEFE}} \color[HTML]{000000} {\cellcolor[HTML]{FEE8DE}} \color[HTML]{000000} 0.1 (0.0-0.2) & {\cellcolor[HTML]{FEFEFE}} \color[HTML]{000000} {\cellcolor[HTML]{FEE7DB}} \color[HTML]{000000} 0.1 (0.0-0.3) & {\cellcolor[HTML]{FEFEFE}} \color[HTML]{000000} {\cellcolor[HTML]{FB7252}} \color[HTML]{F1F1F1} 0.5 (0.3-0.6) & {\cellcolor[HTML]{FEFEFE}} \color[HTML]{000000} {\cellcolor[HTML]{000000}} \color[HTML]{F1F1F1} {\cellcolor{white}}  & {\cellcolor[HTML]{FEFEFE}} \color[HTML]{000000} {\cellcolor[HTML]{FCB79C}} \color[HTML]{000000} 0.3 (0.1-0.5) & {\cellcolor[HTML]{FEFEFE}} \color[HTML]{000000} {\cellcolor[HTML]{FCC4AD}} \color[HTML]{000000} 0.2 (0.1-0.3) & {\cellcolor[HTML]{FEFEFE}} \color[HTML]{000000} {\cellcolor[HTML]{000000}} \color[HTML]{F1F1F1} {\cellcolor{white}} {\cellcolor[HTML]{AAAAAA}}  & {\cellcolor[HTML]{FEFEFE}} \color[HTML]{000000} {\cellcolor[HTML]{FFEDE5}} \color[HTML]{000000} 0.1 (0.0-0.2) & {\cellcolor[HTML]{FEFEFE}} \color[HTML]{000000} {\cellcolor[HTML]{FFF5F0}} \color[HTML]{000000} 0.0 (0.0-0.1) \\
{\cellcolor[HTML]{FEFEFE}} \color[HTML]{000000} 90 & {\cellcolor[HTML]{FEFEFE}} \color[HTML]{000000} $\text{G}^\text{282}_\text{OP2}$-$\text{O}_\text{ph}$ & {\cellcolor[HTML]{FEFEFE}} \color[HTML]{000000} {\cellcolor{white}}  & {\cellcolor[HTML]{FEFEFE}} \color[HTML]{000000}  & {\cellcolor[HTML]{FEFEFE}} \color[HTML]{000000} {\cellcolor[HTML]{000000}} \color[HTML]{F1F1F1} {\cellcolor{white}}  & {\cellcolor[HTML]{FEFEFE}} \color[HTML]{000000} {\cellcolor[HTML]{000000}} \color[HTML]{F1F1F1} {\cellcolor{white}}  & {\cellcolor[HTML]{FEFEFE}} \color[HTML]{000000} {\cellcolor[HTML]{000000}} \color[HTML]{F1F1F1} {\cellcolor{white}}  & {\cellcolor[HTML]{FEFEFE}} \color[HTML]{000000} {\cellcolor[HTML]{000000}} \color[HTML]{F1F1F1} {\cellcolor{white}}  & {\cellcolor[HTML]{FEFEFE}} \color[HTML]{000000} {\cellcolor[HTML]{FEE7DC}} \color[HTML]{000000} 0.1 (0.0-0.3) & {\cellcolor[HTML]{FEFEFE}} \color[HTML]{000000} {\cellcolor[HTML]{FCBBA1}} \color[HTML]{000000} 0.3 (0.1-0.4) & {\cellcolor[HTML]{FEFEFE}} \color[HTML]{000000} {\cellcolor[HTML]{FEE2D5}} \color[HTML]{000000} 0.1 (0.1-0.2) & {\cellcolor[HTML]{FEFEFE}} \color[HTML]{000000} {\cellcolor[HTML]{FEDACA}} \color[HTML]{000000} 0.1 (0.1-0.2) & {\cellcolor[HTML]{FEFEFE}} \color[HTML]{000000} {\cellcolor[HTML]{000000}} \color[HTML]{F1F1F1} {\cellcolor{white}}  & {\cellcolor[HTML]{FEFEFE}} \color[HTML]{000000} {\cellcolor[HTML]{FCC1A8}} \color[HTML]{000000} 0.2 (0.1-0.3) & {\cellcolor[HTML]{FEFEFE}} \color[HTML]{000000} {\cellcolor[HTML]{FCA183}} \color[HTML]{000000} 0.3 (0.2-0.5) & {\cellcolor[HTML]{FEFEFE}} \color[HTML]{000000} {\cellcolor[HTML]{000000}} \color[HTML]{F1F1F1} {\cellcolor{white}} {\cellcolor[HTML]{AAAAAA}}  & {\cellcolor[HTML]{FEFEFE}} \color[HTML]{000000} {\cellcolor[HTML]{000000}} \color[HTML]{F1F1F1} {\cellcolor{white}} {\cellcolor[HTML]{AAAAAA}}  & {\cellcolor[HTML]{FEFEFE}} \color[HTML]{000000} {\cellcolor[HTML]{000000}} \color[HTML]{F1F1F1} {\cellcolor{white}} {\cellcolor[HTML]{AAAAAA}}  \\
{\cellcolor[HTML]{FEFEFE}} \color[HTML]{000000} 91 & {\cellcolor[HTML]{FEFEFE}} \color[HTML]{000000} $\text{G}^\text{220}_\text{OP2}$-$\text{O}_\text{ph}$ & {\cellcolor[HTML]{FEFEFE}} \color[HTML]{000000} {\cellcolor{white}}  & {\cellcolor[HTML]{FEFEFE}} \color[HTML]{000000}  & {\cellcolor[HTML]{FEFEFE}} \color[HTML]{000000} {\cellcolor[HTML]{000000}} \color[HTML]{F1F1F1} {\cellcolor{white}}  & {\cellcolor[HTML]{FEFEFE}} \color[HTML]{000000} {\cellcolor[HTML]{000000}} \color[HTML]{F1F1F1} {\cellcolor{white}}  & {\cellcolor[HTML]{FEFEFE}} \color[HTML]{000000} {\cellcolor[HTML]{000000}} \color[HTML]{F1F1F1} {\cellcolor{white}}  & {\cellcolor[HTML]{FEFEFE}} \color[HTML]{000000} {\cellcolor[HTML]{000000}} \color[HTML]{F1F1F1} {\cellcolor{white}}  & {\cellcolor[HTML]{FEFEFE}} \color[HTML]{000000} {\cellcolor[HTML]{000000}} \color[HTML]{F1F1F1} {\cellcolor{white}}  & {\cellcolor[HTML]{FEFEFE}} \color[HTML]{000000} {\cellcolor[HTML]{FFEFE8}} \color[HTML]{000000} 0.0 (0.0-0.1) & {\cellcolor[HTML]{FEFEFE}} \color[HTML]{000000} {\cellcolor[HTML]{000000}} \color[HTML]{F1F1F1} {\cellcolor{white}}  & {\cellcolor[HTML]{FEFEFE}} \color[HTML]{000000} {\cellcolor[HTML]{FFF0E8}} \color[HTML]{000000} 0.0 (0.0-0.1) & {\cellcolor[HTML]{FEFEFE}} \color[HTML]{000000} {\cellcolor[HTML]{FFF3ED}} \color[HTML]{000000} 0.0 (0.0-0.0) & {\cellcolor[HTML]{FEFEFE}} \color[HTML]{000000} {\cellcolor[HTML]{FFF3ED}} \color[HTML]{000000} 0.0 (0.0-0.0) & {\cellcolor[HTML]{FEFEFE}} \color[HTML]{000000} {\cellcolor[HTML]{FFF0E8}} \color[HTML]{000000} 0.0 (0.0-0.1) & {\cellcolor[HTML]{FEFEFE}} \color[HTML]{000000} {\cellcolor[HTML]{FDCAB5}} \color[HTML]{000000} 0.2 (0.1-0.4) & {\cellcolor[HTML]{FEFEFE}} \color[HTML]{000000} {\cellcolor[HTML]{FC9B7C}} \color[HTML]{000000} 0.4 (0.1-0.8) & {\cellcolor[HTML]{FEFEFE}} \color[HTML]{000000} {\cellcolor[HTML]{FB7A5A}} \color[HTML]{F1F1F1} 0.5 (0.2-0.7) \\
{\cellcolor[HTML]{FEFEFE}} \color[HTML]{000000} 92 & {\cellcolor[HTML]{FEFEFE}} \color[HTML]{000000} $\text{U}^\text{348}_\text{OP1}$-$\text{O}_\text{ph}$ & {\cellcolor[HTML]{FEFEFE}} \color[HTML]{000000} {\cellcolor{white}}  & {\cellcolor[HTML]{FEFEFE}} \color[HTML]{000000}  & {\cellcolor[HTML]{FEFEFE}} \color[HTML]{000000} {\cellcolor[HTML]{000000}} \color[HTML]{F1F1F1} {\cellcolor{white}}  & {\cellcolor[HTML]{FEFEFE}} \color[HTML]{000000} {\cellcolor[HTML]{000000}} \color[HTML]{F1F1F1} {\cellcolor{white}}  & {\cellcolor[HTML]{FEFEFE}} \color[HTML]{000000} {\cellcolor[HTML]{000000}} \color[HTML]{F1F1F1} {\cellcolor{white}}  & {\cellcolor[HTML]{FEFEFE}} \color[HTML]{000000} {\cellcolor[HTML]{000000}} \color[HTML]{F1F1F1} {\cellcolor{white}}  & {\cellcolor[HTML]{FEFEFE}} \color[HTML]{000000} {\cellcolor[HTML]{000000}} \color[HTML]{F1F1F1} {\cellcolor{white}}  & {\cellcolor[HTML]{FEFEFE}} \color[HTML]{000000} {\cellcolor[HTML]{FEEAE0}} \color[HTML]{000000} 0.1 (0.0-0.2) & {\cellcolor[HTML]{FEFEFE}} \color[HTML]{000000} {\cellcolor[HTML]{FCC2AA}} \color[HTML]{000000} 0.2 (0.1-0.4) & {\cellcolor[HTML]{FEFEFE}} \color[HTML]{000000} {\cellcolor[HTML]{FCAD90}} \color[HTML]{000000} 0.3 (0.2-0.4) & {\cellcolor[HTML]{FEFEFE}} \color[HTML]{000000} {\cellcolor[HTML]{000000}} \color[HTML]{F1F1F1} {\cellcolor{white}}  & {\cellcolor[HTML]{FEFEFE}} \color[HTML]{000000} {\cellcolor[HTML]{FEDFD0}} \color[HTML]{000000} 0.1 (0.1-0.2) & {\cellcolor[HTML]{FEFEFE}} \color[HTML]{000000} {\cellcolor[HTML]{FDCBB6}} \color[HTML]{000000} 0.2 (0.1-0.3) & {\cellcolor[HTML]{FEFEFE}} \color[HTML]{000000} {\cellcolor[HTML]{000000}} \color[HTML]{F1F1F1} {\cellcolor{white}} {\cellcolor[HTML]{AAAAAA}}  & {\cellcolor[HTML]{FEFEFE}} \color[HTML]{000000} {\cellcolor[HTML]{FEE5D8}} \color[HTML]{000000} 0.1 (0.0-0.2) & {\cellcolor[HTML]{FEFEFE}} \color[HTML]{000000} {\cellcolor[HTML]{FEE5D8}} \color[HTML]{000000} 0.1 (0.0-0.4) \\
{\cellcolor[HTML]{FEFEFE}} \color[HTML]{000000} 93 & {\cellcolor[HTML]{FEFEFE}} \color[HTML]{000000} $\text{G}^\text{357}_\text{OP2}$-$\text{O}_\text{ph}$ & {\cellcolor[HTML]{FEFEFE}} \color[HTML]{000000} {\cellcolor{white}}  & {\cellcolor[HTML]{FEFEFE}} \color[HTML]{000000}  & {\cellcolor[HTML]{FEFEFE}} \color[HTML]{000000} {\cellcolor[HTML]{000000}} \color[HTML]{F1F1F1} {\cellcolor{white}}  & {\cellcolor[HTML]{FEFEFE}} \color[HTML]{000000} {\cellcolor[HTML]{FEEAE1}} \color[HTML]{000000} 0.1 & {\cellcolor[HTML]{FEFEFE}} \color[HTML]{000000} {\cellcolor[HTML]{000000}} \color[HTML]{F1F1F1} {\cellcolor{white}}  & {\cellcolor[HTML]{FEFEFE}} \color[HTML]{000000} {\cellcolor[HTML]{000000}} \color[HTML]{F1F1F1} {\cellcolor{white}}  & {\cellcolor[HTML]{FEFEFE}} \color[HTML]{000000} {\cellcolor[HTML]{000000}} \color[HTML]{F1F1F1} {\cellcolor{white}}  & {\cellcolor[HTML]{FEFEFE}} \color[HTML]{000000} {\cellcolor[HTML]{FEE7DC}} \color[HTML]{000000} 0.1 (0.0-0.2) & {\cellcolor[HTML]{FEFEFE}} \color[HTML]{000000} {\cellcolor[HTML]{FCB89E}} \color[HTML]{000000} 0.3 (0.1-0.5) & {\cellcolor[HTML]{FEFEFE}} \color[HTML]{000000} {\cellcolor[HTML]{FDCBB6}} \color[HTML]{000000} 0.2 (0.1-0.3) & {\cellcolor[HTML]{FEFEFE}} \color[HTML]{000000} {\cellcolor[HTML]{FDD5C4}} \color[HTML]{000000} 0.2 (0.0-0.4) & {\cellcolor[HTML]{FEFEFE}} \color[HTML]{000000} {\cellcolor[HTML]{FDD1BE}} \color[HTML]{000000} 0.2 (0.1-0.3) & {\cellcolor[HTML]{FEFEFE}} \color[HTML]{000000} {\cellcolor[HTML]{FDD5C4}} \color[HTML]{000000} 0.2 (0.1-0.2) & {\cellcolor[HTML]{FEFEFE}} \color[HTML]{000000} {\cellcolor[HTML]{FFF5F0}} \color[HTML]{000000} 0.0 (0.0-0.0) & {\cellcolor[HTML]{FEFEFE}} \color[HTML]{000000} {\cellcolor[HTML]{FFF5F0}} \color[HTML]{000000} 0.0 (0.0-0.0) & {\cellcolor[HTML]{FEFEFE}} \color[HTML]{000000} {\cellcolor[HTML]{FFF5F0}} \color[HTML]{000000} 0.0 (0.0-0.1) \\
{\cellcolor[HTML]{FEFEFE}} \color[HTML]{000000} 94 & {\cellcolor[HTML]{FEFEFE}} \color[HTML]{000000} $\text{G}^\text{126}_\text{OP1}$-$\text{O}_\text{ph}$ & {\cellcolor[HTML]{FEFEFE}} \color[HTML]{000000} {\cellcolor{white}}  & {\cellcolor[HTML]{FEFEFE}} \color[HTML]{000000}  & {\cellcolor[HTML]{FEFEFE}} \color[HTML]{000000} {\cellcolor[HTML]{000000}} \color[HTML]{F1F1F1} {\cellcolor{white}}  & {\cellcolor[HTML]{FEFEFE}} \color[HTML]{000000} {\cellcolor[HTML]{000000}} \color[HTML]{F1F1F1} {\cellcolor{white}}  & {\cellcolor[HTML]{FEFEFE}} \color[HTML]{000000} {\cellcolor[HTML]{000000}} \color[HTML]{F1F1F1} {\cellcolor{white}}  & {\cellcolor[HTML]{FEFEFE}} \color[HTML]{000000} {\cellcolor[HTML]{000000}} \color[HTML]{F1F1F1} {\cellcolor{white}}  & {\cellcolor[HTML]{FEFEFE}} \color[HTML]{000000} {\cellcolor[HTML]{000000}} \color[HTML]{F1F1F1} {\cellcolor{white}}  & {\cellcolor[HTML]{FEFEFE}} \color[HTML]{000000} {\cellcolor[HTML]{FEE8DD}} \color[HTML]{000000} 0.1 (0.0-0.1) & {\cellcolor[HTML]{FEFEFE}} \color[HTML]{000000} {\cellcolor[HTML]{FCC1A8}} \color[HTML]{000000} 0.2 (0.1-0.4) & {\cellcolor[HTML]{FEFEFE}} \color[HTML]{000000} {\cellcolor[HTML]{FCB89E}} \color[HTML]{000000} 0.3 (0.2-0.4) & {\cellcolor[HTML]{FEFEFE}} \color[HTML]{000000} {\cellcolor[HTML]{000000}} \color[HTML]{F1F1F1} {\cellcolor{white}}  & {\cellcolor[HTML]{FEFEFE}} \color[HTML]{000000} {\cellcolor[HTML]{FEE8DE}} \color[HTML]{000000} 0.1 (0.0-0.2) & {\cellcolor[HTML]{FEFEFE}} \color[HTML]{000000} {\cellcolor[HTML]{FED8C7}} \color[HTML]{000000} 0.2 (0.1-0.2) & {\cellcolor[HTML]{FEFEFE}} \color[HTML]{000000} {\cellcolor[HTML]{000000}} \color[HTML]{F1F1F1} {\cellcolor{white}} {\cellcolor[HTML]{AAAAAA}}  & {\cellcolor[HTML]{FEFEFE}} \color[HTML]{000000} {\cellcolor[HTML]{FCAB8F}} \color[HTML]{000000} 0.3 (0.2-0.5) & {\cellcolor[HTML]{FEFEFE}} \color[HTML]{000000} {\cellcolor[HTML]{FFF5F0}} \color[HTML]{000000} 0.0 (0.0-0.1) \\
{\cellcolor[HTML]{FEFEFE}} \color[HTML]{000000} 95 & {\cellcolor[HTML]{FEFEFE}} \color[HTML]{000000} $\text{A}^\text{94}_\text{OP2}$-$\text{O}_\text{ph}$ & {\cellcolor[HTML]{FEFEFE}} \color[HTML]{000000} {\cellcolor{white}}  & {\cellcolor[HTML]{FEFEFE}} \color[HTML]{000000}  & {\cellcolor[HTML]{FEFEFE}} \color[HTML]{000000} {\cellcolor[HTML]{000000}} \color[HTML]{F1F1F1} {\cellcolor{white}}  & {\cellcolor[HTML]{FEFEFE}} \color[HTML]{000000} {\cellcolor[HTML]{000000}} \color[HTML]{F1F1F1} {\cellcolor{white}}  & {\cellcolor[HTML]{FEFEFE}} \color[HTML]{000000} {\cellcolor[HTML]{000000}} \color[HTML]{F1F1F1} {\cellcolor{white}}  & {\cellcolor[HTML]{FEFEFE}} \color[HTML]{000000} {\cellcolor[HTML]{000000}} \color[HTML]{F1F1F1} {\cellcolor{white}}  & {\cellcolor[HTML]{FEFEFE}} \color[HTML]{000000} {\cellcolor[HTML]{000000}} \color[HTML]{F1F1F1} {\cellcolor{white}}  & {\cellcolor[HTML]{FEFEFE}} \color[HTML]{000000} {\cellcolor[HTML]{FEE4D8}} \color[HTML]{000000} 0.1 (0.1-0.2) & {\cellcolor[HTML]{FEFEFE}} \color[HTML]{000000} {\cellcolor[HTML]{FFEEE7}} \color[HTML]{000000} 0.0 (0.0-0.1) & {\cellcolor[HTML]{FEFEFE}} \color[HTML]{000000} {\cellcolor[HTML]{FCAE92}} \color[HTML]{000000} 0.3 (0.2-0.4) & {\cellcolor[HTML]{FEFEFE}} \color[HTML]{000000} {\cellcolor[HTML]{000000}} \color[HTML]{F1F1F1} {\cellcolor{white}}  & {\cellcolor[HTML]{FEFEFE}} \color[HTML]{000000} {\cellcolor[HTML]{FEDCCD}} \color[HTML]{000000} 0.1 (0.0-0.2) & {\cellcolor[HTML]{FEFEFE}} \color[HTML]{000000} {\cellcolor[HTML]{FDC6B0}} \color[HTML]{000000} 0.2 (0.1-0.3) & {\cellcolor[HTML]{FEFEFE}} \color[HTML]{000000} {\cellcolor[HTML]{000000}} \color[HTML]{F1F1F1} {\cellcolor{white}} {\cellcolor[HTML]{AAAAAA}}  & {\cellcolor[HTML]{FEFEFE}} \color[HTML]{000000} {\cellcolor[HTML]{FFF5F0}} \color[HTML]{000000} 0.0 (0.0-0.1) & {\cellcolor[HTML]{FEFEFE}} \color[HTML]{000000} {\cellcolor[HTML]{FCAB8F}} \color[HTML]{000000} 0.3 (0.1-0.5) \\
{\cellcolor[HTML]{FEFEFE}} \color[HTML]{000000} 96 & {\cellcolor[HTML]{FEFEFE}} \color[HTML]{000000} $\text{A}^\text{308}_\text{OP1}$-$\text{O}_\text{ph}$ & {\cellcolor[HTML]{FEFEFE}} \color[HTML]{000000} {\cellcolor{white}}  & {\cellcolor[HTML]{FEFEFE}} \color[HTML]{000000}  & {\cellcolor[HTML]{FEFEFE}} \color[HTML]{000000} {\cellcolor[HTML]{000000}} \color[HTML]{F1F1F1} {\cellcolor{white}}  & {\cellcolor[HTML]{FEFEFE}} \color[HTML]{000000} {\cellcolor[HTML]{000000}} \color[HTML]{F1F1F1} {\cellcolor{white}}  & {\cellcolor[HTML]{FEFEFE}} \color[HTML]{000000} {\cellcolor[HTML]{000000}} \color[HTML]{F1F1F1} {\cellcolor{white}}  & {\cellcolor[HTML]{FEFEFE}} \color[HTML]{000000} {\cellcolor[HTML]{000000}} \color[HTML]{F1F1F1} {\cellcolor{white}}  & {\cellcolor[HTML]{FEFEFE}} \color[HTML]{000000} {\cellcolor[HTML]{000000}} \color[HTML]{F1F1F1} {\cellcolor{white}}  & {\cellcolor[HTML]{FEFEFE}} \color[HTML]{000000} {\cellcolor[HTML]{000000}} \color[HTML]{F1F1F1} {\cellcolor{white}}  & {\cellcolor[HTML]{FEFEFE}} \color[HTML]{000000} {\cellcolor[HTML]{FFF3ED}} \color[HTML]{000000} 0.0 (0.0-0.1) & {\cellcolor[HTML]{FEFEFE}} \color[HTML]{000000} {\cellcolor[HTML]{FFECE4}} \color[HTML]{000000} 0.1 (0.0-0.1) & {\cellcolor[HTML]{FEFEFE}} \color[HTML]{000000} {\cellcolor[HTML]{000000}} \color[HTML]{F1F1F1} {\cellcolor{white}}  & {\cellcolor[HTML]{FEFEFE}} \color[HTML]{000000} {\cellcolor[HTML]{000000}} \color[HTML]{F1F1F1} {\cellcolor{white}}  & {\cellcolor[HTML]{FEFEFE}} \color[HTML]{000000} {\cellcolor[HTML]{FFF2EC}} \color[HTML]{000000} 0.0 (0.0-0.0) & {\cellcolor[HTML]{FEFEFE}} \color[HTML]{000000} {\cellcolor[HTML]{000000}} \color[HTML]{F1F1F1} {\cellcolor{white}} {\cellcolor[HTML]{AAAAAA}}  & {\cellcolor[HTML]{FEFEFE}} \color[HTML]{000000} {\cellcolor[HTML]{000000}} \color[HTML]{F1F1F1} {\cellcolor{white}} {\cellcolor[HTML]{AAAAAA}}  & {\cellcolor[HTML]{FEFEFE}} \color[HTML]{000000} {\cellcolor[HTML]{67000D}} \color[HTML]{F1F1F1} 1.0 (0.1-1.0) \\
{\cellcolor[HTML]{FEFEFE}} \color[HTML]{000000} 97 & {\cellcolor[HTML]{FEFEFE}} \color[HTML]{000000} $\text{A}^\text{286}_\text{OP2}$-$\text{O}_\text{ph}$ & {\cellcolor[HTML]{FEFEFE}} \color[HTML]{000000} {\cellcolor{white}}  & {\cellcolor[HTML]{FEFEFE}} \color[HTML]{000000}  & {\cellcolor[HTML]{FEFEFE}} \color[HTML]{000000} {\cellcolor[HTML]{000000}} \color[HTML]{F1F1F1} {\cellcolor{white}}  & {\cellcolor[HTML]{FEFEFE}} \color[HTML]{000000} {\cellcolor[HTML]{000000}} \color[HTML]{F1F1F1} {\cellcolor{white}}  & {\cellcolor[HTML]{FEFEFE}} \color[HTML]{000000} {\cellcolor[HTML]{000000}} \color[HTML]{F1F1F1} {\cellcolor{white}}  & {\cellcolor[HTML]{FEFEFE}} \color[HTML]{000000} {\cellcolor[HTML]{000000}} \color[HTML]{F1F1F1} {\cellcolor{white}}  & {\cellcolor[HTML]{FEFEFE}} \color[HTML]{000000} {\cellcolor[HTML]{FEE3D6}} \color[HTML]{000000} 0.1 (0.0-0.3) & {\cellcolor[HTML]{FEFEFE}} \color[HTML]{000000} {\cellcolor[HTML]{FED8C7}} \color[HTML]{000000} 0.2 (0.0-0.3) & {\cellcolor[HTML]{FEFEFE}} \color[HTML]{000000} {\cellcolor[HTML]{FDD7C6}} \color[HTML]{000000} 0.2 (0.0-0.3) & {\cellcolor[HTML]{FEFEFE}} \color[HTML]{000000} {\cellcolor[HTML]{FCA588}} \color[HTML]{000000} 0.3 (0.2-0.4) & {\cellcolor[HTML]{FEFEFE}} \color[HTML]{000000} {\cellcolor[HTML]{000000}} \color[HTML]{F1F1F1} {\cellcolor{white}}  & {\cellcolor[HTML]{FEFEFE}} \color[HTML]{000000} {\cellcolor[HTML]{FCC4AD}} \color[HTML]{000000} 0.2 (0.1-0.3) & {\cellcolor[HTML]{FEFEFE}} \color[HTML]{000000} {\cellcolor[HTML]{FEE3D6}} \color[HTML]{000000} 0.1 (0.1-0.2) & {\cellcolor[HTML]{FEFEFE}} \color[HTML]{000000} {\cellcolor[HTML]{000000}} \color[HTML]{F1F1F1} {\cellcolor{white}} {\cellcolor[HTML]{AAAAAA}}  & {\cellcolor[HTML]{FEFEFE}} \color[HTML]{000000} {\cellcolor[HTML]{000000}} \color[HTML]{F1F1F1} {\cellcolor{white}} {\cellcolor[HTML]{AAAAAA}}  & {\cellcolor[HTML]{FEFEFE}} \color[HTML]{000000} {\cellcolor[HTML]{000000}} \color[HTML]{F1F1F1} {\cellcolor{white}} {\cellcolor[HTML]{AAAAAA}}  \\
{\cellcolor[HTML]{FEFEFE}} \color[HTML]{000000} 98 & {\cellcolor[HTML]{FEFEFE}} \color[HTML]{000000} $\text{G}^\text{50}_\text{OP2}$-$\text{O}_\text{ph}$ & {\cellcolor[HTML]{FEFEFE}} \color[HTML]{000000} {\cellcolor{white}}  & {\cellcolor[HTML]{FEFEFE}} \color[HTML]{000000}  & {\cellcolor[HTML]{FEFEFE}} \color[HTML]{000000} {\cellcolor[HTML]{000000}} \color[HTML]{F1F1F1} {\cellcolor{white}}  & {\cellcolor[HTML]{FEFEFE}} \color[HTML]{000000} {\cellcolor[HTML]{000000}} \color[HTML]{F1F1F1} {\cellcolor{white}}  & {\cellcolor[HTML]{FEFEFE}} \color[HTML]{000000} {\cellcolor[HTML]{000000}} \color[HTML]{F1F1F1} {\cellcolor{white}}  & {\cellcolor[HTML]{FEFEFE}} \color[HTML]{000000} {\cellcolor[HTML]{000000}} \color[HTML]{F1F1F1} {\cellcolor{white}}  & {\cellcolor[HTML]{FEFEFE}} \color[HTML]{000000} {\cellcolor[HTML]{000000}} \color[HTML]{F1F1F1} {\cellcolor{white}}  & {\cellcolor[HTML]{FEFEFE}} \color[HTML]{000000} {\cellcolor[HTML]{FFF0E8}} \color[HTML]{000000} 0.0 (0.0-0.1) & {\cellcolor[HTML]{FEFEFE}} \color[HTML]{000000} {\cellcolor[HTML]{FEE8DD}} \color[HTML]{000000} 0.1 (0.0-0.2) & {\cellcolor[HTML]{FEFEFE}} \color[HTML]{000000} {\cellcolor[HTML]{FEE9DF}} \color[HTML]{000000} 0.1 (0.0-0.1) & {\cellcolor[HTML]{FEFEFE}} \color[HTML]{000000} {\cellcolor[HTML]{000000}} \color[HTML]{F1F1F1} {\cellcolor{white}}  & {\cellcolor[HTML]{FEFEFE}} \color[HTML]{000000} {\cellcolor[HTML]{FEE3D7}} \color[HTML]{000000} 0.1 (0.0-0.2) & {\cellcolor[HTML]{FEFEFE}} \color[HTML]{000000} {\cellcolor[HTML]{FEE8DD}} \color[HTML]{000000} 0.1 (0.0-0.1) & {\cellcolor[HTML]{FEFEFE}} \color[HTML]{000000} {\cellcolor[HTML]{000000}} \color[HTML]{F1F1F1} {\cellcolor{white}} {\cellcolor[HTML]{AAAAAA}}  & {\cellcolor[HTML]{FEFEFE}} \color[HTML]{000000} {\cellcolor[HTML]{FC9B7C}} \color[HTML]{000000} 0.4 (0.2-0.5) & {\cellcolor[HTML]{FEFEFE}} \color[HTML]{000000} {\cellcolor[HTML]{FC9B7C}} \color[HTML]{000000} 0.4 (0.2-0.5) \\
{\cellcolor[HTML]{FEFEFE}} \color[HTML]{000000} 99 & {\cellcolor[HTML]{FEFEFE}} \color[HTML]{000000} $\text{U}^\text{106}_\text{O4}$-$\text{O}_\text{b}$ & {\cellcolor[HTML]{FEFEFE}} \color[HTML]{000000} {\cellcolor{white}}  & {\cellcolor[HTML]{FEFEFE}} \color[HTML]{000000}  & {\cellcolor[HTML]{FEFEFE}} \color[HTML]{000000} {\cellcolor[HTML]{FEEAE1}} \color[HTML]{000000} 0.1 & {\cellcolor[HTML]{FEFEFE}} \color[HTML]{000000} {\cellcolor[HTML]{FEE0D2}} \color[HTML]{000000} 0.1 & {\cellcolor[HTML]{FEFEFE}} \color[HTML]{000000} {\cellcolor[HTML]{FEE0D2}} \color[HTML]{000000} 0.1 & {\cellcolor[HTML]{FEFEFE}} \color[HTML]{000000} {\cellcolor[HTML]{000000}} \color[HTML]{F1F1F1} {\cellcolor{white}}  & {\cellcolor[HTML]{FEFEFE}} \color[HTML]{000000} {\cellcolor[HTML]{FFF5F0}} \color[HTML]{000000} 0.0 (0.0-0.0) & {\cellcolor[HTML]{FEFEFE}} \color[HTML]{000000} {\cellcolor[HTML]{000000}} \color[HTML]{F1F1F1} {\cellcolor{white}}  & {\cellcolor[HTML]{FEFEFE}} \color[HTML]{000000} {\cellcolor[HTML]{000000}} \color[HTML]{F1F1F1} {\cellcolor{white}}  & {\cellcolor[HTML]{FEFEFE}} \color[HTML]{000000} {\cellcolor[HTML]{FFF5F0}} \color[HTML]{000000} 0.0 (0.0-0.0) & {\cellcolor[HTML]{FEFEFE}} \color[HTML]{000000} {\cellcolor[HTML]{FFF5F0}} \color[HTML]{000000} 0.0 (0.0-0.0) & {\cellcolor[HTML]{FEFEFE}} \color[HTML]{000000} {\cellcolor[HTML]{FFF5F0}} \color[HTML]{000000} 0.0 (0.0-0.0) & {\cellcolor[HTML]{FEFEFE}} \color[HTML]{000000} {\cellcolor[HTML]{FFF5F0}} \color[HTML]{000000} 0.0 (0.0-0.0) & {\cellcolor[HTML]{FEFEFE}} \color[HTML]{000000} {\cellcolor[HTML]{FCBBA1}} \color[HTML]{000000} 0.2 (0.1-0.4) & {\cellcolor[HTML]{FEFEFE}} \color[HTML]{000000} {\cellcolor[HTML]{000000}} \color[HTML]{F1F1F1} {\cellcolor{white}} {\cellcolor[HTML]{AAAAAA}}  & {\cellcolor[HTML]{FEFEFE}} \color[HTML]{000000} {\cellcolor[HTML]{FB694A}} \color[HTML]{F1F1F1} 0.5 (0.1-0.6) \\
{\cellcolor[HTML]{FEFEFE}} \color[HTML]{000000} 100 & {\cellcolor[HTML]{FEFEFE}} \color[HTML]{000000} $\text{G}^\text{328}_\text{OP1}$-$\text{O}_\text{ph}$ & {\cellcolor[HTML]{FEFEFE}} \color[HTML]{000000} {\cellcolor{white}}  & {\cellcolor[HTML]{FEFEFE}} \color[HTML]{000000}  & {\cellcolor[HTML]{FEFEFE}} \color[HTML]{000000} {\cellcolor[HTML]{000000}} \color[HTML]{F1F1F1} {\cellcolor{white}}  & {\cellcolor[HTML]{FEFEFE}} \color[HTML]{000000} {\cellcolor[HTML]{000000}} \color[HTML]{F1F1F1} {\cellcolor{white}}  & {\cellcolor[HTML]{FEFEFE}} \color[HTML]{000000} {\cellcolor[HTML]{000000}} \color[HTML]{F1F1F1} {\cellcolor{white}}  & {\cellcolor[HTML]{FEFEFE}} \color[HTML]{000000} {\cellcolor[HTML]{000000}} \color[HTML]{F1F1F1} {\cellcolor{white}}  & {\cellcolor[HTML]{FEFEFE}} \color[HTML]{000000} {\cellcolor[HTML]{000000}} \color[HTML]{F1F1F1} {\cellcolor{white}}  & {\cellcolor[HTML]{FEFEFE}} \color[HTML]{000000} {\cellcolor[HTML]{FFEEE7}} \color[HTML]{000000} 0.0 (0.0-0.1) & {\cellcolor[HTML]{FEFEFE}} \color[HTML]{000000} {\cellcolor[HTML]{FDD0BC}} \color[HTML]{000000} 0.2 (0.1-0.3) & {\cellcolor[HTML]{FEFEFE}} \color[HTML]{000000} {\cellcolor[HTML]{FFF0E8}} \color[HTML]{000000} 0.0 (0.0-0.1) & {\cellcolor[HTML]{FEFEFE}} \color[HTML]{000000} {\cellcolor[HTML]{FFEDE5}} \color[HTML]{000000} 0.0 (0.0-0.1) & {\cellcolor[HTML]{FEFEFE}} \color[HTML]{000000} {\cellcolor[HTML]{FFEEE7}} \color[HTML]{000000} 0.0 (0.0-0.1) & {\cellcolor[HTML]{FEFEFE}} \color[HTML]{000000} {\cellcolor[HTML]{FEE9DF}} \color[HTML]{000000} 0.1 (0.0-0.1) & {\cellcolor[HTML]{FEFEFE}} \color[HTML]{000000} {\cellcolor[HTML]{FC9B7C}} \color[HTML]{000000} 0.4 (0.3-0.5) & {\cellcolor[HTML]{FEFEFE}} \color[HTML]{000000} {\cellcolor[HTML]{FEE5D8}} \color[HTML]{000000} 0.1 (0.0-0.3) & {\cellcolor[HTML]{FEFEFE}} \color[HTML]{000000} {\cellcolor[HTML]{FDCAB5}} \color[HTML]{000000} 0.2 (0.0-0.5) \\
{\cellcolor[HTML]{FEFEFE}} \color[HTML]{000000} 101 & {\cellcolor[HTML]{FEFEFE}} \color[HTML]{000000} $\text{G}^\text{191}_\text{OP2}$-$\text{O}_\text{ph}$ & {\cellcolor[HTML]{FEFEFE}} \color[HTML]{000000} {\cellcolor{white}}  & {\cellcolor[HTML]{FEFEFE}} \color[HTML]{000000}  & {\cellcolor[HTML]{FEFEFE}} \color[HTML]{000000} {\cellcolor[HTML]{000000}} \color[HTML]{F1F1F1} {\cellcolor{white}}  & {\cellcolor[HTML]{FEFEFE}} \color[HTML]{000000} {\cellcolor[HTML]{000000}} \color[HTML]{F1F1F1} {\cellcolor{white}}  & {\cellcolor[HTML]{FEFEFE}} \color[HTML]{000000} {\cellcolor[HTML]{000000}} \color[HTML]{F1F1F1} {\cellcolor{white}}  & {\cellcolor[HTML]{FEFEFE}} \color[HTML]{000000} {\cellcolor[HTML]{000000}} \color[HTML]{F1F1F1} {\cellcolor{white}}  & {\cellcolor[HTML]{FEFEFE}} \color[HTML]{000000} {\cellcolor[HTML]{000000}} \color[HTML]{F1F1F1} {\cellcolor{white}}  & {\cellcolor[HTML]{FEFEFE}} \color[HTML]{000000} {\cellcolor[HTML]{FDC7B2}} \color[HTML]{000000} 0.2 (0.1-0.4) & {\cellcolor[HTML]{FEFEFE}} \color[HTML]{000000} {\cellcolor[HTML]{FEDCCD}} \color[HTML]{000000} 0.1 (0.0-0.3) & {\cellcolor[HTML]{FEFEFE}} \color[HTML]{000000} {\cellcolor[HTML]{FCBBA1}} \color[HTML]{000000} 0.3 (0.1-0.4) & {\cellcolor[HTML]{FEFEFE}} \color[HTML]{000000} {\cellcolor[HTML]{000000}} \color[HTML]{F1F1F1} {\cellcolor{white}}  & {\cellcolor[HTML]{FEFEFE}} \color[HTML]{000000} {\cellcolor[HTML]{FEE4D8}} \color[HTML]{000000} 0.1 (0.0-0.2) & {\cellcolor[HTML]{FEFEFE}} \color[HTML]{000000} {\cellcolor[HTML]{FCB89E}} \color[HTML]{000000} 0.3 (0.2-0.4) & {\cellcolor[HTML]{FEFEFE}} \color[HTML]{000000} {\cellcolor[HTML]{000000}} \color[HTML]{F1F1F1} {\cellcolor{white}} {\cellcolor[HTML]{AAAAAA}}  & {\cellcolor[HTML]{FEFEFE}} \color[HTML]{000000} {\cellcolor[HTML]{FFF5F0}} \color[HTML]{000000} 0.0 (0.0-0.2) & {\cellcolor[HTML]{FEFEFE}} \color[HTML]{000000} {\cellcolor[HTML]{FEE5D8}} \color[HTML]{000000} 0.1 (0.0-0.2) \\
{\cellcolor[HTML]{FEFEFE}} \color[HTML]{000000} 102 & {\cellcolor[HTML]{FEFEFE}} \color[HTML]{000000} $\text{C}^\text{45}_\text{OP1}$-$\text{O}_\text{ph}$ & {\cellcolor[HTML]{FEFEFE}} \color[HTML]{000000} {\cellcolor{white}}  & {\cellcolor[HTML]{FEFEFE}} \color[HTML]{000000}  & {\cellcolor[HTML]{FEFEFE}} \color[HTML]{000000} {\cellcolor[HTML]{000000}} \color[HTML]{F1F1F1} {\cellcolor{white}}  & {\cellcolor[HTML]{FEFEFE}} \color[HTML]{000000} {\cellcolor[HTML]{FEEAE1}} \color[HTML]{000000} 0.1 & {\cellcolor[HTML]{FEFEFE}} \color[HTML]{000000} {\cellcolor[HTML]{000000}} \color[HTML]{F1F1F1} {\cellcolor{white}}  & {\cellcolor[HTML]{FEFEFE}} \color[HTML]{000000} {\cellcolor[HTML]{000000}} \color[HTML]{F1F1F1} {\cellcolor{white}}  & {\cellcolor[HTML]{FEFEFE}} \color[HTML]{000000} {\cellcolor[HTML]{000000}} \color[HTML]{F1F1F1} {\cellcolor{white}}  & {\cellcolor[HTML]{FEFEFE}} \color[HTML]{000000} {\cellcolor[HTML]{FFF3ED}} \color[HTML]{000000} 0.0 (0.0-0.0) & {\cellcolor[HTML]{FEFEFE}} \color[HTML]{000000} {\cellcolor[HTML]{FFECE4}} \color[HTML]{000000} 0.1 (0.0-0.1) & {\cellcolor[HTML]{FEFEFE}} \color[HTML]{000000} {\cellcolor[HTML]{FFEBE2}} \color[HTML]{000000} 0.1 (0.0-0.1) & {\cellcolor[HTML]{FEFEFE}} \color[HTML]{000000} {\cellcolor[HTML]{FFF3ED}} \color[HTML]{000000} 0.0 (0.0-0.0) & {\cellcolor[HTML]{FEFEFE}} \color[HTML]{000000} {\cellcolor[HTML]{FFF3ED}} \color[HTML]{000000} 0.0 (0.0-0.0) & {\cellcolor[HTML]{FEFEFE}} \color[HTML]{000000} {\cellcolor[HTML]{FFF0E8}} \color[HTML]{000000} 0.0 (0.0-0.1) & {\cellcolor[HTML]{FEFEFE}} \color[HTML]{000000} {\cellcolor[HTML]{FB694A}} \color[HTML]{F1F1F1} 0.5 (0.4-0.6) & {\cellcolor[HTML]{FEFEFE}} \color[HTML]{000000} {\cellcolor[HTML]{FCAB8F}} \color[HTML]{000000} 0.3 (0.0-0.6) & {\cellcolor[HTML]{FEFEFE}} \color[HTML]{000000} {\cellcolor[HTML]{FFF5F0}} \color[HTML]{000000} 0.0 (0.0-0.2) \\
{\cellcolor[HTML]{FEFEFE}} \color[HTML]{000000} 103 & {\cellcolor[HTML]{FEFEFE}} \color[HTML]{000000} $\text{U}^\text{310}_\text{OP1}$-$\text{O}_\text{ph}$ & {\cellcolor[HTML]{FEFEFE}} \color[HTML]{000000} {\cellcolor{white}}  & {\cellcolor[HTML]{FEFEFE}} \color[HTML]{000000}  & {\cellcolor[HTML]{FEFEFE}} \color[HTML]{000000} {\cellcolor[HTML]{000000}} \color[HTML]{F1F1F1} {\cellcolor{white}}  & {\cellcolor[HTML]{FEFEFE}} \color[HTML]{000000} {\cellcolor[HTML]{000000}} \color[HTML]{F1F1F1} {\cellcolor{white}}  & {\cellcolor[HTML]{FEFEFE}} \color[HTML]{000000} {\cellcolor[HTML]{000000}} \color[HTML]{F1F1F1} {\cellcolor{white}}  & {\cellcolor[HTML]{FEFEFE}} \color[HTML]{000000} {\cellcolor[HTML]{000000}} \color[HTML]{F1F1F1} {\cellcolor{white}}  & {\cellcolor[HTML]{FEFEFE}} \color[HTML]{000000} {\cellcolor[HTML]{FEE5D9}} \color[HTML]{000000} 0.1 (0.0-0.3) & {\cellcolor[HTML]{FEFEFE}} \color[HTML]{000000} {\cellcolor[HTML]{FEE8DD}} \color[HTML]{000000} 0.1 (0.0-0.2) & {\cellcolor[HTML]{FEFEFE}} \color[HTML]{000000} {\cellcolor[HTML]{FFF4EF}} \color[HTML]{000000} 0.0 (0.0-0.0) & {\cellcolor[HTML]{FEFEFE}} \color[HTML]{000000} {\cellcolor[HTML]{FEDBCC}} \color[HTML]{000000} 0.1 (0.1-0.2) & {\cellcolor[HTML]{FEFEFE}} \color[HTML]{000000} {\cellcolor[HTML]{000000}} \color[HTML]{F1F1F1} {\cellcolor{white}}  & {\cellcolor[HTML]{FEFEFE}} \color[HTML]{000000} {\cellcolor[HTML]{FFF0E8}} \color[HTML]{000000} 0.0 (0.0-0.1) & {\cellcolor[HTML]{FEFEFE}} \color[HTML]{000000} {\cellcolor[HTML]{FEE1D4}} \color[HTML]{000000} 0.1 (0.1-0.2) & {\cellcolor[HTML]{FEFEFE}} \color[HTML]{000000} {\cellcolor[HTML]{F6583E}} \color[HTML]{F1F1F1} 0.6 (0.4-0.7) & {\cellcolor[HTML]{FEFEFE}} \color[HTML]{000000} {\cellcolor[HTML]{000000}} \color[HTML]{F1F1F1} {\cellcolor{white}} {\cellcolor[HTML]{AAAAAA}}  & {\cellcolor[HTML]{FEFEFE}} \color[HTML]{000000} {\cellcolor[HTML]{000000}} \color[HTML]{F1F1F1} {\cellcolor{white}} {\cellcolor[HTML]{AAAAAA}}  \\
{\cellcolor[HTML]{FEFEFE}} \color[HTML]{000000} 104 & {\cellcolor[HTML]{FEFEFE}} \color[HTML]{000000} $\text{G}^\text{327}_\text{OP2}$-$\text{O}_\text{ph}$ & {\cellcolor[HTML]{FEFEFE}} \color[HTML]{000000} {\cellcolor{white}}  & {\cellcolor[HTML]{FEFEFE}} \color[HTML]{000000}  & {\cellcolor[HTML]{FEFEFE}} \color[HTML]{000000} {\cellcolor[HTML]{000000}} \color[HTML]{F1F1F1} {\cellcolor{white}}  & {\cellcolor[HTML]{FEFEFE}} \color[HTML]{000000} {\cellcolor[HTML]{FEEAE1}} \color[HTML]{000000} 0.1 & {\cellcolor[HTML]{FEFEFE}} \color[HTML]{000000} {\cellcolor[HTML]{000000}} \color[HTML]{F1F1F1} {\cellcolor{white}}  & {\cellcolor[HTML]{FEFEFE}} \color[HTML]{000000} {\cellcolor[HTML]{000000}} \color[HTML]{F1F1F1} {\cellcolor{white}}  & {\cellcolor[HTML]{FEFEFE}} \color[HTML]{000000} {\cellcolor[HTML]{000000}} \color[HTML]{F1F1F1} {\cellcolor{white}}  & {\cellcolor[HTML]{FEFEFE}} \color[HTML]{000000} {\cellcolor[HTML]{FEE9DF}} \color[HTML]{000000} 0.1 (0.0-0.2) & {\cellcolor[HTML]{FEFEFE}} \color[HTML]{000000} {\cellcolor[HTML]{FEE5D9}} \color[HTML]{000000} 0.1 (0.0-0.2) & {\cellcolor[HTML]{FEFEFE}} \color[HTML]{000000} {\cellcolor[HTML]{FDCAB5}} \color[HTML]{000000} 0.2 (0.1-0.3) & {\cellcolor[HTML]{FEFEFE}} \color[HTML]{000000} {\cellcolor[HTML]{FEE0D2}} \color[HTML]{000000} 0.1 (0.0-0.4) & {\cellcolor[HTML]{FEFEFE}} \color[HTML]{000000} {\cellcolor[HTML]{FEDFD0}} \color[HTML]{000000} 0.1 (0.0-0.4) & {\cellcolor[HTML]{FEFEFE}} \color[HTML]{000000} {\cellcolor[HTML]{FEDBCC}} \color[HTML]{000000} 0.1 (0.1-0.2) & {\cellcolor[HTML]{FEFEFE}} \color[HTML]{000000} {\cellcolor[HTML]{FFF5F0}} \color[HTML]{000000} 0.0 (0.0-0.0) & {\cellcolor[HTML]{FEFEFE}} \color[HTML]{000000} {\cellcolor[HTML]{FFEDE5}} \color[HTML]{000000} 0.1 (0.0-0.4) & {\cellcolor[HTML]{FEFEFE}} \color[HTML]{000000} {\cellcolor[HTML]{FED9C9}} \color[HTML]{000000} 0.2 (0.0-0.6) \\
{\cellcolor[HTML]{FEFEFE}} \color[HTML]{000000} 105 & {\cellcolor[HTML]{FEFEFE}} \color[HTML]{000000} $\text{G}^\text{279}_\text{OP1}$-$\text{O}_\text{ph}$ & {\cellcolor[HTML]{FEFEFE}} \color[HTML]{000000} {\cellcolor{white}}  & {\cellcolor[HTML]{FEFEFE}} \color[HTML]{000000}  & {\cellcolor[HTML]{FEFEFE}} \color[HTML]{000000} {\cellcolor[HTML]{FEEAE1}} \color[HTML]{000000} 0.1 & {\cellcolor[HTML]{FEFEFE}} \color[HTML]{000000} {\cellcolor[HTML]{000000}} \color[HTML]{F1F1F1} {\cellcolor{white}}  & {\cellcolor[HTML]{FEFEFE}} \color[HTML]{000000} {\cellcolor[HTML]{000000}} \color[HTML]{F1F1F1} {\cellcolor{white}}  & {\cellcolor[HTML]{FEFEFE}} \color[HTML]{000000} {\cellcolor[HTML]{000000}} \color[HTML]{F1F1F1} {\cellcolor{white}}  & {\cellcolor[HTML]{FEFEFE}} \color[HTML]{000000} {\cellcolor[HTML]{000000}} \color[HTML]{F1F1F1} {\cellcolor{white}}  & {\cellcolor[HTML]{FEFEFE}} \color[HTML]{000000} {\cellcolor[HTML]{FFF2EB}} \color[HTML]{000000} 0.0 (0.0-0.1) & {\cellcolor[HTML]{FEFEFE}} \color[HTML]{000000} {\cellcolor[HTML]{FEE7DC}} \color[HTML]{000000} 0.1 (0.0-0.1) & {\cellcolor[HTML]{FEFEFE}} \color[HTML]{000000} {\cellcolor[HTML]{FFF2EC}} \color[HTML]{000000} 0.0 (0.0-0.0) & {\cellcolor[HTML]{FEFEFE}} \color[HTML]{000000} {\cellcolor[HTML]{000000}} \color[HTML]{F1F1F1} {\cellcolor{white}}  & {\cellcolor[HTML]{FEFEFE}} \color[HTML]{000000} {\cellcolor[HTML]{000000}} \color[HTML]{F1F1F1} {\cellcolor{white}}  & {\cellcolor[HTML]{FEFEFE}} \color[HTML]{000000} {\cellcolor[HTML]{FFEEE7}} \color[HTML]{000000} 0.0 (0.0-0.1) & {\cellcolor[HTML]{FEFEFE}} \color[HTML]{000000} {\cellcolor[HTML]{000000}} \color[HTML]{F1F1F1} {\cellcolor{white}} {\cellcolor[HTML]{AAAAAA}}  & {\cellcolor[HTML]{FEFEFE}} \color[HTML]{000000} {\cellcolor[HTML]{BC141A}} \color[HTML]{F1F1F1} 0.8 (0.0-0.9) & {\cellcolor[HTML]{FEFEFE}} \color[HTML]{000000} {\cellcolor[HTML]{000000}} \color[HTML]{F1F1F1} {\cellcolor{white}} {\cellcolor[HTML]{AAAAAA}}  \\
{\cellcolor[HTML]{FEFEFE}} \color[HTML]{000000} 106 & {\cellcolor[HTML]{FEFEFE}} \color[HTML]{000000} $\text{G}^\text{344}_\text{OP2}$-$\text{O}_\text{ph}$ & {\cellcolor[HTML]{FEFEFE}} \color[HTML]{000000} {\cellcolor{white}}  & {\cellcolor[HTML]{FEFEFE}} \color[HTML]{000000}  & {\cellcolor[HTML]{FEFEFE}} \color[HTML]{000000} {\cellcolor[HTML]{000000}} \color[HTML]{F1F1F1} {\cellcolor{white}}  & {\cellcolor[HTML]{FEFEFE}} \color[HTML]{000000} {\cellcolor[HTML]{000000}} \color[HTML]{F1F1F1} {\cellcolor{white}}  & {\cellcolor[HTML]{FEFEFE}} \color[HTML]{000000} {\cellcolor[HTML]{000000}} \color[HTML]{F1F1F1} {\cellcolor{white}}  & {\cellcolor[HTML]{FEFEFE}} \color[HTML]{000000} {\cellcolor[HTML]{FEE0D2}} \color[HTML]{000000} 0.1 & {\cellcolor[HTML]{FEFEFE}} \color[HTML]{000000} {\cellcolor[HTML]{000000}} \color[HTML]{F1F1F1} {\cellcolor{white}}  & {\cellcolor[HTML]{FEFEFE}} \color[HTML]{000000} {\cellcolor[HTML]{000000}} \color[HTML]{F1F1F1} {\cellcolor{white}}  & {\cellcolor[HTML]{FEFEFE}} \color[HTML]{000000} {\cellcolor[HTML]{FFEFE8}} \color[HTML]{000000} 0.0 (0.0-0.1) & {\cellcolor[HTML]{FEFEFE}} \color[HTML]{000000} {\cellcolor[HTML]{FCAD90}} \color[HTML]{000000} 0.3 (0.1-0.5) & {\cellcolor[HTML]{FEFEFE}} \color[HTML]{000000} {\cellcolor[HTML]{000000}} \color[HTML]{F1F1F1} {\cellcolor{white}}  & {\cellcolor[HTML]{FEFEFE}} \color[HTML]{000000} {\cellcolor[HTML]{FDD3C1}} \color[HTML]{000000} 0.2 (0.0-0.3) & {\cellcolor[HTML]{FEFEFE}} \color[HTML]{000000} {\cellcolor[HTML]{FED9C9}} \color[HTML]{000000} 0.2 (0.1-0.2) & {\cellcolor[HTML]{FEFEFE}} \color[HTML]{000000} {\cellcolor[HTML]{000000}} \color[HTML]{F1F1F1} {\cellcolor{white}} {\cellcolor[HTML]{AAAAAA}}  & {\cellcolor[HTML]{FEFEFE}} \color[HTML]{000000} {\cellcolor[HTML]{FFF5F0}} \color[HTML]{000000} 0.0 (0.0-0.1) & {\cellcolor[HTML]{FEFEFE}} \color[HTML]{000000} {\cellcolor[HTML]{FCBBA1}} \color[HTML]{000000} 0.2 (0.0-0.6) \\
{\cellcolor[HTML]{FEFEFE}} \color[HTML]{000000} 107 & {\cellcolor[HTML]{FEFEFE}} \color[HTML]{000000} $\text{A}^\text{31}_\text{OP2}$-$\text{O}_\text{ph}$ & {\cellcolor[HTML]{FEFEFE}} \color[HTML]{000000} {\cellcolor{white}}  & {\cellcolor[HTML]{FEFEFE}} \color[HTML]{000000}  & {\cellcolor[HTML]{FEFEFE}} \color[HTML]{000000} {\cellcolor[HTML]{000000}} \color[HTML]{F1F1F1} {\cellcolor{white}}  & {\cellcolor[HTML]{FEFEFE}} \color[HTML]{000000} {\cellcolor[HTML]{000000}} \color[HTML]{F1F1F1} {\cellcolor{white}}  & {\cellcolor[HTML]{FEFEFE}} \color[HTML]{000000} {\cellcolor[HTML]{000000}} \color[HTML]{F1F1F1} {\cellcolor{white}}  & {\cellcolor[HTML]{FEFEFE}} \color[HTML]{000000} {\cellcolor[HTML]{000000}} \color[HTML]{F1F1F1} {\cellcolor{white}}  & {\cellcolor[HTML]{FEFEFE}} \color[HTML]{000000} {\cellcolor[HTML]{000000}} \color[HTML]{F1F1F1} {\cellcolor{white}}  & {\cellcolor[HTML]{FEFEFE}} \color[HTML]{000000} {\cellcolor[HTML]{FEE3D6}} \color[HTML]{000000} 0.1 (0.1-0.2) & {\cellcolor[HTML]{FEFEFE}} \color[HTML]{000000} {\cellcolor[HTML]{FCA183}} \color[HTML]{000000} 0.3 (0.2-0.6) & {\cellcolor[HTML]{FEFEFE}} \color[HTML]{000000} {\cellcolor[HTML]{FEE9DF}} \color[HTML]{000000} 0.1 (0.0-0.1) & {\cellcolor[HTML]{FEFEFE}} \color[HTML]{000000} {\cellcolor[HTML]{FEE7DB}} \color[HTML]{000000} 0.1 (0.0-0.3) & {\cellcolor[HTML]{FEFEFE}} \color[HTML]{000000} {\cellcolor[HTML]{FEE0D2}} \color[HTML]{000000} 0.1 (0.0-0.3) & {\cellcolor[HTML]{FEFEFE}} \color[HTML]{000000} {\cellcolor[HTML]{FCAE92}} \color[HTML]{000000} 0.3 (0.2-0.4) & {\cellcolor[HTML]{FEFEFE}} \color[HTML]{000000} {\cellcolor[HTML]{000000}} \color[HTML]{F1F1F1} {\cellcolor{white}} {\cellcolor[HTML]{AAAAAA}}  & {\cellcolor[HTML]{FEFEFE}} \color[HTML]{000000} {\cellcolor[HTML]{000000}} \color[HTML]{F1F1F1} {\cellcolor{white}} {\cellcolor[HTML]{AAAAAA}}  & {\cellcolor[HTML]{FEFEFE}} \color[HTML]{000000} {\cellcolor[HTML]{000000}} \color[HTML]{F1F1F1} {\cellcolor{white}} {\cellcolor[HTML]{AAAAAA}}  \\
{\cellcolor[HTML]{FEFEFE}} \color[HTML]{000000} 108 & {\cellcolor[HTML]{FEFEFE}} \color[HTML]{000000} $\text{A}^\text{284}_\text{OP2}$-$\text{O}_\text{ph}$ & {\cellcolor[HTML]{FEFEFE}} \color[HTML]{000000} {\cellcolor{white}}  & {\cellcolor[HTML]{FEFEFE}} \color[HTML]{000000}  & {\cellcolor[HTML]{FEFEFE}} \color[HTML]{000000} {\cellcolor[HTML]{000000}} \color[HTML]{F1F1F1} {\cellcolor{white}}  & {\cellcolor[HTML]{FEFEFE}} \color[HTML]{000000} {\cellcolor[HTML]{000000}} \color[HTML]{F1F1F1} {\cellcolor{white}}  & {\cellcolor[HTML]{FEFEFE}} \color[HTML]{000000} {\cellcolor[HTML]{000000}} \color[HTML]{F1F1F1} {\cellcolor{white}}  & {\cellcolor[HTML]{FEFEFE}} \color[HTML]{000000} {\cellcolor[HTML]{000000}} \color[HTML]{F1F1F1} {\cellcolor{white}}  & {\cellcolor[HTML]{FEFEFE}} \color[HTML]{000000} {\cellcolor[HTML]{000000}} \color[HTML]{F1F1F1} {\cellcolor{white}}  & {\cellcolor[HTML]{FEFEFE}} \color[HTML]{000000} {\cellcolor[HTML]{FDCCB8}} \color[HTML]{000000} 0.2 (0.1-0.3) & {\cellcolor[HTML]{FEFEFE}} \color[HTML]{000000} {\cellcolor[HTML]{000000}} \color[HTML]{F1F1F1} {\cellcolor{white}}  & {\cellcolor[HTML]{FEFEFE}} \color[HTML]{000000} {\cellcolor[HTML]{FCBFA7}} \color[HTML]{000000} 0.2 (0.1-0.4) & {\cellcolor[HTML]{FEFEFE}} \color[HTML]{000000} {\cellcolor[HTML]{FFF0E8}} \color[HTML]{000000} 0.0 (0.0-0.1) & {\cellcolor[HTML]{FEFEFE}} \color[HTML]{000000} {\cellcolor[HTML]{FEE2D5}} \color[HTML]{000000} 0.1 (0.0-0.2) & {\cellcolor[HTML]{FEFEFE}} \color[HTML]{000000} {\cellcolor[HTML]{FCBFA7}} \color[HTML]{000000} 0.2 (0.1-0.3) & {\cellcolor[HTML]{FEFEFE}} \color[HTML]{000000} {\cellcolor[HTML]{FFF5F0}} \color[HTML]{000000} 0.0 (0.0-0.0) & {\cellcolor[HTML]{FEFEFE}} \color[HTML]{000000} {\cellcolor[HTML]{FEE5D8}} \color[HTML]{000000} 0.1 (0.0-0.3) & {\cellcolor[HTML]{FEFEFE}} \color[HTML]{000000} {\cellcolor[HTML]{FEE5D8}} \color[HTML]{000000} 0.1 (0.0-0.2) \\
{\cellcolor[HTML]{FEFEFE}} \color[HTML]{000000} 109 & {\cellcolor[HTML]{FEFEFE}} \color[HTML]{000000} $\text{G}^\text{360}_\text{OP2}$-$\text{O}_\text{ph}$ & {\cellcolor[HTML]{FEFEFE}} \color[HTML]{000000} {\cellcolor{white}}  & {\cellcolor[HTML]{FEFEFE}} \color[HTML]{000000}  & {\cellcolor[HTML]{FEFEFE}} \color[HTML]{000000} {\cellcolor[HTML]{000000}} \color[HTML]{F1F1F1} {\cellcolor{white}}  & {\cellcolor[HTML]{FEFEFE}} \color[HTML]{000000} {\cellcolor[HTML]{000000}} \color[HTML]{F1F1F1} {\cellcolor{white}}  & {\cellcolor[HTML]{FEFEFE}} \color[HTML]{000000} {\cellcolor[HTML]{000000}} \color[HTML]{F1F1F1} {\cellcolor{white}}  & {\cellcolor[HTML]{FEFEFE}} \color[HTML]{000000} {\cellcolor[HTML]{000000}} \color[HTML]{F1F1F1} {\cellcolor{white}}  & {\cellcolor[HTML]{FEFEFE}} \color[HTML]{000000} {\cellcolor[HTML]{000000}} \color[HTML]{F1F1F1} {\cellcolor{white}}  & {\cellcolor[HTML]{FEFEFE}} \color[HTML]{000000} {\cellcolor[HTML]{FEE0D2}} \color[HTML]{000000} 0.1 (0.0-0.4) & {\cellcolor[HTML]{FEFEFE}} \color[HTML]{000000} {\cellcolor[HTML]{FFEFE8}} \color[HTML]{000000} 0.0 (0.0-0.1) & {\cellcolor[HTML]{FEFEFE}} \color[HTML]{000000} {\cellcolor[HTML]{FEE7DB}} \color[HTML]{000000} 0.1 (0.0-0.2) & {\cellcolor[HTML]{FEFEFE}} \color[HTML]{000000} {\cellcolor[HTML]{000000}} \color[HTML]{F1F1F1} {\cellcolor{white}}  & {\cellcolor[HTML]{FEFEFE}} \color[HTML]{000000} {\cellcolor[HTML]{FEE0D2}} \color[HTML]{000000} 0.1 (0.0-0.3) & {\cellcolor[HTML]{FEFEFE}} \color[HTML]{000000} {\cellcolor[HTML]{FCB095}} \color[HTML]{000000} 0.3 (0.1-0.5) & {\cellcolor[HTML]{FEFEFE}} \color[HTML]{000000} {\cellcolor[HTML]{000000}} \color[HTML]{F1F1F1} {\cellcolor{white}} {\cellcolor[HTML]{AAAAAA}}  & {\cellcolor[HTML]{FEFEFE}} \color[HTML]{000000} {\cellcolor[HTML]{FCBBA1}} \color[HTML]{000000} 0.2 (0.2-0.4) & {\cellcolor[HTML]{FEFEFE}} \color[HTML]{000000} {\cellcolor[HTML]{FEE5D8}} \color[HTML]{000000} 0.1 (0.0-0.2) \\
{\cellcolor[HTML]{FEFEFE}} \color[HTML]{000000} 110 & {\cellcolor[HTML]{FEFEFE}} \color[HTML]{000000} $\text{A}^\text{24}_\text{OP1}$-$\text{O}_\text{ph}$ & {\cellcolor[HTML]{FEFEFE}} \color[HTML]{000000} {\cellcolor{white}}  & {\cellcolor[HTML]{FEFEFE}} \color[HTML]{000000}  & {\cellcolor[HTML]{FEFEFE}} \color[HTML]{000000} {\cellcolor[HTML]{000000}} \color[HTML]{F1F1F1} {\cellcolor{white}}  & {\cellcolor[HTML]{FEFEFE}} \color[HTML]{000000} {\cellcolor[HTML]{000000}} \color[HTML]{F1F1F1} {\cellcolor{white}}  & {\cellcolor[HTML]{FEFEFE}} \color[HTML]{000000} {\cellcolor[HTML]{000000}} \color[HTML]{F1F1F1} {\cellcolor{white}}  & {\cellcolor[HTML]{FEFEFE}} \color[HTML]{000000} {\cellcolor[HTML]{FC9272}} \color[HTML]{000000} 0.4 & {\cellcolor[HTML]{FEFEFE}} \color[HTML]{000000} {\cellcolor[HTML]{000000}} \color[HTML]{F1F1F1} {\cellcolor{white}}  & {\cellcolor[HTML]{FEFEFE}} \color[HTML]{000000} {\cellcolor[HTML]{FEE7DC}} \color[HTML]{000000} 0.1 (0.0-0.1) & {\cellcolor[HTML]{FEFEFE}} \color[HTML]{000000} {\cellcolor[HTML]{FEEAE0}} \color[HTML]{000000} 0.1 (0.0-0.2) & {\cellcolor[HTML]{FEFEFE}} \color[HTML]{000000} {\cellcolor[HTML]{FCBCA2}} \color[HTML]{000000} 0.2 (0.2-0.4) & {\cellcolor[HTML]{FEFEFE}} \color[HTML]{000000} {\cellcolor[HTML]{000000}} \color[HTML]{F1F1F1} {\cellcolor{white}}  & {\cellcolor[HTML]{FEFEFE}} \color[HTML]{000000} {\cellcolor[HTML]{FDD4C2}} \color[HTML]{000000} 0.2 (0.1-0.3) & {\cellcolor[HTML]{FEFEFE}} \color[HTML]{000000} {\cellcolor[HTML]{FEEAE1}} \color[HTML]{000000} 0.1 (0.0-0.1) & {\cellcolor[HTML]{FEFEFE}} \color[HTML]{000000} {\cellcolor[HTML]{000000}} \color[HTML]{F1F1F1} {\cellcolor{white}} {\cellcolor[HTML]{AAAAAA}}  & {\cellcolor[HTML]{FEFEFE}} \color[HTML]{000000} {\cellcolor[HTML]{000000}} \color[HTML]{F1F1F1} {\cellcolor{white}} {\cellcolor[HTML]{AAAAAA}}  & {\cellcolor[HTML]{FEFEFE}} \color[HTML]{000000} {\cellcolor[HTML]{000000}} \color[HTML]{F1F1F1} {\cellcolor{white}} {\cellcolor[HTML]{AAAAAA}}  \\
{\cellcolor[HTML]{FEFEFE}} \color[HTML]{000000} 111 & {\cellcolor[HTML]{FEFEFE}} \color[HTML]{000000} $\text{A}^\text{94}_\text{OP1}$-$\text{O}_\text{ph}$ & {\cellcolor[HTML]{FEFEFE}} \color[HTML]{000000} {\cellcolor{white}}  & {\cellcolor[HTML]{FEFEFE}} \color[HTML]{000000}  & {\cellcolor[HTML]{FEFEFE}} \color[HTML]{000000} {\cellcolor[HTML]{000000}} \color[HTML]{F1F1F1} {\cellcolor{white}}  & {\cellcolor[HTML]{FEFEFE}} \color[HTML]{000000} {\cellcolor[HTML]{000000}} \color[HTML]{F1F1F1} {\cellcolor{white}}  & {\cellcolor[HTML]{FEFEFE}} \color[HTML]{000000} {\cellcolor[HTML]{FEEAE1}} \color[HTML]{000000} 0.1 & {\cellcolor[HTML]{FEFEFE}} \color[HTML]{000000} {\cellcolor[HTML]{000000}} \color[HTML]{F1F1F1} {\cellcolor{white}}  & {\cellcolor[HTML]{FEFEFE}} \color[HTML]{000000} {\cellcolor[HTML]{000000}} \color[HTML]{F1F1F1} {\cellcolor{white}}  & {\cellcolor[HTML]{FEFEFE}} \color[HTML]{000000} {\cellcolor[HTML]{000000}} \color[HTML]{F1F1F1} {\cellcolor{white}}  & {\cellcolor[HTML]{FEFEFE}} \color[HTML]{000000} {\cellcolor[HTML]{FFF5F0}} \color[HTML]{000000} 0.0 (0.0-0.0) & {\cellcolor[HTML]{FEFEFE}} \color[HTML]{000000} {\cellcolor[HTML]{FDD3C1}} \color[HTML]{000000} 0.2 (0.0-0.3) & {\cellcolor[HTML]{FEFEFE}} \color[HTML]{000000} {\cellcolor[HTML]{000000}} \color[HTML]{F1F1F1} {\cellcolor{white}}  & {\cellcolor[HTML]{FEFEFE}} \color[HTML]{000000} {\cellcolor[HTML]{FFF5F0}} \color[HTML]{000000} 0.0 (0.0-0.0) & {\cellcolor[HTML]{FEFEFE}} \color[HTML]{000000} {\cellcolor[HTML]{FEEAE0}} \color[HTML]{000000} 0.1 (0.0-0.2) & {\cellcolor[HTML]{FEFEFE}} \color[HTML]{000000} {\cellcolor[HTML]{000000}} \color[HTML]{F1F1F1} {\cellcolor{white}} {\cellcolor[HTML]{AAAAAA}}  & {\cellcolor[HTML]{FEFEFE}} \color[HTML]{000000} {\cellcolor[HTML]{000000}} \color[HTML]{F1F1F1} {\cellcolor{white}} {\cellcolor[HTML]{AAAAAA}}  & {\cellcolor[HTML]{FEFEFE}} \color[HTML]{000000} {\cellcolor[HTML]{D92523}} \color[HTML]{F1F1F1} 0.7 (0.6-0.8) \\
{\cellcolor[HTML]{FEFEFE}} \color[HTML]{000000} 112 & {\cellcolor[HTML]{FEFEFE}} \color[HTML]{000000} $\text{U}^\text{155}_\text{O4}$-$\text{O}_\text{b}$ & {\cellcolor[HTML]{FEFEFE}} \color[HTML]{000000} {\cellcolor{white}}  & {\cellcolor[HTML]{FEFEFE}} \color[HTML]{000000}  & {\cellcolor[HTML]{FEFEFE}} \color[HTML]{000000} {\cellcolor[HTML]{CA181D}} \color[HTML]{F1F1F1} 0.8 & {\cellcolor[HTML]{FEFEFE}} \color[HTML]{000000} {\cellcolor[HTML]{000000}} \color[HTML]{F1F1F1} {\cellcolor{white}}  & {\cellcolor[HTML]{FEFEFE}} \color[HTML]{000000} {\cellcolor[HTML]{FDCDB9}} \color[HTML]{000000} 0.2 & {\cellcolor[HTML]{FEFEFE}} \color[HTML]{000000} {\cellcolor[HTML]{000000}} \color[HTML]{F1F1F1} {\cellcolor{white}}  & {\cellcolor[HTML]{FEFEFE}} \color[HTML]{000000} {\cellcolor[HTML]{FFF4EF}} \color[HTML]{000000} 0.0 (0.0-0.0) & {\cellcolor[HTML]{FEFEFE}} \color[HTML]{000000} {\cellcolor[HTML]{FFF5F0}} \color[HTML]{000000} 0.0 (0.0-0.0) & {\cellcolor[HTML]{FEFEFE}} \color[HTML]{000000} {\cellcolor[HTML]{000000}} \color[HTML]{F1F1F1} {\cellcolor{white}}  & {\cellcolor[HTML]{FEFEFE}} \color[HTML]{000000} {\cellcolor[HTML]{FFF3ED}} \color[HTML]{000000} 0.0 (0.0-0.0) & {\cellcolor[HTML]{FEFEFE}} \color[HTML]{000000} {\cellcolor[HTML]{000000}} \color[HTML]{F1F1F1} {\cellcolor{white}}  & {\cellcolor[HTML]{FEFEFE}} \color[HTML]{000000} {\cellcolor[HTML]{000000}} \color[HTML]{F1F1F1} {\cellcolor{white}}  & {\cellcolor[HTML]{FEFEFE}} \color[HTML]{000000} {\cellcolor[HTML]{000000}} \color[HTML]{F1F1F1} {\cellcolor{white}}  & {\cellcolor[HTML]{FEFEFE}} \color[HTML]{000000} {\cellcolor[HTML]{FFEDE5}} \color[HTML]{000000} 0.1 (0.0-0.1) & {\cellcolor[HTML]{FEFEFE}} \color[HTML]{000000} {\cellcolor[HTML]{FFF5F0}} \color[HTML]{000000} 0.0 (0.0-0.0) & {\cellcolor[HTML]{FEFEFE}} \color[HTML]{000000} {\cellcolor[HTML]{FFF5F0}} \color[HTML]{000000} 0.0 (0.0-0.1) \\
{\cellcolor[HTML]{FEFEFE}} \color[HTML]{000000} 113 & {\cellcolor[HTML]{FEFEFE}} \color[HTML]{000000} $\text{C}^\text{274}_\text{OP1}$-$\text{O}_\text{ph}$ & {\cellcolor[HTML]{FEFEFE}} \color[HTML]{000000} {\cellcolor{white}}  & {\cellcolor[HTML]{FEFEFE}} \color[HTML]{000000}  & {\cellcolor[HTML]{FEFEFE}} \color[HTML]{000000} {\cellcolor[HTML]{000000}} \color[HTML]{F1F1F1} {\cellcolor{white}}  & {\cellcolor[HTML]{FEFEFE}} \color[HTML]{000000} {\cellcolor[HTML]{000000}} \color[HTML]{F1F1F1} {\cellcolor{white}}  & {\cellcolor[HTML]{FEFEFE}} \color[HTML]{000000} {\cellcolor[HTML]{000000}} \color[HTML]{F1F1F1} {\cellcolor{white}}  & {\cellcolor[HTML]{FEFEFE}} \color[HTML]{000000} {\cellcolor[HTML]{000000}} \color[HTML]{F1F1F1} {\cellcolor{white}}  & {\cellcolor[HTML]{FEFEFE}} \color[HTML]{000000} {\cellcolor[HTML]{000000}} \color[HTML]{F1F1F1} {\cellcolor{white}}  & {\cellcolor[HTML]{FEFEFE}} \color[HTML]{000000} {\cellcolor[HTML]{FFEFE8}} \color[HTML]{000000} 0.0 (0.0-0.1) & {\cellcolor[HTML]{FEFEFE}} \color[HTML]{000000} {\cellcolor[HTML]{FFF4EF}} \color[HTML]{000000} 0.0 (0.0-0.0) & {\cellcolor[HTML]{FEFEFE}} \color[HTML]{000000} {\cellcolor[HTML]{FEE8DD}} \color[HTML]{000000} 0.1 (0.0-0.1) & {\cellcolor[HTML]{FEFEFE}} \color[HTML]{000000} {\cellcolor[HTML]{FCC3AB}} \color[HTML]{000000} 0.2 (0.0-0.5) & {\cellcolor[HTML]{FEFEFE}} \color[HTML]{000000} {\cellcolor[HTML]{FCB499}} \color[HTML]{000000} 0.3 (0.1-0.5) & {\cellcolor[HTML]{FEFEFE}} \color[HTML]{000000} {\cellcolor[HTML]{FEDCCD}} \color[HTML]{000000} 0.1 (0.1-0.2) & {\cellcolor[HTML]{FEFEFE}} \color[HTML]{000000} {\cellcolor[HTML]{FFF5F0}} \color[HTML]{000000} 0.0 (0.0-0.3) & {\cellcolor[HTML]{FEFEFE}} \color[HTML]{000000} {\cellcolor[HTML]{FED9C9}} \color[HTML]{000000} 0.2 (0.0-0.3) & {\cellcolor[HTML]{FEFEFE}} \color[HTML]{000000} {\cellcolor[HTML]{FEE5D8}} \color[HTML]{000000} 0.1 (0.0-0.2) \\
{\cellcolor[HTML]{FEFEFE}} \color[HTML]{000000} 114 & {\cellcolor[HTML]{FEFEFE}} \color[HTML]{000000} $\text{G}^\text{40}_\text{OP1}$-$\text{O}_\text{ph}$ & {\cellcolor[HTML]{FEFEFE}} \color[HTML]{000000} {\cellcolor{white}}  & {\cellcolor[HTML]{FEFEFE}} \color[HTML]{000000}  & {\cellcolor[HTML]{FEFEFE}} \color[HTML]{000000} {\cellcolor[HTML]{000000}} \color[HTML]{F1F1F1} {\cellcolor{white}}  & {\cellcolor[HTML]{FEFEFE}} \color[HTML]{000000} {\cellcolor[HTML]{000000}} \color[HTML]{F1F1F1} {\cellcolor{white}}  & {\cellcolor[HTML]{FEFEFE}} \color[HTML]{000000} {\cellcolor[HTML]{000000}} \color[HTML]{F1F1F1} {\cellcolor{white}}  & {\cellcolor[HTML]{FEFEFE}} \color[HTML]{000000} {\cellcolor[HTML]{000000}} \color[HTML]{F1F1F1} {\cellcolor{white}}  & {\cellcolor[HTML]{FEFEFE}} \color[HTML]{000000} {\cellcolor[HTML]{000000}} \color[HTML]{F1F1F1} {\cellcolor{white}}  & {\cellcolor[HTML]{FEFEFE}} \color[HTML]{000000} {\cellcolor[HTML]{FDC6B0}} \color[HTML]{000000} 0.2 (0.1-0.4) & {\cellcolor[HTML]{FEFEFE}} \color[HTML]{000000} {\cellcolor[HTML]{FEE6DA}} \color[HTML]{000000} 0.1 (0.0-0.3) & {\cellcolor[HTML]{FEFEFE}} \color[HTML]{000000} {\cellcolor[HTML]{FCA98C}} \color[HTML]{000000} 0.3 (0.2-0.5) & {\cellcolor[HTML]{FEFEFE}} \color[HTML]{000000} {\cellcolor[HTML]{000000}} \color[HTML]{F1F1F1} {\cellcolor{white}}  & {\cellcolor[HTML]{FEFEFE}} \color[HTML]{000000} {\cellcolor[HTML]{FEE8DD}} \color[HTML]{000000} 0.1 (0.0-0.2) & {\cellcolor[HTML]{FEFEFE}} \color[HTML]{000000} {\cellcolor[HTML]{FDD7C6}} \color[HTML]{000000} 0.2 (0.1-0.2) & {\cellcolor[HTML]{FEFEFE}} \color[HTML]{000000} {\cellcolor[HTML]{000000}} \color[HTML]{F1F1F1} {\cellcolor{white}} {\cellcolor[HTML]{AAAAAA}}  & {\cellcolor[HTML]{FEFEFE}} \color[HTML]{000000} {\cellcolor[HTML]{FED9C9}} \color[HTML]{000000} 0.2 (0.0-0.2) & {\cellcolor[HTML]{FEFEFE}} \color[HTML]{000000} {\cellcolor[HTML]{FFF5F0}} \color[HTML]{000000} 0.0 (0.0-0.1) \\
{\cellcolor[HTML]{FEFEFE}} \color[HTML]{000000} 115 & {\cellcolor[HTML]{FEFEFE}} \color[HTML]{000000} $\text{G}^\text{251}_\text{OP2}$-$\text{O}_\text{ph}$ & {\cellcolor[HTML]{FEFEFE}} \color[HTML]{000000} {\cellcolor{white}}  & {\cellcolor[HTML]{FEFEFE}} \color[HTML]{000000}  & {\cellcolor[HTML]{FEFEFE}} \color[HTML]{000000} {\cellcolor[HTML]{000000}} \color[HTML]{F1F1F1} {\cellcolor{white}}  & {\cellcolor[HTML]{FEFEFE}} \color[HTML]{000000} {\cellcolor[HTML]{000000}} \color[HTML]{F1F1F1} {\cellcolor{white}}  & {\cellcolor[HTML]{FEFEFE}} \color[HTML]{000000} {\cellcolor[HTML]{000000}} \color[HTML]{F1F1F1} {\cellcolor{white}}  & {\cellcolor[HTML]{FEFEFE}} \color[HTML]{000000} {\cellcolor[HTML]{000000}} \color[HTML]{F1F1F1} {\cellcolor{white}}  & {\cellcolor[HTML]{FEFEFE}} \color[HTML]{000000} {\cellcolor[HTML]{000000}} \color[HTML]{F1F1F1} {\cellcolor{white}}  & {\cellcolor[HTML]{FEFEFE}} \color[HTML]{000000} {\cellcolor[HTML]{FFF1EA}} \color[HTML]{000000} 0.0 (0.0-0.1) & {\cellcolor[HTML]{FEFEFE}} \color[HTML]{000000} {\cellcolor[HTML]{FEE0D2}} \color[HTML]{000000} 0.1 (0.0-0.3) & {\cellcolor[HTML]{FEFEFE}} \color[HTML]{000000} {\cellcolor[HTML]{FFECE3}} \color[HTML]{000000} 0.1 (0.0-0.1) & {\cellcolor[HTML]{FEFEFE}} \color[HTML]{000000} {\cellcolor[HTML]{000000}} \color[HTML]{F1F1F1} {\cellcolor{white}}  & {\cellcolor[HTML]{FEFEFE}} \color[HTML]{000000} {\cellcolor[HTML]{FFEEE6}} \color[HTML]{000000} 0.0 (0.0-0.1) & {\cellcolor[HTML]{FEFEFE}} \color[HTML]{000000} {\cellcolor[HTML]{FFEFE8}} \color[HTML]{000000} 0.0 (0.0-0.1) & {\cellcolor[HTML]{FEFEFE}} \color[HTML]{000000} {\cellcolor[HTML]{000000}} \color[HTML]{F1F1F1} {\cellcolor{white}} {\cellcolor[HTML]{AAAAAA}}  & {\cellcolor[HTML]{FEFEFE}} \color[HTML]{000000} {\cellcolor[HTML]{FC9B7C}} \color[HTML]{000000} 0.4 (0.2-0.5) & {\cellcolor[HTML]{FEFEFE}} \color[HTML]{000000} {\cellcolor[HTML]{FC9B7C}} \color[HTML]{000000} 0.4 (0.2-0.5) \\
{\cellcolor[HTML]{FEFEFE}} \color[HTML]{000000} 116 & {\cellcolor[HTML]{FEFEFE}} \color[HTML]{000000} $\text{U}^\text{106}_\text{OP2}$-$\text{O}_\text{ph}$ & {\cellcolor[HTML]{FEFEFE}} \color[HTML]{000000} {\cellcolor{white}}  & {\cellcolor[HTML]{FEFEFE}} \color[HTML]{000000}  & {\cellcolor[HTML]{FEFEFE}} \color[HTML]{000000} {\cellcolor[HTML]{000000}} \color[HTML]{F1F1F1} {\cellcolor{white}}  & {\cellcolor[HTML]{FEFEFE}} \color[HTML]{000000} {\cellcolor[HTML]{000000}} \color[HTML]{F1F1F1} {\cellcolor{white}}  & {\cellcolor[HTML]{FEFEFE}} \color[HTML]{000000} {\cellcolor[HTML]{000000}} \color[HTML]{F1F1F1} {\cellcolor{white}}  & {\cellcolor[HTML]{FEFEFE}} \color[HTML]{000000} {\cellcolor[HTML]{FEE0D2}} \color[HTML]{000000} 0.1 & {\cellcolor[HTML]{FEFEFE}} \color[HTML]{000000} {\cellcolor[HTML]{000000}} \color[HTML]{F1F1F1} {\cellcolor{white}}  & {\cellcolor[HTML]{FEFEFE}} \color[HTML]{000000} {\cellcolor[HTML]{FEE7DC}} \color[HTML]{000000} 0.1 (0.0-0.2) & {\cellcolor[HTML]{FEFEFE}} \color[HTML]{000000} {\cellcolor[HTML]{FFF0E8}} \color[HTML]{000000} 0.0 (0.0-0.1) & {\cellcolor[HTML]{FEFEFE}} \color[HTML]{000000} {\cellcolor[HTML]{FDCBB6}} \color[HTML]{000000} 0.2 (0.1-0.3) & {\cellcolor[HTML]{FEFEFE}} \color[HTML]{000000} {\cellcolor[HTML]{000000}} \color[HTML]{F1F1F1} {\cellcolor{white}}  & {\cellcolor[HTML]{FEFEFE}} \color[HTML]{000000} {\cellcolor[HTML]{FFF0E8}} \color[HTML]{000000} 0.0 (0.0-0.1) & {\cellcolor[HTML]{FEFEFE}} \color[HTML]{000000} {\cellcolor[HTML]{FFEBE2}} \color[HTML]{000000} 0.1 (0.0-0.1) & {\cellcolor[HTML]{FEFEFE}} \color[HTML]{000000} {\cellcolor[HTML]{000000}} \color[HTML]{F1F1F1} {\cellcolor{white}} {\cellcolor[HTML]{AAAAAA}}  & {\cellcolor[HTML]{FEFEFE}} \color[HTML]{000000} {\cellcolor[HTML]{FDCAB5}} \color[HTML]{000000} 0.2 (0.0-0.4) & {\cellcolor[HTML]{FEFEFE}} \color[HTML]{000000} {\cellcolor[HTML]{FCBBA1}} \color[HTML]{000000} 0.2 (0.0-0.5) \\
{\cellcolor[HTML]{FEFEFE}} \color[HTML]{000000} 117 & {\cellcolor[HTML]{FEFEFE}} \color[HTML]{000000} $\text{C}^\text{109}_\text{OP2}$-$\text{O}_\text{ph}$ & {\cellcolor[HTML]{FEFEFE}} \color[HTML]{000000} {\cellcolor{white}}  & {\cellcolor[HTML]{FEFEFE}} \color[HTML]{000000}  & {\cellcolor[HTML]{FEFEFE}} \color[HTML]{000000} {\cellcolor[HTML]{000000}} \color[HTML]{F1F1F1} {\cellcolor{white}}  & {\cellcolor[HTML]{FEFEFE}} \color[HTML]{000000} {\cellcolor[HTML]{FEEAE1}} \color[HTML]{000000} 0.1 & {\cellcolor[HTML]{FEFEFE}} \color[HTML]{000000} {\cellcolor[HTML]{000000}} \color[HTML]{F1F1F1} {\cellcolor{white}}  & {\cellcolor[HTML]{FEFEFE}} \color[HTML]{000000} {\cellcolor[HTML]{000000}} \color[HTML]{F1F1F1} {\cellcolor{white}}  & {\cellcolor[HTML]{FEFEFE}} \color[HTML]{000000} {\cellcolor[HTML]{FFF5F0}} \color[HTML]{000000} 0.0 (0.0-0.0) & {\cellcolor[HTML]{FEFEFE}} \color[HTML]{000000} {\cellcolor[HTML]{FEE9DF}} \color[HTML]{000000} 0.1 (0.0-0.2) & {\cellcolor[HTML]{FEFEFE}} \color[HTML]{000000} {\cellcolor[HTML]{FDD4C2}} \color[HTML]{000000} 0.2 (0.0-0.4) & {\cellcolor[HTML]{FEFEFE}} \color[HTML]{000000} {\cellcolor[HTML]{FFEDE5}} \color[HTML]{000000} 0.0 (0.0-0.1) & {\cellcolor[HTML]{FEFEFE}} \color[HTML]{000000} {\cellcolor[HTML]{000000}} \color[HTML]{F1F1F1} {\cellcolor{white}}  & {\cellcolor[HTML]{FEFEFE}} \color[HTML]{000000} {\cellcolor[HTML]{FEEAE0}} \color[HTML]{000000} 0.1 (0.0-0.2) & {\cellcolor[HTML]{FEFEFE}} \color[HTML]{000000} {\cellcolor[HTML]{FFF4EE}} \color[HTML]{000000} 0.0 (0.0-0.0) & {\cellcolor[HTML]{FEFEFE}} \color[HTML]{000000} {\cellcolor[HTML]{000000}} \color[HTML]{F1F1F1} {\cellcolor{white}} {\cellcolor[HTML]{AAAAAA}}  & {\cellcolor[HTML]{FEFEFE}} \color[HTML]{000000} {\cellcolor[HTML]{FC9B7C}} \color[HTML]{000000} 0.4 (0.2-0.6) & {\cellcolor[HTML]{FEFEFE}} \color[HTML]{000000} {\cellcolor[HTML]{FDCAB5}} \color[HTML]{000000} 0.2 (0.0-0.5) \\
{\cellcolor[HTML]{FEFEFE}} \color[HTML]{000000} 118 & {\cellcolor[HTML]{FEFEFE}} \color[HTML]{000000} $\text{G}^\text{100}_\text{OP2}$-$\text{O}_\text{ph}$ & {\cellcolor[HTML]{FEFEFE}} \color[HTML]{000000} {\cellcolor{white}}  & {\cellcolor[HTML]{FEFEFE}} \color[HTML]{000000}  & {\cellcolor[HTML]{FEFEFE}} \color[HTML]{000000} {\cellcolor[HTML]{000000}} \color[HTML]{F1F1F1} {\cellcolor{white}}  & {\cellcolor[HTML]{FEFEFE}} \color[HTML]{000000} {\cellcolor[HTML]{000000}} \color[HTML]{F1F1F1} {\cellcolor{white}}  & {\cellcolor[HTML]{FEFEFE}} \color[HTML]{000000} {\cellcolor[HTML]{000000}} \color[HTML]{F1F1F1} {\cellcolor{white}}  & {\cellcolor[HTML]{FEFEFE}} \color[HTML]{000000} {\cellcolor[HTML]{000000}} \color[HTML]{F1F1F1} {\cellcolor{white}}  & {\cellcolor[HTML]{FEFEFE}} \color[HTML]{000000} {\cellcolor[HTML]{000000}} \color[HTML]{F1F1F1} {\cellcolor{white}}  & {\cellcolor[HTML]{FEFEFE}} \color[HTML]{000000} {\cellcolor[HTML]{FDC7B2}} \color[HTML]{000000} 0.2 (0.1-0.4) & {\cellcolor[HTML]{FEFEFE}} \color[HTML]{000000} {\cellcolor[HTML]{FCB499}} \color[HTML]{000000} 0.3 (0.2-0.4) & {\cellcolor[HTML]{FEFEFE}} \color[HTML]{000000} {\cellcolor[HTML]{FEE8DD}} \color[HTML]{000000} 0.1 (0.0-0.1) & {\cellcolor[HTML]{FEFEFE}} \color[HTML]{000000} {\cellcolor[HTML]{000000}} \color[HTML]{F1F1F1} {\cellcolor{white}}  & {\cellcolor[HTML]{FEFEFE}} \color[HTML]{000000} {\cellcolor[HTML]{FFEFE8}} \color[HTML]{000000} 0.0 (0.0-0.1) & {\cellcolor[HTML]{FEFEFE}} \color[HTML]{000000} {\cellcolor[HTML]{FC9474}} \color[HTML]{000000} 0.4 (0.2-0.5) & {\cellcolor[HTML]{FEFEFE}} \color[HTML]{000000} {\cellcolor[HTML]{000000}} \color[HTML]{F1F1F1} {\cellcolor{white}} {\cellcolor[HTML]{AAAAAA}}  & {\cellcolor[HTML]{FEFEFE}} \color[HTML]{000000} {\cellcolor[HTML]{FFF5F0}} \color[HTML]{000000} 0.0 (0.0-0.0) & {\cellcolor[HTML]{FEFEFE}} \color[HTML]{000000} {\cellcolor[HTML]{FFF5F0}} \color[HTML]{000000} 0.0 (0.0-0.0) \\
{\cellcolor[HTML]{FEFEFE}} \color[HTML]{000000} 119 & {\cellcolor[HTML]{FEFEFE}} \color[HTML]{000000} $\text{G}^\text{337}_\text{OP2}$-$\text{O}_\text{ph}$ & {\cellcolor[HTML]{FEFEFE}} \color[HTML]{000000} {\cellcolor{white}}  & {\cellcolor[HTML]{FEFEFE}} \color[HTML]{000000}  & {\cellcolor[HTML]{FEFEFE}} \color[HTML]{000000} {\cellcolor[HTML]{000000}} \color[HTML]{F1F1F1} {\cellcolor{white}}  & {\cellcolor[HTML]{FEFEFE}} \color[HTML]{000000} {\cellcolor[HTML]{000000}} \color[HTML]{F1F1F1} {\cellcolor{white}}  & {\cellcolor[HTML]{FEFEFE}} \color[HTML]{000000} {\cellcolor[HTML]{000000}} \color[HTML]{F1F1F1} {\cellcolor{white}}  & {\cellcolor[HTML]{FEFEFE}} \color[HTML]{000000} {\cellcolor[HTML]{000000}} \color[HTML]{F1F1F1} {\cellcolor{white}}  & {\cellcolor[HTML]{FEFEFE}} \color[HTML]{000000} {\cellcolor[HTML]{000000}} \color[HTML]{F1F1F1} {\cellcolor{white}}  & {\cellcolor[HTML]{FEFEFE}} \color[HTML]{000000} {\cellcolor[HTML]{FEE0D2}} \color[HTML]{000000} 0.1 (0.0-0.3) & {\cellcolor[HTML]{FEFEFE}} \color[HTML]{000000} {\cellcolor[HTML]{FDCEBB}} \color[HTML]{000000} 0.2 (0.1-0.4) & {\cellcolor[HTML]{FEFEFE}} \color[HTML]{000000} {\cellcolor[HTML]{FCBCA2}} \color[HTML]{000000} 0.2 (0.1-0.4) & {\cellcolor[HTML]{FEFEFE}} \color[HTML]{000000} {\cellcolor[HTML]{000000}} \color[HTML]{F1F1F1} {\cellcolor{white}}  & {\cellcolor[HTML]{FEFEFE}} \color[HTML]{000000} {\cellcolor[HTML]{FDCDB9}} \color[HTML]{000000} 0.2 (0.1-0.3) & {\cellcolor[HTML]{FEFEFE}} \color[HTML]{000000} {\cellcolor[HTML]{FDC5AE}} \color[HTML]{000000} 0.2 (0.1-0.3) & {\cellcolor[HTML]{FEFEFE}} \color[HTML]{000000} {\cellcolor[HTML]{000000}} \color[HTML]{F1F1F1} {\cellcolor{white}} {\cellcolor[HTML]{AAAAAA}}  & {\cellcolor[HTML]{FEFEFE}} \color[HTML]{000000} {\cellcolor[HTML]{000000}} \color[HTML]{F1F1F1} {\cellcolor{white}} {\cellcolor[HTML]{AAAAAA}}  & {\cellcolor[HTML]{FEFEFE}} \color[HTML]{000000} {\cellcolor[HTML]{000000}} \color[HTML]{F1F1F1} {\cellcolor{white}} {\cellcolor[HTML]{AAAAAA}}  \\
{\cellcolor[HTML]{FEFEFE}} \color[HTML]{000000} 120 & {\cellcolor[HTML]{FEFEFE}} \color[HTML]{000000} $\text{G}^\text{331}_\text{OP1}$-$\text{O}_\text{ph}$ & {\cellcolor[HTML]{FEFEFE}} \color[HTML]{000000} {\cellcolor{white}}  & {\cellcolor[HTML]{FEFEFE}} \color[HTML]{000000}  & {\cellcolor[HTML]{FEFEFE}} \color[HTML]{000000} {\cellcolor[HTML]{000000}} \color[HTML]{F1F1F1} {\cellcolor{white}}  & {\cellcolor[HTML]{FEFEFE}} \color[HTML]{000000} {\cellcolor[HTML]{000000}} \color[HTML]{F1F1F1} {\cellcolor{white}}  & {\cellcolor[HTML]{FEFEFE}} \color[HTML]{000000} {\cellcolor[HTML]{000000}} \color[HTML]{F1F1F1} {\cellcolor{white}}  & {\cellcolor[HTML]{FEFEFE}} \color[HTML]{000000} {\cellcolor[HTML]{000000}} \color[HTML]{F1F1F1} {\cellcolor{white}}  & {\cellcolor[HTML]{FEFEFE}} \color[HTML]{000000} {\cellcolor[HTML]{000000}} \color[HTML]{F1F1F1} {\cellcolor{white}}  & {\cellcolor[HTML]{FEFEFE}} \color[HTML]{000000} {\cellcolor[HTML]{FFF1EA}} \color[HTML]{000000} 0.0 (0.0-0.1) & {\cellcolor[HTML]{FEFEFE}} \color[HTML]{000000} {\cellcolor[HTML]{FC9C7D}} \color[HTML]{000000} 0.3 (0.2-0.5) & {\cellcolor[HTML]{FEFEFE}} \color[HTML]{000000} {\cellcolor[HTML]{FEDBCC}} \color[HTML]{000000} 0.1 (0.1-0.2) & {\cellcolor[HTML]{FEFEFE}} \color[HTML]{000000} {\cellcolor[HTML]{000000}} \color[HTML]{F1F1F1} {\cellcolor{white}}  & {\cellcolor[HTML]{FEFEFE}} \color[HTML]{000000} {\cellcolor[HTML]{FFECE3}} \color[HTML]{000000} 0.1 (0.0-0.1) & {\cellcolor[HTML]{FEFEFE}} \color[HTML]{000000} {\cellcolor[HTML]{FCBEA5}} \color[HTML]{000000} 0.2 (0.1-0.4) & {\cellcolor[HTML]{FEFEFE}} \color[HTML]{000000} {\cellcolor[HTML]{000000}} \color[HTML]{F1F1F1} {\cellcolor{white}} {\cellcolor[HTML]{AAAAAA}}  & {\cellcolor[HTML]{FEFEFE}} \color[HTML]{000000} {\cellcolor[HTML]{FFEDE5}} \color[HTML]{000000} 0.1 (0.0-0.1) & {\cellcolor[HTML]{FEFEFE}} \color[HTML]{000000} {\cellcolor[HTML]{FEE5D8}} \color[HTML]{000000} 0.1 (0.1-0.2) \\
{\cellcolor[HTML]{FEFEFE}} \color[HTML]{000000} 121 & {\cellcolor[HTML]{FEFEFE}} \color[HTML]{000000} $\text{U}^\text{253}_\text{OP2}$-$\text{O}_\text{ph}$ & {\cellcolor[HTML]{FEFEFE}} \color[HTML]{000000} {\cellcolor{white}}  & {\cellcolor[HTML]{FEFEFE}} \color[HTML]{000000}  & {\cellcolor[HTML]{FEFEFE}} \color[HTML]{000000} {\cellcolor[HTML]{000000}} \color[HTML]{F1F1F1} {\cellcolor{white}}  & {\cellcolor[HTML]{FEFEFE}} \color[HTML]{000000} {\cellcolor[HTML]{000000}} \color[HTML]{F1F1F1} {\cellcolor{white}}  & {\cellcolor[HTML]{FEFEFE}} \color[HTML]{000000} {\cellcolor[HTML]{000000}} \color[HTML]{F1F1F1} {\cellcolor{white}}  & {\cellcolor[HTML]{FEFEFE}} \color[HTML]{000000} {\cellcolor[HTML]{FEE0D2}} \color[HTML]{000000} 0.1 & {\cellcolor[HTML]{FEFEFE}} \color[HTML]{000000} {\cellcolor[HTML]{000000}} \color[HTML]{F1F1F1} {\cellcolor{white}}  & {\cellcolor[HTML]{FEFEFE}} \color[HTML]{000000} {\cellcolor[HTML]{FFF5F0}} \color[HTML]{000000} 0.0 (0.0-0.0) & {\cellcolor[HTML]{FEFEFE}} \color[HTML]{000000} {\cellcolor[HTML]{FEE7DC}} \color[HTML]{000000} 0.1 (0.0-0.1) & {\cellcolor[HTML]{FEFEFE}} \color[HTML]{000000} {\cellcolor[HTML]{FFF4EF}} \color[HTML]{000000} 0.0 (0.0-0.0) & {\cellcolor[HTML]{FEFEFE}} \color[HTML]{000000} {\cellcolor[HTML]{000000}} \color[HTML]{F1F1F1} {\cellcolor{white}}  & {\cellcolor[HTML]{FEFEFE}} \color[HTML]{000000} {\cellcolor[HTML]{FFEEE6}} \color[HTML]{000000} 0.0 (0.0-0.1) & {\cellcolor[HTML]{FEFEFE}} \color[HTML]{000000} {\cellcolor[HTML]{FEE6DA}} \color[HTML]{000000} 0.1 (0.0-0.2) & {\cellcolor[HTML]{FEFEFE}} \color[HTML]{000000} {\cellcolor[HTML]{000000}} \color[HTML]{F1F1F1} {\cellcolor{white}} {\cellcolor[HTML]{AAAAAA}}  & {\cellcolor[HTML]{FEFEFE}} \color[HTML]{000000} {\cellcolor[HTML]{FCAB8F}} \color[HTML]{000000} 0.3 (0.2-0.4) & {\cellcolor[HTML]{FEFEFE}} \color[HTML]{000000} {\cellcolor[HTML]{FCAB8F}} \color[HTML]{000000} 0.3 (0.2-0.4) \\
{\cellcolor[HTML]{FEFEFE}} \color[HTML]{000000} 122 & {\cellcolor[HTML]{FEFEFE}} \color[HTML]{000000} $\text{A}^\text{380}_\text{OP2}$-$\text{O}_\text{ph}$ & {\cellcolor[HTML]{FEFEFE}} \color[HTML]{000000} {\cellcolor{white}}  & {\cellcolor[HTML]{FEFEFE}} \color[HTML]{000000}  & {\cellcolor[HTML]{FEFEFE}} \color[HTML]{000000} {\cellcolor[HTML]{000000}} \color[HTML]{F1F1F1} {\cellcolor{white}}  & {\cellcolor[HTML]{FEFEFE}} \color[HTML]{000000} {\cellcolor[HTML]{000000}} \color[HTML]{F1F1F1} {\cellcolor{white}}  & {\cellcolor[HTML]{FEFEFE}} \color[HTML]{000000} {\cellcolor[HTML]{000000}} \color[HTML]{F1F1F1} {\cellcolor{white}}  & {\cellcolor[HTML]{FEFEFE}} \color[HTML]{000000} {\cellcolor[HTML]{000000}} \color[HTML]{F1F1F1} {\cellcolor{white}}  & {\cellcolor[HTML]{FEFEFE}} \color[HTML]{000000} {\cellcolor[HTML]{000000}} \color[HTML]{F1F1F1} {\cellcolor{white}}  & {\cellcolor[HTML]{FEFEFE}} \color[HTML]{000000} {\cellcolor[HTML]{FEE7DC}} \color[HTML]{000000} 0.1 (0.0-0.2) & {\cellcolor[HTML]{FEFEFE}} \color[HTML]{000000} {\cellcolor[HTML]{FEE9DF}} \color[HTML]{000000} 0.1 (0.0-0.1) & {\cellcolor[HTML]{FEFEFE}} \color[HTML]{000000} {\cellcolor[HTML]{FB7151}} \color[HTML]{F1F1F1} 0.5 (0.3-0.6) & {\cellcolor[HTML]{FEFEFE}} \color[HTML]{000000} {\cellcolor[HTML]{FEEAE1}} \color[HTML]{000000} 0.1 (0.0-0.2) & {\cellcolor[HTML]{FEFEFE}} \color[HTML]{000000} {\cellcolor[HTML]{FFF4EE}} \color[HTML]{000000} 0.0 (0.0-0.0) & {\cellcolor[HTML]{FEFEFE}} \color[HTML]{000000} {\cellcolor[HTML]{FCBDA4}} \color[HTML]{000000} 0.2 (0.1-0.4) & {\cellcolor[HTML]{FEFEFE}} \color[HTML]{000000} {\cellcolor[HTML]{000000}} \color[HTML]{F1F1F1} {\cellcolor{white}} {\cellcolor[HTML]{AAAAAA}}  & {\cellcolor[HTML]{FEFEFE}} \color[HTML]{000000} {\cellcolor[HTML]{000000}} \color[HTML]{F1F1F1} {\cellcolor{white}} {\cellcolor[HTML]{AAAAAA}}  & {\cellcolor[HTML]{FEFEFE}} \color[HTML]{000000} {\cellcolor[HTML]{000000}} \color[HTML]{F1F1F1} {\cellcolor{white}} {\cellcolor[HTML]{AAAAAA}}  \\
{\cellcolor[HTML]{FEFEFE}} \color[HTML]{000000} 123 & {\cellcolor[HTML]{FEFEFE}} \color[HTML]{000000} $\text{G}^\text{181}_\text{OP2}$-$\text{O}_\text{ph}$ & {\cellcolor[HTML]{FEFEFE}} \color[HTML]{000000} {\cellcolor{white}}  & {\cellcolor[HTML]{FEFEFE}} \color[HTML]{000000}  & {\cellcolor[HTML]{FEFEFE}} \color[HTML]{000000} {\cellcolor[HTML]{000000}} \color[HTML]{F1F1F1} {\cellcolor{white}}  & {\cellcolor[HTML]{FEFEFE}} \color[HTML]{000000} {\cellcolor[HTML]{000000}} \color[HTML]{F1F1F1} {\cellcolor{white}}  & {\cellcolor[HTML]{FEFEFE}} \color[HTML]{000000} {\cellcolor[HTML]{000000}} \color[HTML]{F1F1F1} {\cellcolor{white}}  & {\cellcolor[HTML]{FEFEFE}} \color[HTML]{000000} {\cellcolor[HTML]{000000}} \color[HTML]{F1F1F1} {\cellcolor{white}}  & {\cellcolor[HTML]{FEFEFE}} \color[HTML]{000000} {\cellcolor[HTML]{000000}} \color[HTML]{F1F1F1} {\cellcolor{white}}  & {\cellcolor[HTML]{FEFEFE}} \color[HTML]{000000} {\cellcolor[HTML]{000000}} \color[HTML]{F1F1F1} {\cellcolor{white}}  & {\cellcolor[HTML]{FEFEFE}} \color[HTML]{000000} {\cellcolor[HTML]{FFEEE7}} \color[HTML]{000000} 0.0 (0.0-0.1) & {\cellcolor[HTML]{FEFEFE}} \color[HTML]{000000} {\cellcolor[HTML]{FCAA8D}} \color[HTML]{000000} 0.3 (0.2-0.5) & {\cellcolor[HTML]{FEFEFE}} \color[HTML]{000000} {\cellcolor[HTML]{FEE4D8}} \color[HTML]{000000} 0.1 (0.0-0.3) & {\cellcolor[HTML]{FEFEFE}} \color[HTML]{000000} {\cellcolor[HTML]{FFECE4}} \color[HTML]{000000} 0.1 (0.0-0.2) & {\cellcolor[HTML]{FEFEFE}} \color[HTML]{000000} {\cellcolor[HTML]{FCAA8D}} \color[HTML]{000000} 0.3 (0.2-0.4) & {\cellcolor[HTML]{FEFEFE}} \color[HTML]{000000} {\cellcolor[HTML]{FEE5D8}} \color[HTML]{000000} 0.1 (0.1-0.2) & {\cellcolor[HTML]{FEFEFE}} \color[HTML]{000000} {\cellcolor[HTML]{FFEDE5}} \color[HTML]{000000} 0.1 (0.0-0.2) & {\cellcolor[HTML]{FEFEFE}} \color[HTML]{000000} {\cellcolor[HTML]{FFF5F0}} \color[HTML]{000000} 0.0 (0.0-0.2) \\
{\cellcolor[HTML]{FEFEFE}} \color[HTML]{000000} 124 & {\cellcolor[HTML]{FEFEFE}} \color[HTML]{000000} $\text{C}^\text{383}_\text{OP2}$-$\text{O}_\text{ph}$ & {\cellcolor[HTML]{FEFEFE}} \color[HTML]{000000} {\cellcolor{white}}  & {\cellcolor[HTML]{FEFEFE}} \color[HTML]{000000}  & {\cellcolor[HTML]{FEFEFE}} \color[HTML]{000000} {\cellcolor[HTML]{000000}} \color[HTML]{F1F1F1} {\cellcolor{white}}  & {\cellcolor[HTML]{FEFEFE}} \color[HTML]{000000} {\cellcolor[HTML]{000000}} \color[HTML]{F1F1F1} {\cellcolor{white}}  & {\cellcolor[HTML]{FEFEFE}} \color[HTML]{000000} {\cellcolor[HTML]{000000}} \color[HTML]{F1F1F1} {\cellcolor{white}}  & {\cellcolor[HTML]{FEFEFE}} \color[HTML]{000000} {\cellcolor[HTML]{000000}} \color[HTML]{F1F1F1} {\cellcolor{white}}  & {\cellcolor[HTML]{FEFEFE}} \color[HTML]{000000} {\cellcolor[HTML]{FEE8DD}} \color[HTML]{000000} 0.1 (0.0-0.2) & {\cellcolor[HTML]{FEFEFE}} \color[HTML]{000000} {\cellcolor[HTML]{FEE0D2}} \color[HTML]{000000} 0.1 (0.0-0.2) & {\cellcolor[HTML]{FEFEFE}} \color[HTML]{000000} {\cellcolor[HTML]{000000}} \color[HTML]{F1F1F1} {\cellcolor{white}}  & {\cellcolor[HTML]{FEFEFE}} \color[HTML]{000000} {\cellcolor[HTML]{FB7A5A}} \color[HTML]{F1F1F1} 0.5 (0.3-0.6) & {\cellcolor[HTML]{FEFEFE}} \color[HTML]{000000} {\cellcolor[HTML]{000000}} \color[HTML]{F1F1F1} {\cellcolor{white}}  & {\cellcolor[HTML]{FEFEFE}} \color[HTML]{000000} {\cellcolor[HTML]{FFF1EA}} \color[HTML]{000000} 0.0 (0.0-0.1) & {\cellcolor[HTML]{FEFEFE}} \color[HTML]{000000} {\cellcolor[HTML]{FCB79C}} \color[HTML]{000000} 0.3 (0.1-0.4) & {\cellcolor[HTML]{FEFEFE}} \color[HTML]{000000} {\cellcolor[HTML]{000000}} \color[HTML]{F1F1F1} {\cellcolor{white}} {\cellcolor[HTML]{AAAAAA}}  & {\cellcolor[HTML]{FEFEFE}} \color[HTML]{000000} {\cellcolor[HTML]{000000}} \color[HTML]{F1F1F1} {\cellcolor{white}} {\cellcolor[HTML]{AAAAAA}}  & {\cellcolor[HTML]{FEFEFE}} \color[HTML]{000000} {\cellcolor[HTML]{000000}} \color[HTML]{F1F1F1} {\cellcolor{white}} {\cellcolor[HTML]{AAAAAA}}  \\
{\cellcolor[HTML]{FEFEFE}} \color[HTML]{000000} 125 & {\cellcolor[HTML]{FEFEFE}} \color[HTML]{000000} $\text{G}^\text{212}_\text{OP1}$-$\text{O}_\text{ph}$ & {\cellcolor[HTML]{FEFEFE}} \color[HTML]{000000} {\cellcolor{white}}  & {\cellcolor[HTML]{FEFEFE}} \color[HTML]{000000}  & {\cellcolor[HTML]{FEFEFE}} \color[HTML]{000000} {\cellcolor[HTML]{000000}} \color[HTML]{F1F1F1} {\cellcolor{white}}  & {\cellcolor[HTML]{FEFEFE}} \color[HTML]{000000} {\cellcolor[HTML]{000000}} \color[HTML]{F1F1F1} {\cellcolor{white}}  & {\cellcolor[HTML]{FEFEFE}} \color[HTML]{000000} {\cellcolor[HTML]{000000}} \color[HTML]{F1F1F1} {\cellcolor{white}}  & {\cellcolor[HTML]{FEFEFE}} \color[HTML]{000000} {\cellcolor[HTML]{000000}} \color[HTML]{F1F1F1} {\cellcolor{white}}  & {\cellcolor[HTML]{FEFEFE}} \color[HTML]{000000} {\cellcolor[HTML]{000000}} \color[HTML]{F1F1F1} {\cellcolor{white}}  & {\cellcolor[HTML]{FEFEFE}} \color[HTML]{000000} {\cellcolor[HTML]{FEE2D5}} \color[HTML]{000000} 0.1 (0.0-0.2) & {\cellcolor[HTML]{FEFEFE}} \color[HTML]{000000} {\cellcolor[HTML]{FEE5D8}} \color[HTML]{000000} 0.1 (0.0-0.1) & {\cellcolor[HTML]{FEFEFE}} \color[HTML]{000000} {\cellcolor[HTML]{FDD5C4}} \color[HTML]{000000} 0.2 (0.1-0.2) & {\cellcolor[HTML]{FEFEFE}} \color[HTML]{000000} {\cellcolor[HTML]{FFEEE7}} \color[HTML]{000000} 0.0 (0.0-0.1) & {\cellcolor[HTML]{FEFEFE}} \color[HTML]{000000} {\cellcolor[HTML]{FFF5F0}} \color[HTML]{000000} 0.0 (0.0-0.0) & {\cellcolor[HTML]{FEFEFE}} \color[HTML]{000000} {\cellcolor[HTML]{FEE3D7}} \color[HTML]{000000} 0.1 (0.1-0.2) & {\cellcolor[HTML]{FEFEFE}} \color[HTML]{000000} {\cellcolor[HTML]{FFF5F0}} \color[HTML]{000000} 0.0 (0.0-0.1) & {\cellcolor[HTML]{FEFEFE}} \color[HTML]{000000} {\cellcolor[HTML]{FFF5F0}} \color[HTML]{000000} 0.0 (0.0-0.1) & {\cellcolor[HTML]{FEFEFE}} \color[HTML]{000000} {\cellcolor[HTML]{FC8A6A}} \color[HTML]{F1F1F1} 0.4 (0.1-0.6) \\
{\cellcolor[HTML]{FEFEFE}} \color[HTML]{000000} 126 & {\cellcolor[HTML]{FEFEFE}} \color[HTML]{000000} $\text{A}^\text{268}_\text{OP2}$-$\text{O}_\text{ph}$ & {\cellcolor[HTML]{FEFEFE}} \color[HTML]{000000} {\cellcolor{white}}  & {\cellcolor[HTML]{FEFEFE}} \color[HTML]{000000}  & {\cellcolor[HTML]{FEFEFE}} \color[HTML]{000000} {\cellcolor[HTML]{FEE0D2}} \color[HTML]{000000} 0.1 & {\cellcolor[HTML]{FEFEFE}} \color[HTML]{000000} {\cellcolor[HTML]{FDCDB9}} \color[HTML]{000000} 0.2 & {\cellcolor[HTML]{FEFEFE}} \color[HTML]{000000} {\cellcolor[HTML]{FEEAE1}} \color[HTML]{000000} 0.1 & {\cellcolor[HTML]{FEFEFE}} \color[HTML]{000000} {\cellcolor[HTML]{000000}} \color[HTML]{F1F1F1} {\cellcolor{white}}  & {\cellcolor[HTML]{FEFEFE}} \color[HTML]{000000} {\cellcolor[HTML]{000000}} \color[HTML]{F1F1F1} {\cellcolor{white}}  & {\cellcolor[HTML]{FEFEFE}} \color[HTML]{000000} {\cellcolor[HTML]{FEE3D6}} \color[HTML]{000000} 0.1 (0.0-0.4) & {\cellcolor[HTML]{FEFEFE}} \color[HTML]{000000} {\cellcolor[HTML]{000000}} \color[HTML]{F1F1F1} {\cellcolor{white}}  & {\cellcolor[HTML]{FEFEFE}} \color[HTML]{000000} {\cellcolor[HTML]{FEDECF}} \color[HTML]{000000} 0.1 (0.0-0.3) & {\cellcolor[HTML]{FEFEFE}} \color[HTML]{000000} {\cellcolor[HTML]{FCBBA1}} \color[HTML]{000000} 0.2 (0.0-0.5) & {\cellcolor[HTML]{FEFEFE}} \color[HTML]{000000} {\cellcolor[HTML]{FFF5F0}} \color[HTML]{000000} 0.0 (0.0-0.0) & {\cellcolor[HTML]{FEFEFE}} \color[HTML]{000000} {\cellcolor[HTML]{FFF0E8}} \color[HTML]{000000} 0.0 (0.0-0.1) & {\cellcolor[HTML]{FEFEFE}} \color[HTML]{000000} {\cellcolor[HTML]{000000}} \color[HTML]{F1F1F1} {\cellcolor{white}} {\cellcolor[HTML]{AAAAAA}}  & {\cellcolor[HTML]{FEFEFE}} \color[HTML]{000000} {\cellcolor[HTML]{000000}} \color[HTML]{F1F1F1} {\cellcolor{white}} {\cellcolor[HTML]{AAAAAA}}  & {\cellcolor[HTML]{FEFEFE}} \color[HTML]{000000} {\cellcolor[HTML]{000000}} \color[HTML]{F1F1F1} {\cellcolor{white}} {\cellcolor[HTML]{AAAAAA}}  \\
{\cellcolor[HTML]{FEFEFE}} \color[HTML]{000000} 127 & {\cellcolor[HTML]{FEFEFE}} \color[HTML]{000000} $\text{A}^\text{369}_\text{OP2}$,$\text{A}^\text{370}_\text{OP2}$-\textit{cis}-2$\text{O}_\text{ph}$ & {\cellcolor[HTML]{FEFEFE}} \color[HTML]{000000} {\cellcolor{white}}  & {\cellcolor[HTML]{FEFEFE}} \color[HTML]{000000}  & {\cellcolor[HTML]{FEFEFE}} \color[HTML]{000000} {\cellcolor[HTML]{000000}} \color[HTML]{F1F1F1} {\cellcolor{white}}  & {\cellcolor[HTML]{FEFEFE}} \color[HTML]{000000} {\cellcolor[HTML]{000000}} \color[HTML]{F1F1F1} {\cellcolor{white}}  & {\cellcolor[HTML]{FEFEFE}} \color[HTML]{000000} {\cellcolor[HTML]{000000}} \color[HTML]{F1F1F1} {\cellcolor{white}}  & {\cellcolor[HTML]{FEFEFE}} \color[HTML]{000000} {\cellcolor[HTML]{000000}} \color[HTML]{F1F1F1} {\cellcolor{white}}  & {\cellcolor[HTML]{FEFEFE}} \color[HTML]{000000} {\cellcolor[HTML]{000000}} \color[HTML]{F1F1F1} {\cellcolor{white}}  & {\cellcolor[HTML]{FEFEFE}} \color[HTML]{000000} {\cellcolor[HTML]{FFECE4}} \color[HTML]{000000} 0.1 (0.0-0.1) & {\cellcolor[HTML]{FEFEFE}} \color[HTML]{000000} {\cellcolor[HTML]{000000}} \color[HTML]{F1F1F1} {\cellcolor{white}}  & {\cellcolor[HTML]{FEFEFE}} \color[HTML]{000000} {\cellcolor[HTML]{FCAB8F}} \color[HTML]{000000} 0.3 (0.1-0.5) & {\cellcolor[HTML]{FEFEFE}} \color[HTML]{000000} {\cellcolor[HTML]{000000}} \color[HTML]{F1F1F1} {\cellcolor{white}}  & {\cellcolor[HTML]{FEFEFE}} \color[HTML]{000000} {\cellcolor[HTML]{FEE0D2}} \color[HTML]{000000} 0.1 (0.0-0.3) & {\cellcolor[HTML]{FEFEFE}} \color[HTML]{000000} {\cellcolor[HTML]{FC8262}} \color[HTML]{F1F1F1} 0.4 (0.2-0.6) & {\cellcolor[HTML]{FEFEFE}} \color[HTML]{000000} {\cellcolor[HTML]{000000}} \color[HTML]{F1F1F1} {\cellcolor{white}} {\cellcolor[HTML]{AAAAAA}}  & {\cellcolor[HTML]{FEFEFE}} \color[HTML]{000000} {\cellcolor[HTML]{000000}} \color[HTML]{F1F1F1} {\cellcolor{white}} {\cellcolor[HTML]{AAAAAA}}  & {\cellcolor[HTML]{FEFEFE}} \color[HTML]{000000} {\cellcolor[HTML]{000000}} \color[HTML]{F1F1F1} {\cellcolor{white}} {\cellcolor[HTML]{AAAAAA}}  \\
{\cellcolor[HTML]{FEFEFE}} \color[HTML]{000000} 128 & {\cellcolor[HTML]{FEFEFE}} \color[HTML]{000000} $\text{A}^\text{151}_\text{OP1}$-$\text{O}_\text{ph}$ & {\cellcolor[HTML]{FEFEFE}} \color[HTML]{000000} {\cellcolor{white}}  & {\cellcolor[HTML]{FEFEFE}} \color[HTML]{000000}  & {\cellcolor[HTML]{FEFEFE}} \color[HTML]{000000} {\cellcolor[HTML]{000000}} \color[HTML]{F1F1F1} {\cellcolor{white}}  & {\cellcolor[HTML]{FEFEFE}} \color[HTML]{000000} {\cellcolor[HTML]{000000}} \color[HTML]{F1F1F1} {\cellcolor{white}}  & {\cellcolor[HTML]{FEFEFE}} \color[HTML]{000000} {\cellcolor[HTML]{000000}} \color[HTML]{F1F1F1} {\cellcolor{white}}  & {\cellcolor[HTML]{FEFEFE}} \color[HTML]{000000} {\cellcolor[HTML]{000000}} \color[HTML]{F1F1F1} {\cellcolor{white}}  & {\cellcolor[HTML]{FEFEFE}} \color[HTML]{000000} {\cellcolor[HTML]{000000}} \color[HTML]{F1F1F1} {\cellcolor{white}}  & {\cellcolor[HTML]{FEFEFE}} \color[HTML]{000000} {\cellcolor[HTML]{FEDBCC}} \color[HTML]{000000} 0.1 (0.0-0.3) & {\cellcolor[HTML]{FEFEFE}} \color[HTML]{000000} {\cellcolor[HTML]{FEDCCD}} \color[HTML]{000000} 0.1 (0.1-0.2) & {\cellcolor[HTML]{FEFEFE}} \color[HTML]{000000} {\cellcolor[HTML]{FEE2D5}} \color[HTML]{000000} 0.1 (0.1-0.2) & {\cellcolor[HTML]{FEFEFE}} \color[HTML]{000000} {\cellcolor[HTML]{FFEDE5}} \color[HTML]{000000} 0.0 (0.0-0.1) & {\cellcolor[HTML]{FEFEFE}} \color[HTML]{000000} {\cellcolor[HTML]{FEE5D9}} \color[HTML]{000000} 0.1 (0.0-0.1) & {\cellcolor[HTML]{FEFEFE}} \color[HTML]{000000} {\cellcolor[HTML]{FDC7B2}} \color[HTML]{000000} 0.2 (0.2-0.3) & {\cellcolor[HTML]{FEFEFE}} \color[HTML]{000000} {\cellcolor[HTML]{FFEDE5}} \color[HTML]{000000} 0.1 (0.0-0.2) & {\cellcolor[HTML]{FEFEFE}} \color[HTML]{000000} {\cellcolor[HTML]{FFEDE5}} \color[HTML]{000000} 0.1 (0.0-0.3) & {\cellcolor[HTML]{FEFEFE}} \color[HTML]{000000} {\cellcolor[HTML]{FFEDE5}} \color[HTML]{000000} 0.1 (0.0-0.2) \\
{\cellcolor[HTML]{FEFEFE}} \color[HTML]{000000} 129 & {\cellcolor[HTML]{FEFEFE}} \color[HTML]{000000} $\text{A}^\text{306}_\text{OP1}$-$\text{O}_\text{ph}$ & {\cellcolor[HTML]{FEFEFE}} \color[HTML]{000000} {\cellcolor{white}}  & {\cellcolor[HTML]{FEFEFE}} \color[HTML]{000000}  & {\cellcolor[HTML]{FEFEFE}} \color[HTML]{000000} {\cellcolor[HTML]{000000}} \color[HTML]{F1F1F1} {\cellcolor{white}}  & {\cellcolor[HTML]{FEFEFE}} \color[HTML]{000000} {\cellcolor[HTML]{FDCDB9}} \color[HTML]{000000} 0.2 & {\cellcolor[HTML]{FEFEFE}} \color[HTML]{000000} {\cellcolor[HTML]{000000}} \color[HTML]{F1F1F1} {\cellcolor{white}}  & {\cellcolor[HTML]{FEFEFE}} \color[HTML]{000000} {\cellcolor[HTML]{FCBBA1}} \color[HTML]{000000} 0.2 & {\cellcolor[HTML]{FEFEFE}} \color[HTML]{000000} {\cellcolor[HTML]{000000}} \color[HTML]{F1F1F1} {\cellcolor{white}}  & {\cellcolor[HTML]{FEFEFE}} \color[HTML]{000000} {\cellcolor[HTML]{000000}} \color[HTML]{F1F1F1} {\cellcolor{white}}  & {\cellcolor[HTML]{FEFEFE}} \color[HTML]{000000} {\cellcolor[HTML]{FFF5F0}} \color[HTML]{000000} 0.0 (0.0-0.0) & {\cellcolor[HTML]{FEFEFE}} \color[HTML]{000000} {\cellcolor[HTML]{FFF4EE}} \color[HTML]{000000} 0.0 (0.0-0.0) & {\cellcolor[HTML]{FEFEFE}} \color[HTML]{000000} {\cellcolor[HTML]{FCBBA1}} \color[HTML]{000000} 0.2 (0.0-0.6) & {\cellcolor[HTML]{FEFEFE}} \color[HTML]{000000} {\cellcolor[HTML]{FFEFE8}} \color[HTML]{000000} 0.0 (0.0-0.1) & {\cellcolor[HTML]{FEFEFE}} \color[HTML]{000000} {\cellcolor[HTML]{FDD7C6}} \color[HTML]{000000} 0.2 (0.0-0.4) & {\cellcolor[HTML]{FEFEFE}} \color[HTML]{000000} {\cellcolor[HTML]{FFF5F0}} \color[HTML]{000000} 0.0 (0.0-0.0) & {\cellcolor[HTML]{FEFEFE}} \color[HTML]{000000} {\cellcolor[HTML]{FFF5F0}} \color[HTML]{000000} 0.0 (0.0-0.0) & {\cellcolor[HTML]{FEFEFE}} \color[HTML]{000000} {\cellcolor[HTML]{FFF5F0}} \color[HTML]{000000} 0.0 (0.0-0.0) \\
{\cellcolor[HTML]{FEFEFE}} \color[HTML]{000000} 130 & {\cellcolor[HTML]{FEFEFE}} \color[HTML]{000000} $\text{G}^\text{329}_\text{OP2}$-$\text{O}_\text{ph}$ & {\cellcolor[HTML]{FEFEFE}} \color[HTML]{000000} {\cellcolor{white}}  & {\cellcolor[HTML]{FEFEFE}} \color[HTML]{000000}  & {\cellcolor[HTML]{FEFEFE}} \color[HTML]{000000} {\cellcolor[HTML]{000000}} \color[HTML]{F1F1F1} {\cellcolor{white}}  & {\cellcolor[HTML]{FEFEFE}} \color[HTML]{000000} {\cellcolor[HTML]{000000}} \color[HTML]{F1F1F1} {\cellcolor{white}}  & {\cellcolor[HTML]{FEFEFE}} \color[HTML]{000000} {\cellcolor[HTML]{FEEAE1}} \color[HTML]{000000} 0.1 & {\cellcolor[HTML]{FEFEFE}} \color[HTML]{000000} {\cellcolor[HTML]{FEE0D2}} \color[HTML]{000000} 0.1 & {\cellcolor[HTML]{FEFEFE}} \color[HTML]{000000} {\cellcolor[HTML]{000000}} \color[HTML]{F1F1F1} {\cellcolor{white}}  & {\cellcolor[HTML]{FEFEFE}} \color[HTML]{000000} {\cellcolor[HTML]{FFF2EB}} \color[HTML]{000000} 0.0 (0.0-0.1) & {\cellcolor[HTML]{FEFEFE}} \color[HTML]{000000} {\cellcolor[HTML]{FFF3ED}} \color[HTML]{000000} 0.0 (0.0-0.0) & {\cellcolor[HTML]{FEFEFE}} \color[HTML]{000000} {\cellcolor[HTML]{FFEBE2}} \color[HTML]{000000} 0.1 (0.0-0.2) & {\cellcolor[HTML]{FEFEFE}} \color[HTML]{000000} {\cellcolor[HTML]{000000}} \color[HTML]{F1F1F1} {\cellcolor{white}}  & {\cellcolor[HTML]{FEFEFE}} \color[HTML]{000000} {\cellcolor[HTML]{FEE5D8}} \color[HTML]{000000} 0.1 (0.0-0.3) & {\cellcolor[HTML]{FEFEFE}} \color[HTML]{000000} {\cellcolor[HTML]{FFEBE2}} \color[HTML]{000000} 0.1 (0.0-0.1) & {\cellcolor[HTML]{FEFEFE}} \color[HTML]{000000} {\cellcolor[HTML]{000000}} \color[HTML]{F1F1F1} {\cellcolor{white}} {\cellcolor[HTML]{AAAAAA}}  & {\cellcolor[HTML]{FEFEFE}} \color[HTML]{000000} {\cellcolor[HTML]{FCAB8F}} \color[HTML]{000000} 0.3 (0.2-0.4) & {\cellcolor[HTML]{FEFEFE}} \color[HTML]{000000} {\cellcolor[HTML]{FED9C9}} \color[HTML]{000000} 0.2 (0.0-0.2) \\
{\cellcolor[HTML]{FEFEFE}} \color[HTML]{000000} 131 & {\cellcolor[HTML]{FEFEFE}} \color[HTML]{000000} $\text{G}^\text{313}_\text{OP1}$-$\text{O}_\text{ph}$ & {\cellcolor[HTML]{FEFEFE}} \color[HTML]{000000} {\cellcolor{white}}  & {\cellcolor[HTML]{FEFEFE}} \color[HTML]{000000}  & {\cellcolor[HTML]{FEFEFE}} \color[HTML]{000000} {\cellcolor[HTML]{000000}} \color[HTML]{F1F1F1} {\cellcolor{white}}  & {\cellcolor[HTML]{FEFEFE}} \color[HTML]{000000} {\cellcolor[HTML]{FEEAE1}} \color[HTML]{000000} 0.1 & {\cellcolor[HTML]{FEFEFE}} \color[HTML]{000000} {\cellcolor[HTML]{000000}} \color[HTML]{F1F1F1} {\cellcolor{white}}  & {\cellcolor[HTML]{FEFEFE}} \color[HTML]{000000} {\cellcolor[HTML]{000000}} \color[HTML]{F1F1F1} {\cellcolor{white}}  & {\cellcolor[HTML]{FEFEFE}} \color[HTML]{000000} {\cellcolor[HTML]{000000}} \color[HTML]{F1F1F1} {\cellcolor{white}}  & {\cellcolor[HTML]{FEFEFE}} \color[HTML]{000000} {\cellcolor[HTML]{FFECE4}} \color[HTML]{000000} 0.1 (0.0-0.1) & {\cellcolor[HTML]{FEFEFE}} \color[HTML]{000000} {\cellcolor[HTML]{FFF1EA}} \color[HTML]{000000} 0.0 (0.0-0.0) & {\cellcolor[HTML]{FEFEFE}} \color[HTML]{000000} {\cellcolor[HTML]{FFECE3}} \color[HTML]{000000} 0.1 (0.0-0.1) & {\cellcolor[HTML]{FEFEFE}} \color[HTML]{000000} {\cellcolor[HTML]{FEE9DF}} \color[HTML]{000000} 0.1 (0.0-0.2) & {\cellcolor[HTML]{FEFEFE}} \color[HTML]{000000} {\cellcolor[HTML]{FEE3D7}} \color[HTML]{000000} 0.1 (0.0-0.2) & {\cellcolor[HTML]{FEFEFE}} \color[HTML]{000000} {\cellcolor[HTML]{FEE3D6}} \color[HTML]{000000} 0.1 (0.1-0.2) & {\cellcolor[HTML]{FEFEFE}} \color[HTML]{000000} {\cellcolor[HTML]{FFF5F0}} \color[HTML]{000000} 0.0 (0.0-0.0) & {\cellcolor[HTML]{FEFEFE}} \color[HTML]{000000} {\cellcolor[HTML]{FEE5D8}} \color[HTML]{000000} 0.1 (0.0-0.3) & {\cellcolor[HTML]{FEFEFE}} \color[HTML]{000000} {\cellcolor[HTML]{FCAB8F}} \color[HTML]{000000} 0.3 (0.0-0.5) \\
{\cellcolor[HTML]{FEFEFE}} \color[HTML]{000000} 132 & {\cellcolor[HTML]{FEFEFE}} \color[HTML]{000000} $\text{C}^\text{216}_\text{OP2}$-$\text{O}_\text{ph}$ & {\cellcolor[HTML]{FEFEFE}} \color[HTML]{000000} {\cellcolor{white}}  & {\cellcolor[HTML]{FEFEFE}} \color[HTML]{000000}  & {\cellcolor[HTML]{FEFEFE}} \color[HTML]{000000} {\cellcolor[HTML]{000000}} \color[HTML]{F1F1F1} {\cellcolor{white}}  & {\cellcolor[HTML]{FEFEFE}} \color[HTML]{000000} {\cellcolor[HTML]{FEEAE1}} \color[HTML]{000000} 0.1 & {\cellcolor[HTML]{FEFEFE}} \color[HTML]{000000} {\cellcolor[HTML]{000000}} \color[HTML]{F1F1F1} {\cellcolor{white}}  & {\cellcolor[HTML]{FEFEFE}} \color[HTML]{000000} {\cellcolor[HTML]{000000}} \color[HTML]{F1F1F1} {\cellcolor{white}}  & {\cellcolor[HTML]{FEFEFE}} \color[HTML]{000000} {\cellcolor[HTML]{000000}} \color[HTML]{F1F1F1} {\cellcolor{white}}  & {\cellcolor[HTML]{FEFEFE}} \color[HTML]{000000} {\cellcolor[HTML]{000000}} \color[HTML]{F1F1F1} {\cellcolor{white}}  & {\cellcolor[HTML]{FEFEFE}} \color[HTML]{000000} {\cellcolor[HTML]{FEE5D8}} \color[HTML]{000000} 0.1 (0.0-0.2) & {\cellcolor[HTML]{FEFEFE}} \color[HTML]{000000} {\cellcolor[HTML]{FFF0E9}} \color[HTML]{000000} 0.0 (0.0-0.1) & {\cellcolor[HTML]{FEFEFE}} \color[HTML]{000000} {\cellcolor[HTML]{FEE0D2}} \color[HTML]{000000} 0.1 (0.0-0.4) & {\cellcolor[HTML]{FEFEFE}} \color[HTML]{000000} {\cellcolor[HTML]{FEE0D2}} \color[HTML]{000000} 0.1 (0.0-0.3) & {\cellcolor[HTML]{FEFEFE}} \color[HTML]{000000} {\cellcolor[HTML]{FFEEE6}} \color[HTML]{000000} 0.0 (0.0-0.1) & {\cellcolor[HTML]{FEFEFE}} \color[HTML]{000000} {\cellcolor[HTML]{FED9C9}} \color[HTML]{000000} 0.2 (0.0-0.3) & {\cellcolor[HTML]{FEFEFE}} \color[HTML]{000000} {\cellcolor[HTML]{FEE5D8}} \color[HTML]{000000} 0.1 (0.0-0.2) & {\cellcolor[HTML]{FEFEFE}} \color[HTML]{000000} {\cellcolor[HTML]{FED9C9}} \color[HTML]{000000} 0.2 (0.0-0.3) \\
{\cellcolor[HTML]{FEFEFE}} \color[HTML]{000000} 133 & {\cellcolor[HTML]{FEFEFE}} \color[HTML]{000000} $\text{C}^\text{98}_\text{OP2}$-$\text{O}_\text{ph}$ & {\cellcolor[HTML]{FEFEFE}} \color[HTML]{000000} {\cellcolor{white}}  & {\cellcolor[HTML]{FEFEFE}} \color[HTML]{000000}  & {\cellcolor[HTML]{FEFEFE}} \color[HTML]{000000} {\cellcolor[HTML]{000000}} \color[HTML]{F1F1F1} {\cellcolor{white}}  & {\cellcolor[HTML]{FEFEFE}} \color[HTML]{000000} {\cellcolor[HTML]{000000}} \color[HTML]{F1F1F1} {\cellcolor{white}}  & {\cellcolor[HTML]{FEFEFE}} \color[HTML]{000000} {\cellcolor[HTML]{000000}} \color[HTML]{F1F1F1} {\cellcolor{white}}  & {\cellcolor[HTML]{FEFEFE}} \color[HTML]{000000} {\cellcolor[HTML]{000000}} \color[HTML]{F1F1F1} {\cellcolor{white}}  & {\cellcolor[HTML]{FEFEFE}} \color[HTML]{000000} {\cellcolor[HTML]{FEEAE1}} \color[HTML]{000000} 0.1 (0.0-0.2) & {\cellcolor[HTML]{FEFEFE}} \color[HTML]{000000} {\cellcolor[HTML]{FCA183}} \color[HTML]{000000} 0.3 (0.2-0.6) & {\cellcolor[HTML]{FEFEFE}} \color[HTML]{000000} {\cellcolor[HTML]{FEE7DC}} \color[HTML]{000000} 0.1 (0.0-0.2) & {\cellcolor[HTML]{FEFEFE}} \color[HTML]{000000} {\cellcolor[HTML]{FC9373}} \color[HTML]{000000} 0.4 (0.2-0.5) & {\cellcolor[HTML]{FEFEFE}} \color[HTML]{000000} {\cellcolor[HTML]{000000}} \color[HTML]{F1F1F1} {\cellcolor{white}}  & {\cellcolor[HTML]{FEFEFE}} \color[HTML]{000000} {\cellcolor[HTML]{FFF0E9}} \color[HTML]{000000} 0.0 (0.0-0.1) & {\cellcolor[HTML]{FEFEFE}} \color[HTML]{000000} {\cellcolor[HTML]{FFF3ED}} \color[HTML]{000000} 0.0 (0.0-0.0) & {\cellcolor[HTML]{FEFEFE}} \color[HTML]{000000} {\cellcolor[HTML]{000000}} \color[HTML]{F1F1F1} {\cellcolor{white}} {\cellcolor[HTML]{AAAAAA}}  & {\cellcolor[HTML]{FEFEFE}} \color[HTML]{000000} {\cellcolor[HTML]{000000}} \color[HTML]{F1F1F1} {\cellcolor{white}} {\cellcolor[HTML]{AAAAAA}}  & {\cellcolor[HTML]{FEFEFE}} \color[HTML]{000000} {\cellcolor[HTML]{000000}} \color[HTML]{F1F1F1} {\cellcolor{white}} {\cellcolor[HTML]{AAAAAA}}  \\
{\cellcolor[HTML]{FEFEFE}} \color[HTML]{000000} 134 & {\cellcolor[HTML]{FEFEFE}} \color[HTML]{000000} $\text{G}^\text{23}_\text{OP1}$-$\text{O}_\text{ph}$ & {\cellcolor[HTML]{FEFEFE}} \color[HTML]{000000} {\cellcolor{white}}  & {\cellcolor[HTML]{FEFEFE}} \color[HTML]{000000}  & {\cellcolor[HTML]{FEFEFE}} \color[HTML]{000000} {\cellcolor[HTML]{000000}} \color[HTML]{F1F1F1} {\cellcolor{white}}  & {\cellcolor[HTML]{FEFEFE}} \color[HTML]{000000} {\cellcolor[HTML]{000000}} \color[HTML]{F1F1F1} {\cellcolor{white}}  & {\cellcolor[HTML]{FEFEFE}} \color[HTML]{000000} {\cellcolor[HTML]{FEEAE1}} \color[HTML]{000000} 0.1 & {\cellcolor[HTML]{FEFEFE}} \color[HTML]{000000} {\cellcolor[HTML]{000000}} \color[HTML]{F1F1F1} {\cellcolor{white}}  & {\cellcolor[HTML]{FEFEFE}} \color[HTML]{000000} {\cellcolor[HTML]{000000}} \color[HTML]{F1F1F1} {\cellcolor{white}}  & {\cellcolor[HTML]{FEFEFE}} \color[HTML]{000000} {\cellcolor[HTML]{FEE2D5}} \color[HTML]{000000} 0.1 (0.0-0.3) & {\cellcolor[HTML]{FEFEFE}} \color[HTML]{000000} {\cellcolor[HTML]{FEE4D8}} \color[HTML]{000000} 0.1 (0.0-0.2) & {\cellcolor[HTML]{FEFEFE}} \color[HTML]{000000} {\cellcolor[HTML]{FDCBB6}} \color[HTML]{000000} 0.2 (0.1-0.3) & {\cellcolor[HTML]{FEFEFE}} \color[HTML]{000000} {\cellcolor[HTML]{000000}} \color[HTML]{F1F1F1} {\cellcolor{white}}  & {\cellcolor[HTML]{FEFEFE}} \color[HTML]{000000} {\cellcolor[HTML]{FDD4C2}} \color[HTML]{000000} 0.2 (0.1-0.3) & {\cellcolor[HTML]{FEFEFE}} \color[HTML]{000000} {\cellcolor[HTML]{FCC1A8}} \color[HTML]{000000} 0.2 (0.1-0.4) & {\cellcolor[HTML]{FEFEFE}} \color[HTML]{000000} {\cellcolor[HTML]{000000}} \color[HTML]{F1F1F1} {\cellcolor{white}} {\cellcolor[HTML]{AAAAAA}}  & {\cellcolor[HTML]{FEFEFE}} \color[HTML]{000000} {\cellcolor[HTML]{000000}} \color[HTML]{F1F1F1} {\cellcolor{white}} {\cellcolor[HTML]{AAAAAA}}  & {\cellcolor[HTML]{FEFEFE}} \color[HTML]{000000} {\cellcolor[HTML]{000000}} \color[HTML]{F1F1F1} {\cellcolor{white}} {\cellcolor[HTML]{AAAAAA}}  \\
{\cellcolor[HTML]{FEFEFE}} \color[HTML]{000000} 135 & {\cellcolor[HTML]{FEFEFE}} \color[HTML]{000000} $\text{G}^\text{73}_\text{OP1}$-$\text{O}_\text{ph}$ & {\cellcolor[HTML]{FEFEFE}} \color[HTML]{000000} {\cellcolor{white}}  & {\cellcolor[HTML]{FEFEFE}} \color[HTML]{000000}  & {\cellcolor[HTML]{FEFEFE}} \color[HTML]{000000} {\cellcolor[HTML]{000000}} \color[HTML]{F1F1F1} {\cellcolor{white}}  & {\cellcolor[HTML]{FEFEFE}} \color[HTML]{000000} {\cellcolor[HTML]{000000}} \color[HTML]{F1F1F1} {\cellcolor{white}}  & {\cellcolor[HTML]{FEFEFE}} \color[HTML]{000000} {\cellcolor[HTML]{000000}} \color[HTML]{F1F1F1} {\cellcolor{white}}  & {\cellcolor[HTML]{FEFEFE}} \color[HTML]{000000} {\cellcolor[HTML]{000000}} \color[HTML]{F1F1F1} {\cellcolor{white}}  & {\cellcolor[HTML]{FEFEFE}} \color[HTML]{000000} {\cellcolor[HTML]{000000}} \color[HTML]{F1F1F1} {\cellcolor{white}}  & {\cellcolor[HTML]{FEFEFE}} \color[HTML]{000000} {\cellcolor[HTML]{FEE8DD}} \color[HTML]{000000} 0.1 (0.0-0.2) & {\cellcolor[HTML]{FEFEFE}} \color[HTML]{000000} {\cellcolor[HTML]{FFF4EE}} \color[HTML]{000000} 0.0 (0.0-0.0) & {\cellcolor[HTML]{FEFEFE}} \color[HTML]{000000} {\cellcolor[HTML]{FC7F5F}} \color[HTML]{F1F1F1} 0.4 (0.3-0.6) & {\cellcolor[HTML]{FEFEFE}} \color[HTML]{000000} {\cellcolor[HTML]{000000}} \color[HTML]{F1F1F1} {\cellcolor{white}}  & {\cellcolor[HTML]{FEFEFE}} \color[HTML]{000000} {\cellcolor[HTML]{FEE2D5}} \color[HTML]{000000} 0.1 (0.0-0.3) & {\cellcolor[HTML]{FEFEFE}} \color[HTML]{000000} {\cellcolor[HTML]{FCBFA7}} \color[HTML]{000000} 0.2 (0.2-0.3) & {\cellcolor[HTML]{FEFEFE}} \color[HTML]{000000} {\cellcolor[HTML]{000000}} \color[HTML]{F1F1F1} {\cellcolor{white}} {\cellcolor[HTML]{AAAAAA}}  & {\cellcolor[HTML]{FEFEFE}} \color[HTML]{000000} {\cellcolor[HTML]{000000}} \color[HTML]{F1F1F1} {\cellcolor{white}} {\cellcolor[HTML]{AAAAAA}}  & {\cellcolor[HTML]{FEFEFE}} \color[HTML]{000000} {\cellcolor[HTML]{000000}} \color[HTML]{F1F1F1} {\cellcolor{white}} {\cellcolor[HTML]{AAAAAA}}  \\
{\cellcolor[HTML]{FEFEFE}} \color[HTML]{000000} 136 & {\cellcolor[HTML]{FEFEFE}} \color[HTML]{000000} $\text{U}^\text{168}_\text{OP2}$-$\text{O}_\text{ph}$ & {\cellcolor[HTML]{FEFEFE}} \color[HTML]{000000} {\cellcolor{white}}  & {\cellcolor[HTML]{FEFEFE}} \color[HTML]{000000}  & {\cellcolor[HTML]{FEFEFE}} \color[HTML]{000000} {\cellcolor[HTML]{000000}} \color[HTML]{F1F1F1} {\cellcolor{white}}  & {\cellcolor[HTML]{FEFEFE}} \color[HTML]{000000} {\cellcolor[HTML]{000000}} \color[HTML]{F1F1F1} {\cellcolor{white}}  & {\cellcolor[HTML]{FEFEFE}} \color[HTML]{000000} {\cellcolor[HTML]{000000}} \color[HTML]{F1F1F1} {\cellcolor{white}}  & {\cellcolor[HTML]{FEFEFE}} \color[HTML]{000000} {\cellcolor[HTML]{FCBBA1}} \color[HTML]{000000} 0.2 & {\cellcolor[HTML]{FEFEFE}} \color[HTML]{000000} {\cellcolor[HTML]{000000}} \color[HTML]{F1F1F1} {\cellcolor{white}}  & {\cellcolor[HTML]{FEFEFE}} \color[HTML]{000000} {\cellcolor[HTML]{000000}} \color[HTML]{F1F1F1} {\cellcolor{white}}  & {\cellcolor[HTML]{FEFEFE}} \color[HTML]{000000} {\cellcolor[HTML]{FEE8DE}} \color[HTML]{000000} 0.1 (0.0-0.2) & {\cellcolor[HTML]{FEFEFE}} \color[HTML]{000000} {\cellcolor[HTML]{FFEEE7}} \color[HTML]{000000} 0.0 (0.0-0.1) & {\cellcolor[HTML]{FEFEFE}} \color[HTML]{000000} {\cellcolor[HTML]{FFF2EC}} \color[HTML]{000000} 0.0 (0.0-0.1) & {\cellcolor[HTML]{FEFEFE}} \color[HTML]{000000} {\cellcolor[HTML]{FEE3D6}} \color[HTML]{000000} 0.1 (0.0-0.4) & {\cellcolor[HTML]{FEFEFE}} \color[HTML]{000000} {\cellcolor[HTML]{FCB296}} \color[HTML]{000000} 0.3 (0.1-0.5) & {\cellcolor[HTML]{FEFEFE}} \color[HTML]{000000} {\cellcolor[HTML]{FEE5D8}} \color[HTML]{000000} 0.1 (0.0-0.2) & {\cellcolor[HTML]{FEFEFE}} \color[HTML]{000000} {\cellcolor[HTML]{FFF5F0}} \color[HTML]{000000} 0.0 (0.0-0.0) & {\cellcolor[HTML]{FEFEFE}} \color[HTML]{000000} {\cellcolor[HTML]{FFF5F0}} \color[HTML]{000000} 0.0 (0.0-0.0) \\
{\cellcolor[HTML]{FEFEFE}} \color[HTML]{000000} 137 & {\cellcolor[HTML]{FEFEFE}} \color[HTML]{000000} $\text{C}^\text{332}_\text{OP1}$-$\text{O}_\text{ph}$ & {\cellcolor[HTML]{FEFEFE}} \color[HTML]{000000} {\cellcolor{white}}  & {\cellcolor[HTML]{FEFEFE}} \color[HTML]{000000}  & {\cellcolor[HTML]{FEFEFE}} \color[HTML]{000000} {\cellcolor[HTML]{000000}} \color[HTML]{F1F1F1} {\cellcolor{white}}  & {\cellcolor[HTML]{FEFEFE}} \color[HTML]{000000} {\cellcolor[HTML]{000000}} \color[HTML]{F1F1F1} {\cellcolor{white}}  & {\cellcolor[HTML]{FEFEFE}} \color[HTML]{000000} {\cellcolor[HTML]{000000}} \color[HTML]{F1F1F1} {\cellcolor{white}}  & {\cellcolor[HTML]{FEFEFE}} \color[HTML]{000000} {\cellcolor[HTML]{000000}} \color[HTML]{F1F1F1} {\cellcolor{white}}  & {\cellcolor[HTML]{FEFEFE}} \color[HTML]{000000} {\cellcolor[HTML]{FFECE3}} \color[HTML]{000000} 0.1 (0.0-0.2) & {\cellcolor[HTML]{FEFEFE}} \color[HTML]{000000} {\cellcolor[HTML]{FEE3D7}} \color[HTML]{000000} 0.1 (0.0-0.2) & {\cellcolor[HTML]{FEFEFE}} \color[HTML]{000000} {\cellcolor[HTML]{FEDFD0}} \color[HTML]{000000} 0.1 (0.0-0.3) & {\cellcolor[HTML]{FEFEFE}} \color[HTML]{000000} {\cellcolor[HTML]{FCA183}} \color[HTML]{000000} 0.3 (0.2-0.5) & {\cellcolor[HTML]{FEFEFE}} \color[HTML]{000000} {\cellcolor[HTML]{000000}} \color[HTML]{F1F1F1} {\cellcolor{white}}  & {\cellcolor[HTML]{FEFEFE}} \color[HTML]{000000} {\cellcolor[HTML]{FEE5D9}} \color[HTML]{000000} 0.1 (0.0-0.2) & {\cellcolor[HTML]{FEFEFE}} \color[HTML]{000000} {\cellcolor[HTML]{FEDCCD}} \color[HTML]{000000} 0.1 (0.1-0.2) & {\cellcolor[HTML]{FEFEFE}} \color[HTML]{000000} {\cellcolor[HTML]{000000}} \color[HTML]{F1F1F1} {\cellcolor{white}} {\cellcolor[HTML]{AAAAAA}}  & {\cellcolor[HTML]{FEFEFE}} \color[HTML]{000000} {\cellcolor[HTML]{000000}} \color[HTML]{F1F1F1} {\cellcolor{white}} {\cellcolor[HTML]{AAAAAA}}  & {\cellcolor[HTML]{FEFEFE}} \color[HTML]{000000} {\cellcolor[HTML]{000000}} \color[HTML]{F1F1F1} {\cellcolor{white}} {\cellcolor[HTML]{AAAAAA}}  \\
{\cellcolor[HTML]{FEFEFE}} \color[HTML]{000000} 138 & {\cellcolor[HTML]{FEFEFE}} \color[HTML]{000000} $\text{A}^\text{122}_\text{OP1}$,$\text{A}^\text{123}_\text{OP1}$-\textit{cis}-2$\text{O}_\text{ph}$ & {\cellcolor[HTML]{FEFEFE}} \color[HTML]{000000} {\cellcolor{white}}  & {\cellcolor[HTML]{FEFEFE}} \color[HTML]{000000}  & {\cellcolor[HTML]{FEFEFE}} \color[HTML]{000000} {\cellcolor[HTML]{000000}} \color[HTML]{F1F1F1} {\cellcolor{white}}  & {\cellcolor[HTML]{FEFEFE}} \color[HTML]{000000} {\cellcolor[HTML]{000000}} \color[HTML]{F1F1F1} {\cellcolor{white}}  & {\cellcolor[HTML]{FEFEFE}} \color[HTML]{000000} {\cellcolor[HTML]{000000}} \color[HTML]{F1F1F1} {\cellcolor{white}}  & {\cellcolor[HTML]{FEFEFE}} \color[HTML]{000000} {\cellcolor[HTML]{000000}} \color[HTML]{F1F1F1} {\cellcolor{white}}  & {\cellcolor[HTML]{FEFEFE}} \color[HTML]{000000} {\cellcolor[HTML]{000000}} \color[HTML]{F1F1F1} {\cellcolor{white}}  & {\cellcolor[HTML]{FEFEFE}} \color[HTML]{000000} {\cellcolor[HTML]{FEE4D8}} \color[HTML]{000000} 0.1 (0.0-0.3) & {\cellcolor[HTML]{FEFEFE}} \color[HTML]{000000} {\cellcolor[HTML]{000000}} \color[HTML]{F1F1F1} {\cellcolor{white}}  & {\cellcolor[HTML]{FEFEFE}} \color[HTML]{000000} {\cellcolor[HTML]{FDC5AE}} \color[HTML]{000000} 0.2 (0.1-0.4) & {\cellcolor[HTML]{FEFEFE}} \color[HTML]{000000} {\cellcolor[HTML]{000000}} \color[HTML]{F1F1F1} {\cellcolor{white}}  & {\cellcolor[HTML]{FEFEFE}} \color[HTML]{000000} {\cellcolor[HTML]{FEE5D9}} \color[HTML]{000000} 0.1 (0.0-0.3) & {\cellcolor[HTML]{FEFEFE}} \color[HTML]{000000} {\cellcolor[HTML]{FCAD90}} \color[HTML]{000000} 0.3 (0.2-0.4) & {\cellcolor[HTML]{FEFEFE}} \color[HTML]{000000} {\cellcolor[HTML]{000000}} \color[HTML]{F1F1F1} {\cellcolor{white}} {\cellcolor[HTML]{AAAAAA}}  & {\cellcolor[HTML]{FEFEFE}} \color[HTML]{000000} {\cellcolor[HTML]{FFEDE5}} \color[HTML]{000000} 0.1 (0.0-0.1) & {\cellcolor[HTML]{FEFEFE}} \color[HTML]{000000} {\cellcolor[HTML]{FEE5D8}} \color[HTML]{000000} 0.1 (0.0-0.2) \\
{\cellcolor[HTML]{FEFEFE}} \color[HTML]{000000} 139 & {\cellcolor[HTML]{FEFEFE}} \color[HTML]{000000} $\text{A}^\text{283}_\text{OP2}$-$\text{O}_\text{ph}$ & {\cellcolor[HTML]{FEFEFE}} \color[HTML]{000000} {\cellcolor{white}}  & {\cellcolor[HTML]{FEFEFE}} \color[HTML]{000000}  & {\cellcolor[HTML]{FEFEFE}} \color[HTML]{000000} {\cellcolor[HTML]{000000}} \color[HTML]{F1F1F1} {\cellcolor{white}}  & {\cellcolor[HTML]{FEFEFE}} \color[HTML]{000000} {\cellcolor[HTML]{000000}} \color[HTML]{F1F1F1} {\cellcolor{white}}  & {\cellcolor[HTML]{FEFEFE}} \color[HTML]{000000} {\cellcolor[HTML]{000000}} \color[HTML]{F1F1F1} {\cellcolor{white}}  & {\cellcolor[HTML]{FEFEFE}} \color[HTML]{000000} {\cellcolor[HTML]{000000}} \color[HTML]{F1F1F1} {\cellcolor{white}}  & {\cellcolor[HTML]{FEFEFE}} \color[HTML]{000000} {\cellcolor[HTML]{000000}} \color[HTML]{F1F1F1} {\cellcolor{white}}  & {\cellcolor[HTML]{FEFEFE}} \color[HTML]{000000} {\cellcolor[HTML]{FEE2D5}} \color[HTML]{000000} 0.1 (0.0-0.2) & {\cellcolor[HTML]{FEFEFE}} \color[HTML]{000000} {\cellcolor[HTML]{FEE5D8}} \color[HTML]{000000} 0.1 (0.0-0.2) & {\cellcolor[HTML]{FEFEFE}} \color[HTML]{000000} {\cellcolor[HTML]{FED9C9}} \color[HTML]{000000} 0.2 (0.1-0.3) & {\cellcolor[HTML]{FEFEFE}} \color[HTML]{000000} {\cellcolor[HTML]{000000}} \color[HTML]{F1F1F1} {\cellcolor{white}}  & {\cellcolor[HTML]{FEFEFE}} \color[HTML]{000000} {\cellcolor[HTML]{FEE8DD}} \color[HTML]{000000} 0.1 (0.0-0.1) & {\cellcolor[HTML]{FEFEFE}} \color[HTML]{000000} {\cellcolor[HTML]{FDCBB6}} \color[HTML]{000000} 0.2 (0.1-0.3) & {\cellcolor[HTML]{FEFEFE}} \color[HTML]{000000} {\cellcolor[HTML]{000000}} \color[HTML]{F1F1F1} {\cellcolor{white}} {\cellcolor[HTML]{AAAAAA}}  & {\cellcolor[HTML]{FEFEFE}} \color[HTML]{000000} {\cellcolor[HTML]{FFEDE5}} \color[HTML]{000000} 0.1 (0.0-0.4) & {\cellcolor[HTML]{FEFEFE}} \color[HTML]{000000} {\cellcolor[HTML]{FED9C9}} \color[HTML]{000000} 0.2 (0.0-0.3) \\
{\cellcolor[HTML]{FEFEFE}} \color[HTML]{000000} 140 & {\cellcolor[HTML]{FEFEFE}} \color[HTML]{000000} $\text{A}^\text{122}_\text{OP1}$-$\text{O}_\text{ph}$ & {\cellcolor[HTML]{FEFEFE}} \color[HTML]{000000} {\cellcolor{white}}  & {\cellcolor[HTML]{FEFEFE}} \color[HTML]{000000}  & {\cellcolor[HTML]{FEFEFE}} \color[HTML]{000000} {\cellcolor[HTML]{000000}} \color[HTML]{F1F1F1} {\cellcolor{white}}  & {\cellcolor[HTML]{FEFEFE}} \color[HTML]{000000} {\cellcolor[HTML]{000000}} \color[HTML]{F1F1F1} {\cellcolor{white}}  & {\cellcolor[HTML]{FEFEFE}} \color[HTML]{000000} {\cellcolor[HTML]{000000}} \color[HTML]{F1F1F1} {\cellcolor{white}}  & {\cellcolor[HTML]{FEFEFE}} \color[HTML]{000000} {\cellcolor[HTML]{000000}} \color[HTML]{F1F1F1} {\cellcolor{white}}  & {\cellcolor[HTML]{FEFEFE}} \color[HTML]{000000} {\cellcolor[HTML]{FFF5F0}} \color[HTML]{000000} 0.0 (0.0-0.0) & {\cellcolor[HTML]{FEFEFE}} \color[HTML]{000000} {\cellcolor[HTML]{FDD4C2}} \color[HTML]{000000} 0.2 (0.1-0.3) & {\cellcolor[HTML]{FEFEFE}} \color[HTML]{000000} {\cellcolor[HTML]{FEE5D8}} \color[HTML]{000000} 0.1 (0.0-0.2) & {\cellcolor[HTML]{FEFEFE}} \color[HTML]{000000} {\cellcolor[HTML]{FDD2BF}} \color[HTML]{000000} 0.2 (0.1-0.2) & {\cellcolor[HTML]{FEFEFE}} \color[HTML]{000000} {\cellcolor[HTML]{000000}} \color[HTML]{F1F1F1} {\cellcolor{white}}  & {\cellcolor[HTML]{FEFEFE}} \color[HTML]{000000} {\cellcolor[HTML]{FDC9B3}} \color[HTML]{000000} 0.2 (0.1-0.3) & {\cellcolor[HTML]{FEFEFE}} \color[HTML]{000000} {\cellcolor[HTML]{FDCCB8}} \color[HTML]{000000} 0.2 (0.1-0.3) & {\cellcolor[HTML]{FEFEFE}} \color[HTML]{000000} {\cellcolor[HTML]{FFF5F0}} \color[HTML]{000000} 0.0 (0.0-0.0) & {\cellcolor[HTML]{FEFEFE}} \color[HTML]{000000} {\cellcolor[HTML]{000000}} \color[HTML]{F1F1F1} {\cellcolor{white}} {\cellcolor[HTML]{AAAAAA}}  & {\cellcolor[HTML]{FEFEFE}} \color[HTML]{000000} {\cellcolor[HTML]{000000}} \color[HTML]{F1F1F1} {\cellcolor{white}} {\cellcolor[HTML]{AAAAAA}}  \\
{\cellcolor[HTML]{FEFEFE}} \color[HTML]{000000} 141 & {\cellcolor[HTML]{FEFEFE}} \color[HTML]{000000} $\text{G}^\text{272}_\text{OP2}$-$\text{O}_\text{ph}$ & {\cellcolor[HTML]{FEFEFE}} \color[HTML]{000000} {\cellcolor{white}}  & {\cellcolor[HTML]{FEFEFE}} \color[HTML]{000000}  & {\cellcolor[HTML]{FEFEFE}} \color[HTML]{000000} {\cellcolor[HTML]{000000}} \color[HTML]{F1F1F1} {\cellcolor{white}}  & {\cellcolor[HTML]{FEFEFE}} \color[HTML]{000000} {\cellcolor[HTML]{FEEAE1}} \color[HTML]{000000} 0.1 & {\cellcolor[HTML]{FEFEFE}} \color[HTML]{000000} {\cellcolor[HTML]{000000}} \color[HTML]{F1F1F1} {\cellcolor{white}}  & {\cellcolor[HTML]{FEFEFE}} \color[HTML]{000000} {\cellcolor[HTML]{FEE0D2}} \color[HTML]{000000} 0.1 & {\cellcolor[HTML]{FEFEFE}} \color[HTML]{000000} {\cellcolor[HTML]{000000}} \color[HTML]{F1F1F1} {\cellcolor{white}}  & {\cellcolor[HTML]{FEFEFE}} \color[HTML]{000000} {\cellcolor[HTML]{000000}} \color[HTML]{F1F1F1} {\cellcolor{white}}  & {\cellcolor[HTML]{FEFEFE}} \color[HTML]{000000} {\cellcolor[HTML]{FFEEE7}} \color[HTML]{000000} 0.0 (0.0-0.1) & {\cellcolor[HTML]{FEFEFE}} \color[HTML]{000000} {\cellcolor[HTML]{FFF5F0}} \color[HTML]{000000} 0.0 (0.0-0.0) & {\cellcolor[HTML]{FEFEFE}} \color[HTML]{000000} {\cellcolor[HTML]{FEE0D2}} \color[HTML]{000000} 0.1 (0.0-0.4) & {\cellcolor[HTML]{FEFEFE}} \color[HTML]{000000} {\cellcolor[HTML]{FFF0E9}} \color[HTML]{000000} 0.0 (0.0-0.1) & {\cellcolor[HTML]{FEFEFE}} \color[HTML]{000000} {\cellcolor[HTML]{FFECE4}} \color[HTML]{000000} 0.1 (0.0-0.1) & {\cellcolor[HTML]{FEFEFE}} \color[HTML]{000000} {\cellcolor[HTML]{FDCAB5}} \color[HTML]{000000} 0.2 (0.2-0.3) & {\cellcolor[HTML]{FEFEFE}} \color[HTML]{000000} {\cellcolor[HTML]{FFF5F0}} \color[HTML]{000000} 0.0 (0.0-0.1) & {\cellcolor[HTML]{FEFEFE}} \color[HTML]{000000} {\cellcolor[HTML]{FDCAB5}} \color[HTML]{000000} 0.2 (0.2-0.4) \\
{\cellcolor[HTML]{FEFEFE}} \color[HTML]{000000} 142 & {\cellcolor[HTML]{FEFEFE}} \color[HTML]{000000} $\text{C}^\text{166}_\text{OP1}$-$\text{O}_\text{ph}$ & {\cellcolor[HTML]{FEFEFE}} \color[HTML]{000000} {\cellcolor{white}}  & {\cellcolor[HTML]{FEFEFE}} \color[HTML]{000000}  & {\cellcolor[HTML]{FEFEFE}} \color[HTML]{000000} {\cellcolor[HTML]{000000}} \color[HTML]{F1F1F1} {\cellcolor{white}}  & {\cellcolor[HTML]{FEFEFE}} \color[HTML]{000000} {\cellcolor[HTML]{000000}} \color[HTML]{F1F1F1} {\cellcolor{white}}  & {\cellcolor[HTML]{FEFEFE}} \color[HTML]{000000} {\cellcolor[HTML]{000000}} \color[HTML]{F1F1F1} {\cellcolor{white}}  & {\cellcolor[HTML]{FEFEFE}} \color[HTML]{000000} {\cellcolor[HTML]{FEE0D2}} \color[HTML]{000000} 0.1 & {\cellcolor[HTML]{FEFEFE}} \color[HTML]{000000} {\cellcolor[HTML]{FFF5F0}} \color[HTML]{000000} 0.0 (0.0-0.0) & {\cellcolor[HTML]{FEFEFE}} \color[HTML]{000000} {\cellcolor[HTML]{FEE9DF}} \color[HTML]{000000} 0.1 (0.0-0.2) & {\cellcolor[HTML]{FEFEFE}} \color[HTML]{000000} {\cellcolor[HTML]{FCBBA1}} \color[HTML]{000000} 0.3 (0.1-0.5) & {\cellcolor[HTML]{FEFEFE}} \color[HTML]{000000} {\cellcolor[HTML]{FFECE3}} \color[HTML]{000000} 0.1 (0.0-0.1) & {\cellcolor[HTML]{FEFEFE}} \color[HTML]{000000} {\cellcolor[HTML]{000000}} \color[HTML]{F1F1F1} {\cellcolor{white}}  & {\cellcolor[HTML]{FEFEFE}} \color[HTML]{000000} {\cellcolor[HTML]{FFF2EC}} \color[HTML]{000000} 0.0 (0.0-0.1) & {\cellcolor[HTML]{FEFEFE}} \color[HTML]{000000} {\cellcolor[HTML]{FFF4EF}} \color[HTML]{000000} 0.0 (0.0-0.0) & {\cellcolor[HTML]{FEFEFE}} \color[HTML]{000000} {\cellcolor[HTML]{FFEDE5}} \color[HTML]{000000} 0.1 (0.0-0.1) & {\cellcolor[HTML]{FEFEFE}} \color[HTML]{000000} {\cellcolor[HTML]{FEE5D8}} \color[HTML]{000000} 0.1 (0.0-0.3) & {\cellcolor[HTML]{FEFEFE}} \color[HTML]{000000} {\cellcolor[HTML]{FED9C9}} \color[HTML]{000000} 0.2 (0.0-0.4) \\
{\cellcolor[HTML]{FEFEFE}} \color[HTML]{000000} 143 & {\cellcolor[HTML]{FEFEFE}} \color[HTML]{000000} $\text{A}^\text{69}_\text{OP2}$-$\text{O}_\text{ph}$ & {\cellcolor[HTML]{FEFEFE}} \color[HTML]{000000} {\cellcolor{white}}  & {\cellcolor[HTML]{FEFEFE}} \color[HTML]{000000}  & {\cellcolor[HTML]{FEFEFE}} \color[HTML]{000000} {\cellcolor[HTML]{000000}} \color[HTML]{F1F1F1} {\cellcolor{white}}  & {\cellcolor[HTML]{FEFEFE}} \color[HTML]{000000} {\cellcolor[HTML]{000000}} \color[HTML]{F1F1F1} {\cellcolor{white}}  & {\cellcolor[HTML]{FEFEFE}} \color[HTML]{000000} {\cellcolor[HTML]{000000}} \color[HTML]{F1F1F1} {\cellcolor{white}}  & {\cellcolor[HTML]{FEFEFE}} \color[HTML]{000000} {\cellcolor[HTML]{000000}} \color[HTML]{F1F1F1} {\cellcolor{white}}  & {\cellcolor[HTML]{FEFEFE}} \color[HTML]{000000} {\cellcolor[HTML]{000000}} \color[HTML]{F1F1F1} {\cellcolor{white}}  & {\cellcolor[HTML]{FEFEFE}} \color[HTML]{000000} {\cellcolor[HTML]{FCAD90}} \color[HTML]{000000} 0.3 (0.2-0.4) & {\cellcolor[HTML]{FEFEFE}} \color[HTML]{000000} {\cellcolor[HTML]{FFECE3}} \color[HTML]{000000} 0.1 (0.0-0.1) & {\cellcolor[HTML]{FEFEFE}} \color[HTML]{000000} {\cellcolor[HTML]{FCC1A8}} \color[HTML]{000000} 0.2 (0.1-0.3) & {\cellcolor[HTML]{FEFEFE}} \color[HTML]{000000} {\cellcolor[HTML]{000000}} \color[HTML]{F1F1F1} {\cellcolor{white}}  & {\cellcolor[HTML]{FEFEFE}} \color[HTML]{000000} {\cellcolor[HTML]{FEEAE1}} \color[HTML]{000000} 0.1 (0.0-0.1) & {\cellcolor[HTML]{FEFEFE}} \color[HTML]{000000} {\cellcolor[HTML]{FDCEBB}} \color[HTML]{000000} 0.2 (0.1-0.3) & {\cellcolor[HTML]{FEFEFE}} \color[HTML]{000000} {\cellcolor[HTML]{000000}} \color[HTML]{F1F1F1} {\cellcolor{white}} {\cellcolor[HTML]{AAAAAA}}  & {\cellcolor[HTML]{FEFEFE}} \color[HTML]{000000} {\cellcolor[HTML]{000000}} \color[HTML]{F1F1F1} {\cellcolor{white}} {\cellcolor[HTML]{AAAAAA}}  & {\cellcolor[HTML]{FEFEFE}} \color[HTML]{000000} {\cellcolor[HTML]{000000}} \color[HTML]{F1F1F1} {\cellcolor{white}} {\cellcolor[HTML]{AAAAAA}}  \\
{\cellcolor[HTML]{FEFEFE}} \color[HTML]{000000} 144 & {\cellcolor[HTML]{FEFEFE}} \color[HTML]{000000} $\text{G}^\text{118}_\text{OP2}$-$\text{O}_\text{ph}$ & {\cellcolor[HTML]{FEFEFE}} \color[HTML]{000000} {\cellcolor{white}}  & {\cellcolor[HTML]{FEFEFE}} \color[HTML]{000000}  & {\cellcolor[HTML]{FEFEFE}} \color[HTML]{000000} {\cellcolor[HTML]{000000}} \color[HTML]{F1F1F1} {\cellcolor{white}}  & {\cellcolor[HTML]{FEFEFE}} \color[HTML]{000000} {\cellcolor[HTML]{000000}} \color[HTML]{F1F1F1} {\cellcolor{white}}  & {\cellcolor[HTML]{FEFEFE}} \color[HTML]{000000} {\cellcolor[HTML]{000000}} \color[HTML]{F1F1F1} {\cellcolor{white}}  & {\cellcolor[HTML]{FEFEFE}} \color[HTML]{000000} {\cellcolor[HTML]{000000}} \color[HTML]{F1F1F1} {\cellcolor{white}}  & {\cellcolor[HTML]{FEFEFE}} \color[HTML]{000000} {\cellcolor[HTML]{000000}} \color[HTML]{F1F1F1} {\cellcolor{white}}  & {\cellcolor[HTML]{FEFEFE}} \color[HTML]{000000} {\cellcolor[HTML]{FEE5D8}} \color[HTML]{000000} 0.1 (0.0-0.2) & {\cellcolor[HTML]{FEFEFE}} \color[HTML]{000000} {\cellcolor[HTML]{FEE3D6}} \color[HTML]{000000} 0.1 (0.0-0.3) & {\cellcolor[HTML]{FEFEFE}} \color[HTML]{000000} {\cellcolor[HTML]{FCC3AB}} \color[HTML]{000000} 0.2 (0.1-0.3) & {\cellcolor[HTML]{FEFEFE}} \color[HTML]{000000} {\cellcolor[HTML]{000000}} \color[HTML]{F1F1F1} {\cellcolor{white}}  & {\cellcolor[HTML]{FEFEFE}} \color[HTML]{000000} {\cellcolor[HTML]{FEE8DE}} \color[HTML]{000000} 0.1 (0.0-0.2) & {\cellcolor[HTML]{FEFEFE}} \color[HTML]{000000} {\cellcolor[HTML]{FCA588}} \color[HTML]{000000} 0.3 (0.2-0.5) & {\cellcolor[HTML]{FEFEFE}} \color[HTML]{000000} {\cellcolor[HTML]{000000}} \color[HTML]{F1F1F1} {\cellcolor{white}} {\cellcolor[HTML]{AAAAAA}}  & {\cellcolor[HTML]{FEFEFE}} \color[HTML]{000000} {\cellcolor[HTML]{000000}} \color[HTML]{F1F1F1} {\cellcolor{white}} {\cellcolor[HTML]{AAAAAA}}  & {\cellcolor[HTML]{FEFEFE}} \color[HTML]{000000} {\cellcolor[HTML]{000000}} \color[HTML]{F1F1F1} {\cellcolor{white}} {\cellcolor[HTML]{AAAAAA}}  \\
{\cellcolor[HTML]{FEFEFE}} \color[HTML]{000000} 145 & {\cellcolor[HTML]{FEFEFE}} \color[HTML]{000000} $\text{C}^\text{216}_\text{OP1}$-$\text{O}_\text{ph}$ & {\cellcolor[HTML]{FEFEFE}} \color[HTML]{000000} {\cellcolor{white}}  & {\cellcolor[HTML]{FEFEFE}} \color[HTML]{000000}  & {\cellcolor[HTML]{FEFEFE}} \color[HTML]{000000} {\cellcolor[HTML]{000000}} \color[HTML]{F1F1F1} {\cellcolor{white}}  & {\cellcolor[HTML]{FEFEFE}} \color[HTML]{000000} {\cellcolor[HTML]{000000}} \color[HTML]{F1F1F1} {\cellcolor{white}}  & {\cellcolor[HTML]{FEFEFE}} \color[HTML]{000000} {\cellcolor[HTML]{000000}} \color[HTML]{F1F1F1} {\cellcolor{white}}  & {\cellcolor[HTML]{FEFEFE}} \color[HTML]{000000} {\cellcolor[HTML]{000000}} \color[HTML]{F1F1F1} {\cellcolor{white}}  & {\cellcolor[HTML]{FEFEFE}} \color[HTML]{000000} {\cellcolor[HTML]{FEE1D3}} \color[HTML]{000000} 0.1 (0.0-0.4) & {\cellcolor[HTML]{FEFEFE}} \color[HTML]{000000} {\cellcolor[HTML]{FFF5F0}} \color[HTML]{000000} 0.0 (0.0-0.0) & {\cellcolor[HTML]{FEFEFE}} \color[HTML]{000000} {\cellcolor[HTML]{FFF0E8}} \color[HTML]{000000} 0.0 (0.0-0.1) & {\cellcolor[HTML]{FEFEFE}} \color[HTML]{000000} {\cellcolor[HTML]{FFF3ED}} \color[HTML]{000000} 0.0 (0.0-0.0) & {\cellcolor[HTML]{FEFEFE}} \color[HTML]{000000} {\cellcolor[HTML]{000000}} \color[HTML]{F1F1F1} {\cellcolor{white}}  & {\cellcolor[HTML]{FEFEFE}} \color[HTML]{000000} {\cellcolor[HTML]{FFF2EB}} \color[HTML]{000000} 0.0 (0.0-0.0) & {\cellcolor[HTML]{FEFEFE}} \color[HTML]{000000} {\cellcolor[HTML]{FFF2EC}} \color[HTML]{000000} 0.0 (0.0-0.0) & {\cellcolor[HTML]{FEFEFE}} \color[HTML]{000000} {\cellcolor[HTML]{FFF5F0}} \color[HTML]{000000} 0.0 (0.0-0.0) & {\cellcolor[HTML]{FEFEFE}} \color[HTML]{000000} {\cellcolor[HTML]{FCBBA1}} \color[HTML]{000000} 0.2 (0.1-0.3) & {\cellcolor[HTML]{FEFEFE}} \color[HTML]{000000} {\cellcolor[HTML]{FC9B7C}} \color[HTML]{000000} 0.4 (0.1-0.5) \\
{\cellcolor[HTML]{FEFEFE}} \color[HTML]{000000} 146 & {\cellcolor[HTML]{FEFEFE}} \color[HTML]{000000} $\text{G}^\text{312}_\text{OP2}$-$\text{O}_\text{ph}$ & {\cellcolor[HTML]{FEFEFE}} \color[HTML]{000000} {\cellcolor{white}}  & {\cellcolor[HTML]{FEFEFE}} \color[HTML]{000000}  & {\cellcolor[HTML]{FEFEFE}} \color[HTML]{000000} {\cellcolor[HTML]{000000}} \color[HTML]{F1F1F1} {\cellcolor{white}}  & {\cellcolor[HTML]{FEFEFE}} \color[HTML]{000000} {\cellcolor[HTML]{000000}} \color[HTML]{F1F1F1} {\cellcolor{white}}  & {\cellcolor[HTML]{FEFEFE}} \color[HTML]{000000} {\cellcolor[HTML]{FEEAE1}} \color[HTML]{000000} 0.1 & {\cellcolor[HTML]{FEFEFE}} \color[HTML]{000000} {\cellcolor[HTML]{000000}} \color[HTML]{F1F1F1} {\cellcolor{white}}  & {\cellcolor[HTML]{FEFEFE}} \color[HTML]{000000} {\cellcolor[HTML]{000000}} \color[HTML]{F1F1F1} {\cellcolor{white}}  & {\cellcolor[HTML]{FEFEFE}} \color[HTML]{000000} {\cellcolor[HTML]{FDCBB6}} \color[HTML]{000000} 0.2 (0.1-0.3) & {\cellcolor[HTML]{FEFEFE}} \color[HTML]{000000} {\cellcolor[HTML]{FEE1D4}} \color[HTML]{000000} 0.1 (0.0-0.2) & {\cellcolor[HTML]{FEFEFE}} \color[HTML]{000000} {\cellcolor[HTML]{FEE0D2}} \color[HTML]{000000} 0.1 (0.1-0.2) & {\cellcolor[HTML]{FEFEFE}} \color[HTML]{000000} {\cellcolor[HTML]{FFF3ED}} \color[HTML]{000000} 0.0 (0.0-0.0) & {\cellcolor[HTML]{FEFEFE}} \color[HTML]{000000} {\cellcolor[HTML]{FEDFD0}} \color[HTML]{000000} 0.1 (0.1-0.2) & {\cellcolor[HTML]{FEFEFE}} \color[HTML]{000000} {\cellcolor[HTML]{FDD5C4}} \color[HTML]{000000} 0.2 (0.1-0.3) & {\cellcolor[HTML]{FEFEFE}} \color[HTML]{000000} {\cellcolor[HTML]{000000}} \color[HTML]{F1F1F1} {\cellcolor{white}} {\cellcolor[HTML]{AAAAAA}}  & {\cellcolor[HTML]{FEFEFE}} \color[HTML]{000000} {\cellcolor[HTML]{000000}} \color[HTML]{F1F1F1} {\cellcolor{white}} {\cellcolor[HTML]{AAAAAA}}  & {\cellcolor[HTML]{FEFEFE}} \color[HTML]{000000} {\cellcolor[HTML]{FFF5F0}} \color[HTML]{000000} 0.0 (0.0-0.2) \\
{\cellcolor[HTML]{FEFEFE}} \color[HTML]{000000} 147 & {\cellcolor[HTML]{FEFEFE}} \color[HTML]{000000} $\text{U}^\text{202}_\text{OP1}$-$\text{O}_\text{ph}$ & {\cellcolor[HTML]{FEFEFE}} \color[HTML]{000000} {\cellcolor{white}}  & {\cellcolor[HTML]{FEFEFE}} \color[HTML]{000000}  & {\cellcolor[HTML]{FEFEFE}} \color[HTML]{000000} {\cellcolor[HTML]{000000}} \color[HTML]{F1F1F1} {\cellcolor{white}}  & {\cellcolor[HTML]{FEFEFE}} \color[HTML]{000000} {\cellcolor[HTML]{000000}} \color[HTML]{F1F1F1} {\cellcolor{white}}  & {\cellcolor[HTML]{FEFEFE}} \color[HTML]{000000} {\cellcolor[HTML]{000000}} \color[HTML]{F1F1F1} {\cellcolor{white}}  & {\cellcolor[HTML]{FEFEFE}} \color[HTML]{000000} {\cellcolor[HTML]{000000}} \color[HTML]{F1F1F1} {\cellcolor{white}}  & {\cellcolor[HTML]{FEFEFE}} \color[HTML]{000000} {\cellcolor[HTML]{FFF3ED}} \color[HTML]{000000} 0.0 (0.0-0.0) & {\cellcolor[HTML]{FEFEFE}} \color[HTML]{000000} {\cellcolor[HTML]{FEE7DC}} \color[HTML]{000000} 0.1 (0.0-0.2) & {\cellcolor[HTML]{FEFEFE}} \color[HTML]{000000} {\cellcolor[HTML]{FFEEE6}} \color[HTML]{000000} 0.0 (0.0-0.1) & {\cellcolor[HTML]{FEFEFE}} \color[HTML]{000000} {\cellcolor[HTML]{FEE8DE}} \color[HTML]{000000} 0.1 (0.0-0.1) & {\cellcolor[HTML]{FEFEFE}} \color[HTML]{000000} {\cellcolor[HTML]{000000}} \color[HTML]{F1F1F1} {\cellcolor{white}}  & {\cellcolor[HTML]{FEFEFE}} \color[HTML]{000000} {\cellcolor[HTML]{FFF0E9}} \color[HTML]{000000} 0.0 (0.0-0.1) & {\cellcolor[HTML]{FEFEFE}} \color[HTML]{000000} {\cellcolor[HTML]{FEEAE1}} \color[HTML]{000000} 0.1 (0.0-0.1) & {\cellcolor[HTML]{FEFEFE}} \color[HTML]{000000} {\cellcolor[HTML]{FCBBA1}} \color[HTML]{000000} 0.2 (0.2-0.4) & {\cellcolor[HTML]{FEFEFE}} \color[HTML]{000000} {\cellcolor[HTML]{FED9C9}} \color[HTML]{000000} 0.2 (0.1-0.2) & {\cellcolor[HTML]{FEFEFE}} \color[HTML]{000000} {\cellcolor[HTML]{FEE5D8}} \color[HTML]{000000} 0.1 (0.0-0.2) \\
{\cellcolor[HTML]{FEFEFE}} \color[HTML]{000000} 148 & {\cellcolor[HTML]{FEFEFE}} \color[HTML]{000000} $\text{U}^\text{356}_\text{OP2}$-$\text{O}_\text{ph}$ & {\cellcolor[HTML]{FEFEFE}} \color[HTML]{000000} {\cellcolor{white}}  & {\cellcolor[HTML]{FEFEFE}} \color[HTML]{000000}  & {\cellcolor[HTML]{FEFEFE}} \color[HTML]{000000} {\cellcolor[HTML]{000000}} \color[HTML]{F1F1F1} {\cellcolor{white}}  & {\cellcolor[HTML]{FEFEFE}} \color[HTML]{000000} {\cellcolor[HTML]{000000}} \color[HTML]{F1F1F1} {\cellcolor{white}}  & {\cellcolor[HTML]{FEFEFE}} \color[HTML]{000000} {\cellcolor[HTML]{000000}} \color[HTML]{F1F1F1} {\cellcolor{white}}  & {\cellcolor[HTML]{FEFEFE}} \color[HTML]{000000} {\cellcolor[HTML]{000000}} \color[HTML]{F1F1F1} {\cellcolor{white}}  & {\cellcolor[HTML]{FEFEFE}} \color[HTML]{000000} {\cellcolor[HTML]{000000}} \color[HTML]{F1F1F1} {\cellcolor{white}}  & {\cellcolor[HTML]{FEFEFE}} \color[HTML]{000000} {\cellcolor[HTML]{FEE8DE}} \color[HTML]{000000} 0.1 (0.0-0.1) & {\cellcolor[HTML]{FEFEFE}} \color[HTML]{000000} {\cellcolor[HTML]{FFF4EF}} \color[HTML]{000000} 0.0 (0.0-0.0) & {\cellcolor[HTML]{FEFEFE}} \color[HTML]{000000} {\cellcolor[HTML]{FDD4C2}} \color[HTML]{000000} 0.2 (0.1-0.3) & {\cellcolor[HTML]{FEFEFE}} \color[HTML]{000000} {\cellcolor[HTML]{000000}} \color[HTML]{F1F1F1} {\cellcolor{white}}  & {\cellcolor[HTML]{FEFEFE}} \color[HTML]{000000} {\cellcolor[HTML]{FFEEE7}} \color[HTML]{000000} 0.0 (0.0-0.1) & {\cellcolor[HTML]{FEFEFE}} \color[HTML]{000000} {\cellcolor[HTML]{FEEAE0}} \color[HTML]{000000} 0.1 (0.0-0.1) & {\cellcolor[HTML]{FEFEFE}} \color[HTML]{000000} {\cellcolor[HTML]{000000}} \color[HTML]{F1F1F1} {\cellcolor{white}} {\cellcolor[HTML]{AAAAAA}}  & {\cellcolor[HTML]{FEFEFE}} \color[HTML]{000000} {\cellcolor[HTML]{FB7A5A}} \color[HTML]{F1F1F1} 0.5 (0.1-0.7) & {\cellcolor[HTML]{FEFEFE}} \color[HTML]{000000} {\cellcolor[HTML]{000000}} \color[HTML]{F1F1F1} {\cellcolor{white}} {\cellcolor[HTML]{AAAAAA}}  \\
{\cellcolor[HTML]{FEFEFE}} \color[HTML]{000000} 149 & {\cellcolor[HTML]{FEFEFE}} \color[HTML]{000000} $\text{U}^\text{307}_\text{OP2}$-$\text{O}_\text{ph}$ & {\cellcolor[HTML]{FEFEFE}} \color[HTML]{000000} {\cellcolor{white}}  & {\cellcolor[HTML]{FEFEFE}} \color[HTML]{000000}  & {\cellcolor[HTML]{FEFEFE}} \color[HTML]{000000} {\cellcolor[HTML]{000000}} \color[HTML]{F1F1F1} {\cellcolor{white}}  & {\cellcolor[HTML]{FEFEFE}} \color[HTML]{000000} {\cellcolor[HTML]{FEEAE1}} \color[HTML]{000000} 0.1 & {\cellcolor[HTML]{FEFEFE}} \color[HTML]{000000} {\cellcolor[HTML]{000000}} \color[HTML]{F1F1F1} {\cellcolor{white}}  & {\cellcolor[HTML]{FEFEFE}} \color[HTML]{000000} {\cellcolor[HTML]{000000}} \color[HTML]{F1F1F1} {\cellcolor{white}}  & {\cellcolor[HTML]{FEFEFE}} \color[HTML]{000000} {\cellcolor[HTML]{000000}} \color[HTML]{F1F1F1} {\cellcolor{white}}  & {\cellcolor[HTML]{FEFEFE}} \color[HTML]{000000} {\cellcolor[HTML]{FFEDE5}} \color[HTML]{000000} 0.0 (0.0-0.1) & {\cellcolor[HTML]{FEFEFE}} \color[HTML]{000000} {\cellcolor[HTML]{FFF4EF}} \color[HTML]{000000} 0.0 (0.0-0.0) & {\cellcolor[HTML]{FEFEFE}} \color[HTML]{000000} {\cellcolor[HTML]{FFECE3}} \color[HTML]{000000} 0.1 (0.0-0.1) & {\cellcolor[HTML]{FEFEFE}} \color[HTML]{000000} {\cellcolor[HTML]{FFF1EA}} \color[HTML]{000000} 0.0 (0.0-0.1) & {\cellcolor[HTML]{FEFEFE}} \color[HTML]{000000} {\cellcolor[HTML]{FEE7DC}} \color[HTML]{000000} 0.1 (0.0-0.2) & {\cellcolor[HTML]{FEFEFE}} \color[HTML]{000000} {\cellcolor[HTML]{FEE9DF}} \color[HTML]{000000} 0.1 (0.0-0.2) & {\cellcolor[HTML]{FEFEFE}} \color[HTML]{000000} {\cellcolor[HTML]{FFF5F0}} \color[HTML]{000000} 0.0 (0.0-0.0) & {\cellcolor[HTML]{FEFEFE}} \color[HTML]{000000} {\cellcolor[HTML]{FFF5F0}} \color[HTML]{000000} 0.0 (0.0-0.2) & {\cellcolor[HTML]{FEFEFE}} \color[HTML]{000000} {\cellcolor[HTML]{FB7A5A}} \color[HTML]{F1F1F1} 0.5 (0.2-0.7) \\
\bottomrule
\end{xltabular}